\documentclass[journal]{new-aiaa}
\usepackage[utf8]{inputenc}
\usepackage{textcomp}

\usepackage{graphicx}
\usepackage{amsmath}
\usepackage[version=4]{mhchem}
\usepackage{siunitx}
\usepackage{longtable,tabularx}
\usepackage{bm}
\usepackage{subcaption}
\usepackage{threeparttable}
\usepackage[normalem]{ulem}
\title{State-Specific Kinetic Modeling of Atomic H \\ for H$\bf _2$/ He Entry Flows}

\author{Alex T. Carroll\footnote{PhD Candidate, Department of Mechanical and Civil Engineering, AIAA Student Member. Contact: amoricar@caltech.edu.} and Guillaume Blanquart\footnote{Professor, Department of Mechanical and Civil Engineering.}}
\affil{California Institute of Technology, Pasadena, CA, 91125}
\author{Aaron M. Brandis\footnote{Senior Research Scientist, Aerothermodynamics Branch, AIAA Associate Fellow.}}
\affil{NASA Ames Research Center, Mountain View, CA, 94035}
\author{Brett A. Cruden\footnote{Senior Research Scientist, Aerothermodynamics Branch, AIAA Associate Fellow.}}
\affil{Analytical Mechanics Associates, Inc. at NASA Ames Research Center, Mountain View, CA, 94035}

\begin{document}

\maketitle

\begin{abstract}

An 11-species thermochemical model for $\bf H_2$/ He mixtures with state-specific kinetics for atomic H is developed and used to simulate 1-D shocks at conditions relevant for ice and gas giant entry flows. To implement this kinetic model, a literature review of the state-specific excitation and ionization rate constants of atomic H is first performed. While electron-impact rate constants from various sources are found to be in good agreement, large discrepancies are found in the limited data available on heavy-particle-impact rate constants. To validate the kinetic model, 1-D steady shocks are simulated using a space-marching code that explicitly accounts for shock tube boundary layer effects. The resulting radiance profiles are compared to experimental data from the NASA Ames Electric Arc Shock Tube (EAST) facility, and are found to reproduce the measured values reasonably accurately while capturing the distinct induction zone behavior observed in the experiments. A sensitivity analysis of the kinetic rates and boundary layer treatment reveals avenues for further improvement of the model. Finally, a comparison to alternate models from the literature underscores the improved accuracy of the present model in predicting ionization and radiation profiles.

\end{abstract}

\section*{Nomenclature}
\label{sec:nomen}
{\renewcommand\arraystretch{1.0}
\noindent\begin{longtable*}{@{}l @{\quad=\quad} l@{}}
$a_0$ & Bohr radius, $5.2918\times10^{-11}$ [m] \\
$A_{n\rightarrow n'}$ & bound-bound Einstein coefficient for spontaneous emission [$\rm s^{-1}$] \\
$A_{\infty\rightarrow n}$ & free-bound radiative transition rate constant [$\rm m^3$/s] \\
$A_{n}$, $B_{n}$ & Bethe coefficients in Johnson's~\cite{Johnson1972} formulation [-] \\
$C(n\rightarrow n')$ & Arrhenius pre-exponential constant for state-specific electronic excitation [$\rm m^3/s/K^{\gamma(n\rightarrow n')}$] \\
$C(n\rightarrow \infty)$ & Arrhenius pre-exponential constant for state-specific ionization [$\rm m^3/s/K^{\gamma(n\rightarrow \infty)}$] \\
$d$ & shock tube driven section diameter [m] \\
$e$ & specific energy [J/kg] \\
$e_{\rm rv}$ & specific rovibrational energy [J/kg] \\
$e_{\rm e}$ & specific electronic energy [J/kg] \\
$E$ & relative translational/ collision energy [J] \\
$E_{{\rm H}(n)}$ & electronic energy of H($n$) relative to the ground state [J] \\
$E_{\rm H^+}$ & ionization energy of H relative to the ground state, $2.1787\times10^{-18}$ [J] \\
$E_{n \rightarrow n'}$ & threshold energy for state-specific electronic excitation [J] \\
$E_{n \rightarrow \infty}$ & threshold energy for state-specific ionization [J] \\
$g$ & degeneracy [-] \\
$g_J$ & statistical weight due to nuclear spin-splitting [-] \\
$h$ & specific enthalpy [J/kg] \\
$h_{\rm P}$ & Planck constant, $6.6261\times10^{-34}$ [J$\cdot$s] \\
$j$ & diffusive mass flux [kg/$\rm m^2$/s] \\
$J$ & rotational quantum number [-] \\
$k_{\rm AI}$ & associative ionization rate constant [$\rm m^3$/s] \\
$k_{\rm B}$ & Boltzmann constant, $1.3806\times10^{-23}$ [J/K] \\
$k_{\rm d}$ & macroscopic dissociation rate constant [$\rm m^3$/s] \\
$k(n \rightarrow n')$ & state-specific electronic excitation rate constant [$\rm m^3$/s] \\
$k(n \rightarrow \infty)$ & state-specific ionization rate constant [$\rm m^3$/s]  \\
$k(\infty \rightarrow n)$ & state-specific recombination rate constant [$\rm m^6$/s]  \\
$K_{\rm eq}$ & macroscopic equilibrium constant [$\rm m^{-3}$] \\
$L$ & shock tube test slug length [m] \\
$m$ & mass [kg] \\
$\bar{m}$ & reduced mass [kg] \\
$n$ & principal quantum number [-] \\
$N$ & number density [$\rm m^{-3}$] \\
$P$ & pressure [Pa] \\
$q$ & heat flux [J/$\rm m^2$/s] \\
$Q$ & internal partition function [-] \\
$Q_{\rm t}$ & translational partition function [$\rm m^{-3}$] \\
$r$ & radial spatial coordinate [m] \\
$r_{n}$, $y_{n}$, $z_{n}$ & normalized threshold energy terms in Johnson's~\cite{Johnson1972} formulation [-] \\
$t$ & time [s] \\
$T_{\rm t}$ & translational temperature [K] \\
$T_{\rm rv}$ & rovibrational temperature [K] \\
$T_{\rm e}$ & electronic temperature [K] \\
$u$ & axial velocity [m/s] \\
$u_{\rm r}$ & radial velocity [m/s] \\
$x$ & axial spatial coordinate [m] \\
$Y$ & mass fraction [-] \\
$\alpha$ & degree of dissociation [-] \\
$\beta$ & Mirels's~\cite{Mirels1963} boundary layer thickness parameter [-] \\
$\gamma(n\rightarrow n')$ & Arrhenius temperature exponent for state-specific electronic excitation [-] \\
$\gamma(n\rightarrow \infty)$ & Arrhenius temperature exponent for state-specific ionization [-] \\
$\eta$ & pre-QSS correction factor [-] \\
$\theta_{\rm rv}$ & characteristic rovibrational temperature relative to the ground state [K] \\
$\Lambda_{n\rightarrow n'}$ & bound-bound radiative transition escape factor [-] \\
$\Lambda_{\infty\rightarrow n}$ & free-bound radiative transition escape factor [-] \\
$\mu$ & dynamic viscosity [kg/m/s] \\
$\nu$ & vibrational quantum number [-] \\
$\rho$ & density [kg/$\rm m^3$] \\
$\sigma(n\rightarrow n')$ & state-specific electronic excitation cross section [$\rm m^2$] \\
$\bar{\sigma}$ & thermally-averaged elastic-scattering cross section [$\rm m^2$] \\
$\tau$ & viscous stress tensor [Pa] \\
$\dot{\omega}$ & species source term [kg/$\rm m^3$/s] \\
$\dot{\Omega}$ & energy source term [J/$\rm m^3$/s] \\

\multicolumn{2}{@{}l}{Subscripts} \\
0 & initial/ pre-shock condition \\
eff & effective \\
f & post-shock frozen state \\
h & heavy (non-electron) species \\
M & through collisions with the third-body M \\
nr & non-recombining limit of the QSS regime \\
pre-QSS & pre-QSS regime \\
rad & radiative processes \\
th & thermal equilibrium limit \\
w & wall properties \\

\end{longtable*}}

\section{Introduction}
\label{sec:intro}

Following the prioritization of Saturn and Uranus probe missions in the latest planetary science decadal surveys~\cite{NAP13117,NAP26522}, there have been a number of studies investigating the aerothermal environments for ice and gas giant entry flows~\cite{Palmer2014,Cruden2017,Higdon2018,Colonna2020,Liu2020,Liu2022,Hansson2021,Carroll2023_conv,Coelho2023,Scoggins2024,Steer2024,Steuer2024,Rataczak2025}. The atmospheres of all of the giant planets (Jupiter, Saturn, Uranus, and Neptune) are primarily composed of molecular hydrogen ($\rm H_2$) and helium ($\rm He$). During atmospheric entry, a shock wave forms in front of the probe. At the high temperatures found in the post-shock region, $\rm H_2$ molecules dissociate, H atoms get electronically excited, and ionization ultimately occurs forming protons ($\rm H^+$) and electrons ($\rm e^-$). Since the heat loads encountered by an entry probe are controlled by the state and composition of the surrounding gas, modeling each of these thermochemical processes accurately is critical for the mass-efficient design of thermal protection systems.

There are two potential mechanisms of heating for an entry vehicle: convection and radiation. While convective heating is often dominant in many entry flows of interest, the analysis of Palmer et al.~\cite{Palmer2014} suggested that there is currently a large associated uncertainty for $\rm H_2$/ He radiation in state-of-the-art computational fluid dynamics (CFD) codes. To better characterize this radiative heating component for $\rm H_2$/ He shocks, Cruden and Bogdanoff~\cite{Cruden2017} conducted experiments in the Electric Arc Shock Tube (EAST) facility for a range of Saturn and Uranus entry flow conditions. A representative gas mixture of 89\% $\rm H_2$ and 11\% $\rm He$ (by mole) was used, and radiance was measured from the vacuum-ultraviolet through the near-infrared for shots with freestream velocities ranging from approximately 25 to 30 km/s. For most shot conditions, a delayed rise in radiance or induction zone was observed in the post-shock region, along with equilibration lengths on the order of several centimeters. Evidence of non-Boltzmann state distributions for atomic H was also observed.

Radiation for $\rm H_2$/ He shocks was also investigated by Liu et al.~\cite{Liu2020,Liu2022} in the X2 expansion tube facility. In these experiments, the gas substitution technique proposed by Stalker and Edwards~\cite{Stalker1998} was used with a 20\% $\rm H_2$ and 80\% $\rm He$ (by mole) mixture. This technique was used to achieve flight-representative post-shock temperatures at a relatively lower freestream velocity of approximately 18 km/s. Consistent with the results of Cruden and Bogdanoff, a slow ionization/ equilibration process was observed in the post-shock flow.

Several computational studies have tried to reproduce the measured radiance and/ or inferred quantities of interest from either set of these experiments. Higdon et al.~\cite{Higdon2018} used the direct simulation Monte Carlo (DSMC) method with the rate constants of Leibowitz~\cite{Leibowitz1973,Leibowitz1976} and a Boltzmann assumption for the radiation to simulate two of the lower pressure shot conditions from Cruden and Bogdanoff. While some qualitative agreement was obtained, in general, the simulated radiance significantly overpredicted the experimental values (likely due to the Boltzmann assumption), and the post-shock induction behavior was not captured. Hansson et al.~\cite{Hansson2021} performed CFD simulations corresponding to the experiments of Liu et al.~\cite{Liu2020}, and found that the rate constants of Leibowitz, Park~\cite{Park2001}, and Boyd~\cite{Boyd1997} predicted vastly different ionization profiles. Moreover, none of these models reproduced the experimental radiance values accurately, even with the use of a non-Boltzmann quasi-steady-state (QSS) solver for radiation. Liu et al.~\cite{Liu2022} performed accompanying CFD simulations for their experiments, and did find good agreement for electron number densities, but only after applying an ad hoc modification factor to the ionization rate constants of Leibowitz. Finally, Colonna et al.~\cite{Colonna2020} performed inviscid 1-D shock calculations with a detailed State-to-State (StS) kinetic model which explicitly treats the non-Boltzmann distributions for the vibrational mode of $\rm H_2$, the electronic modes of $\rm H_2$, H, and He, and a non-Maxwellian electron translational energy. With this model, Colonna et al. were only able to obtain reasonable estimates of electron number densities and H electronic temperatures (when compared to the experiments of Cruden and Bogdanoff) by also including the trace species $\rm H_3^+$ and $\rm H^-$, as these species allowed for an alternate pathway to ionization earlier in the shock layer. Ultimately, it is not clear from these previous investigations if the discrepancies between experiments and simulations are due to ``missing'' reactions in heritage kinetic models or simply inaccurate rate constants.

In light of these issues, the objective of the present work is twofold. First, to determine which rate constants should be used to model the critical processes of H excitation and ionization. Second, to determine if the conventional approach of modeling just $\rm H_2$ dissociation and H ionization is sufficient for the accurate prediction of radiation in $\rm H_2$/ He shocks, or if the inclusion of additional minority species (e.g., $\rm H_2^+$, $\rm H_3^+$, and $\rm H^-$) and their associated kinetics~\cite{Janev1987,Janev1993,Janev2003} is necessary. The paper is organized as follows. In section~\ref{sec:revH}, the electronic excitation and ionization of H by both electron and heavy-particle-impact are reviewed. Next, in section~\ref{sec:methods}, an 11-species kinetic model is constructed using the best estimates of the H excitation and ionization rate constants along with the $\rm H_2$ dissociation model developed recently by the present authors~\cite{Carroll2026_diss1,Carroll2026_diss2}. This kinetic model is then implemented in a quasi-1-D framework that explicitly accounts for shock tube boundary layer effects. Finally, in section~\ref{sec:results}, the resulting radiance predictions from this model are compared to the EAST measurements of Cruden and Bogdanoff to assess the model's accuracy and identify critical sensitivities for future improvements.

\section{Rate Constant Review for the Excitation and Ionization of H}
\label{sec:revH}

Modeling atomic ionization is complicated by the fact that the processes of thermal and chemical non-equilibrium, i.e., electronic energy relaxation and ionization, are inherently coupled.
To capture this coupling explicitly, state-specific rate constants for the electronic excitation and ionization of H are considered in the present work. The state-specific excitation and ionization of H can be written as
\begin{equation}
    \rm H({\it n}) + M \leftrightarrow H({\it n'}) + M
    \label{eqn:Hexc}
\end{equation}
and
\begin{equation}
    \rm H({\it n}) + M \leftrightarrow H^+ + M + e^-,
    \label{eqn:Hion}
\end{equation}
respectively. In these reactions, the electronic states of H are binned by principal quantum number, $n$, and the underlying angular momentum states are assumed to follow a statistical distribution. As a consequence, all electronic state degeneracies and energies are given by $g_{{\rm H}(n)} = 2n^2$ and the Rydberg formula, $E_{{\rm H}(n)} = E_{\rm H^+}(1-n^{-2})$, respectively. Section~\ref{sec:revEI} reviews the available rate constants in the literature for electron-impact ($\rm M = e^-$) excitation and ionization, while section \ref{sec:revHI} discusses rate constants for heavy-particle-impact ($\rm M \neq e^-$) excitation and ionization.

\subsection{Electron-Impact}
\label{sec:revEI}

\subsubsection{Excitation}
\label{sec:revEIE}

For the electron-impact excitation of H, rate constants are fortunately available from several different sources. The majority of the available data can be split into three categories: experimental data, semi-empirical formulations, and ab initio calculations. Experimental data mostly consists of cross section measurements for specific transitions at select collision energies, while the semi-empirical formulations combine experimental data with theory to construct analytical formulas for cross sections and rate constants for arbitrary transitions/ quantum numbers. Since the majority of the experimental data of interest has been considered and/ or incorporated into the semi-empirical formulations, experimental data are not included separately in the present review. The semi-empirical formulations of Mansbach and Keck~\cite{Mansbach1969}, Park~\cite{Park1971}, Johnson~\cite{Johnson1972}, and Vriens and Smeets~\cite{Vriens1980} are considered in the present work.

In contrast to the experimental and semi-empirical approaches, ab initio calculations use numerical methods to solve the close-coupling equations and compute cross sections and rate constants. For a more detailed discussion of the relevant ab initio methods, the reader is referred to the works by Bray and Stelbovics~\cite{Bray1995} and Bartschat~\cite{Bartschat2013}. The ab initio rate constants of Scholz et al.~\cite{Scholz1990}, Aggarwal et al.~\cite{Aggarwal1991}, Callaway~\cite{Callaway1994}, Anderson et al.~\cite{Anderson2000,Anderson2002}, and Przybilla and Butler~\cite{Przybilla2004} are considered in the present work. In general, the ab initio calculations are believed to be more accurate than the semi-empirical formulations, as they resolve more of the details/ resonances of the cross sections, especially at the near-threshold energies to which the thermally-averaged rate constants are most sensitive.

As an additional point of comparison, the recommended cross sections of Janev and Smith~\cite{Janev1993} (based on a review of experimental, semi-empirical, and ab initio data available at the time) are presented here as well. These cross sections have been integrated over a Maxwellian electron distribution as
\begin{equation}
    k_{\rm M=e^-}(n \rightarrow n') = \sqrt{\frac{8k_{\rm B}T_{\rm t,e^-}}{\pi m_{\rm e^-}}} \int_{\frac{E_{n\rightarrow n'}}{k_{\rm B}T_{\rm t,e^-}}}^{\infty} \sigma_{\rm M = e^-}(n \rightarrow n') \frac{E}{k_{\rm B}T_{\rm t,e^-}} \exp \left( -\frac{E}{k_{\rm B}T_{\rm t,e^-}} \right) {\rm d}\left( \frac{E}{k_{\rm B}T_{\rm t,e^-}} \right),
\label{eqn:maxwell}
\end{equation}
where $E_{n\rightarrow n'}\equiv E_{{\rm H}(n')}-E_{{\rm H}(n)}$, to obtain corresponding thermally-averaged rate constants.

\begin{figure}[hbt!]
     \centering
     \begin{subfigure}[b]{0.33\textwidth}
         \centering
         \includegraphics[width=\textwidth,trim={0cm 0cm 1cm 0.5cm},clip]{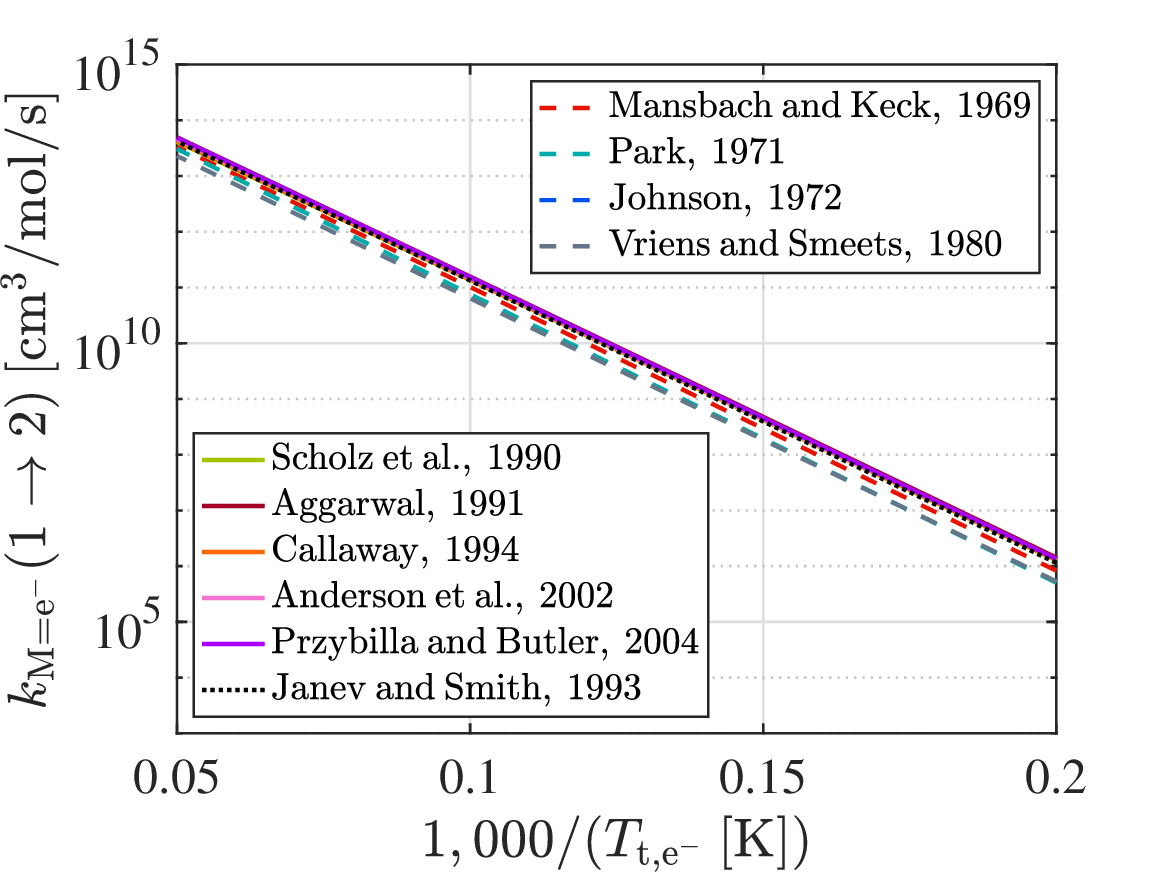}
     \end{subfigure}
     \hfill
     \begin{subfigure}[b]{0.33\textwidth}
         \centering
         \includegraphics[width=\textwidth,trim={0cm 0cm 1cm 0.5cm},clip]{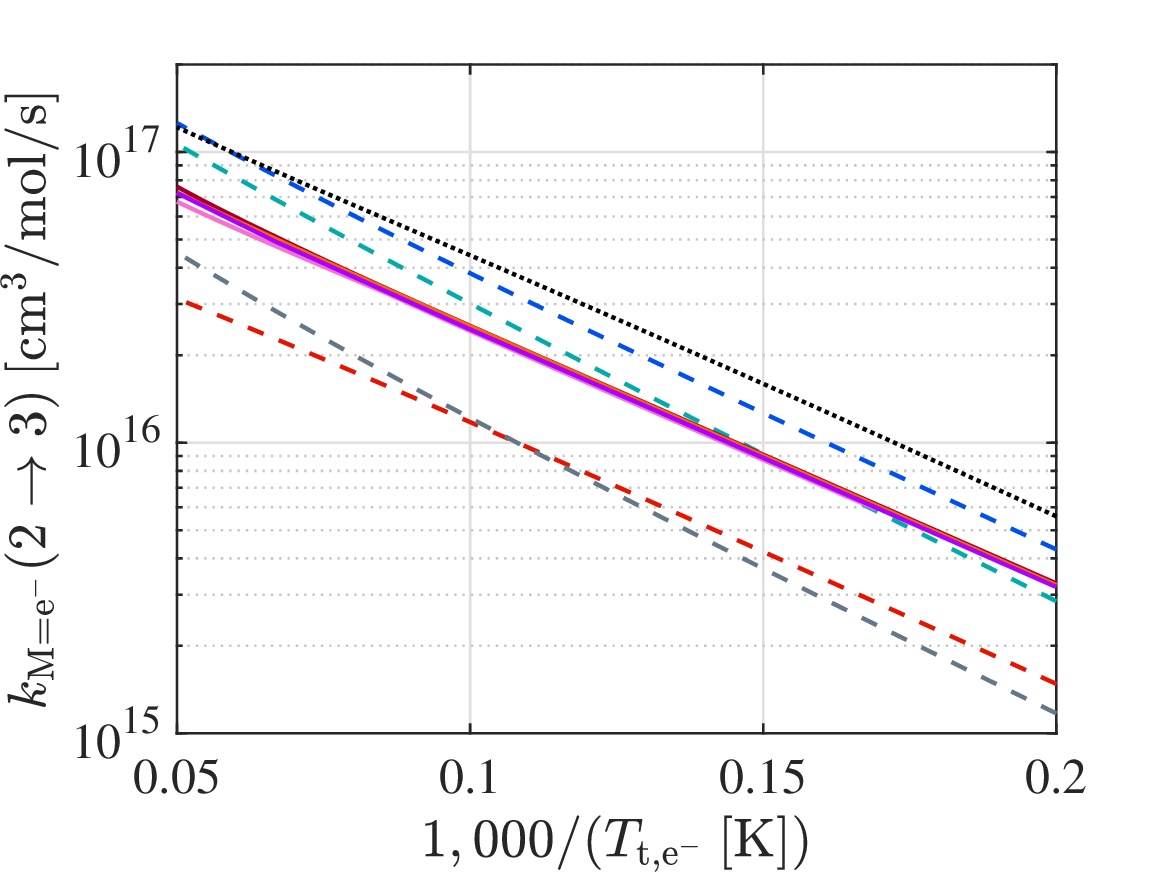}
     \end{subfigure}
     \hfill
     \begin{subfigure}[b]{0.33\textwidth}
         \centering
         \includegraphics[width=\textwidth,trim={0cm 0cm 1cm 0.5cm},clip]{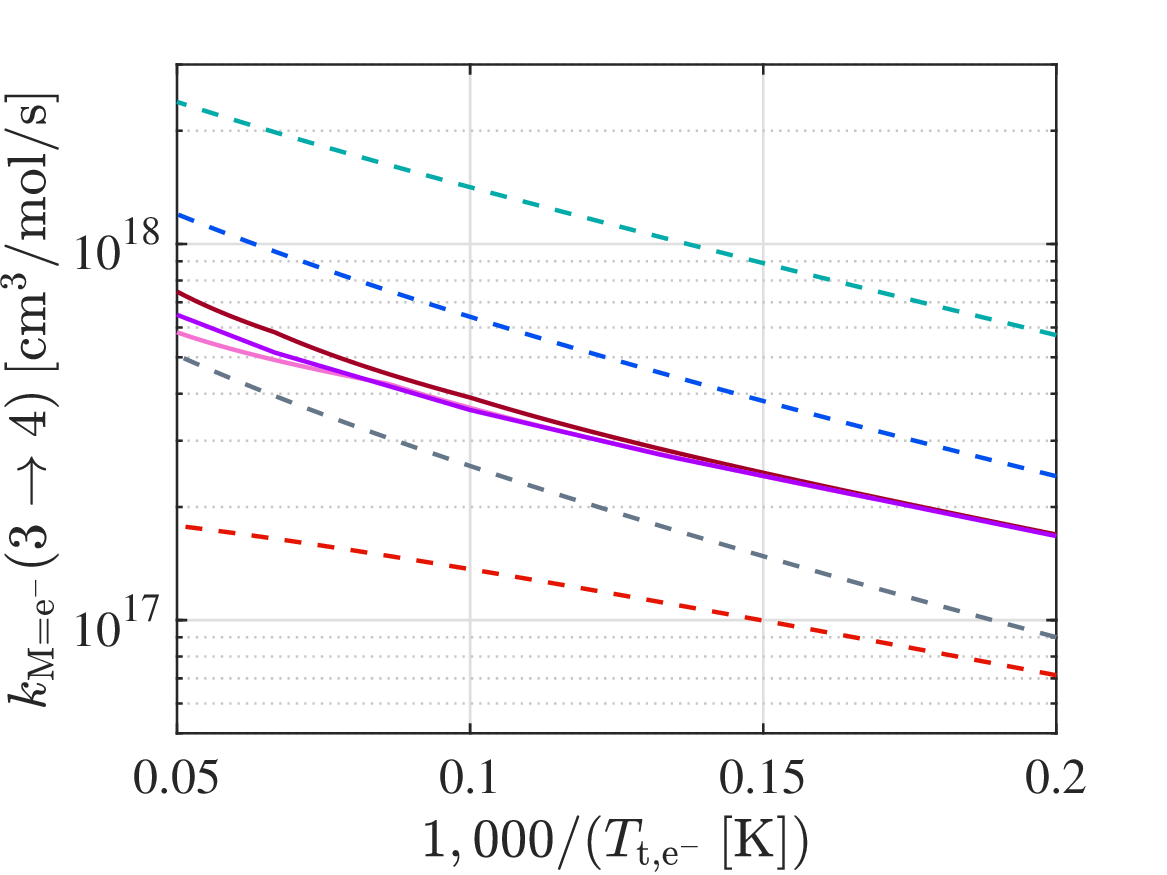}
     \end{subfigure}
     \begin{subfigure}[b]{0.33\textwidth}
         \centering
         \includegraphics[width=\textwidth,trim={0cm 0cm 1cm 0.5cm},clip]{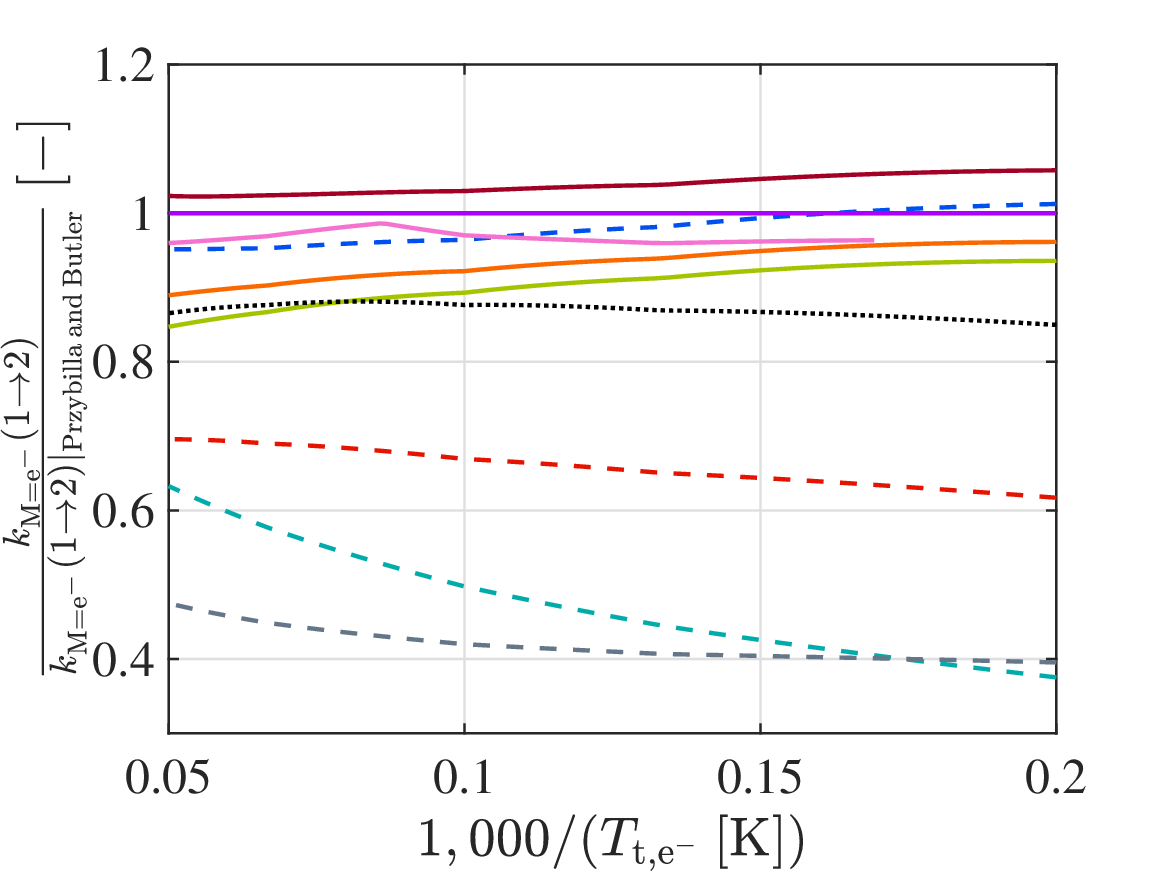}
         \caption{$\bm{n = 1\rightarrow2}$}
     \end{subfigure}
     \hfill
     \begin{subfigure}[b]{0.33\textwidth}
         \centering
         \includegraphics[width=\textwidth,trim={0cm 0cm 1cm 0.5cm},clip]{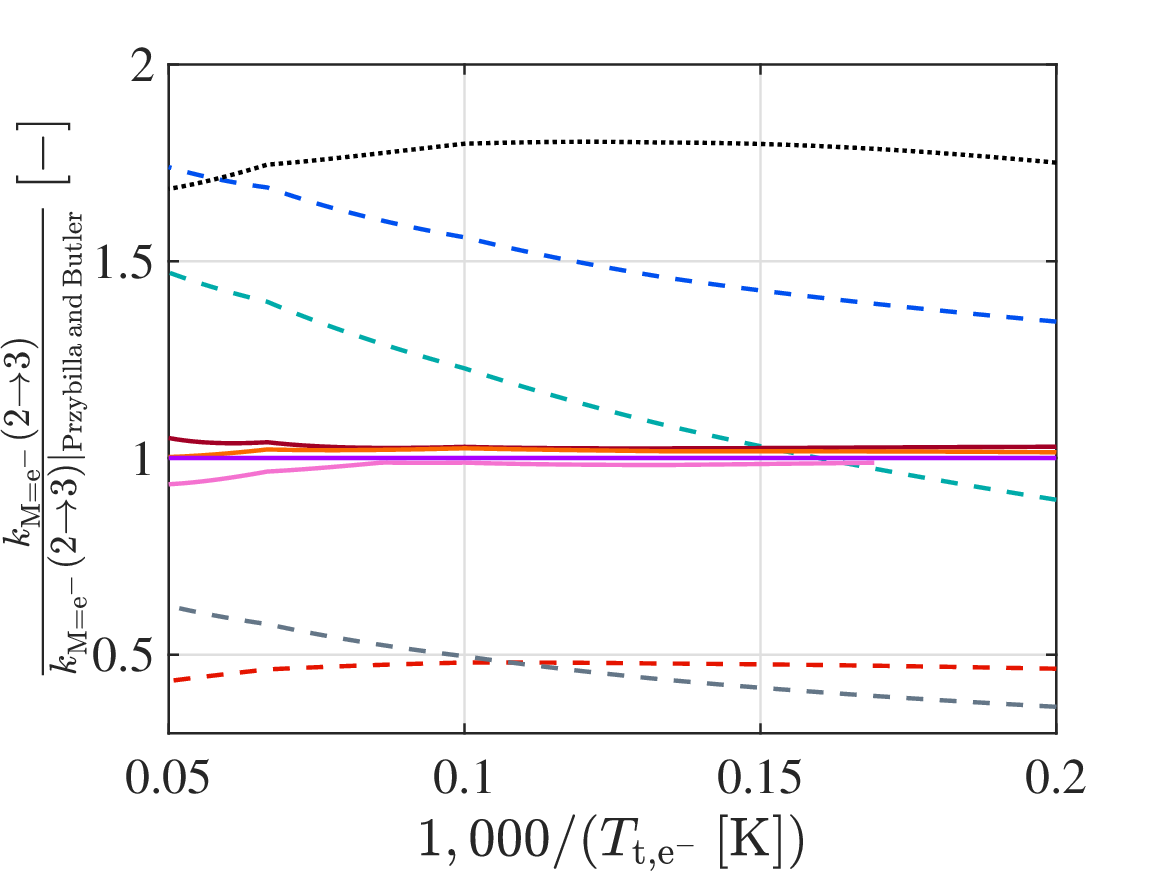}
         \caption{$\bm{n = 2\rightarrow3}$}
     \end{subfigure}
     \hfill
     \begin{subfigure}[b]{0.33\textwidth}
         \centering
         \includegraphics[width=\textwidth,trim={0cm 0cm 1cm 0.5cm},clip]{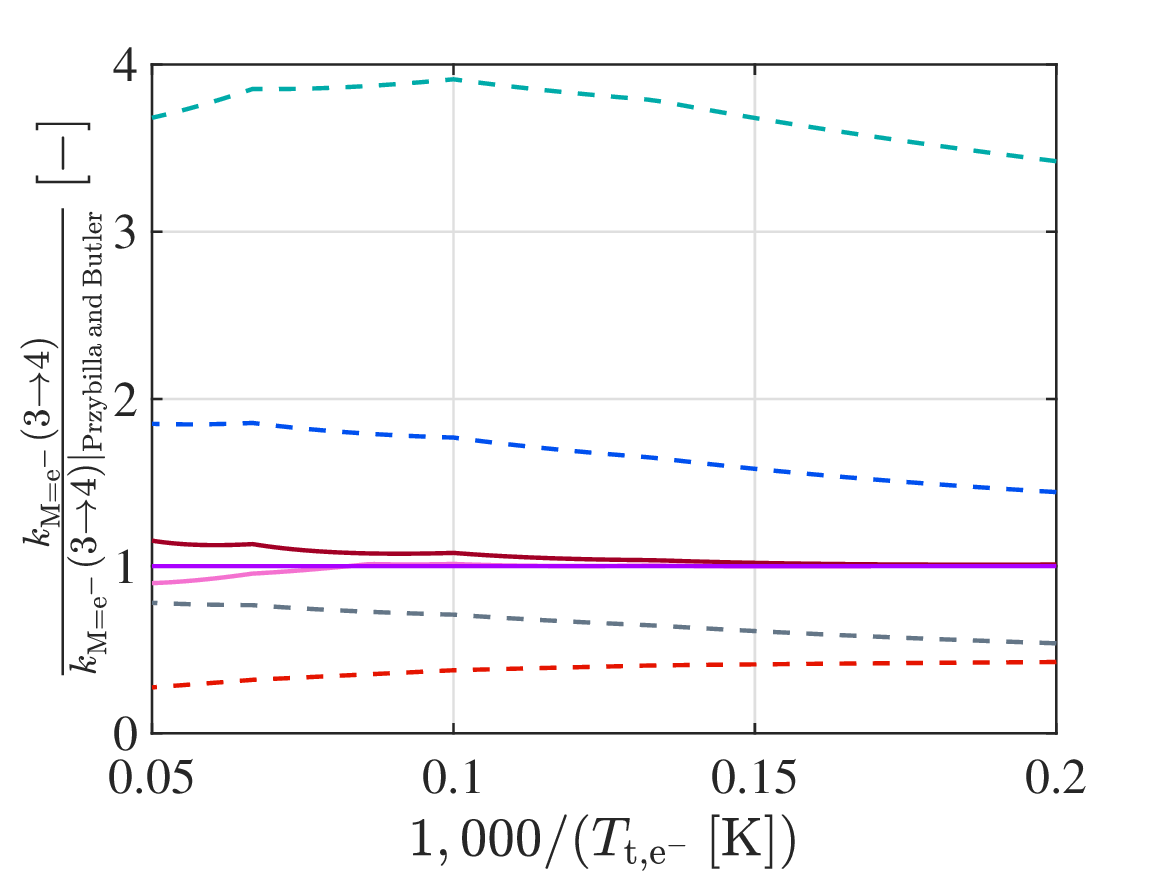}
         \caption{$\bm{n = 3\rightarrow4}$}
     \end{subfigure}
    \caption{Electron-impact excitation rate constants (top) and the same rate constants normalized by those of Przybilla and Butler~\cite{Przybilla2004} (bottom) for the $\bm{n = 1\rightarrow2}$ (left), $\bm{n = 2\rightarrow3}$ (middle), and $\bm{n = 3\rightarrow4}$ (right) transitions.}
    \label{fig:keexc}
\end{figure}

Figure~\ref{fig:keexc} shows the excitation rate constants for the dominant mono-quantum transitions, $n = 1\rightarrow2$, $n = 2\rightarrow3$, and $n = 3\rightarrow4$. The same plots normalized by the ab initio rate constants of Przybilla and Butler are also shown. The $n = 1\rightarrow2$ rate constants show a large variation with $T_{\rm t,e^-}$, as they have a large associated threshold energy of $E_{1\rightarrow2}=10.2$ eV. As the quantum numbers increase, this variation with temperature diminishes, such that for the $n = 3\rightarrow4$ transition, the rate constants only vary by a factor of approximately four over 5,000 to 20,000 K. In general, all of the ab initio rate constants (given by the solid lines) are in excellent agreement with each other, with an estimated uncertainty of approximately $\pm 15\%$. The semi-empirical rate constants on the other hand (given by the dashed lines) fail to reproduce the ab initio rate constants. For the $n = 1\rightarrow2$ transition, all of the semi-empirical rate constants except that of Johnson consistently underpredict the ab initio results; for the $n = 2\rightarrow3$ and $n = 3\rightarrow4$ transitions, the rate constants of Mansbach and Keck and Vriens and Smeets underpredict, while the Park and Johnson rate constants overpredict the ab initio results.

Based on the consistency between the ab initio rate constants, any of these datasets would likely give accurate estimates of the electron-impact excitation rate constants. In the present work, the rate constants of Przybilla and Butler are used, as they provide rate constants for the largest range of quantum numbers ($n \leq 7$) among the reviewed ab initio datasets.

\subsubsection{Ionization}
\label{sec:revEII}

\begin{figure}[hbt!]
     \centering
     \begin{subfigure}[b]{0.33\textwidth}
         \centering
         \includegraphics[width=\textwidth,trim={0cm 0cm 1cm 0.5cm},clip]{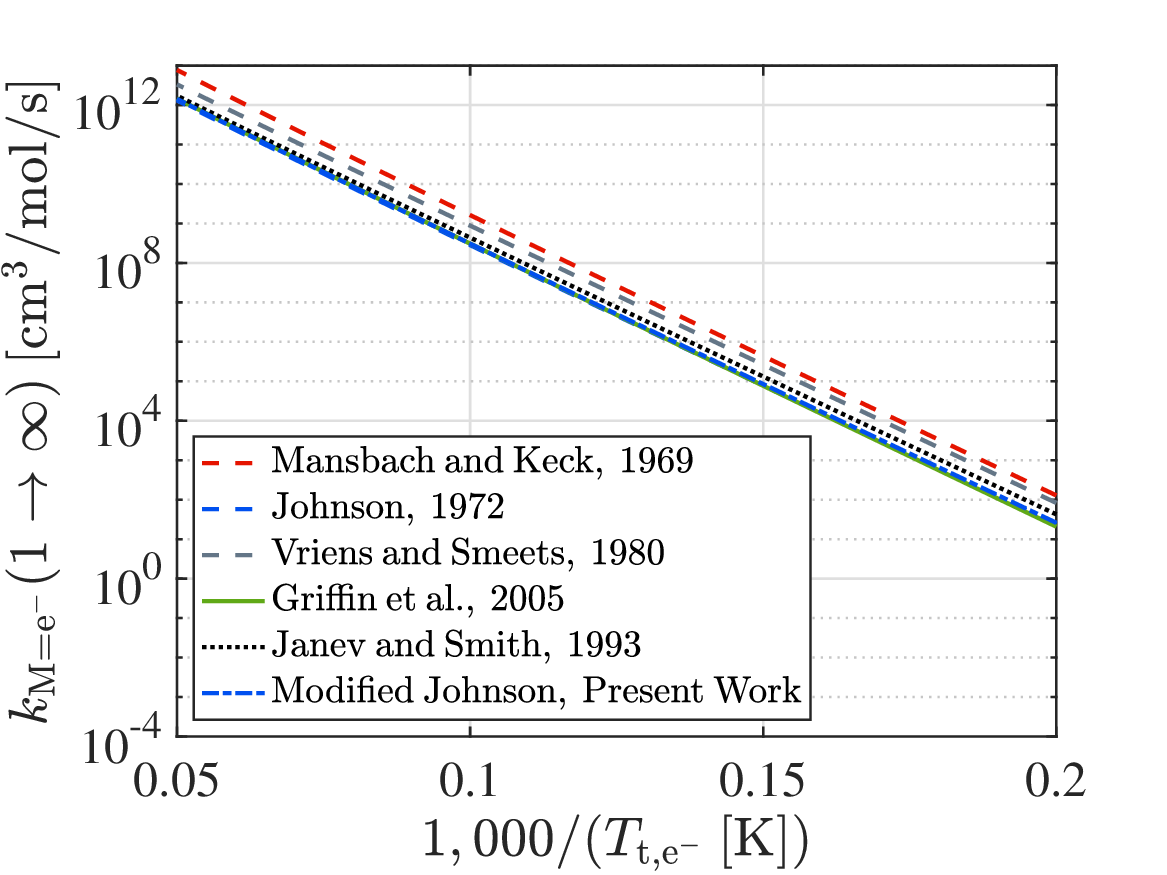}
     \end{subfigure}
     \hfill
     \begin{subfigure}[b]{0.33\textwidth}
         \centering
         \includegraphics[width=\textwidth,trim={0cm 0cm 1cm 0.5cm},clip]{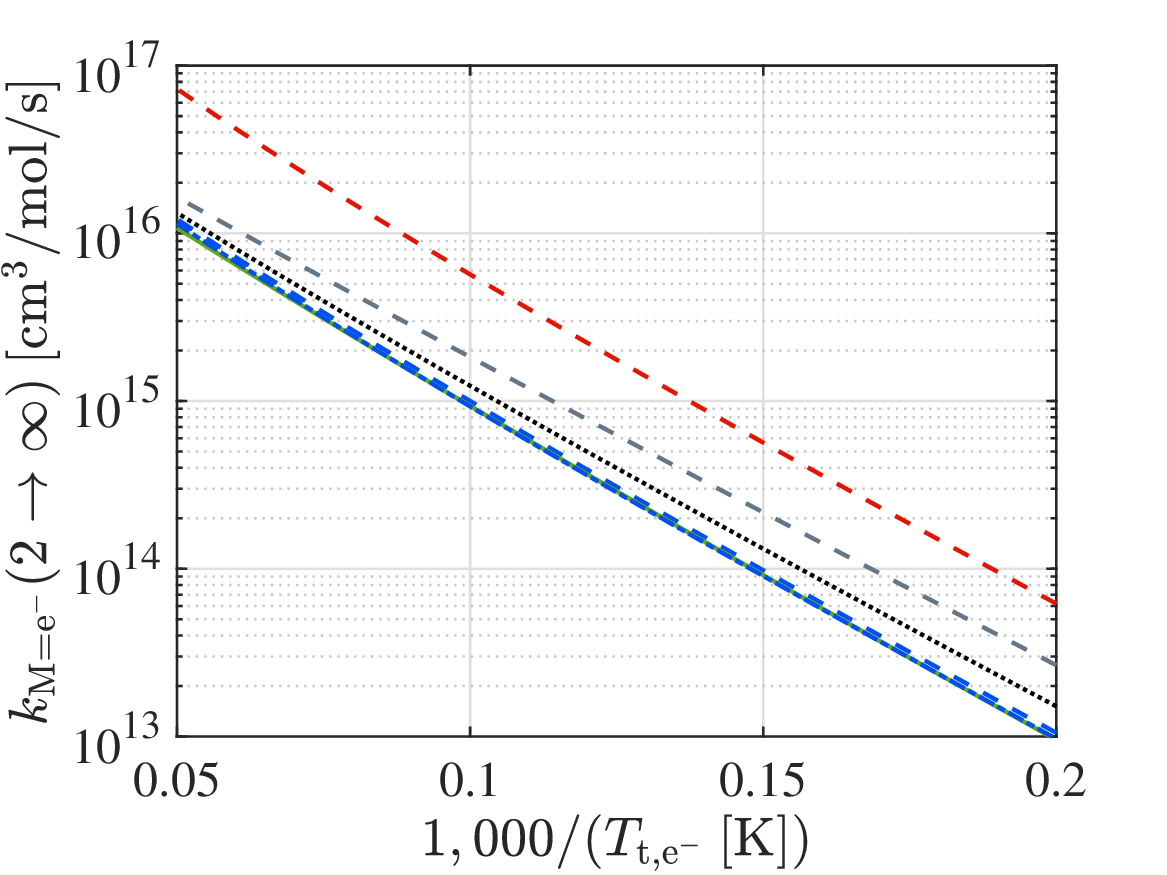}
     \end{subfigure}
     \hfill
     \begin{subfigure}[b]{0.33\textwidth}
         \centering
         \includegraphics[width=\textwidth,trim={0cm 0cm 1cm 0.5cm},clip]{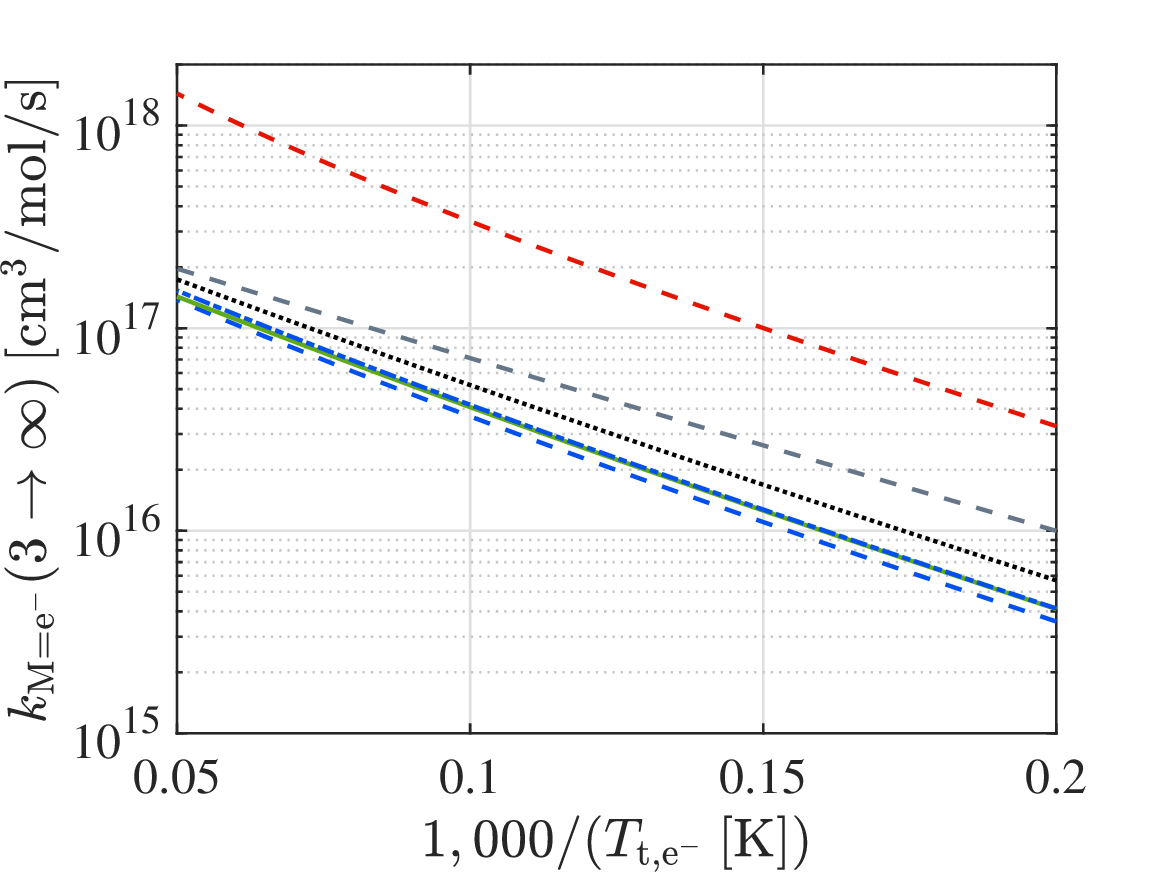}
     \end{subfigure}
     \begin{subfigure}[b]{0.33\textwidth}
         \centering
         \includegraphics[width=\textwidth,trim={0cm 0cm 1cm 0.5cm},clip]{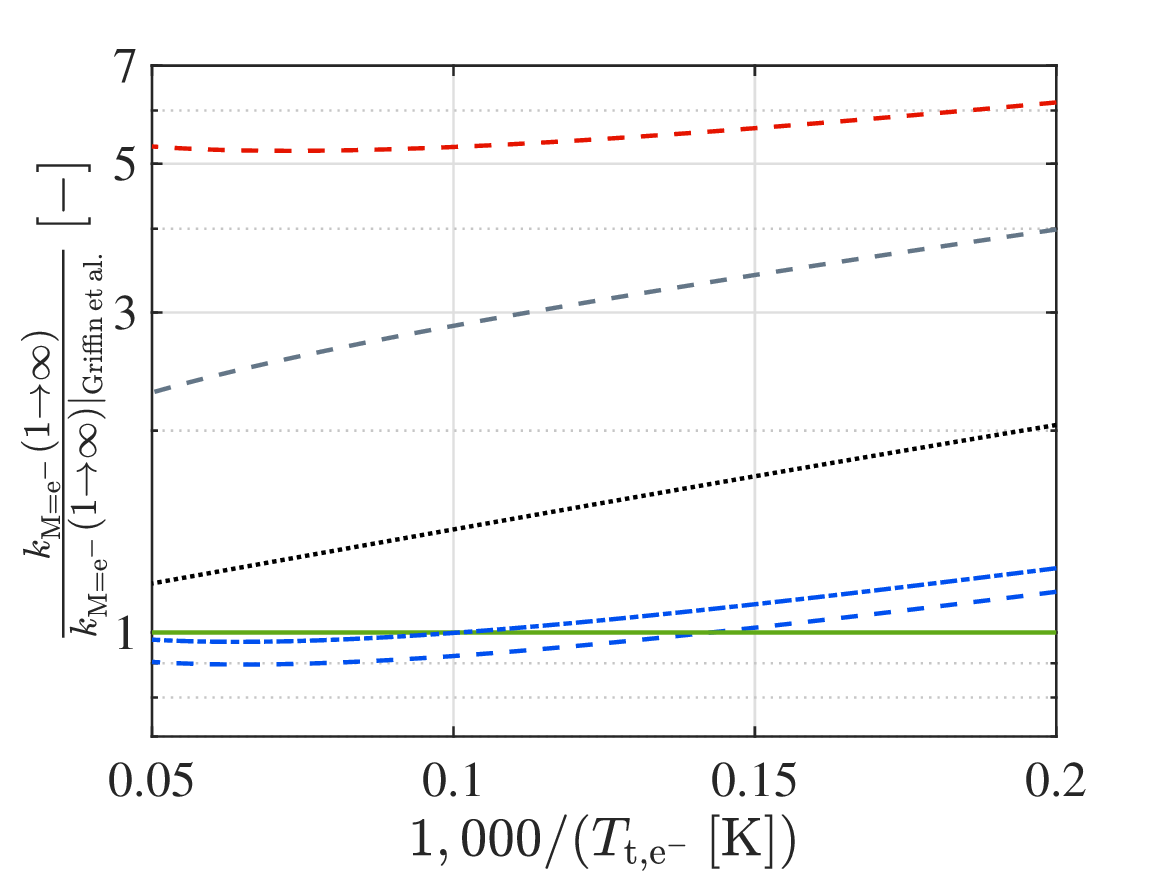}
         \caption{$\bm{n = 1\rightarrow\infty}$}
     \end{subfigure}
     \hfill
     \begin{subfigure}[b]{0.33\textwidth}
         \centering
         \includegraphics[width=\textwidth,trim={0cm 0cm 1cm 0.5cm},clip]{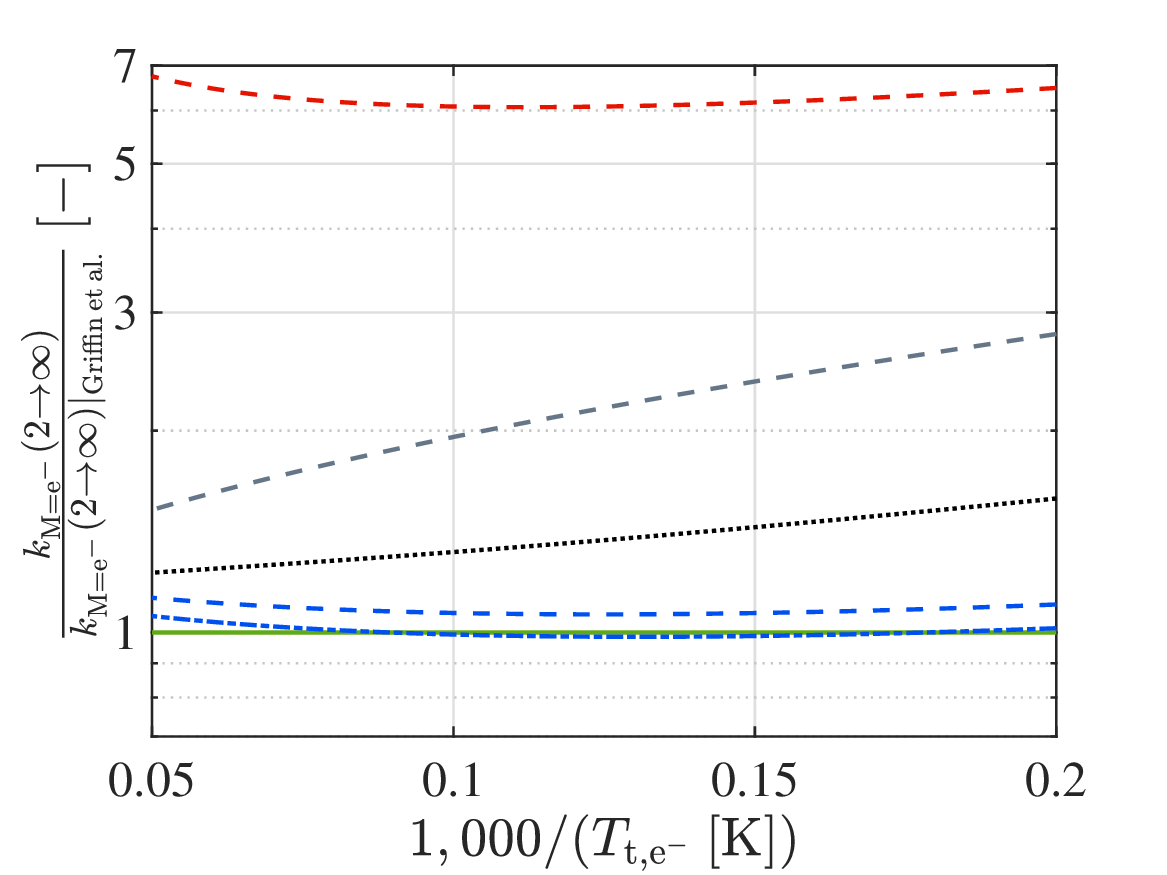}
         \caption{$\bm{n = 2\rightarrow\infty}$}
     \end{subfigure}
     \hfill
     \begin{subfigure}[b]{0.33\textwidth}
         \centering
         \includegraphics[width=\textwidth,trim={0cm 0cm 1cm 0.5cm},clip]{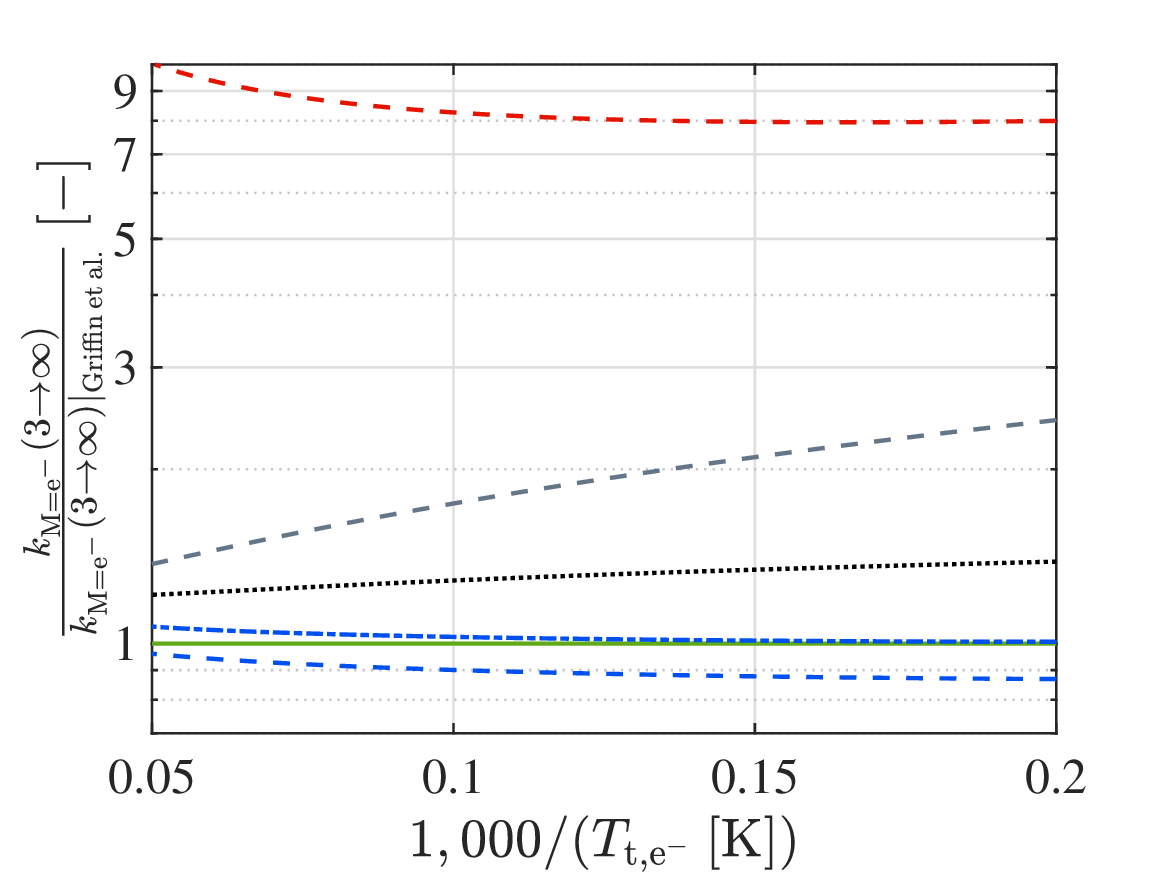}
         \caption{$\bm{n = 3\rightarrow\infty}$}
     \end{subfigure}
    \caption{Electron-impact ionization rate constants (top) and the same rate constants normalized by those of Griffin et al.~\cite{Griffin2005} (bottom) for the $\bm{n = 1\rightarrow\infty}$ (left), $\bm{n = 2\rightarrow\infty}$ (middle), and $\bm{n = 3\rightarrow\infty}$ (right) reactions.}
    \label{fig:keion}
\end{figure}

For the electron-impact ionization of H, the semi-empirical formulations of Mansbach and Keck~\cite{Mansbach1969}, Johnson~\cite{Johnson1972}, and Vriens and Smeets~\cite{Vriens1980} are considered, along with the recommended cross sections of Janev and Smith~\cite{Janev1993} and the ab initio cross sections of Griffin et al.~\cite{Griffin2005}. The cross sections of Janev and Smith and Griffin et al. have been integrated over a Maxwellian electron distribution using Eq.~\eqref{eqn:maxwell} as before, but with the threshold energy given instead by $E_{n\rightarrow \infty} \equiv E_{\rm H^+}-E_{{\rm H}(n)}$. The ionization rate constants for the $n$ = 1, 2, and 3 states along with the same plots normalized by the rate constants of Griffin et al. are shown in Fig.~\ref{fig:keion}. In general, the rate constants of Mansbach and Keck and Vriens and Smeets overpredict the ab initio rate constants of Griffin et al., while the rate constants of Johnson and Janev and Smith are consistently within a factor of two of the ab initio results. Since the $\rm 1s\rightarrow\infty$ and $\rm 2s\rightarrow\infty$ ionization cross sections of Griffin et al. have been validated extensively against other computational~\cite{Pindzola1996,Bartschat1996,Witthoeft2004} and experimental~\cite{Shah1987,Defrance1981} results, and as discussed, ab initio calculations are generally believed to more accurate than semi-empirical formulations, the rate constants of Griffin et al. are used throughout the present work where possible.

Unfortunately, the rate constants of Griffin et al. are limited to quantum numbers $n \leq 4$.  Hence, the use of a semi-empirical formulation is still necessary to compute ionization rate constants for the $n > 4$ states. For this purpose, a modified version of the Johnson rate constants are used. Johnson's formulation is based on high-energy approximations (Born-Bethe) with an empirical correction applied at low collision energies. The rate constants are given by
\begin{equation}
    k_{\rm M=e^-}(n \rightarrow \infty) = \sqrt{\frac{8k_{\rm B}T_{\rm t,e^-}}{\pi m_{\rm e^-}}} 2n^2\pi a_0^2 y_{n}^2 \biggr[ A_{n} \left( \frac{{\rm Ei}_1(y_{n})}{y_{n}} - \frac{{\rm Ei}_1(z_{n})}{z_{n}} \right) + \left(B_{n} - A_{n} \ln \left(2n^2\right)\right) \left(\xi(y_{n}) - \xi(z_{n}) \right) \biggr],
\label{eqn:johnsonion}
\end{equation}
where
\begin{equation}
    y_{n} \equiv\frac{E_{n\rightarrow \infty}}{k_{\rm B}T_{\rm t,e^-}}
    \quad {\rm and} \quad
    z_{n} \equiv r_{n} + \frac{E_{n\rightarrow \infty}}{k_{\rm B}T_{\rm t,e^-}},
\label{eqn:yz}
\end{equation}
and
\begin{equation}
    {\rm Ei}_{i}(y_n) \equiv \int_1^{\infty} \frac{\exp(-y_n\ell)}{\ell^i}{\rm d}\ell \quad (i=0,1,2,...)
\end{equation}
is the generalized exponential integral with
\begin{equation}
    \xi(y_{n}) \equiv {\rm Ei}_0(y_{n})-2{\rm Ei}_1(y_{n})+{\rm Ei}_2(y_{n}).
\end{equation}
In this formulation, the only free parameter is the empirical correction factor, $r_{n}$. For this, Johnson proposed the fit $r_{n=1}=0.45$ and $r_{n\geq2}=1.94n^{-1.57}$. To obtain a better agreement with the ab initio rate constants of Griffin et al., this fit is modified in the present work to be $r_{n=1}=0.50$ and $r_{n\geq2}=0.97n^{-0.75}$. As shown in Fig.~\ref{fig:keion}, these modified Johnson rate constants reproduce the ab initio results more accurately than any of the other semi-empirical formulations.

\begin{figure}[hbt!]
     \centering
     \hspace*{\fill}
     \begin{subfigure}[b]{0.4\textwidth}
         \centering
         \includegraphics[width=\textwidth,trim={0cm 0cm 1cm 0.5cm},clip]{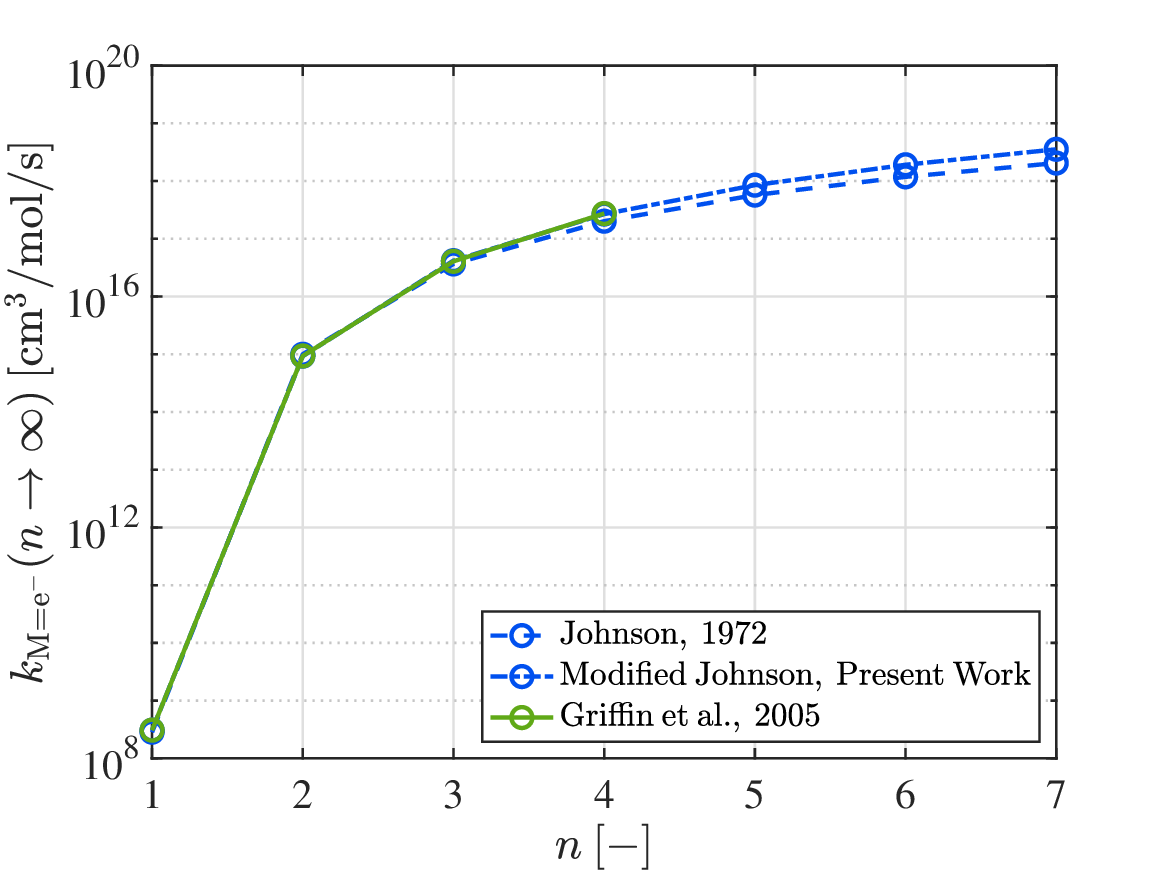}
     \end{subfigure}
     \hfill
     \begin{subfigure}[b]{0.4\textwidth}
         \centering
         \includegraphics[width=\textwidth,trim={0cm 0cm 1cm 0.5cm},clip]{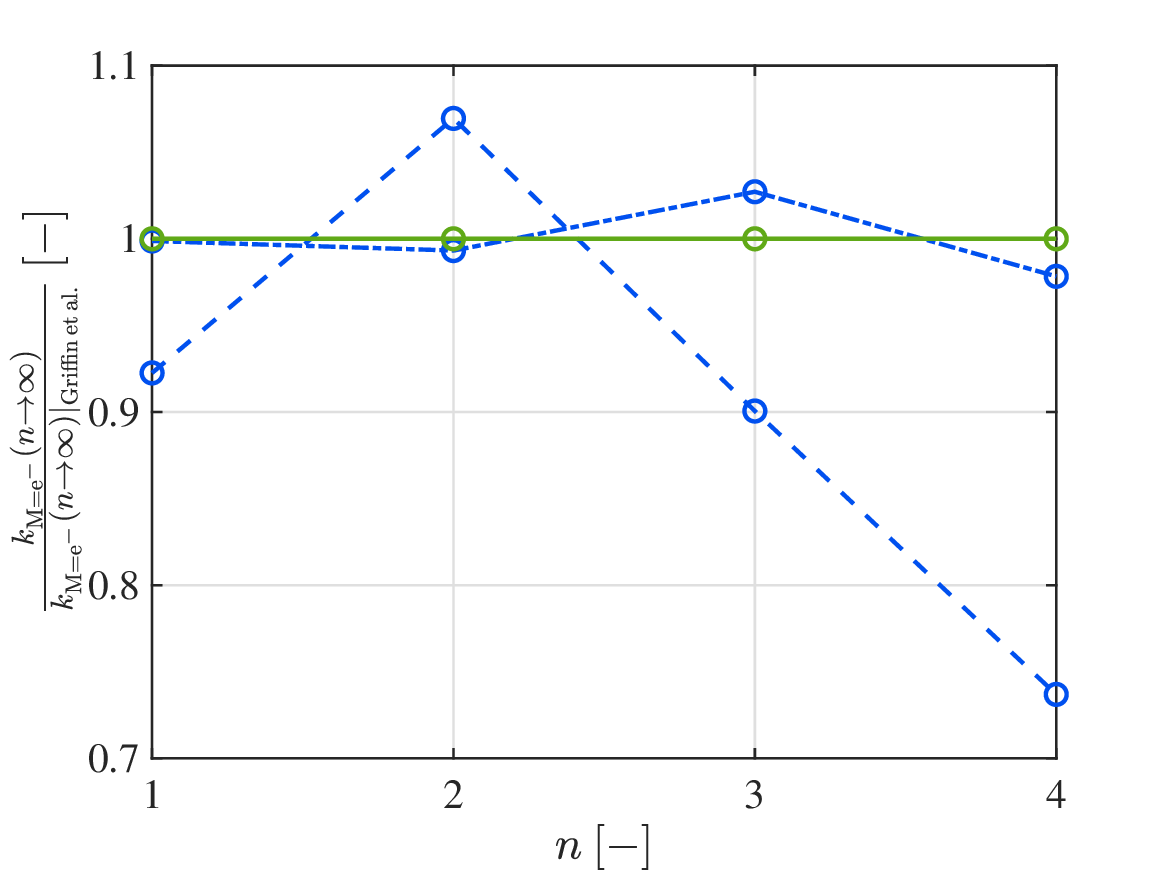}
     \end{subfigure}
     \hspace*{\fill}
    \caption{Electron-impact ionization rate constants as a function of $\bm n$ (left) and the same rate constants normalized by those of Griffin et al.~\cite{Griffin2005} (right) at $\bm{T_{\rm t,e^-}}$ = 10,000 K.}
    \label{fig:kenc}
\end{figure}

Figure~\ref{fig:kenc} shows the same ionization rate constants of Johnson and Griffin et al., but now plotted as a function of $n$ at a fixed temperature of $T_{\rm t,e^-}$ = 10,000 K. Both Johnson and Griffin et al. predict a smooth trend in the rate constants for $n\geq2$ resembling a power law, but with slightly different coefficients. After the modification, the Johnson rate constants are within 5\% of the rate constants of Griffin et al. for $n\leq4$, and lead to a more consistent extension of the rate constants for higher $n$.

\subsection{Heavy-Particle-Impact}
\label{sec:revHI}

\subsubsection{Datasets with Multiple Transitions}

As reviewed by Barklem~\cite{Barklem2007}, there are unfortunately very limited kinetic data available in the literature for the heavy-particle-impact excitation and ionization of H. More specifically, while a plethora of experimental and computational cross section data does exist for all third-bodies of interest~\cite{Hill1979,Hill1980,Hunter1990,Riley1999,AlAtawneh2021}, nearly all of this data is in the keV collision energy range, and therefore does not resolve the near-threshold behavior that is relevant for the present work.

In light of this, many computational studies investigating weakly-ionized hydrogen plasmas, including the StS model of Colonna et al.~\cite{Colonna2017,Colonna2018,Colonna2020}, use the classical rate constants of Drawin~\cite{Drawin1969,Drawin1973}. The Drawin formulation extends the formula of Thomson~\cite{Thomson1912} for electron-impact to the heavy-particle-impact process by assuming that the efficiency of the energy transfer for atom-atom collisions is the same as that for atom-electron collisions. In other words, the interaction is treated classically, and the impact of the nucleus of the perturbing atom is neglected~\cite{Barklem2011}.

Despite their widespread use, the validity of the Drawin rate constants is questionable. The works by Fleischmann and Dehmel~\cite{Fleischmann1972}, Steenbock and Holweger~\cite{Steenbock1984}, and Lambert~\cite{Lambert1993} all suggest that there is an error for the threshold energy in the derivation by Drawin due to the confusion of the center-of-mass and laboratory frames. Additionally, while ab initio calculations have not yet been performed for the H + H system to the authors' knowledge, they have been performed for several other systems of astrophysical relevance, including Li + H~\cite{Belyaev2003,Barklem2003}, Na + H~\cite{Belyaev2010,Barklem2010}, and Mg + H~\cite{Guitou2011,Barklem2012}; in each of these cases, the Drawin formulation has not been able to reproduce the ab initio results~\cite{Barklem2011}.

In contrast to the theoretical approach of Drawin, Park~\cite{Park2012} extracted heavy-particle-impact rate constants using an empirical/ inverse approach. Namely, Park simulated $\rm H_2$/ He shocks in a 1-D CFD code and tuned rate constant parameters to match the measured equilibration distances from the experiments by Livingston and Poon~\cite{Livingston1976} and Leibowitz~\cite{Leibowitz1973}. To do this, the functional form given by
\begin{equation}
    k_{\rm M=H}(n\rightarrow n') = \left(\frac{12000}{T_{\rm t,h}}\right)^{\gamma_{\rm M=H}}\sqrt{\frac{8k_{\rm B}T_{\rm t,h}}{\pi \bar{m}_{\rm H,H}}} \sigma_{\rm M=H} \exp \left(-\frac{E_{n\rightarrow n'}}{k_{\rm B}T_{\rm t,h}} \right)
    \label{eqn:Parkhvy_n12}
\end{equation}
was assumed for all excitation rate constants, where $\bar{m}_{\rm H,H}=m_{\rm H}^2/(m_{\rm H}+m_{\rm H})=m_{\rm H}/2$, and $\gamma_{\rm M=H}=0$ for all transitions except $n = 1\rightarrow2$. This functional form comes from assuming that all transitions can be described using the same constant/ energy-independent cross section, $\sigma_{\rm M=H}$, and then applying an ad hoc, temperature-dependent modification factor for just the $n=1\rightarrow2$ transition. Park argued that the energy/ temperature-independent cross sections were justified for all $n \geq 2$ transitions, as the relevant collision energies are substantially larger than the threshold energies for these transitions, which leads to a Maxwellian-integrated $\sigma_{\rm M=H}$ value with effectively no temperature dependence. Park also assumed that the same cross section and temperature exponent could be used for computing rate constants with M = He as a collision partner, i.e., with $\sigma_{\rm M=He} = \sigma_{\rm M=H}$, $\gamma_{\rm M=He} = \gamma_{\rm M=H}$, and with the reduced mass adjusted accordingly. The values $\gamma_{\rm M=H} = 6$ and $\sigma_{\rm M=H} = 4\times 10^{-21}$ $\rm m^2$ were found to give the best agreement with the experimental data. While Park did also implement state-specific ionization rate constants, the form of these rate constants were not stated explicitly. For the present work, the ionization rate constants of Park are interpreted as also following Eq.~\eqref{eqn:Parkhvy_n12}, but with the threshold energy, $E_{n\rightarrow n'}$, replaced accordingly with $E_{n\rightarrow \infty}$.

The most relevant post-shock temperatures investigated in Park's study were in the range of 15,000 to 20,000 K. As will be shown later in section~\ref{sec:sens}, equilibration distances are most sensitive to the heavy-particle-impact $n=1\rightarrow2$ transition, as it has the largest associated threshold energy of all of the mono-quantum transitions, and thus acts as a rate-limiting step for H ionization.  Therefore, the accuracy of Park's proposed rate constants in all other temperature ranges and for all other transitions is uncertain, due to the inherently reduced sensitivity to them in an inverse approach based on equilibration distances.

Finally, semi-classical rate constants have also been computed by Mihajlov et al.~\cite{Mihajlov2004} and Dimitrijević et al.~\cite{Dimitrijevic2021} for excitation, and by Mihajlov et al.~\cite{Mihajlov2011} and Srećković et al.~\cite{Sreckovic2018} for ionization. For simplicity, this entire set of rate constants will be referred to as the Mihajlov et al. rate constants throughout the present work. These calculations are based on the quasi-resonant energy transfer mechanism of Janev and Mihajlov~\cite{Janev1979}, and are believed to be more accurate than the classical theory of Drawin due to the more sophisticated theoretical treatment of the H + H interaction~\cite{Barklem2007}. However, due to the fundamental assumption of the quasi-resonant energy transfer mechanism of treating the H + H system as the interaction between an outer electron and the internal $\rm H^+$ + H(1s) subsystem, rate constants were only computed for quantum numbers $n \geq 4$ for excitation and $n \geq 5$ for ionization\footnote{Mihajlov et al.~\cite{Mihajlov2011} and Srećković et al.~\cite{Sreckovic2018} do provide estimates of the ionization rate constants for the $n$ = 2, 3, and 4 states; however these are based on extrapolations, and are therefore not considered in the present work.}.

\subsubsection{Data for Individual Transitions}

\begin{figure}[hbt!]
     \centering
     \begin{subfigure}[b]{0.4\textwidth}
         \centering
         \includegraphics[width=\textwidth,trim={0cm 0cm 1cm 0.5cm},clip]{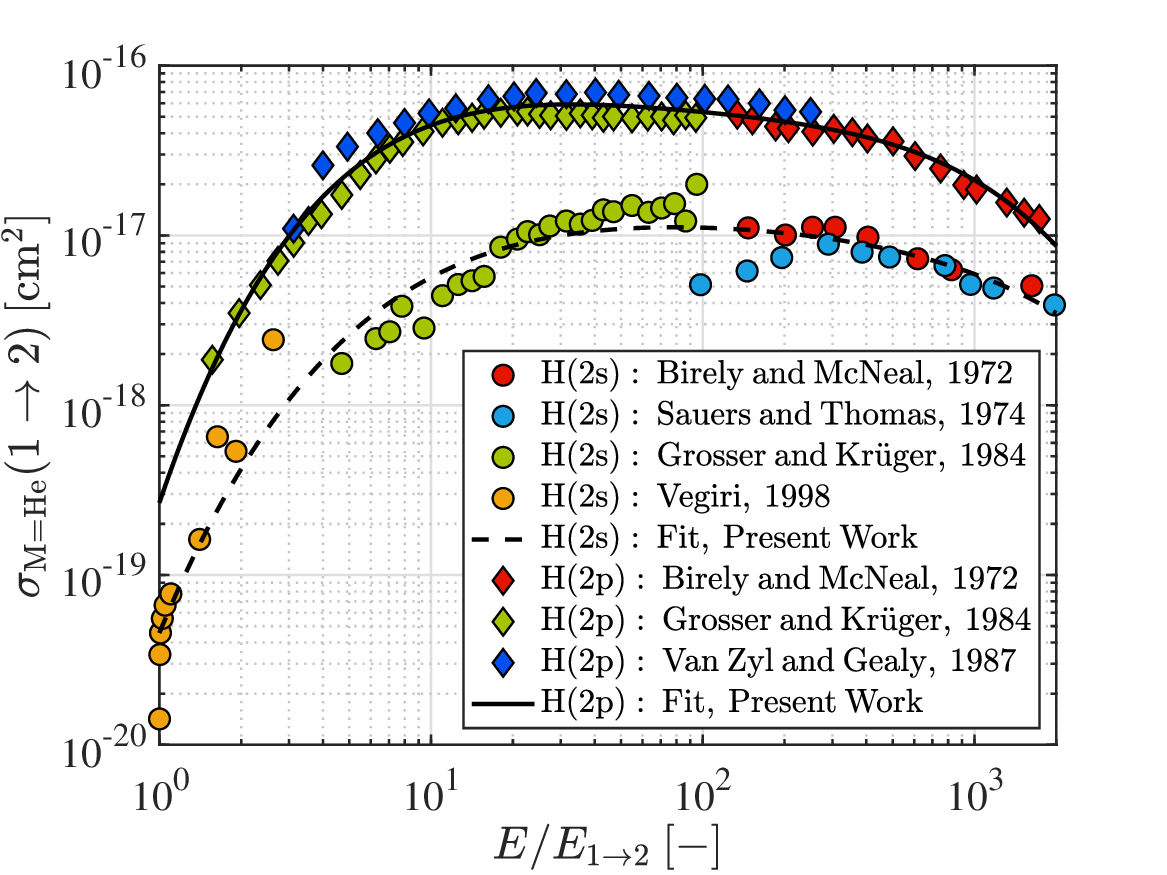}
     \end{subfigure}
    \caption{He-impact excitation cross sections for the $\bm{n = 1\rightarrow2}$ transition.}
    \label{fig:sigmaMHe12}
\end{figure}

While limited, there are also some data available for individual transitions of interest. For the $n=1\rightarrow 2$ transition with M = He, experimental measurements of the cross section near threshold energies have been obtained by Birely and McNeal~\cite{Birely1972}, Sauers and Thomas~\cite{Sauers1974}, Grosser and Krüger~\cite{Grosser1984}, and Van Zyl and Gealy~\cite{VanZyl1987}. Additionally, estimates of the $\rm 2s\rightarrow1s$ de-excitation cross section have been obtained via ab initio methods by Vegiri~\cite{Vegiri1998}. The cross section data from all of these sources are plotted in Fig.~\ref{fig:sigmaMHe12}. The cross section of Vegiri has been converted to a $\rm 1s\rightarrow 2s$ excitation cross section by applying micro-reversibility. In general, there is good agreement between the different datasets, especially for the dominant excitation to the H(2p) state.

To obtain a corresponding rate constant, separate fits for the $\rm 1s\rightarrow 2s$ and $\rm 1s\rightarrow 2p$ cross sections are first computed using a threshold-energy-normalized version of the polynomial functional form of Janev et al.~\cite{Janev1987} given by
\begin{equation}
    \ln(\sigma_{\rm M=He}({\rm 1s\rightarrow2s,2p})) = \sum_{i=0}^4 p_i\left(\ln\left(\frac{E}{E_{1\rightarrow 2}}\right)\right)^i,
    \label{eqn:sigmaMHe12}
\end{equation}
with fitting coefficients provided in Table~\ref{tab:sigmaMHe12}. These cross sections are then integrated over a Maxwellian distribution to obtain a thermally-averaged rate constant as
\begin{equation}
    k_{\rm M=He}(1 \rightarrow 2) = \sqrt{\frac{8k_{\rm B}T_{\rm t,h}}{\pi\bar{m}_{\rm H,He}}} \int_{\frac{E_{1\rightarrow 2}}{k_{\rm B}T_{\rm t,h}}}^{\infty} (\sigma_{\rm M=He}({\rm 1s \rightarrow 2s})+\sigma_{\rm M=He}({\rm 1s \rightarrow 2p})) \frac{E}{k_{\rm B}T_{\rm t,h}} \exp \left( -\frac{E}{k_{\rm B}T_{\rm t,h}} \right) {\rm d}\left( \frac{E}{k_{\rm B}T_{\rm t,h}} \right).
\end{equation}

\begin{table}[hbt!]
\small
\centering
\caption{\label{tab:sigmaMHe12}Fitting Coefficients for Eq.~\eqref{eqn:sigmaMHe12}}
\begin{tabular}{lccccc}
\hline \hline
$p_i$ [$\ln(\rm m^2)$] & $i=0$ & $i=1$ & $i=2$ & $i=3$ & $i=4$ \\\hline
1s $\rightarrow$ 2s & $-4.453\times10^{1}$ & $3.855\times10^{0}$ & $-1.017\times10^{0}$ & $1.235\times10^{-1}$ & $-6.128\times10^{-3}$ \\
1s $\rightarrow$ 2p & $-4.277\times10^{1}$ & $4.706\times10^{0}$ & $-1.508\times10^{0}$ & $2.136\times10^{-1}$ & $-1.167\times10^{-2}$ \\
\hline \hline
\end{tabular}
\end{table}

Separately, the rate constant for the $n=3\rightarrow2$ transition with M = H has been computed by Bates and Lewis~\cite{Bates1955} using the Landau-Zener model. In their work, the ionic-curve-crossing mechanism was used to compute cross sections and rate constants for the mutual-neutralization reaction,
\begin{equation}
\begin{split}
    \rm H^- + H^+ & \rm \rightarrow H({\it n}=1) + H({\it n}=2) \\
                  & \rm \rightarrow H({\it n}=1) + H({\it n}=3).
\label{eqn:mutneut}
\end{split}
\end{equation}
These cross sections are in reasonable agreement with more recent experimental~\cite{Szucs1984} and computational results~\cite{Stenrup2009}.
Then, treating $n=3\rightarrow2$ as the double transition of reaction~\eqref{eqn:mutneut} (i.e., $\rm H({\it n}=1) + H({\it n}=3) \rightarrow H^- + H^+ \rightarrow H({\it n}=1) + H({\it n}=2)$), a corresponding rate constant was obtained for the $n=3\rightarrow2$ transition.

\subsubsection{Comparison of Rate Constants}

\begin{figure}[hbt!]
     \centering
     \begin{subfigure}[b]{0.33\textwidth}
         \centering
         \includegraphics[width=\textwidth,trim={0cm 0cm 1cm 0.5cm},clip]{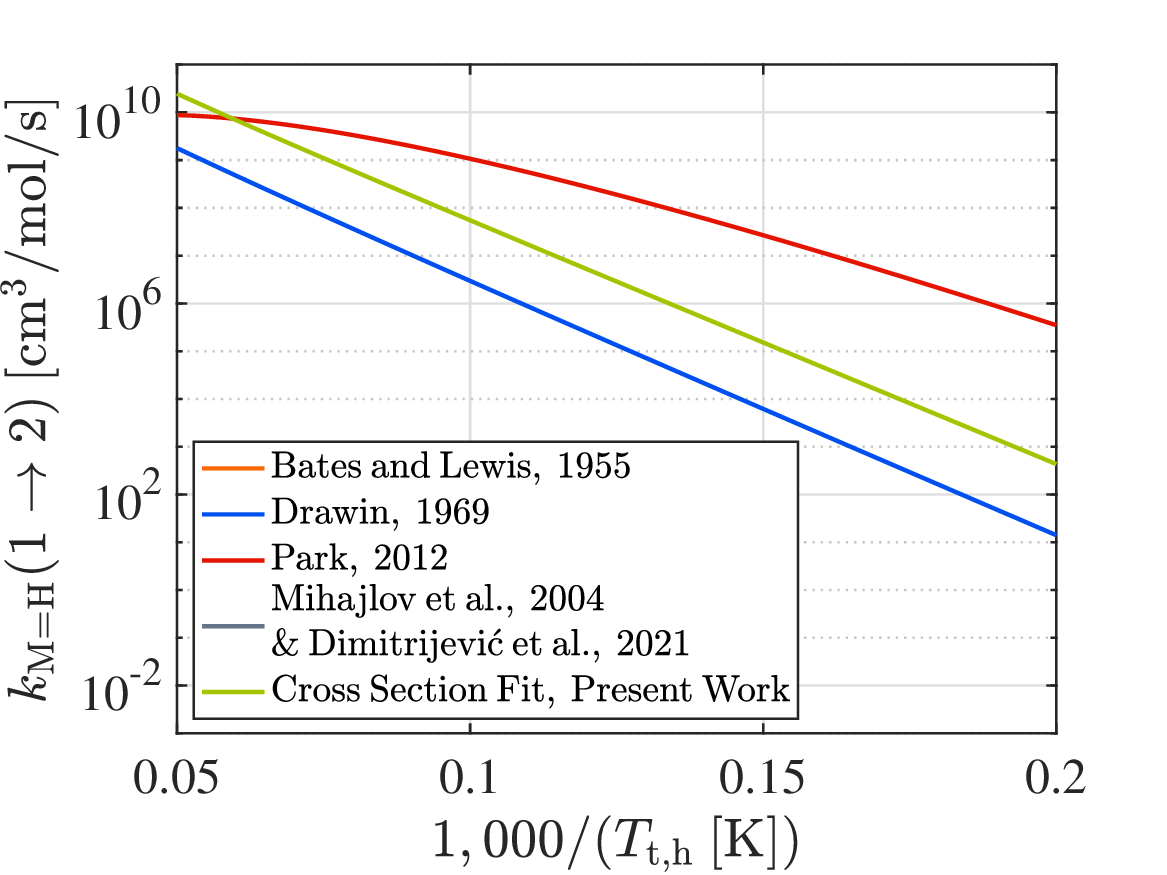}
         \caption{$\bm{n = 1\rightarrow2}$}
     \end{subfigure}
     \hfill
     \begin{subfigure}[b]{0.33\textwidth}
         \centering
         \includegraphics[width=\textwidth,trim={0cm 0cm 1cm 0.5cm},clip]{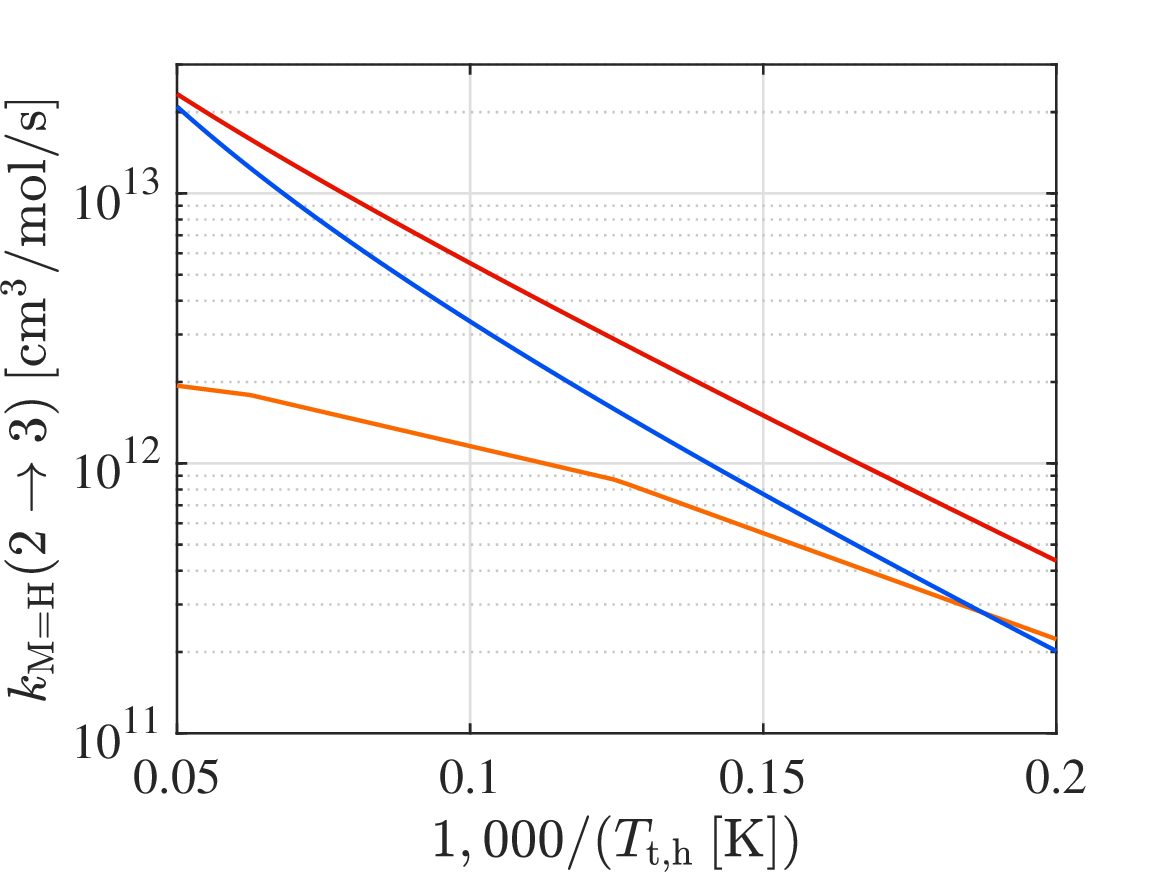}
         \caption{$\bm{n = 2\rightarrow3}$}
     \end{subfigure}
     \hfill
     \begin{subfigure}[b]{0.33\textwidth}
         \centering
         \includegraphics[width=\textwidth,trim={0cm 0cm 1cm 0.5cm},clip]{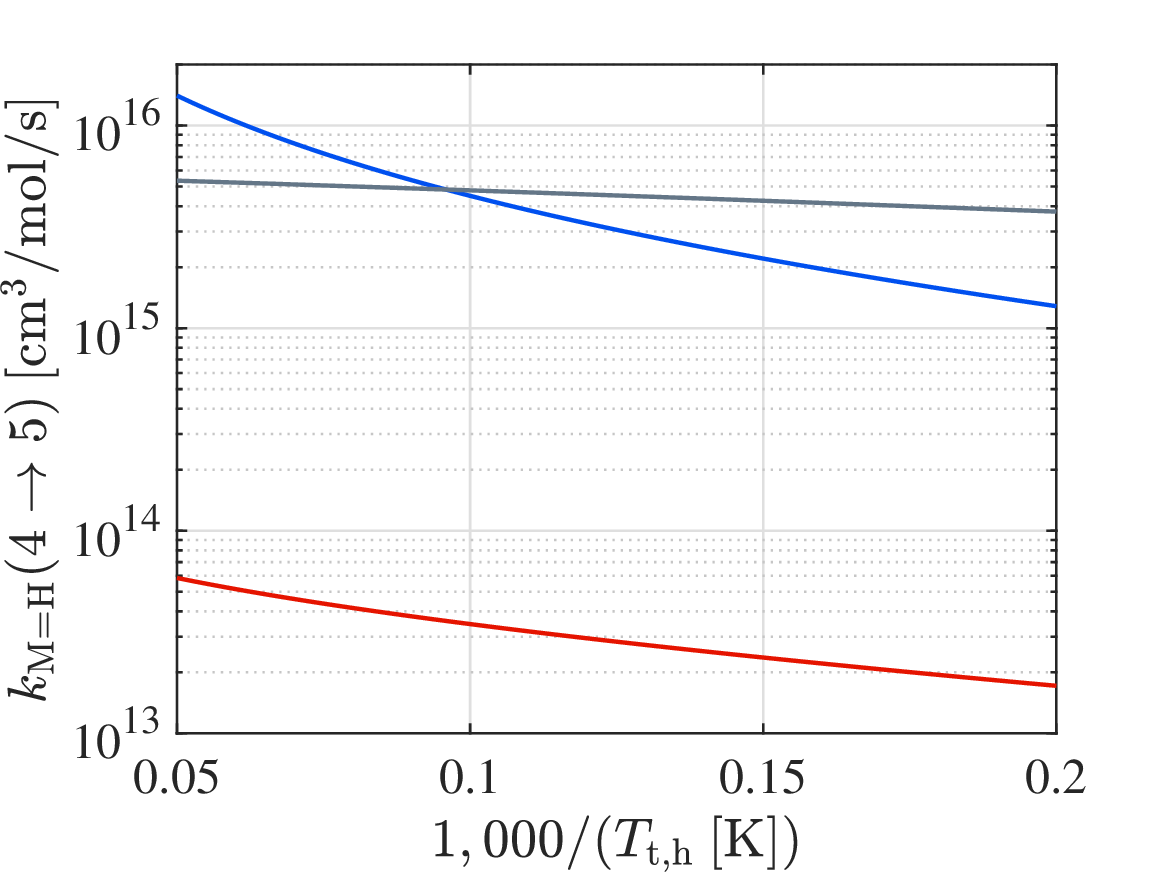}
         \caption{$\bm{n = 4\rightarrow5}$}
     \end{subfigure}
    \caption{H-impact excitation rate constants for the $\bm{n = 1\rightarrow2}$ (left), $\bm{n = 2\rightarrow3}$ (middle), and $\bm{n = 4\rightarrow5}$ (right) transitions.}
    \label{fig:kHexc}
\end{figure}

Figure~\ref{fig:kHexc} shows the H-impact excitation rate constants for select transitions. For the cross-section-fitted rate constant computed from Fig.~\ref{fig:sigmaMHe12}, it has been assumed that $\sigma_{\rm M=H}=\sigma_{\rm M=He}$, such that $k_{\rm M=H}=\sqrt{\bar{m}_{\rm H,He}/\bar{m}_{\rm H,H}}k_{\rm M=He}$. In general, there is very little agreement between the plotted rate constants, with discrepancies on the order of several magnitudes. For the $n=1\rightarrow2$ transition, the Drawin and cross-section-fitted rate constants show similar temperature dependencies. In particular, the temperature dependence of the cross-section-fitted rate constant is effectively characterized by the Arrhenius exponential factor, $\exp(-E_{1\rightarrow2}/(k_{\rm B}T_{\rm t,h}))$. In contrast, the rate constant of Park displays a much shallower temperature dependence, due to the $(12000/T_{\rm t,h})^6$ modification factor applied for this specific transition. Interestingly, near 17,000 K where Park's $n=1\rightarrow2$  rate constant is expected to be most reliable, there is good agreement with the cross-section-fitted rate constant. For the $n=2\rightarrow3$ transition, while there is reasonable agreement between the Drawin and Park rate constants, the rate constant of Bates and Lewis has a shallower temperature dependence, resulting in approximately a magnitude of difference at 20,000 K. Finally, as $n$ increases further, the rate constants of Drawin increase in magnitude faster than those of Park, such that for all $n\geq3$ transitions, Drawin's rate constants are larger than Park's. The apparent agreement for the $n=4\rightarrow5$ transition between the rate constants of Mihajlov et al. and Drawin is coincidental; they diverge from each other as $n$ increases further (not shown).

\begin{figure}[hbt!]
     \centering
     \begin{subfigure}[b]{0.33\textwidth}
         \centering
         \includegraphics[width=\textwidth,trim={0cm 0cm 1cm 0.5cm},clip]{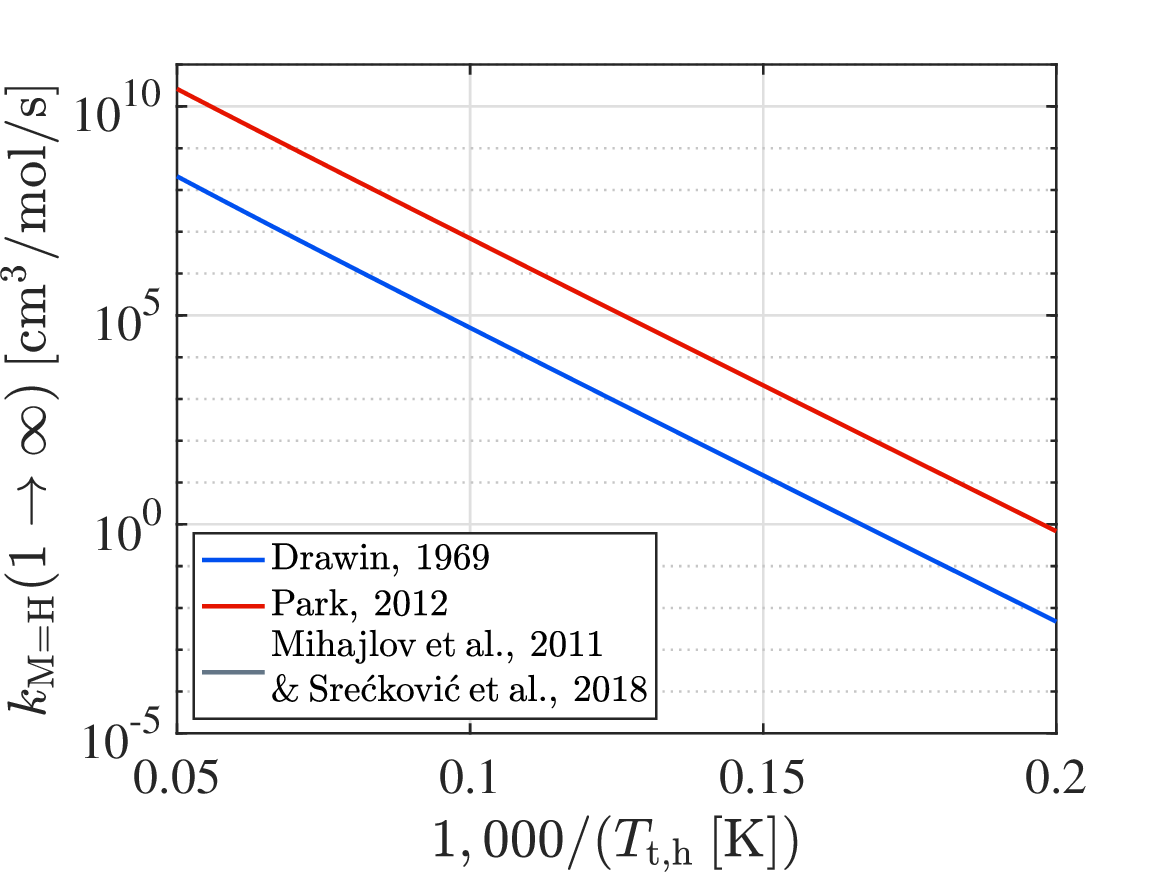}
         \caption{$\bm{n = 1\rightarrow\infty}$}
     \end{subfigure}
     \hfill
     \begin{subfigure}[b]{0.33\textwidth}
         \centering
         \includegraphics[width=\textwidth,trim={0cm 0cm 1cm 0.5cm},clip]{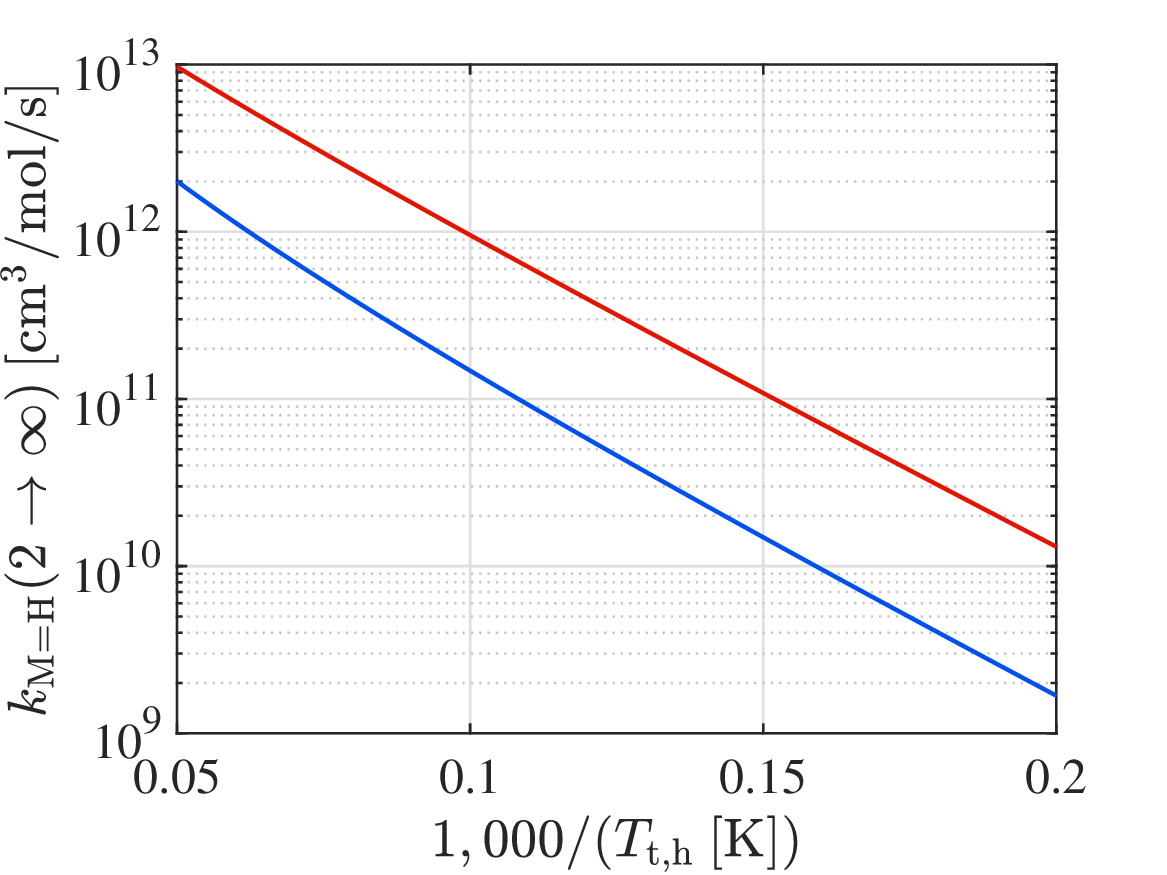}
         \caption{$\bm{n = 2\rightarrow\infty}$}
     \end{subfigure}
     \hfill
     \begin{subfigure}[b]{0.33\textwidth}
         \centering
         \includegraphics[width=\textwidth,trim={0cm 0cm 1cm 0.5cm},clip]{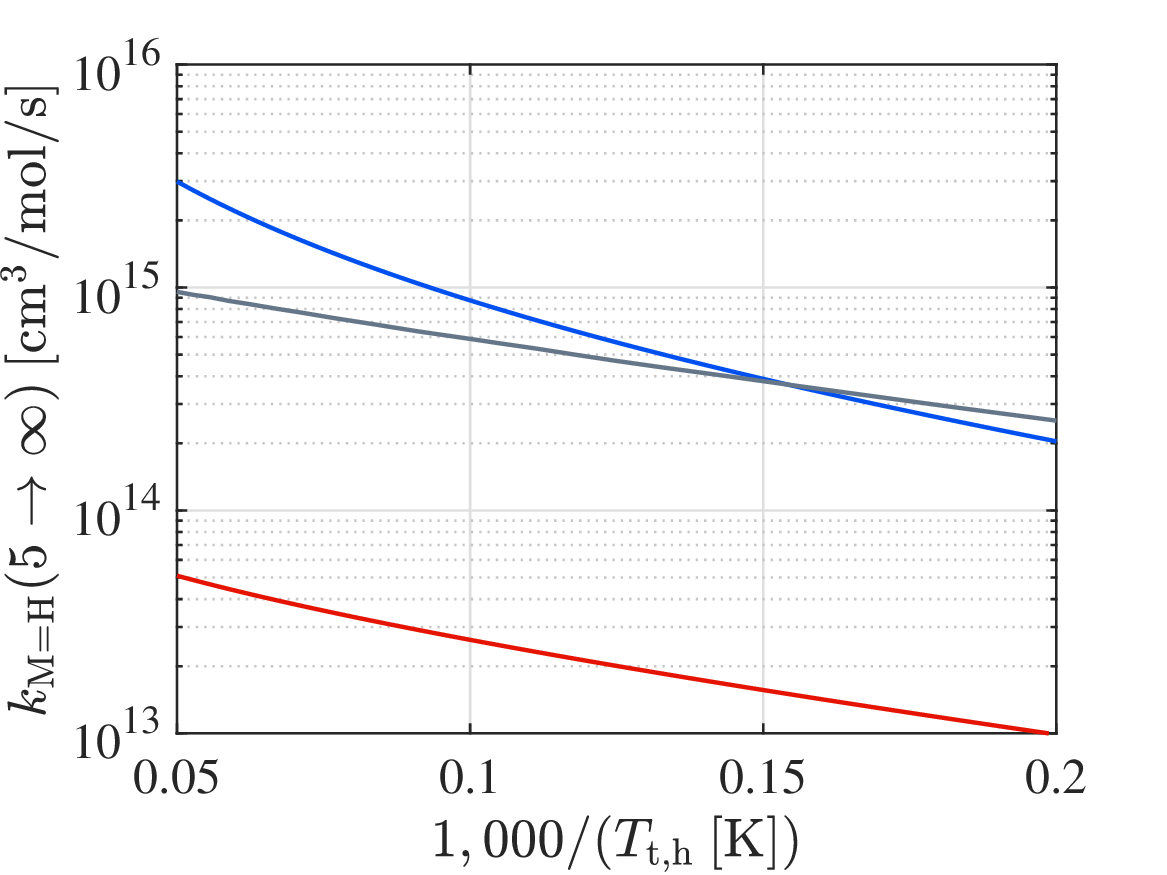}
         \caption{$\bm{n = 5\rightarrow\infty}$}
     \end{subfigure}
    \caption{H-impact ionization rate constants for the $\bm{n = 1\rightarrow\infty}$ (left), $\bm{n = 2\rightarrow\infty}$ (middle), and $\bm{n = 5\rightarrow\infty}$ (right) reactions.}
    \label{fig:kHion}
\end{figure}

The H-impact ionization rate constants for the $n$ = 1, 2, and 5 states are shown in Fig.~\ref{fig:kHion}. In general, similar trends are observed as for the excitation rate constants. Namely, the rate constants of Park are greater than those of Drawin for $n\leq2$, and lower for $n>2$. While there is some agreement between the rate constants of Mihajlov et al. and Drawin for the $n=5\rightarrow\infty$ reaction, they diverge from each other for higher $n$.

\subsubsection{Proposed Rate Constants}
\label{sec:prophvy}

The review presented in the previous sections shows that the cross-section-fitted and Mihajlov et al. rate constants are the most reliable of the available heavy-particle-impact rate constants. However, the cross-section-fitted and Mihajlov et al. datasets alone are not enough to specify rate constants for all transitions of interest; it is still necessary to specify rate constants for the $n=1\rightarrow n'\geq3$, $n=2\rightarrow n'\geq3$, $n=3\rightarrow n'\geq4$, and $n\leq4\rightarrow\infty$ transitions separately, as well as the relevant third-body efficiencies.

\begin{figure}[hbt!]
     \centering
     \begin{subfigure}[b]{0.4\textwidth}
         \centering
         \includegraphics[width=\textwidth,trim={0cm 0cm 1cm 0.5cm},clip]{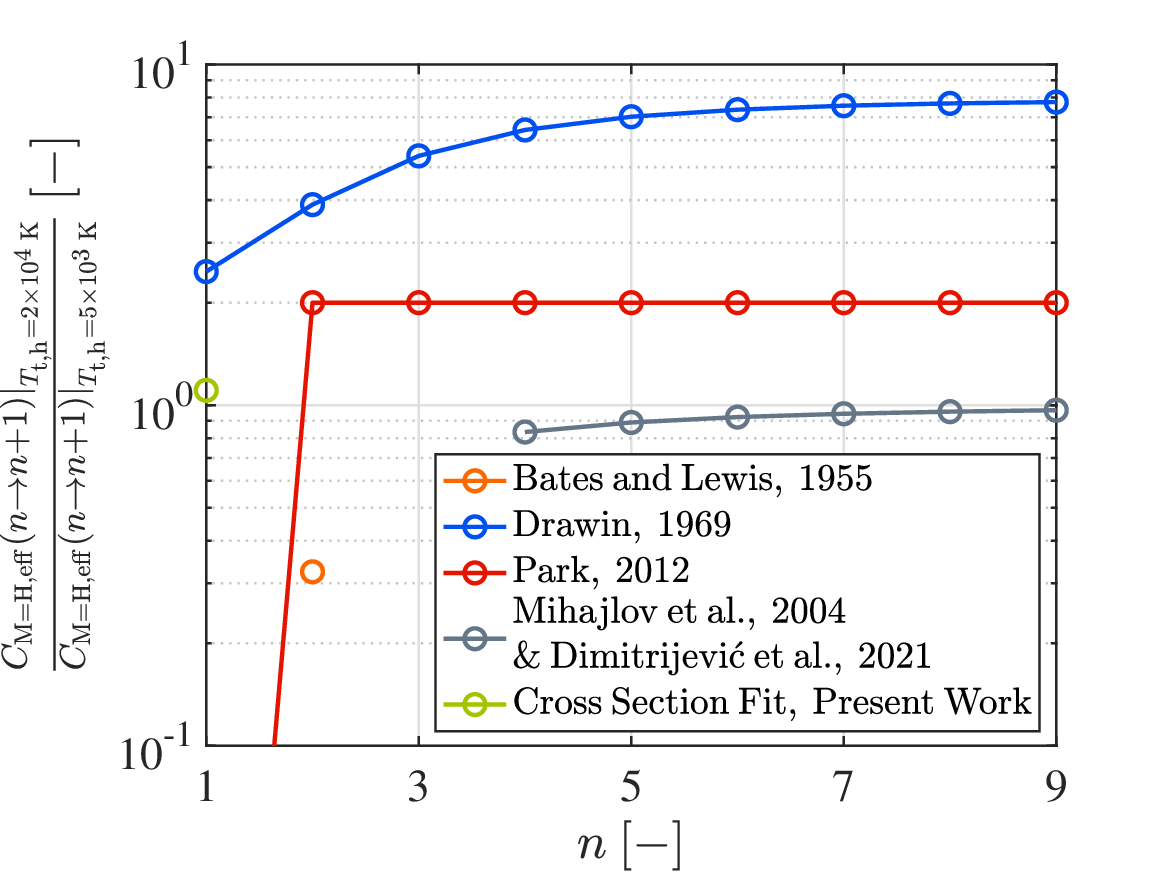}
     \end{subfigure}
    \caption{Ratio of the effective Arrhenius pre-exponential factor evaluated at 20,000 K versus 5,000 K. For Park~\cite{Park2012} at $\bm{n=1}$, this ratio has a value of $\bm{4.9\times10^{-4}}$ (not shown).}
    \label{fig:hvyscale_Eth}
\end{figure}

To specify these remaining rate constants, two assumptions are first made about the functional form of the heavy-particle-impact rate constants. First, it is assumed that for any given excitation transition or ionization reaction, the cross sections are the same for all heavy third-bodies, such that $k_{\rm M\neq e^-}=\sqrt{\bar{m}_{\rm H,H}/\bar{m}_{\rm H,M\neq e^-}}k_{\rm M=H}$. Second, it is assumed that the rate constants for all excitation transitions are described by the Arrhenius functional form, $k_{\rm M=H}(n\rightarrow n')=C_{\rm M=H}(n\rightarrow n')\exp(-E_{n\rightarrow n'}/(k_{\rm B}T_{\rm t,h}))$, where $C_{\rm M=H}(n\rightarrow n')$ is not a function of temperature. This assumption is motivated by Fig.~\ref{fig:hvyscale_Eth}, which shows the ``effective'' $C_{\rm M=H}(n\rightarrow n+1)$ for all mono-quantum transitions, computed as $k_{\rm M=H}(n\rightarrow n+1)/\exp(-E_{n\rightarrow n+1}/(k_{\rm B}T_{\rm t,h}))$, evaluated at 20,000 K and normalized by the same quantity evaluated at 5,000 K. The plotted ratio is very close to one for the cross-section-fitted and Mihajlov et al. rate constants, indicating that the assumption of a temperature-independent $C_{\rm M=H}(n\rightarrow n')$ is reasonable for the heavy-particle-impact excitation rate constants.

\begin{figure}[hbt!]
     \centering
     \begin{subfigure}[b]{0.4\textwidth}
         \centering
         \includegraphics[width=\textwidth,trim={0cm 0cm 1cm 0cm},clip]{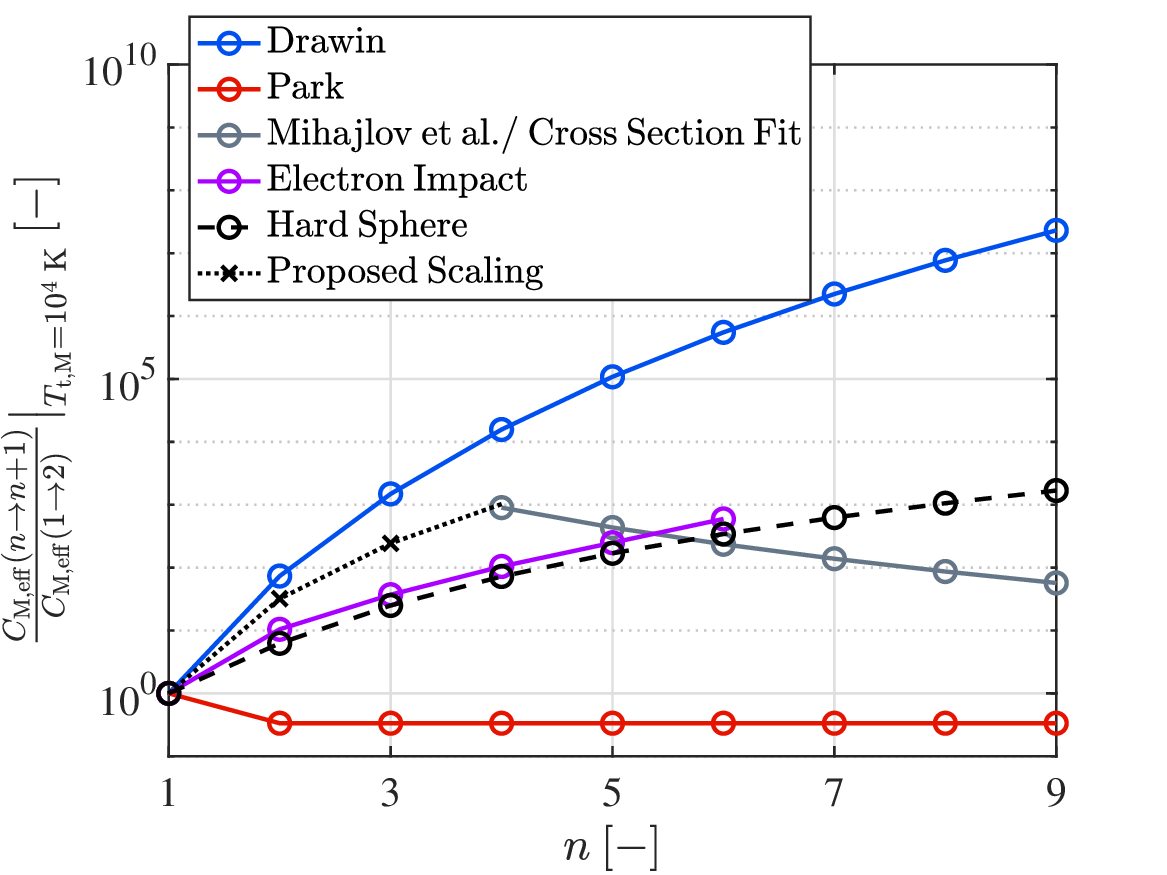}
     \end{subfigure}
    \caption{Scalings of effective Arrhenius pre-exponential factors for mono-quantum $\bm{n\rightarrow n+1}$ transitions.}
    \label{fig:hvyscale_monoq}
\end{figure}

Next, three additional assumptions are made about the scaling of the heavy-particle-impact rate constants with $n$ and $n'$. For the $n$ = 2 and 3 mono-quantum transitions, it is assumed that the pre-exponential factors can be described by the power law, $C_{\rm M=H}(n\rightarrow n+1)/C_{\rm M=H}(1\rightarrow 2)=n^5$. This assumption is based on Fig.~\ref{fig:hvyscale_monoq}, which shows the scalings, $C_{\rm M,eff}(n\rightarrow n+1)/C_{\rm M,eff}(1\rightarrow 2)$, from various sources alongside the proposed scaling. In this plot, the electron-impact scaling is computed using the proposed rate constants from section~\ref{sec:revEI}, and the hard-sphere scaling is computed assuming the radius of H scales with the Bohr radius, $a_0n^2$, such that $C_{\rm M}(n\rightarrow n+1)/C_{\rm M}(1\rightarrow 2)=(n^2+1)^2/4$. Notably, the electron-impact scaling closely follows the hard-sphere scaling. The proposed scaling is chosen to follow the power-law-type dependence of the Drawin, electron-impact, and hard-sphere scalings, while still matching the scaling of the Mihajlov et al./ cross-section-fitted rate constants at $n$ = 4.

\begin{figure}[hbt!]
     \centering
     \begin{subfigure}[b]{0.33\textwidth}
         \centering
         \includegraphics[width=\textwidth,trim={0cm 0cm 1cm 0.5cm},clip]{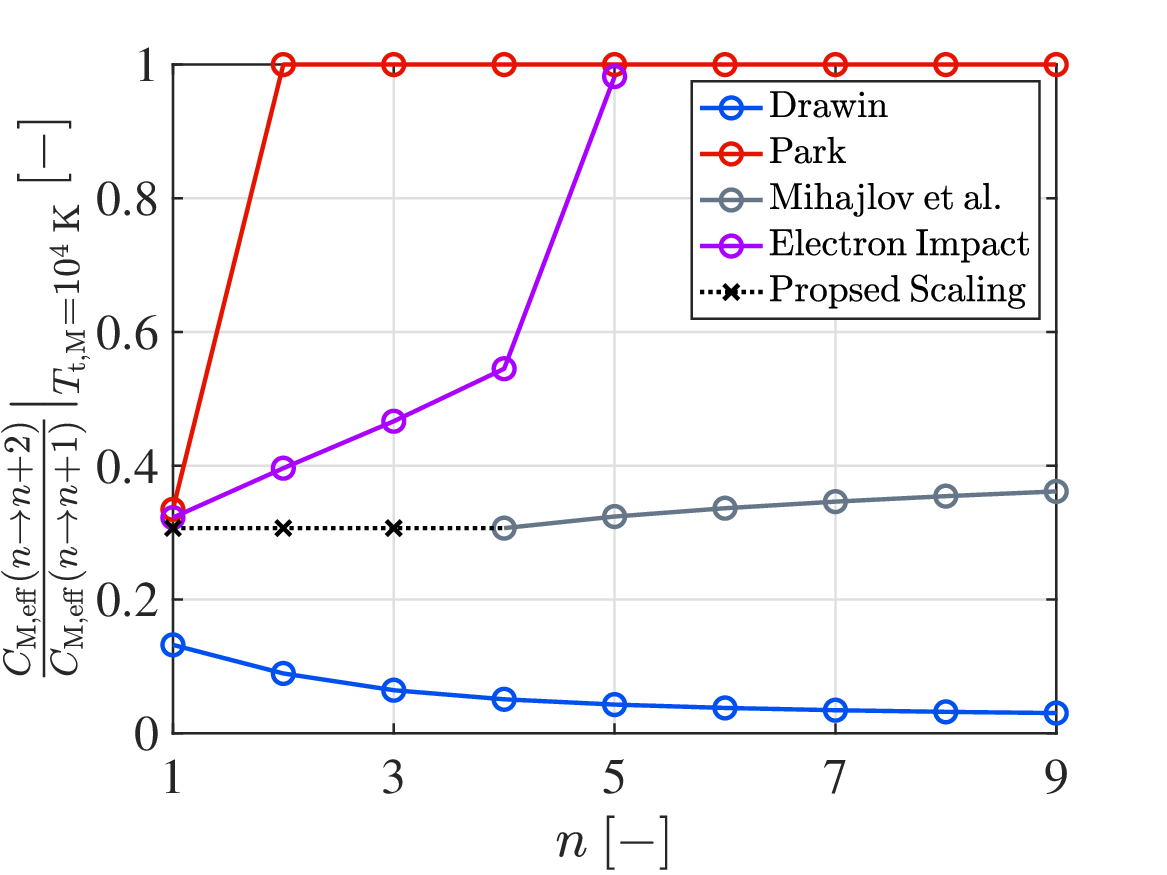}
         \caption{$\bm{n \rightarrow n+2}$}
     \end{subfigure}
     \hfill
     \begin{subfigure}[b]{0.33\textwidth}
         \centering
         \includegraphics[width=\textwidth,trim={0cm 0cm 1cm 0.5cm},clip]{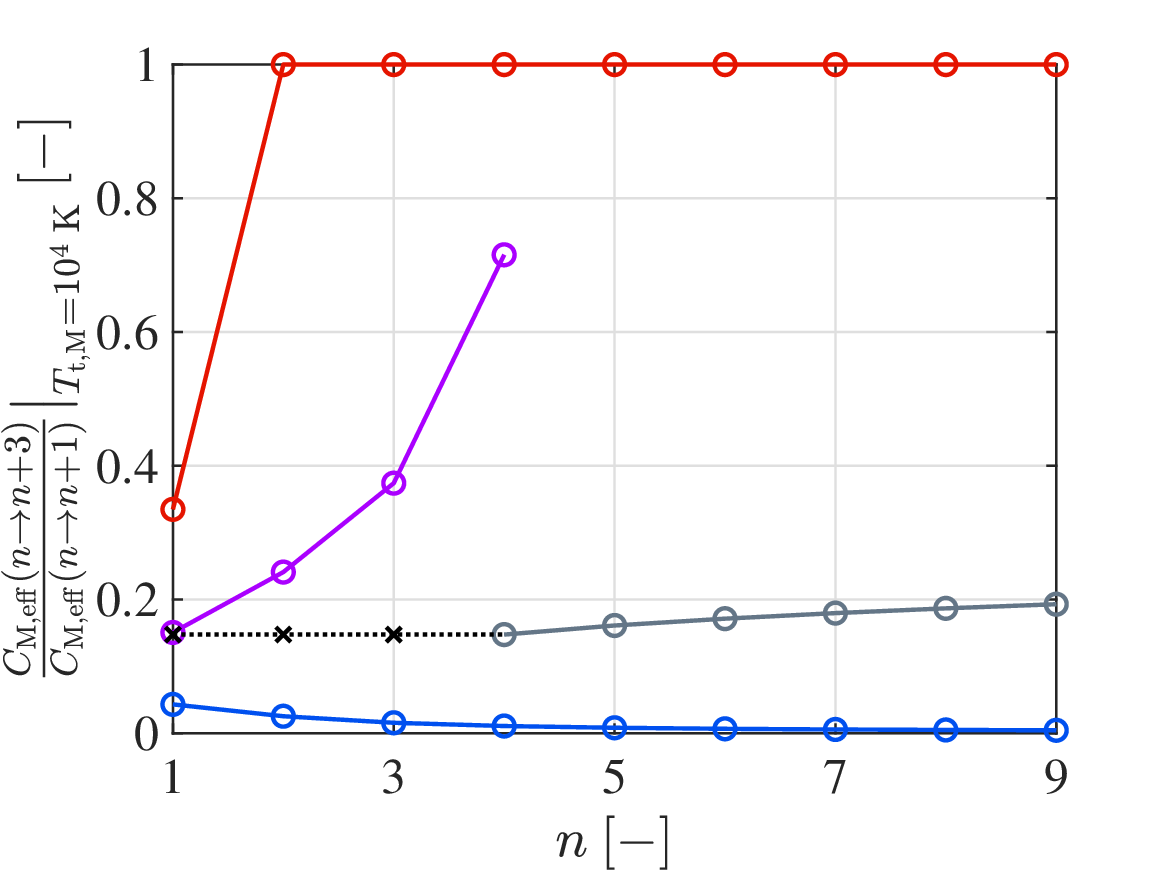}
         \caption{$\bm{n \rightarrow n+3}$}
     \end{subfigure}
     \hfill
     \begin{subfigure}[b]{0.33\textwidth}
         \centering
         \includegraphics[width=\textwidth,trim={0cm 0cm 1cm 0.5cm},clip]{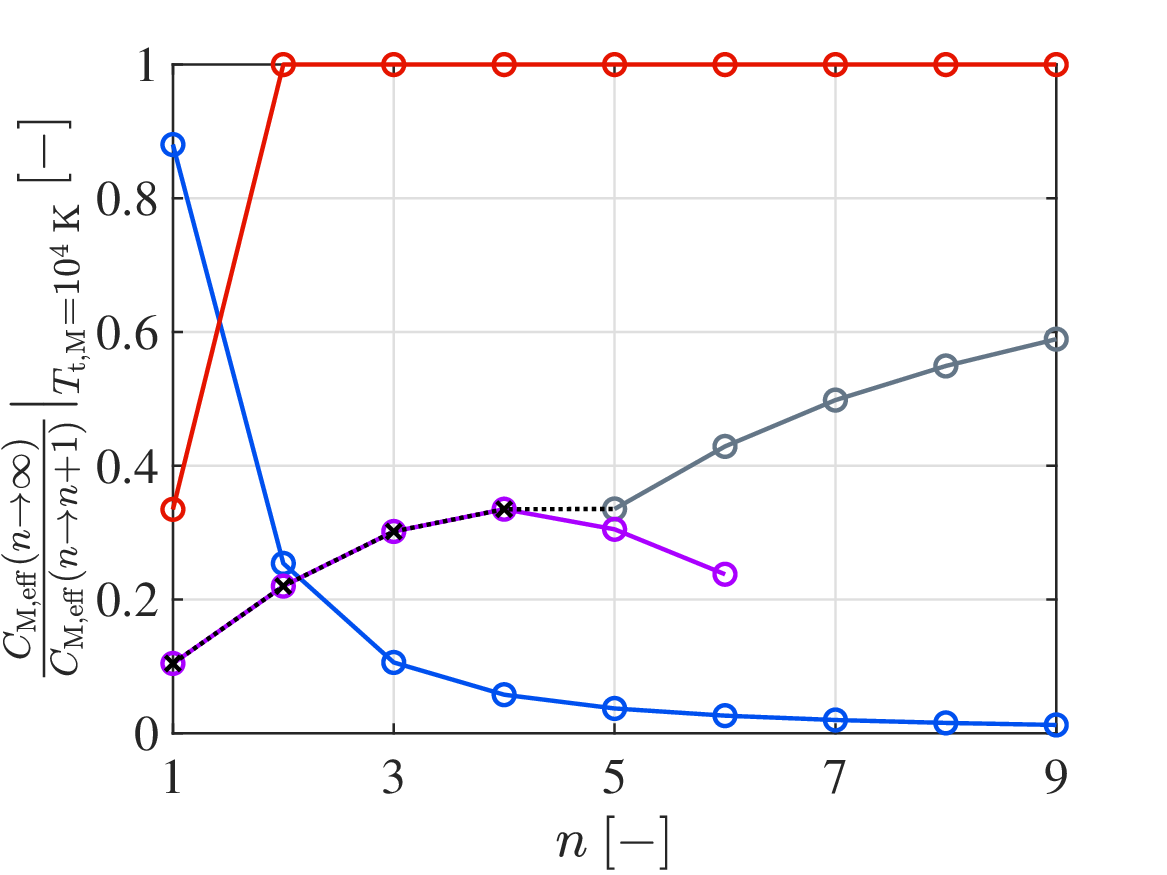}
         \caption{$\bm{n \rightarrow \infty}$}
     \end{subfigure}
    \caption{Scalings of effective Arrhenius pre-exponential factors for the $\bm{n\rightarrow n+2}$ (left), $\bm{n\rightarrow n+3}$ (middle), and $\bm{n\rightarrow \infty}$ (right) transitions relative to the mono-quantum $\bm{n\rightarrow n+1}$ transitions.}
    \label{fig:hvyscale_multiq}
\end{figure}

For the remaining multi-quantum transitions, $n\rightarrow n+\Delta n$ where $n$ = 1, 2, and 3 and $\Delta n\geq2$, it is assumed that $C_{\rm M=H}(n\rightarrow n+\Delta n)/C_{\rm M=H}(n\rightarrow n+1)=C_{\rm M=H}(4\rightarrow 4+\Delta n)/C_{\rm M=H}(4\rightarrow 5)$. This assumption is based on Fig.~\ref{fig:hvyscale_multiq} (a) and (b), which show that the scalings of Mihajlov et al. for the multi-quantum transitions are nearly constant with $n$. Interestingly, the proposed scaling also nearly matches that of the electron-impact rate constants at $n$ = 1. Finally, the remaining ionization rate constants for the $n$ = 1 to 4 states are assumed to follow the electron-impact scaling, as this leads to a more consistent extension to low $n$ when compared to the Drawin and Park scalings, as evidenced in Fig.~\ref{fig:hvyscale_multiq} (c). It is worth noting that while Fig.~\ref{fig:hvyscale_monoq} and Fig.~\ref{fig:hvyscale_multiq} only show the rate constant scalings at 10,000 K, similar trends were observed for all relevant temperatures.

\section{Methodology}
\label{sec:methods}

In the present work, simulations of 1-D normal shocks in a $\rm H_2$/ He mixture are performed using a space-marching CFD code. The resulting flowfield solutions are used to compute radiance values which are compared to the measurements from the EAST experiments of Cruden and Bogdanoff~\cite{Cruden2017}. The governing equations for the flowfield calculations are first discussed in section~\ref{sec:goveq}. Then, the implementation of the thermochemical model and radiance calculations are described in sections~\ref{sec:thermochem} and \ref{sec:neqair}, respectively. Finally, details of the simulated shot conditions are discussed in section~\ref{sec:shots}.

\subsection{Governing Equations}
\label{sec:goveq}

\subsubsection{Previous Work}

Most previous studies in the literature investigating $\rm H_2$/ He flows in shock tubes have assumed that the post-shock flow can be modeled adequately using the 1-D steady Euler equations in the shock-fixed frame (referred to as the Rankine-Hugoniot approach in the present work). This includes the previous studies by Park~\cite{Park2012} and Furudate~\cite{Furudate2009}, which simulated the experiments of Livingston and Poon~\cite{Livingston1976} and Leibowitz~\cite{Leibowitz1973}, as well as the more recent study by Colonna et al.~\cite{Colonna2020} which simulated the experiments of Cruden and Bogdanoff~\cite{Cruden2017}. Such an approach is often used because post-shock flow solutions can be obtained efficiently using a space-marching numerical scheme. However, this approach does not allow for the modeling of real shock tube flow non-uniformities, namely boundary layer effects.

In an idealized shock described by the Rankine-Hugoniot relations, the post-shock speed is nearly constant. Consequently, in the shock frame-of-reference, the post-shock flow does not stagnate. However, in an actual shock tube, the presence of a boundary layer behind the shock displaces mass away from the core flow, eventually resulting in a steady-state configuration with stagnated flow at the contact surface~\cite{Mirels1963,Mirels1966}. This difference in post-shock speed profiles results in a spatial ``distortion'' of the thermochemical processes and therefore the radiance seen in shock tube experiments.

To address this shortcoming of the Rankine-Hugoniot approach, Clarke et al.~\cite{Clarke2023} proposed a simple post-shock spatial transformation based on the theory of Mirels~\cite{Mirels1963}, which can be applied a posteriori to a Rankine-Hugoniot flow solution. However, as noted by Clarke et al., this spatial transformation alone does not account for the associated compression effect that would result from a stagnated post-shock flow. In a subsequent study, Clarke et al.~\cite{Clarke2024} proposed an alternative approach, which captures the effect of the spatial transformation and the coupled compression effect simultaneously. Unfortunately, the governing equations for this alternative approach are elliptic, and therefore cannot be solved in a simple space-marching manner.

A third approach is proposed here which accounts for the spatial transformation and compression effect in a quasi-1-D framework, but neglects viscous effects in the core flow such that solutions can still be obtained rapidly using a space-marching method. The derivation of this method is described in the following sections.

\subsubsection{Navier-Stokes Equations}

In the general case, the compressible Navier-Stokes equations for reacting flows are given by
\begin{equation}
    \frac{\partial\rho}{\partial t}+\frac{\partial}{\partial x_{\rm i}}(\rho u_{\rm i}) = 0
    \label{eqn:continuity}
\end{equation}
for continuity,
\begin{equation}
    \frac{\partial}{\partial t}(\rho Y_{\rm s})+\frac{\partial}{\partial x_{\rm i}}(\rho u_{\rm i}Y_{\rm s}) = -\frac{\partial j_{\rm s,i}}{\partial x_{\rm i}} + \dot{\omega}_{\rm s}
    \label{eqn:species}
\end{equation}
for continuity of species s,
\begin{equation}
    \frac{\partial}{\partial t}(\rho u_{\rm i})+\frac{\partial}{\partial x_{\rm j}}(\rho u_{\rm i}u_{\rm j}) = -\frac{\partial P}{\partial x_{\rm i}} + \frac{\partial\tau_{\rm ij}}{\partial x_{\rm j}}
    \label{eqn:momentum}
\end{equation}
for momentum, and
\begin{equation}
    \frac{\partial}{\partial t}\left(\rho\left(e+\frac{u_{\rm i}^2}{2}\right)\right) + \frac{\partial}{\partial x_{\rm j}}\left(\rho u_{\rm i}\left(h+\frac{u_{\rm i}^2}{2}\right)\right) = -\frac{\partial q_{\rm i}}{\partial x_{\rm i}} + \frac{\partial}{\partial x_{\rm i}}(\tau_{\rm ij}u_{\rm j})
    \label{eqn:energy}
\end{equation}
for total energy. Here, the subscripts i and j indicate spatial directions, and the specific energy and specific enthalpy are related as $h=e+P/\rho$, where $h = \sum_{\rm s} Y_{\rm s} h_{\rm s}$. To simplify calculations, radiative processes are assumed to be negligible in the flowfield governing equations, i.e., only collisional processes are considered in the kinetic model. The validity of this assumption is discussed in Appendix A.

As will be discussed in section~\ref{sec:thermochem}, the non-equilibrium kinetic mechanism for $\rm H_2$ dissociation used in the present work does not rely on the solution of separate rotational and/ or vibrational energy equations. However, the electron translational mode is not assumed to be in equilibrium with the heavy gas translational mode, and hence the solution of a separate electron energy equation is still required. Assuming a Maxwellian electron distribution, the electron energy equation is given by
\begin{equation}
     \frac{\partial}{\partial t}(\rho e_{\rm e^-}) + \frac{\partial}{\partial x_{\rm i}}(\rho u_{\rm i} e_{\rm e^-}) = -\frac{\partial q_{\rm e^-,i}}{\partial x_{\rm i}} + \dot{\Omega}_{\rm e^-},
    \label{eqn:electron}
\end{equation}
where $e_{\rm e^-} = 3Y_{\rm e^-}k_{\rm B}T_{\rm t,e^-}/(2m_{\rm e^-})$. Assuming the work done on electrons by the electron-pressure-gradient-induced electric field is negligible, the source term of Eq.~\eqref{eqn:electron} is computed as the sum of the contributions from elastic and inelastic electron-impact collisions, i.e.,
\begin{equation}
    \dot{\Omega}_{\rm e^-} = \dot{\Omega}_{\rm e^-,elastic} + \dot{\Omega}_{\rm e^-,inelastic}.
    \label{eqn:omegae}
\end{equation}
Following the formulation of Appleton and Bray~\cite{Appleton1964}, the elastic term is computed as
\begin{equation}
    \dot{\Omega}_{\rm e^-,elastic} = 3\rho Y_{\rm e^-}k_{\rm B}(T_{\rm t,h}-T_{\rm t,e^-}) \sqrt{\frac{8k_{\rm B}T_{\rm t,e^-}}{\pi m_{\rm e^-}}} \left(\sum_{\rm s \neq e^-} \frac{N_{\rm s}}{m_{\rm s}} \bar{\sigma}_{\rm e^-,s} \right),
    \label{eqn:omegaeel}
\end{equation}
where $\bar{\sigma}_{\rm e^-,s}$ is computed by averaging the cross sections from the IST-Lisbon LXCAT database~\cite{Alves2014} over a Maxwellian electron distribution for s = H and He, and by similarly averaging the cross section of Zammit et al.~\cite{Zammit2016} for s = $\rm H_2$. The resulting fits for $\bar{\sigma}_{\rm e^-,s}$ as a function of $T_{\rm t,e^-}$ are provided in Table~\ref{tab:elasticXS}. For the case where s is an ion, the Coulomb cross section with the Debye cutoff approximation from Park~\cite{Park1989NonequilibriumHA} is used. Assuming that all of the energy for electron-impact excitation and ionization comes from the translational energy of electrons, the inelastic term is computed as
\begin{equation}
\begin{split}
    \dot{\Omega}_{\rm e^-,inelastic} = N_{\rm e^-}\bigg[
    & - \sum_{n=1}^{n_{\rm max}-1}\sum_{\rm n'>n}^{n_{\rm max}} N_{\rm H\it(n)} k_{\rm M=e^-}(n\rightarrow n') E_{n\rightarrow n'}
    + \sum_{n=1}^{n_{\rm max}-1}\sum_{\rm n'>n}^{n_{\rm max}} N_{\rm H\it(n')} k_{\rm M=e^-}(n'\rightarrow n) E_{n\rightarrow n'} \\
    & - \sum_{n=1}^{n_{\rm max}} N_{\rm H\it(n)}k_{\rm M=e^-}(n\rightarrow \infty) E_{n\rightarrow\infty}
    + N_{\rm e^-}N_{\rm H^+}\sum_{n=1}^{n_{\rm max}} k_{\rm M=e^-}(\infty\rightarrow n) E_{n\rightarrow\infty}\bigg].
    \label{eqn:omegaeinel}
\end{split}
\end{equation}

\begin{table}[hbt!]
\small
\centering
\caption{\label{tab:elasticXS}Fits of Thermally-Averaged Elastic-Scattering Cross Sections}
\begin{tabular}{lc}
\hline \hline
s & $\bar{\sigma}_{\rm e^-,s}$ [$\rm m^2$] \\\hline
$\rm H_2$ & $-1.463\times10^{-21}[\ln(T_{\rm t,e^-})]^2+4.229\times10^{-20}[\ln(T_{\rm t,e^-})]-1.255\times10^{-19}$ \\
H & \hspace{2mm}$1.111\times10^{-21}[\ln(T_{\rm t,e^-})]^2-1.193\times10^{-19}[\ln(T_{\rm t,e^-})]+1.239\times10^{-18}$ \\
He & $-3.621\times10^{-21}[\ln(T_{\rm t,e^-})]^2+6.699\times10^{-20}[\ln(T_{\rm t,e^-})]-2.423\times10^{-19}$ \\
\hline \hline
\end{tabular}
\end{table}

\subsubsection{Proposed Quasi-1-D Formulation}
\label{sec:Q1D}

To derive the relevant governing equations, Eq.~\eqref{eqn:continuity}-\eqref{eqn:electron} are applied to the shock tube core flow in the shock-fixed frame. It is assumed that the core flow is steady, inviscid, and quasi-1-D (all quantities except $u_{\rm r}$ only vary with $x$). Under these assumptions, the governing equations in cylindrical coordinates are given by
\begin{equation}
    \frac{\partial}{\partial x}(\rho u)+\frac{1}{r}\frac{\partial}{\partial r}(r \rho u_{\rm r}) = 0,
    \label{eqn:continuity_core}
\end{equation}
\begin{equation}
    \frac{\partial}{\partial x}(\rho u Y_{\rm s})+\frac{1}{r}\frac{\partial}{\partial r}(r \rho u_{\rm r} Y_{\rm s}) = \dot{\omega}_{\rm s},
    \label{eqn:species_core}
\end{equation}
\begin{equation}
    \frac{\partial}{\partial x}(\rho u^2)+\frac{1}{r}\frac{\partial}{\partial r}(r \rho u_{\rm r} u) = -\frac{\partial P}{\partial x},
    \label{eqn:momentum_core}
\end{equation}
\begin{equation}
    \frac{\partial}{\partial x}\left(\rho u \left(h+\frac{u^2}{2}\right) \right) + \frac{1}{r}\frac{\partial}{\partial r}\left(r \rho u_{\rm r} \left(h+\frac{u^2}{2}\right) \right) = 0,
    \label{eqn:energy_core}
\end{equation}
and
\begin{equation}
    \frac{\partial}{\partial x}(\rho ue_{\rm e^-})+\frac{1}{r}\frac{\partial}{\partial r}(r\rho u_{\rm r}e_{\rm e^-}) = \dot{\Omega}_{\rm e^-}.
    \label{eqn:electron_core}
\end{equation}
Mirels's~\cite{Mirels1963} theory gives that the axial momentum profile at any post-shock location in the core flow is described by
\begin{equation}
    \rho u = \rho_0 u_0 \left( 1 - \sqrt{\frac{x}{L}} \right).
    \label{eqn:rhoux}
\end{equation}
The estimation of test slug lengths, $L$, is discussed separately in section~\ref{sec:Lmirels}. Substituting Eq.~\eqref{eqn:rhoux} into Eq.~\eqref{eqn:continuity_core} yields the corresponding radial momentum profile as
\begin{equation}
    \rho u_{\rm r} = r \frac{\rho_0 u_0}{4\sqrt{xL}}.
    \label{eqn:rhour}
\end{equation}
Equations~\eqref{eqn:continuity_core}-\eqref{eqn:rhour} are consistent with the formulation of Clarke et al.~\cite{Clarke2024} when the transport terms (diffusion, viscosity, and conduction) are neglected.

Substituting both Eq.~\eqref{eqn:rhoux} and \eqref{eqn:rhour} into Eq.~\eqref{eqn:species_core}-\eqref{eqn:electron_core} gives the final set of governing equations as
\begin{equation}
    \frac{{\rm d}Y_{\rm s}}{{\rm d}x} = \frac{\dot{\omega}_{\rm s}}{\rho_0 u_0 \left(1-\sqrt{x/L}\right)},
    \label{eqn:species_march}
\end{equation}
\begin{equation}
    \frac{\rm d}{{\rm d}x}(P + \rho u^2) = - \frac{\rho_0 u_0 u}{2\sqrt{xL}},
    \label{eqn:momentum_march}
\end{equation}
\begin{equation}
    h+\frac{u^2}{2} = h_0 +\frac{u_0^2}{2},
    \label{eqn:energy_march}
\end{equation}
and
\begin{equation}
   \frac{{\rm d}e_{\rm e^-}}{{\rm d}x} = \frac{\dot{\Omega}_{\rm e^-}}{\rho_0 u_0 \left(1-\sqrt{x/L}\right)}.
   \label{eqn:electron_march}
\end{equation}
As expected, in the limit where $L \rightarrow \infty$, the standard Rankine-Hugoniot shock invariants are recovered. The system of equations is closed by using the ideal gas equation of state,
\begin{equation}
    P = \rho k_{\rm B}\left( \sum_{\rm s\neq e^-} \frac{Y_{\rm s}}{m_{\rm s}}T_{\rm t,h} + \frac{Y_{\rm e^-}}{m_{\rm e^-}}T_{\rm t,e^-} \right).
    \label{eqn:eos}
\end{equation}

To compute the flowfield solution, these equations are solved numerically in a space-marching manner as follows. Starting from a given grid point in the post-shock, $x_i$, Eq.~\eqref{eqn:species_march} and \eqref{eqn:electron_march} are first integrated to $x_{i+1}$ to obtain $Y_{\rm s}(x_{i+1})$ and $e_{\rm e^-}(x_{i+1})$ using a modified version of the implicit method of Beardsell and Blanquart~\cite{Beardsell2020}. Equation~\eqref{eqn:momentum_march} is similarly integrated using the explicit RK4 method to obtain $(P+\rho u^2)|_{x_{i+1}}$. These quantities along with Eq.~\eqref{eqn:energy_march} and \eqref{eqn:eos} are then solved iteratively to compute the remaining flowfield variables: $P(x_{i+1})$, $\rho(x_{i+1})$, $u(x_{i+1})$, $h(x_{i+1})$, and $T_{\rm t,h}(x_{i+1})$. This entire procedure is then repeated at each point in the post-shock domain length of interest.

\subsection{Thermochemical Model}
\label{sec:thermochem}

\subsubsection{Thermodynamics}

Four bulk species, $\rm H_2$, He, $\rm H^+$, and $\rm e^-$, and seven electronic states of H($n$=1-7) are included in the thermochemical model for a total of 11 species. Including seven electronic states of H effectively covers the list of relevant spectral features (discussed in section~\ref{sec:neqair}) while minimizing computational cost\footnote{Calculations were also performed with up to 20 electronic states of H, but the difference in results was found to be negligible.}. The impact of considering the additional species $\rm H_2^+$ is discussed separately in Appendix B.

The thermodynamic properties of $\rm H_2$ are computed using the non-equilibrium expression proposed by Carroll et al.~\cite{Carroll2026_diss1}. In the work of Carroll et al., the rovibrational master equations were used to derive macroscopic chemical source term and rovibrational energy expressions that hold in the three key limits/ regimes of dissociation-dominated flows: the thermal equilibrium limit, the quasi-steady-state (QSS) regime, and the pre-QSS regime. In contrast to the widely used multi-temperature~\cite{Park1989,Chaudhry2020,Singh2020_1} and coarse-graining~\cite{Magin2012,Liu2015,Venturi2020} methods, this new method does not rely on the solution of additional energy and/ or species equations, and is ultimately only a function of the heavy-gas translational temperature, $T_{\rm t,h}$, and the degree of dissociation, $\alpha\equiv N_{\rm H({\it n}=1)}/(N_{\rm H({\it n}=1)}+2N_{\rm H_2})$. The macroscopic rovibrational energy of $\rm H_2$ in this formulation is given by
\begin{equation}
    e_{\rm rv,H_2} = \left(1-\frac{N_{\rm H}^2}{N_{\rm H_2}}K_{\rm eq}^{-1}\right)e_{\rm rv,H_2,pre-QSS}+\left(\frac{N_{\rm H}^2}{N_{\rm H_2}}K_{\rm eq}^{-1}\right)e_{\rm rv,H_2,th},
\end{equation}
where
\begin{equation}
    e_{\rm rv,H_2,pre-QSS} = e_{\rm rv,H_2,nr}\left[1 - \left(1-\frac{e_{\rm rv,H_2,0}}{e_{\rm rv,H_2,nr}}\right) \exp\left( -\sqrt{-\frac{\ln(1-\alpha)}{\eta}} \right) \right].
    \label{eqn:erv_preQSS}
\end{equation}
Here, $K_{\rm eq}(T_{\rm t,h})$ is the macroscopic equilibrium constant for $\rm H_2$ dissociation, $\eta(T_{\rm t,h})$ is a pre-QSS correction factor, and $e_{\rm rv,H_2,nr}(T_{\rm t,h})$ and $e_{\rm rv,H_2,th}(T_{\rm t,h})$ are the rovibrational energy of $\rm H_2$ in the QSS regime when recombination is neglected and in the thermal equilibrium limit, respectively. For further details on the implementation of this model, the reader is referred to section III.D of~\cite{Carroll2026_diss1}.

For He, the lowest-lying electronically excited state and the ionized state have energies of 19.8 eV and 24.6 eV, respectively. At the post-shock temperatures relevant for the present work, these energies are sufficiently high such that the electronic excitation and ionization of He can be neglected. Therefore, only the translational partition function is considered when computing thermodynamic properties for He. Similarly for $\rm H^+$ and $\rm e^-$, there are no internal energy modes to consider, so thermodynamic properties are computed directly from the translational partition functions.

Finally, thermodynamic properties of H($n$=1-7) are evaluated on a state-specific basis. The internal partition function for each individual H($n$) state is computed as
\begin{equation}
    Q_{{\rm H}(n)}(T) = g_{{\rm H}(n)}\exp\left(-\frac{E_{{\rm H}(n)}}{k_{\rm B}T}\right).
\end{equation}
The corresponding thermodynamic properties are computed from the standard statistical mechanics expressions~\cite{vincenti1965}.

\subsubsection{Chemical Kinetics}
\label{sec:kinetics}

Table~\ref{tab:kinetics} summarizes the reactions and corresponding sources of rate constants for the present kinetic model. Consistent with the thermodynamics, the formulation of Carroll et al.~\cite{Carroll2026_diss1} is used for modeling the dissociation of $\rm H_2$. In this formulation, the chemical source term for $\rm H_2$ dissociation with the third-body M is given by
\begin{equation}
    \frac{{\rm d}N_{\rm H}}{{\rm d}t} = 2N_{\rm M}N_{\rm H_2}k_{\rm d,pre-QSS} -2N_{\rm M}N_{\rm H}^2k_{\rm d,pre-QSS}K_{\rm eq}^{-1},
\end{equation}
where
\begin{equation}
    k_{\rm d,pre-QSS} = k_{\rm d,nr}\left[1 - \left(1-\frac{k_{\rm d,0}}{k_{\rm d,nr}}\right) \exp\left( -\sqrt{-\frac{\ln(1-\alpha)}{\eta}} \right) \right].
    \label{eqn:preQSSkd}
\end{equation}
Here, $\eta(T_{\rm t,h})$ is the same pre-QSS correction factor as from Eq.~\eqref{eqn:erv_preQSS}, and $k_{\rm d,nr}(T_{\rm t,h})$ is the macroscopic dissociation rate constant in the QSS regime when recombination is neglected. Fits of $k_{\rm d,nr}(T_{\rm t,h})$ were evaluated by Carroll et al.~\cite{Carroll2026_diss2} for M = $\rm H_2$, H, and He by reviewing data from a wide range of experimental and computational sources. These fits are used throughout the present work.

\begin{table}[hbt!]
\centering
\caption{Reactions Included in the Kinetic Model}
\begin{tabular}{lccc}
\hline \hline
Type         & Reaction                                                     & Third-bodies, M             & Source                            \\\hline
Dissociation & $\rm H_2$ + M $\leftrightarrow$ 2H($n$=1) + M                & $\rm H_2$, H, He            & Carroll et al.~\cite{Carroll2026_diss1,Carroll2026_diss2} \\
Excitation   & H($n$) + $\rm e^-$ $\leftrightarrow$ H($n'$) + $\rm e^-$     &                             & Przybilla and Butler~\cite{Przybilla2004} \\
             & H($n$) + M $\leftrightarrow$ H($n'$) + M                     & $\rm H_2$, H, He, $\rm H^+$ & Present work for $n\leq3$ \\
             &                                                              &                             & Mihajlov et al.~\cite{Mihajlov2004,Dimitrijevic2021} for $n>3$ \\
Ionization   & H($n$) + $\rm e^-$ $\leftrightarrow$ $\rm H^+$ + 2$\rm e^-$  &                             & Griffin et al.~\cite{Griffin2005} for $n\leq4$ \\
             &                                                              &                             & Present work for $n>4$ \\
             & H($n$) + M $\leftrightarrow$ $\rm H^+$ + $\rm e^-$ + M       & $\rm H_2$, H, He, $\rm H^+$ & Present work for $n\leq4$ \\
             &                                                              &                             & Mihajlov et al.~\cite{Mihajlov2011,Sreckovic2018} for $n>4$ \\
\hline \hline
\end{tabular}
\label{tab:kinetics}
\end{table}

The proposed rate constants from section~\ref{sec:revH} are used for the electron and heavy-particle-impact excitation and ionization reactions of H($n$). These rate constants have been fit to the modified-Arrhenius functional form and are reported in the supplementary material. The associated reverse reaction rate constants are computed via micro-reversibility~\cite{Oxenius1986} as
\begin{equation}
    \frac{k_{\rm M=e^-}(n\rightarrow n')}{k_{\rm M=e^-}(n'\rightarrow n)} = \frac{Q_{{\rm H}(n')}(T_{\rm t,e^-})}{Q_{{\rm H}(n)}(T_{\rm t,e^-})} = \frac{g_{{\rm H}(n')}}{g_{{\rm H}(n)}} \exp\left(-\frac{E_{n\rightarrow n'}}{k_{\rm B}T_{\rm t,e^-}} \right),
    \label{eqn:micro_e}
\end{equation}
\begin{equation}
    \frac{k_{\rm M\neq e^-}(n\rightarrow n')}{k_{\rm M\neq e^-}(n'\rightarrow n)} = \frac{Q_{{\rm H}(n')}(T_{\rm t,h})}{Q_{{\rm H}(n)}(T_{\rm t,h})} = \frac{g_{{\rm H}(n')}}{g_{{\rm H}(n)}} \exp\left(-\frac{E_{n\rightarrow n'}}{k_{\rm B}T_{\rm t,h}} \right),
    \label{eqn:micro_h}
\end{equation}
\begin{equation}
\begin{split}
    \frac{k_{\rm M=e^-}(n\rightarrow \infty)}{k_{\rm M=e^-}(\infty\rightarrow n)} & = 
    \frac{Q_{\rm t,e^-}(T_{\rm t,e^-})g_{\rm e^-}Q_{\rm t,H^+}(T_{\rm t,e^-})g_{\rm H^+}}{Q_{{\rm t,H}(n)}(T_{\rm t,e^-})Q_{{\rm H}(n)}(T_{\rm t,e^-})}\exp\left(-\frac{E_{\rm H^+}}{k_{\rm B}T_{\rm t,e^-}}\right) \\
    & = \frac{g_{\rm e^-}g_{\rm H^+}}{g_{{\rm H}(n)}}\left(\frac{2\pi m_{\rm e^-}k_{\rm B}T_{\rm t,e^-}}{h_{\rm P}^2}\right)^{3/2}\exp\left(-\frac{E_{n\rightarrow \infty}}{k_{\rm B}T_{\rm t,e^-}} \right),
\end{split}
\label{eqn:saha_e}
\end{equation}
and
\begin{equation}
\begin{split}
    \frac{k_{\rm M\neq e^-}(n\rightarrow \infty)}{k_{\rm M\neq e^-}(\infty\rightarrow n)} & =
    \frac{Q_{\rm t,e^-}(T_{\rm t,e^-})g_{\rm e^-}Q_{\rm t,H^+}(T_{\rm t,h})g_{\rm H^+}}{Q_{{\rm t,H}(n)}(T_{\rm t,h})Q_{{\rm H}(n)}(T_{\rm t,h})}\exp\left(-\frac{E_{\rm H^+}}{k_{\rm B}T_{\rm t,h}}\right) \\
    & = \frac{g_{\rm e^-}g_{\rm H^+}}{g_{{\rm H}(n)}}\left(\frac{2\pi m_{\rm e^-}k_{\rm B}T_{\rm t,e^-}}{h_{\rm P}^2}\right)^{3/2}\exp\left(-\frac{E_{n\rightarrow \infty}}{k_{\rm B}T_{\rm t,h}} \right),
\end{split}
\label{eqn:saha_h}
\end{equation}
where $g_{\rm e^-}=2$, $g_{\rm H^+}=1$, and $Q_{\rm t,s} = (2\pi m_{\rm s} k_{\rm B} T_{\rm t}/h_{\rm P}^2)^{3/2}$. In the second line of Eq.~\eqref{eqn:saha_e} and \eqref{eqn:saha_h}, it has been assumed that $m_{\rm H}\approx m_{\rm H^+}$ such that $Q_{{\rm t,H}(n)}\approx Q_{\rm t,H^+}$. Unlike Eq.~\eqref{eqn:micro_e}-\eqref{eqn:saha_e}, Eq.~\eqref{eqn:saha_h} is unique in that the reverse reaction rate constant is controlled by both $T_{\rm t,e^-}$ and $T_{\rm t,h}$~\cite{Annaloro2017}.

\subsection{Radiance Calculations}
\label{sec:neqair}

Radiance calculations are performed using version 15.3 of the radiation code, NEQAIR~\cite{Cruden2014Neqair,Brandis2019Neqair}. Radiance is computed for the 117 to 173 nm, 323 to 497 nm, and 653 to 659 nm wavelength ranges, corresponding to the vacuum-ultraviolet (VUV), blue, and red spectral ranges respectively from the experiments of Cruden and Bogdanoff~\cite{Cruden2017}. The spectral features of interest are summarized in Table \ref{tab:spec}. While measurements were also obtained in the near-infrared (NIR) range for a select number of shots, these are not discussed in the present work, as the NIR contribution to the total radiance is negligible compared to the VUV, blue, and red contributions.

\begin{table}[hbt!]
\centering
\caption{Spectral Features for Radiance Calculations}
\begin{tabular}{lccccc}
\hline \hline
        & Spectral range [nm] & Species   & Spectral feature      & Transition                     & Wavelength [nm] \\\hline
VUV     & 117 to 173          & H         & Lyman $\rm \alpha$    & $n$ = 2 $\rightarrow$ 1        & 121.57          \\
        &                     & $\rm H_2$ & Lyman band            & $\rm B^1 \Sigma - X^1 \Sigma$  & 100 to 160      \\
Blue    & 323 to 497          & H         & Balmer $\rm \beta$    & $n$ = 4 $\rightarrow$ 2        & 486.14          \\
        &                     &           & Balmer $\rm \gamma$   & $n$ = 5 $\rightarrow$ 2        & 434.05          \\
        &                     &           & Balmer $\rm \delta$   & $n$ = 6 $\rightarrow$ 2        & 410.17          \\
        &                     &           & Balmer $\rm \epsilon$ & $n$ = 7 $\rightarrow$ 2        & 397.00          \\
        &                     &           & Balmer continuum      & $n$ = $\infty \rightarrow$ 2   & <364.6          \\
Red     & 653 to 659          & H         & Balmer $\rm \alpha$   & $n$ = 3 $\rightarrow$ 2        & 656.28          \\
\hline \hline
\end{tabular}
\label{tab:spec}
\end{table}

The number densities of the explicitly treated H($n$) states are passed directly to NEQAIR, while the internal energy states of $\rm H_2$ are assumed to follow Boltzmann distributions about separate rovibrational and electronic temperatures. As discussed in section~\ref{sec:thermochem}, the evolution of $e_{\rm rv,H_2}$ is tracked explicitly in the dissociation formulation of Carroll et al.~\cite{Carroll2026_diss1}. Therefore, an energy-equivalent rovibrational temperature, $T_{\rm rv,H_2}$, is computed by inverting the relation
\begin{equation}
    e_{\rm rv,H_2} = \frac{k_{\rm B}}{m_{\rm H_2}} \frac{\sum_{\nu=0}^{\nu_{\rm max}}\sum_{J=0}^{J_{\rm max}(\nu)} \theta_{\rm rv,H_2\it(J,\nu)} Q_{\rm H_2\it(J,\nu)}(T_{\rm rv,H_2})}{\sum_{\nu=0}^{\nu_{\rm max}}\sum_{J=0}^{J_{\rm max}(\nu)} Q_{\rm H_2\it(J,\nu)}(T_{\rm rv,H_2})}. 
    \label{eqn:erv}
\end{equation}
Here, $Q_{{\rm H_2}(J,\nu)}(T_{\rm rv,H_2}) = g_J (2J+1) \exp( -\theta_{\rm rv, H_2 \it(J,\nu)} / T_{\rm rv,H_2} )$, where $g_J=1/4$ for all even $J$ (para-hydrogen) and $g_J=3/4$ for all odd $J$ (ortho-hydrogen)~\cite{Colonna2012,Popovas2016}. The electronic temperature of $\rm H_2$ is assumed to be equal to $T_{\rm t,e^-}$.

The computed signals are convolved with instrument line shapes (ILS) in wavelength space and spatial resolution functions (SRF) in physical space to capture the effects of camera resolution and shock motion from the experiments~\cite{Cruden2014AbsoluteRM}. The ILSs are parametrized as a linearly weighted average of Gaussian and Lorentzian functions. The SRFs are composed of three functions: a triangular optical function, a square root of a Voigt camera function, and a square temporal function. The parameters used to specify all of these functions per camera and shot condition are given in the supplemental test data of~\cite{Cruden2017} (test campaign number 56).

\subsection{Shot Conditions}
\label{sec:shots}

\subsubsection{Freestream Conditions and Shock Locations}
\label{sec:freestream}

The simulations performed in the present work correspond to various shot conditions from the experiments of Cruden and Bogdanoff~\cite{Cruden2017}. The freestream pressures and velocities for the considered shots are summarized in Table~\ref{tab:shots}. The freestream composition and temperature are taken to be 89\% $\rm H_2$/ 11\% He (by mole) and 300 K, respectively, for all shots.

\begin{table}[hbt!]
\centering
\caption{Shot Conditions}
\begin{threeparttable}
\begin{tabular}{lccccc}
\hline \hline
Shot no. & $P_0$ [torr] & $u_0$ [km/s] & Shift in shock location [cm] & $L$ [cm] & Contaminant length [cm] \\\hline
7        & 0.50         & 27.34        & + 0.5\hspace{1.5mm}          & 22.6     & 6.90                    \\
11       & 0.10         & 28.68        & + 0.9\hspace{1.5mm}          & 4.29     & 3.46                    \\
17       & 0.10         & 27.40        & + 1.1\hspace{1.5mm}          & 4.50     & 3.78                    \\
20       & 0.20         & 26.34        & + 1.2\hspace{1.5mm}          & 9.38     & 6.13                    \\
22       & 0.51         & 27.66        & + 1.0\hspace{1.5mm}          & 22.7     & 7.71                    \\
23       & 0.51         & 27.63        & + 0.6\footnotemark[1]        & 22.7     & 8.66                    \\
25       & 0.20         & 27.80        & + 1.2\hspace{1.5mm}          & 8.86     & 6.57                    \\
\hline \hline
\end{tabular}
\begin{tablenotes}
\footnotesize
    \item[*] This shift was used for the VUV and Red cameras. The Blue camera was instead shifted by + 1.1 cm.
\end{tablenotes}
\end{threeparttable}
\label{tab:shots}
\end{table}

To compare between experimental and simulated profiles, the shock location is consistently defined as the intersection point between the freestream pressure and the tangent line to the maximum gradient in the pressure profile. For all shock profiles presented, the shock location is set at $x$ = 0 cm, such that all negative values of $x$ correspond to the pre-shock region, and all positive values of $x$ correspond to the post-shock region.

\begin{figure}[hbt!]
     \centering
     \begin{subfigure}[b]{0.4\textwidth}
         \centering
         \includegraphics[width=\textwidth]{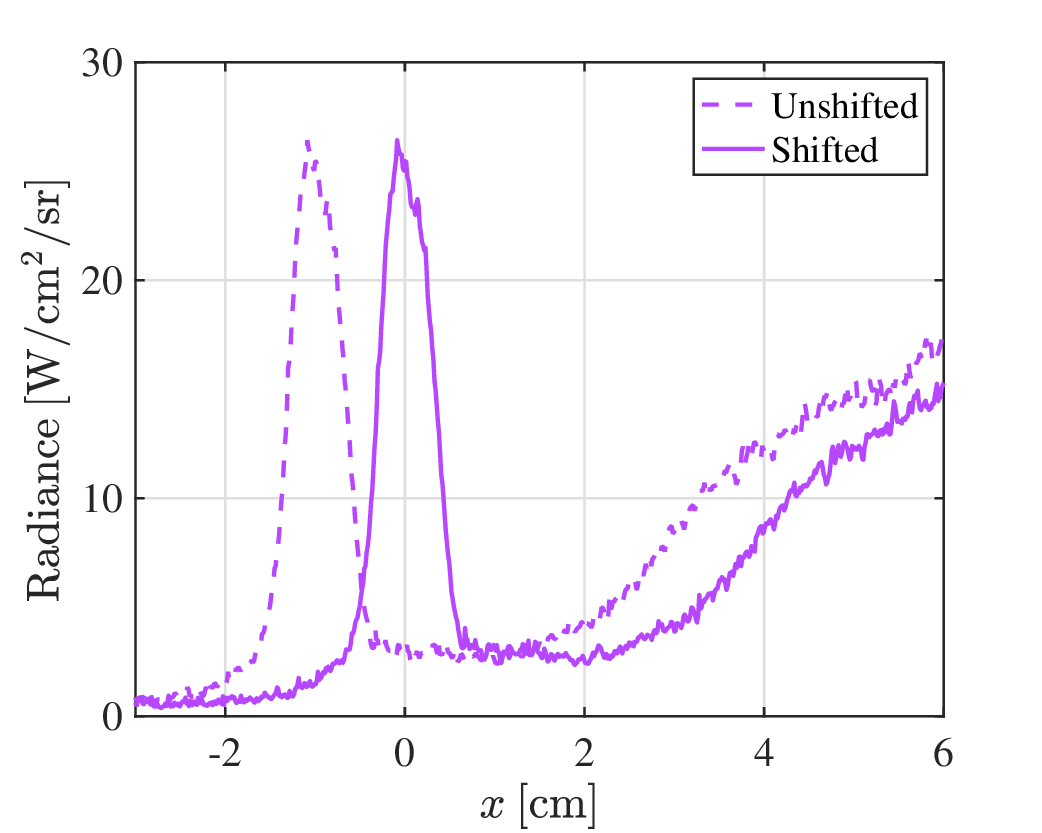}
     \end{subfigure}
        \caption{Unshifted and shifted integrated radiance profiles for the shot 22 VUV camera (117 to 173 nm).}
        \label{fig:eastshift}
\end{figure}

Despite this consistent definition, there is still an uncertainty associated with the shock location in the experimental profiles due to potential time lags in the pressure transducers. At lower speeds, the uncertainty introduced by this delay time would be negligible. However, at shock speeds in excess of 26 km/s as investigated here, a delay time of 0.5 $\mu$s~\cite{Cruden2024} can lead to a discrepancy in the shock location by up to 26 km/s $\times$ 0.5 $\mu$s = 1.3 cm. Fortunately, an estimate of the shift in the shock locations can be obtained from the measured radiance profiles. In particular, for shots 17, 20, 22, and 23 a distinct peak was captured in the VUV profiles at the shock front (due to the $\rm H_2$ Lyman band feature) as illustrated in Fig.~\ref{fig:eastshift}. The required shift in the shock location is estimated as the shift required to place this feature at $x$ = 0 cm. For shots 7, 11, and 25 for which the onset of this feature was not captured, the same approach is used but based on the Balmer $\beta$ feature. For shot 23 alone, the shock location as estimated from the Lyman band and Balmer $\beta$ feature is inconsistent, so a separate shift is used for the VUV/ Red and Blue cameras. The estimated shifts in shock locations are summarized in the fourth column of Table~\ref{tab:shots}. All of the experimental results to be presented in section~\ref{sec:results} are shifted from the originally reported shock locations by these values.

\subsubsection{Test Slug Lengths}
\label{sec:Lmirels}

As discussed in section~\ref{sec:Q1D}, it is necessary to specify test slug lengths to solve the governing equations Eq.~\eqref{eqn:species_march}-\eqref{eqn:electron_march}. An estimate of these values can be computed using the relation derived by Mirels~\cite{Mirels1963}, namely
\begin{equation}
    L = \frac{d^2}{16\beta^2}\left(\frac{\rho_{\rm f}}{\rho_{\rm w}}\right)^2\frac{u_{\rm f}}{u_{\rm w}-u_{\rm f}}\frac{\rho_{\rm w}u_{\rm f}}{\mu_{\rm w}},
    \label{eqn:L}
\end{equation}
where $d$ = 10.16 cm is the driven section diameter of EAST, the subscript w denotes wall properties, and the subscript f denotes the post-shock state, taken here to be the post-shock frozen state. $\beta$ is computed using Mirels's proposed correlation for strong non-ideal shocks\footnote{Equation~\eqref{eqn:beta} was proposed by Mirels for the case with a Prandtl number of 0.72. It is assumed that this correlation is also appropriate for the $\rm H_2$/ He shocks considered here, as the frozen Prandtl number is estimated to be approximately 0.68 for all shot conditions (based on the collision integral fits of Palmer et al.~\cite{Palmer2014}).},
\begin{equation}
    \beta = 1.59\left(\frac{\rho_{\rm f}\mu_{\rm f}}{\rho_{\rm w}\mu_{\rm w}}\right)^{0.37}\left(1+\frac{1.796+0.802(u_{\rm w}/u_{\rm f})}{(u_{\rm w}/u_{\rm f})^2-1}\right),
    \label{eqn:beta}
\end{equation}
where $u_{\rm w}$ is the shock speed in the laboratory frame of reference. Viscosities are evaluated using the collision integral fits of Palmer et al.~\cite{Palmer2014}. The computed values of $L$ for each shot condition are summarized in the fifth column of Table~\ref{tab:shots}.

Practically, these estimates of $L$ need to satisfy the constraint imposed by measured contamination lengths from the experiments. Namely, since the driver gas in the experiments was composed of pure He, the onset of contamination should either precede or coincide with the contact surface. Therefore, the reported contamination lengths give lower bound values for $L$. The contamination lengths reported by Cruden and Bogdanoff~\cite{Cruden2017} (with the shock location shifts accounted for) are given in the final column of Table~\ref{tab:shots}. As expected, the estimated test slug lengths are greater than the reported contamination lengths for all shot conditions. The impact of uncertainties in the test slug lengths on the computed radiance profiles will be discussed in section~\ref{sec:sens}.

\section{Results}
\label{sec:results}

The general features of the flowfield solutions are first discussed in section~\ref{sec:flowfield}. Then, the computed radiance profiles are compared to the EAST measurements of Cruden and Bogdanoff~\cite{Cruden2017} in section~\ref{sec:rad}, and a corresponding sensitivity analysis for the kinetic rates and boundary layer treatment is presented in section~\ref{sec:sens}. Finally, comparisons are made to the alternate kinetic models of Colonna et al.~\cite{Colonna2020} and Liu et al.~\cite{Liu2020} in section~\ref{sec:altmodels}.

\subsection{Flowfield Solution}
\label{sec:flowfield}

\begin{figure}[hbt!]
     \centering
     \begin{subfigure}[b]{0.33\textwidth}
         \centering
         \includegraphics[width=\textwidth,trim={0cm 0cm 1.2cm 0cm},clip]{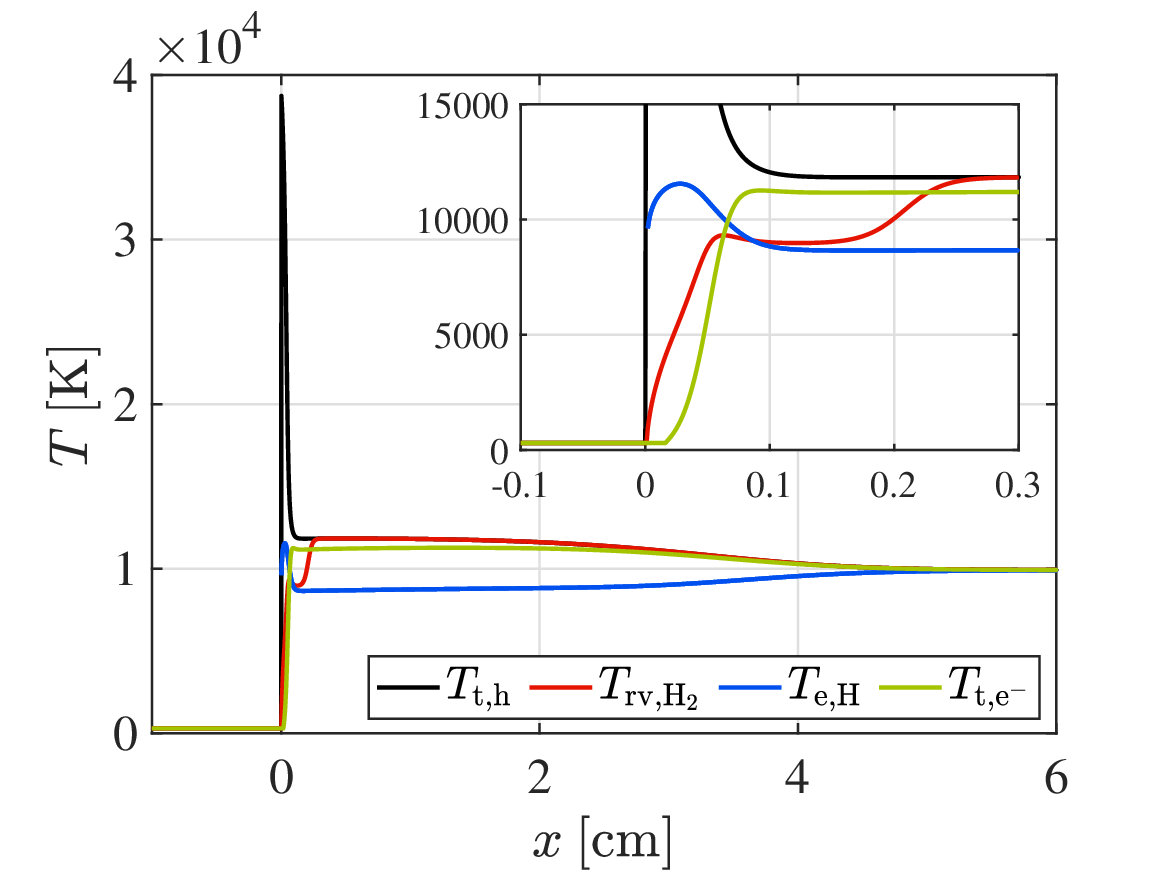}
     \end{subfigure}
     \hfill
     \begin{subfigure}[b]{0.33\textwidth}
         \centering
         \includegraphics[width=\textwidth,trim={0cm 0cm 1.2cm 0cm},clip]{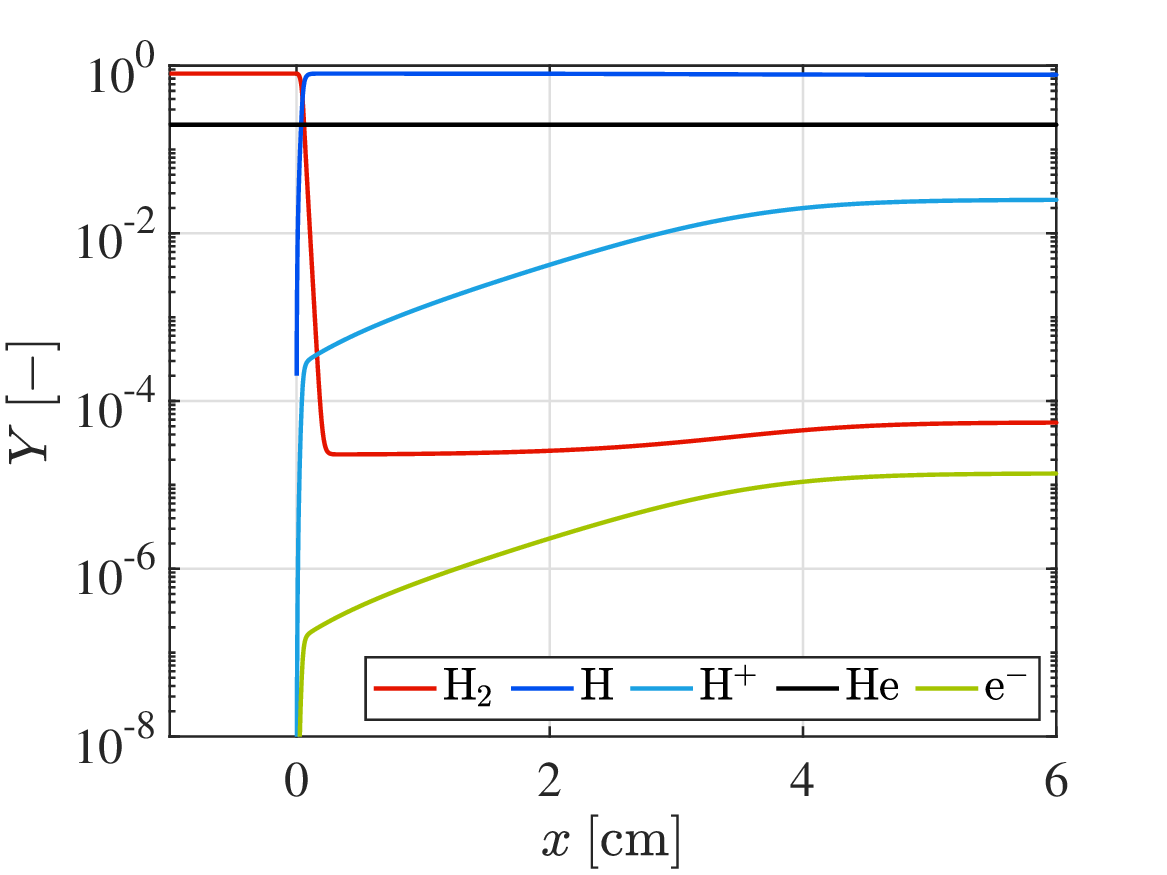}
     \end{subfigure}
     \hfill
     \begin{subfigure}[b]{0.33\textwidth}
         \centering
         \includegraphics[width=\textwidth,trim={0cm 0cm 1.2cm 0cm},clip]{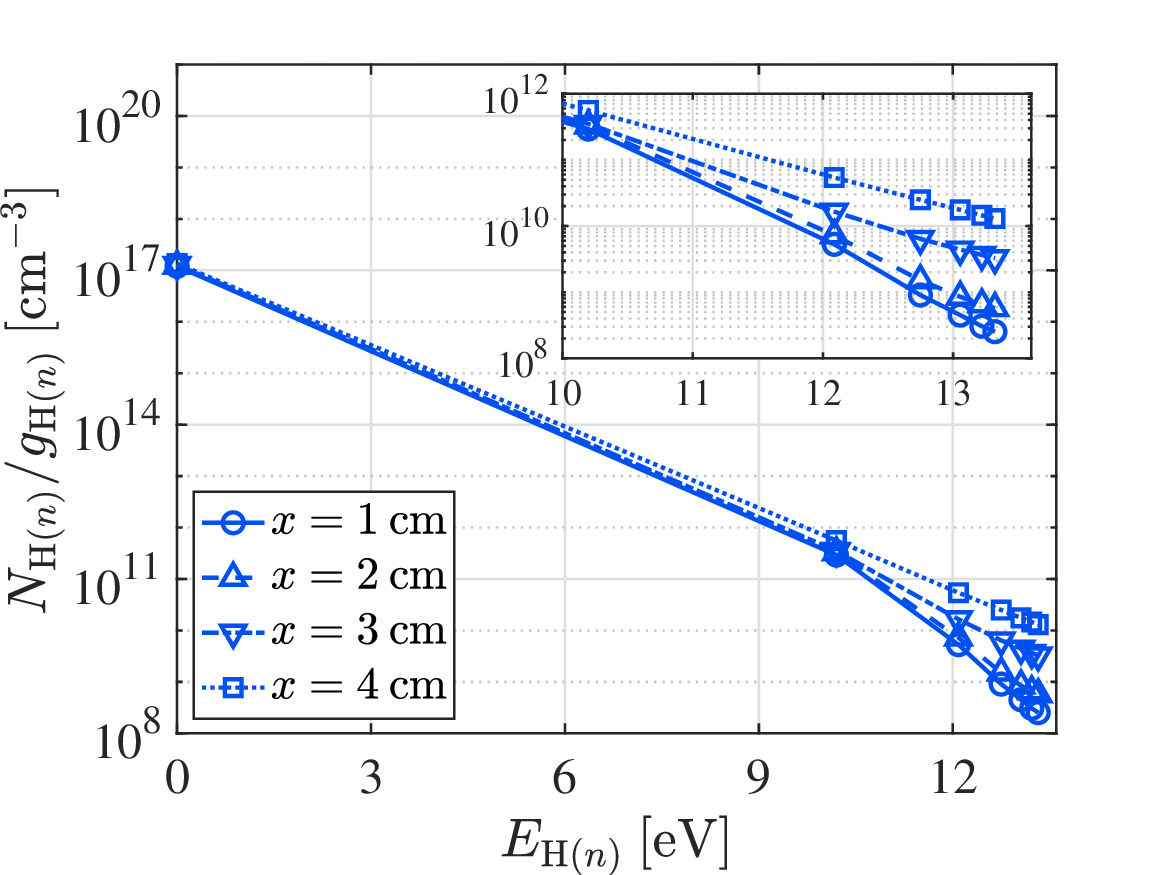}
     \end{subfigure}
    \caption{Simulated temperatures (left), mass fractions (middle), and H($\bm n$) distributions (right) for shot 23 ($\bm{P_0}$ = 0.51 torr and $\bm{u_0}$ = 27.63 km/s).}
    \label{fig:flowfield}
\end{figure}

Figure~\ref{fig:flowfield} shows the simulated temperatures, mass fractions, and H({\it n}) distributions (Boltzmann plots) for shot 23 from the EAST experiments. In the temperature plot, $T_{\rm rv,H_2}$ is computed by inverting Eq.~\eqref{eqn:erv}, and $T_{\rm e,H}$ is computed by inverting the analogous equation for the electronic energy of H,
\begin{equation}
    e_{\rm e,H} = m_{\rm H}^{-1}\frac{\sum_{n=1}^{n_{\rm max}} E_{\rm H\it(n)} N_{\rm H\it(n)}}{\sum_{n=1}^{n_{\rm max}} N_{\rm H\it(n)}} = m_{\rm H}^{-1} \frac{\sum_{n=1}^{n_{\rm max}} E_{\rm H\it(n)} Q_{\rm H\it(n)}(T_{\rm e,H})}{\sum_{n=1}^{n_{\rm max}} Q_{\rm H\it(n)}(T_{\rm e,H})}.
    \label{eqn:eeH}
\end{equation}

In the post-shock region, the rovibrational energy of $\rm H_2$, the electronic energy of H, and the electron translational energy all begin in a state of non-equilibrium with the heavy-gas translational mode. Then, each of these modes follow the same general steps to reach equilibrium; namely, an initial rise from the pre-shock condition, followed by a QSS region, and finally a relaxation to equilibrium with the heavy-gas translational mode. These QSS conditions arise from the competition between the relevant production and consumption processes for each energy mode. For the rovibrational energy of $\rm H_2$ and the electronic energy of H, these are the excitation and dissociation/ ionization of internal energy states; for the electron translational mode, these are the elastic and inelastic collision processes.

For the rovibrational energy of $\rm H_2$, the QSS plateau (characterized by an approximate 3,000 K difference between $T_{\rm t,h}$ and $T_{\rm rv,H_2}$) is reached around $x\approx 0.05$ cm, and equilibrium with the heavy-gas translational mode is reached by $x\approx$ 0.25 cm. For the electronic mode of H, $T_{\rm e,H}$ establishes a QSS condition by $x\approx 0.1$ cm, but does not equilibrate until further downstream at $x\approx$ 4 to 5 cm. Similarly, for the electron translational mode, the QSS plateau (with an approximate 500 K difference between $T_{\rm t,h}$ and $T_{\rm t,e^-}$) is reached around $x\approx 0.1$ cm, but equilibrium with the heavy-gas translational mode is not reached until $x\approx$ 4 to 5 cm, when H, $\rm H^+$, and $\rm e^-$ have reached the Saha equilibrium condition. Together, these results suggest that while $\rm H_2$ rovibrational relaxation and dissociation occur rapidly in the post-shock flow, H electronic relaxation and ionization occur over a significantly longer timescale.

These trends are also reflected in the mass fraction and H({\it n}) Boltzmann plots. Namely, the mass fraction plot shows the rapid dissociation of $\rm H_2$, in contrast to the relatively slow ionization of H and production of $\rm H^+$ and $\rm e^-$. Consistent with this slow ionization trend, the Boltzmann plots show non-Boltzmann H($\it n$) distributions (characterized by depleted tails for the {\it n} > 2 states), which do not relax to a Boltzmann distribution until $x\approx$ 4 cm. Similar trends of the rapid dissociation of $\rm H_2$ within $x\lesssim$ 0.5 cm of the shock front and the slow ionization of H over several centimeters were observed for all shot conditions.

\subsection{Radiation}
\label{sec:rad}

\subsubsection{Atomic H Features}
\label{sec:Hrad}

\begin{figure}[hbt!]
     \centering
     \hspace*{\fill}
     \begin{subfigure}[b]{0.4\textwidth}
         \centering
         \includegraphics[width=\textwidth,trim={0cm 0cm 1.2cm 0.5cm},clip]{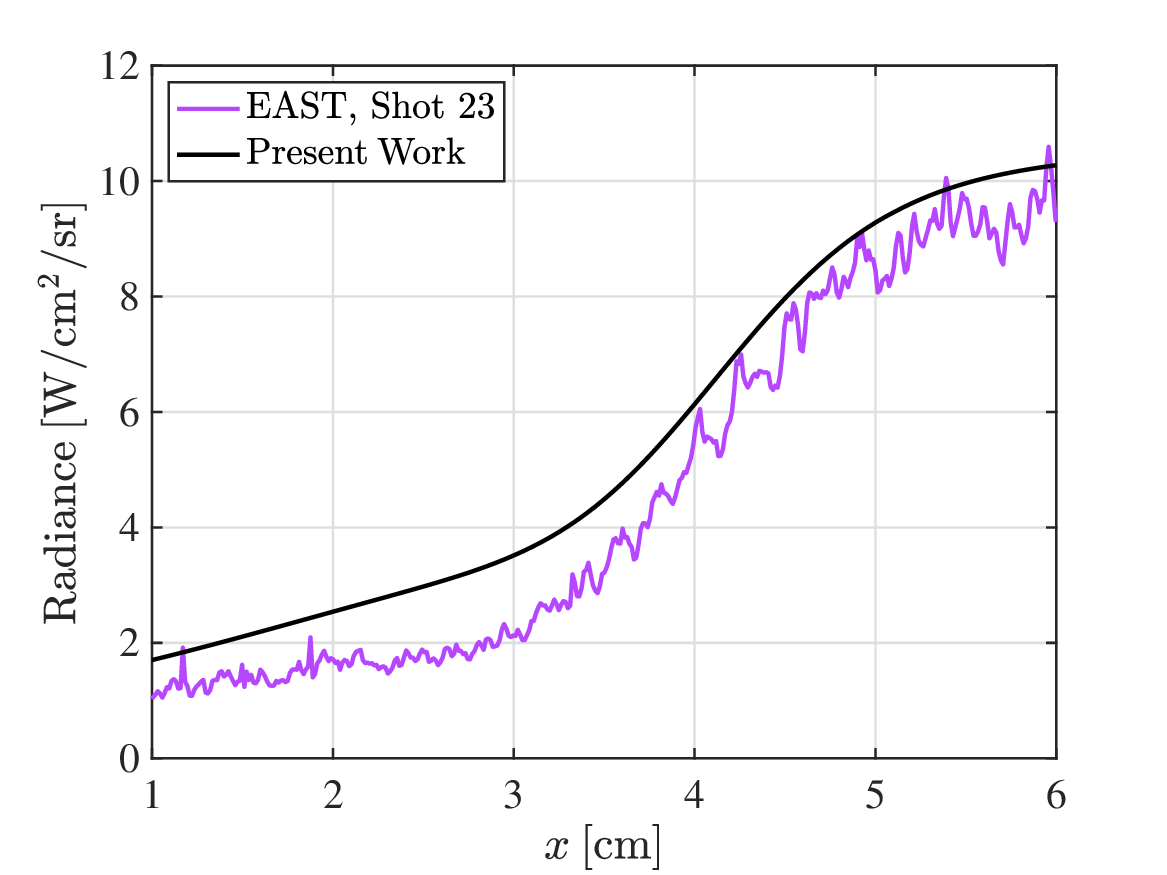}
     \end{subfigure}
     \hfill
     \begin{subfigure}[b]{0.4\textwidth}
         \centering
         \includegraphics[width=\textwidth,trim={0cm 0cm 1.2cm 0.5cm},clip]{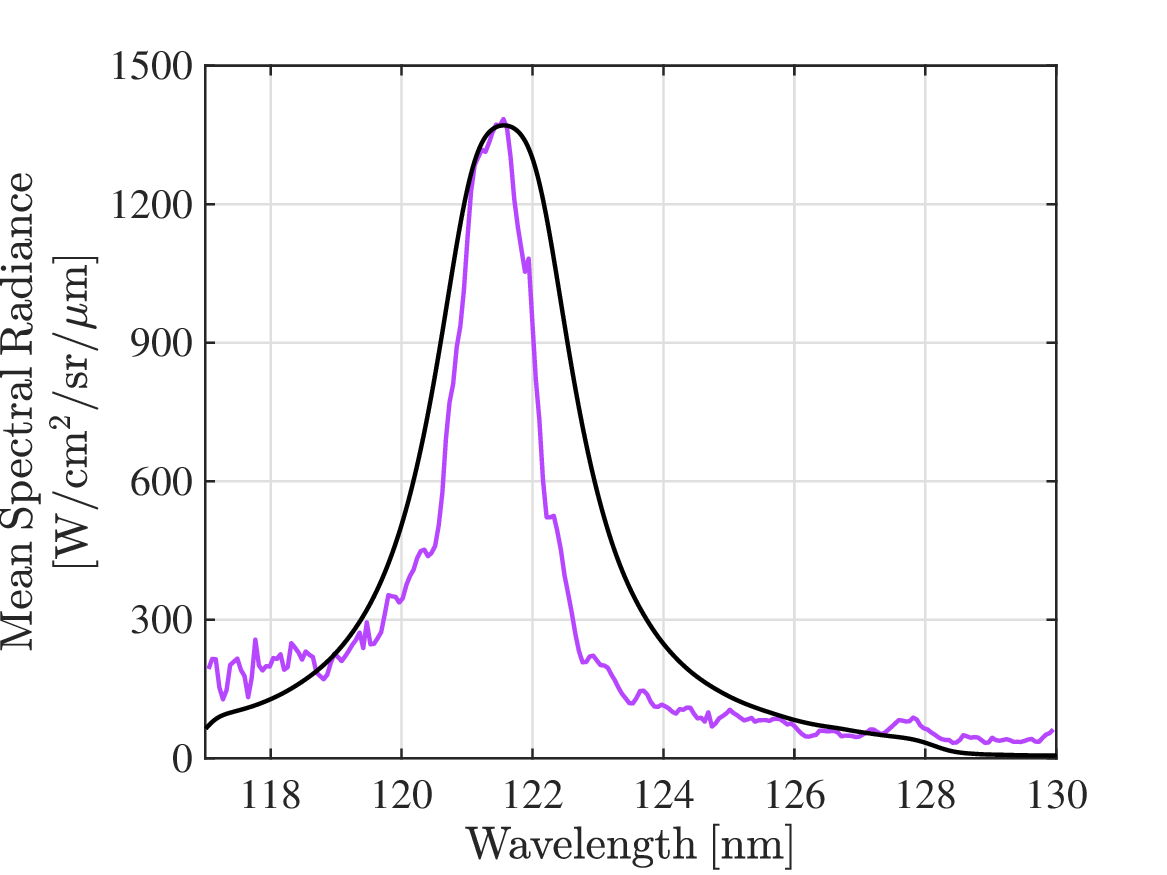}
     \end{subfigure}
     \hspace*{\fill}
    \caption{Integrated radiance over 117 to 130 nm (left) and mean spectral radiance over $\bm x$ = 3 to 4 cm (right) for the Lyman-$\bm \alpha$ feature.}
    \label{fig:S23_Lya}
\end{figure}

\begin{figure}[hbt!]
     \centering
     \hspace*{\fill}
     \begin{subfigure}[b]{0.4\textwidth}
         \centering
         \includegraphics[width=\textwidth,trim={0cm 0cm 1.2cm 0.5cm},clip]{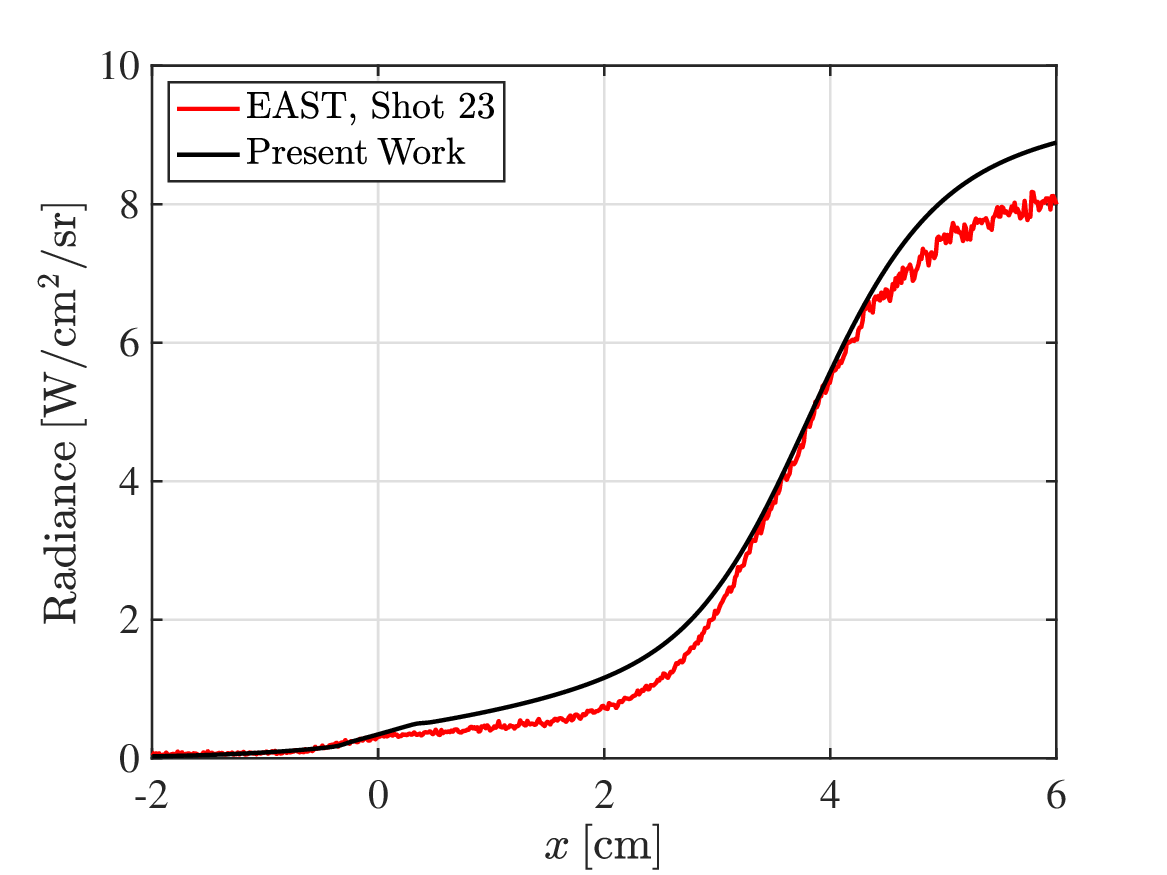}
     \end{subfigure}
     \hfill
     \begin{subfigure}[b]{0.4\textwidth}
         \centering
         \includegraphics[width=\textwidth,trim={0cm 0cm 1.2cm 0.5cm},clip]{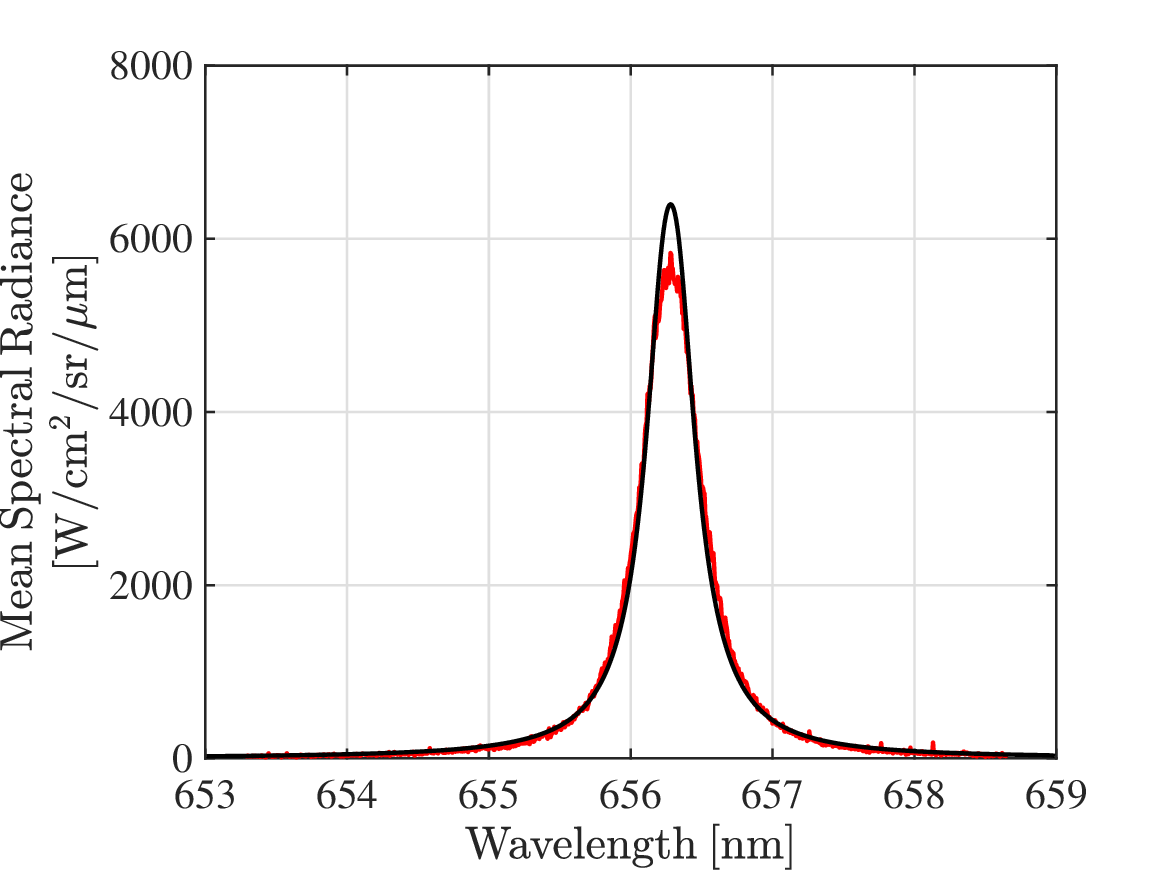}
     \end{subfigure}
     \hspace*{\fill}
    \caption{Integrated radiance over 653 to 659 nm (left) and mean spectral radiance over $\bm x$ = 3 to 4 cm (right) for the Balmer-$\bm \alpha$ feature.}
    \label{fig:S23_Baa}
\end{figure}

Figures~\ref{fig:S23_Lya} and \ref{fig:S23_Baa} show the radiance profiles for shot 23 for the dominant atomic H features, i.e., the Lyman-$\alpha$ ($n=2\rightarrow1$) and Balmer-$\alpha$ ($n=3\rightarrow2$) lines, respectively. The integrated radiance for the Lyman-$\alpha$ feature is only shown for $x\geq$ 1 cm, as there is an overlap with the $\rm H_2$ Lyman band for earlier post-shock locations. In general, both of these features are predicted accurately by the computational model. While there is a slight overprediction of the radiance in the induction zone from $x\approx$ 0 to 3 cm, the slope in the subsequent rise for the $x\gtrsim$ 3 cm region are captured well for both features. The spectral plot for the Balmer-$\alpha$ feature in particular shows an excellent agreement in both shape and magnitude in the post-induction zone.

\begin{figure}[hbt!]
     \centering
     \begin{subfigure}[b]{0.33\textwidth}
         \centering
         \includegraphics[width=\textwidth,trim={0cm 0cm 1.2cm 0.5cm},clip]{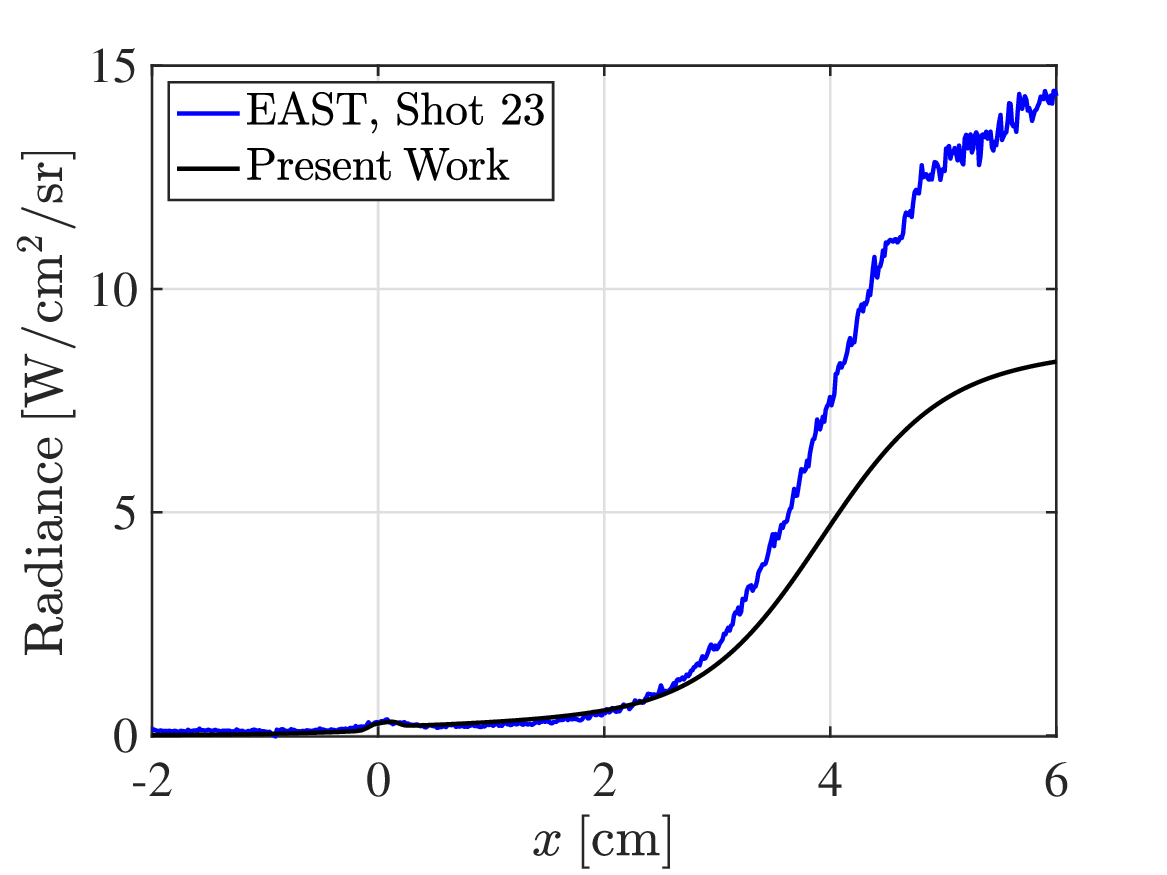}
     \end{subfigure}
     \hfill
     \begin{subfigure}[b]{0.33\textwidth}
         \centering
         \includegraphics[width=\textwidth,trim={0cm 0cm 1.2cm 0.5cm},clip]{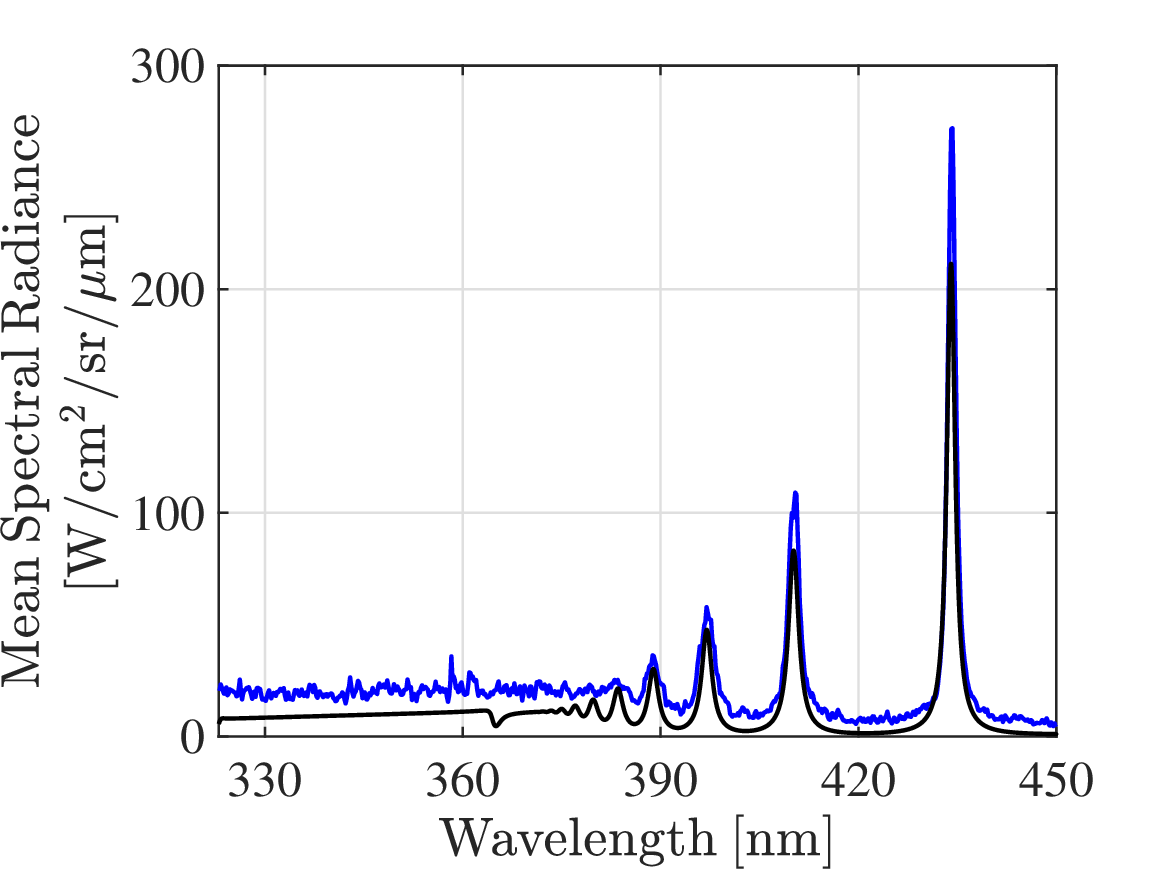}
     \end{subfigure}
     \hfill
     \begin{subfigure}[b]{0.33\textwidth}
         \centering
         \includegraphics[width=\textwidth,trim={0cm 0cm 1.2cm 0.5cm},clip]{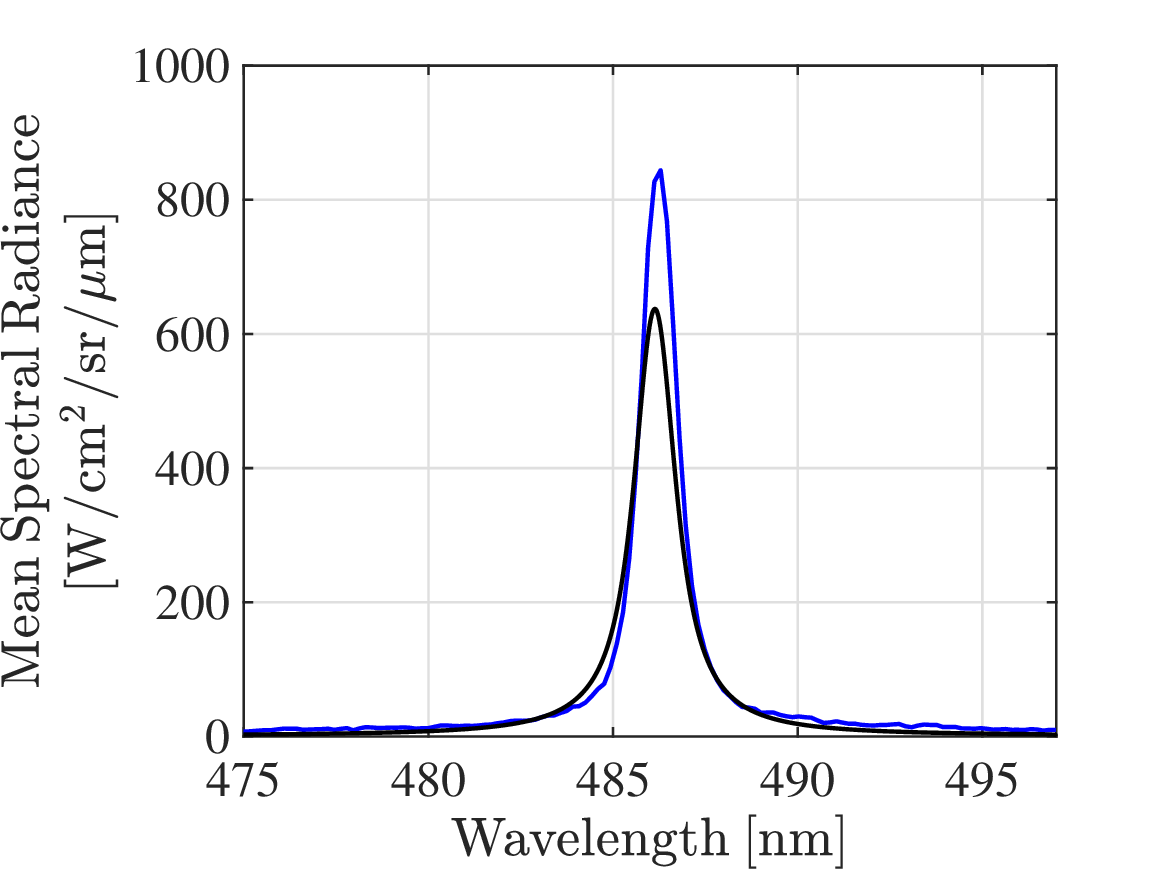}
     \end{subfigure}
    \caption{Integrated radiance over 323 to 497 nm (left) and mean spectral radiance over $\bm x$ = 3 to 4 cm for the Balmer-$\bm \gamma$ through continuum (middle) and Balmer-$\bm \beta$ (right) features.
    }
    \label{fig:S23_Bas}
\end{figure}

In contrast, the integrated radiance for the higher-order Balmer series features shown in Figure~\ref{fig:S23_Bas} is predicted accurately for $x\lesssim$ 2 cm, but is then underpredicted in the far post-shock region. Notably, the predicted radiance is approximately a factor of two lower than the experimental values at $x$ = 6 cm, where the gas is expected to be near or at equilibrium. The corresponding mean spectral radiance plots show that this is due to an underprediction of the magnitudes of all of the features in this spectral range.

\begin{figure}[hbt!]
     \centering
     \begin{subfigure}[b]{0.33\textwidth}
         \centering
         \includegraphics[width=\textwidth,trim={0cm 0cm 1.2cm 0.5cm},clip]{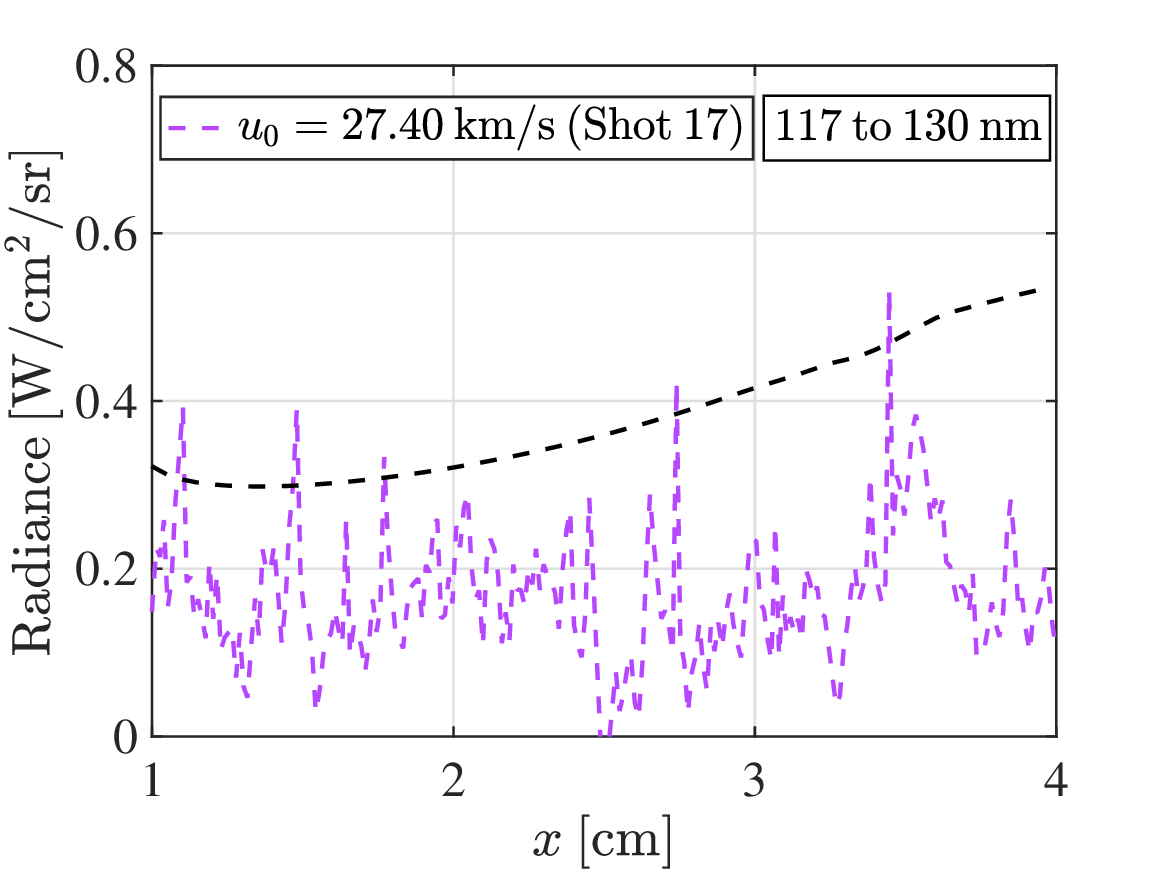}
     \end{subfigure}
     \hfill
     \begin{subfigure}[b]{0.33\textwidth}
         \centering
         \includegraphics[width=\textwidth,trim={0cm 0cm 1.2cm 0.5cm},clip]{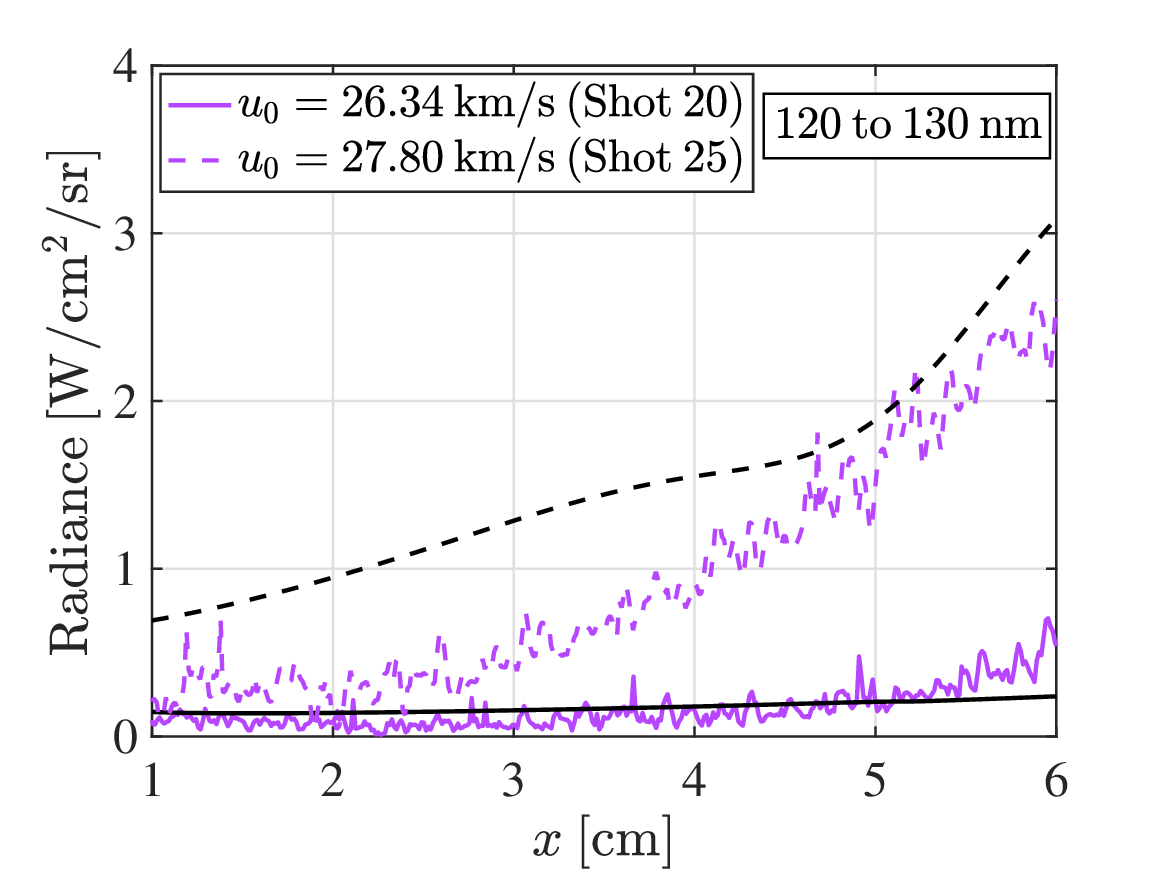}
     \end{subfigure}
     \hfill
     \begin{subfigure}[b]{0.33\textwidth}
         \centering
         \includegraphics[width=\textwidth,trim={0cm 0cm 1.2cm 0.5cm},clip]{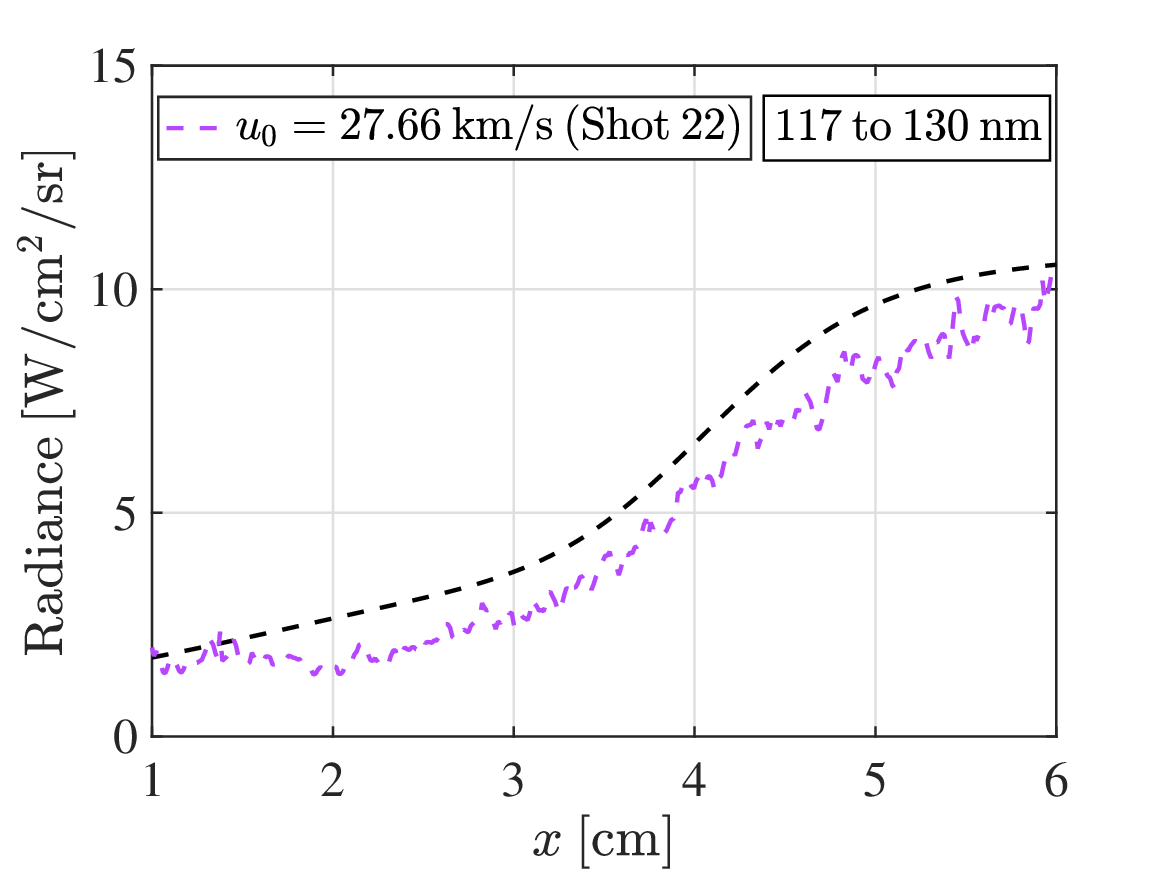}
     \end{subfigure}
     \begin{subfigure}[b]{0.33\textwidth}
         \centering
         \includegraphics[width=\textwidth,trim={0cm 0cm 1.2cm 0.5cm},clip]{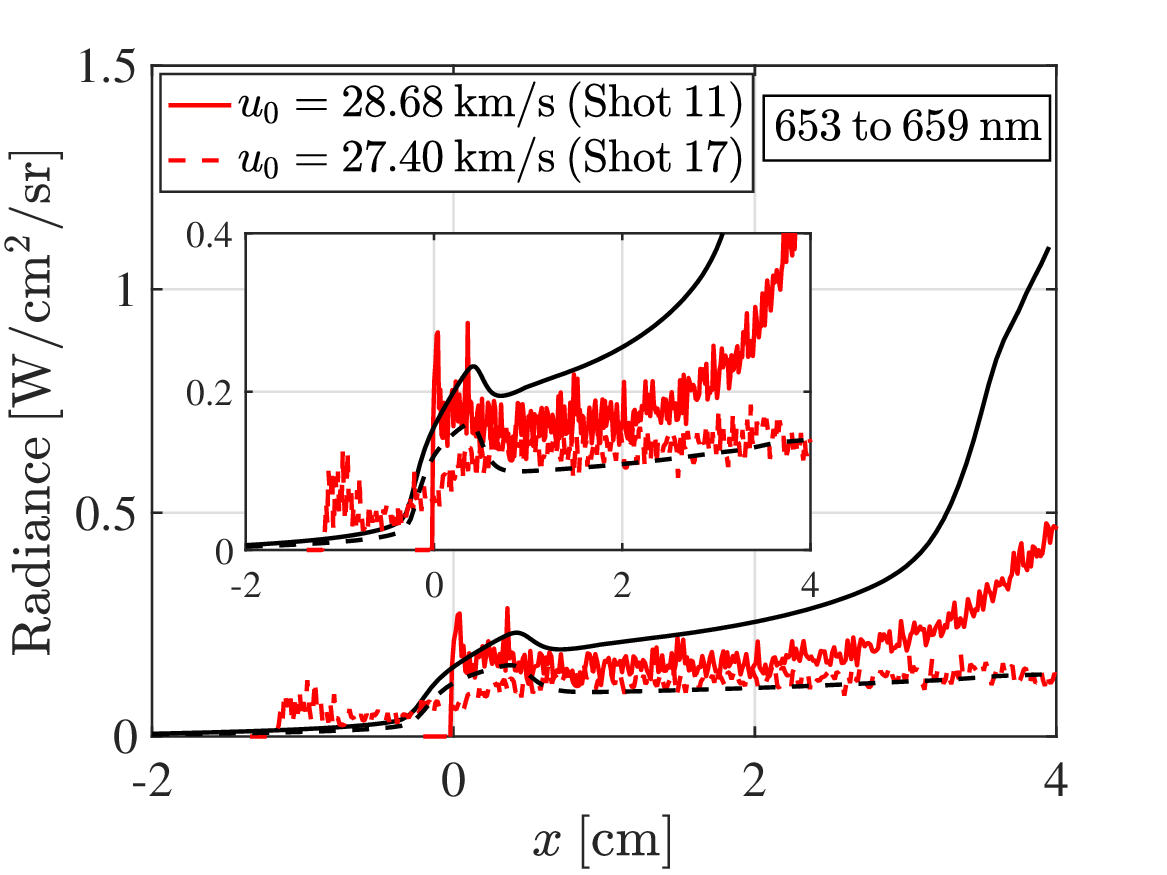}
     \end{subfigure}
     \hfill
     \begin{subfigure}[b]{0.33\textwidth}
         \centering
         \includegraphics[width=\textwidth,trim={0cm 0cm 1.2cm 0.5cm},clip]{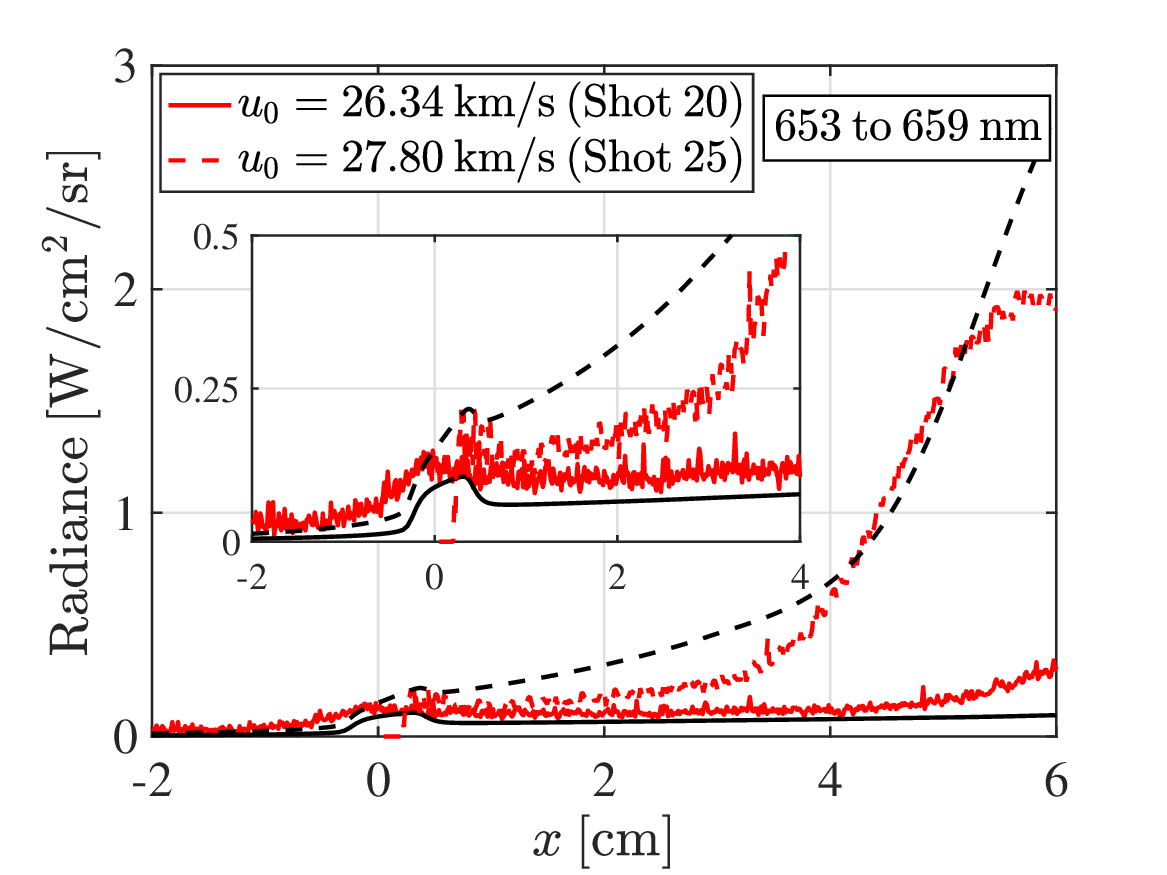}
     \end{subfigure}
     \hfill
     \begin{subfigure}[b]{0.33\textwidth}
         \centering
         \includegraphics[width=\textwidth,trim={0cm 0cm 1.2cm 0.5cm},clip]{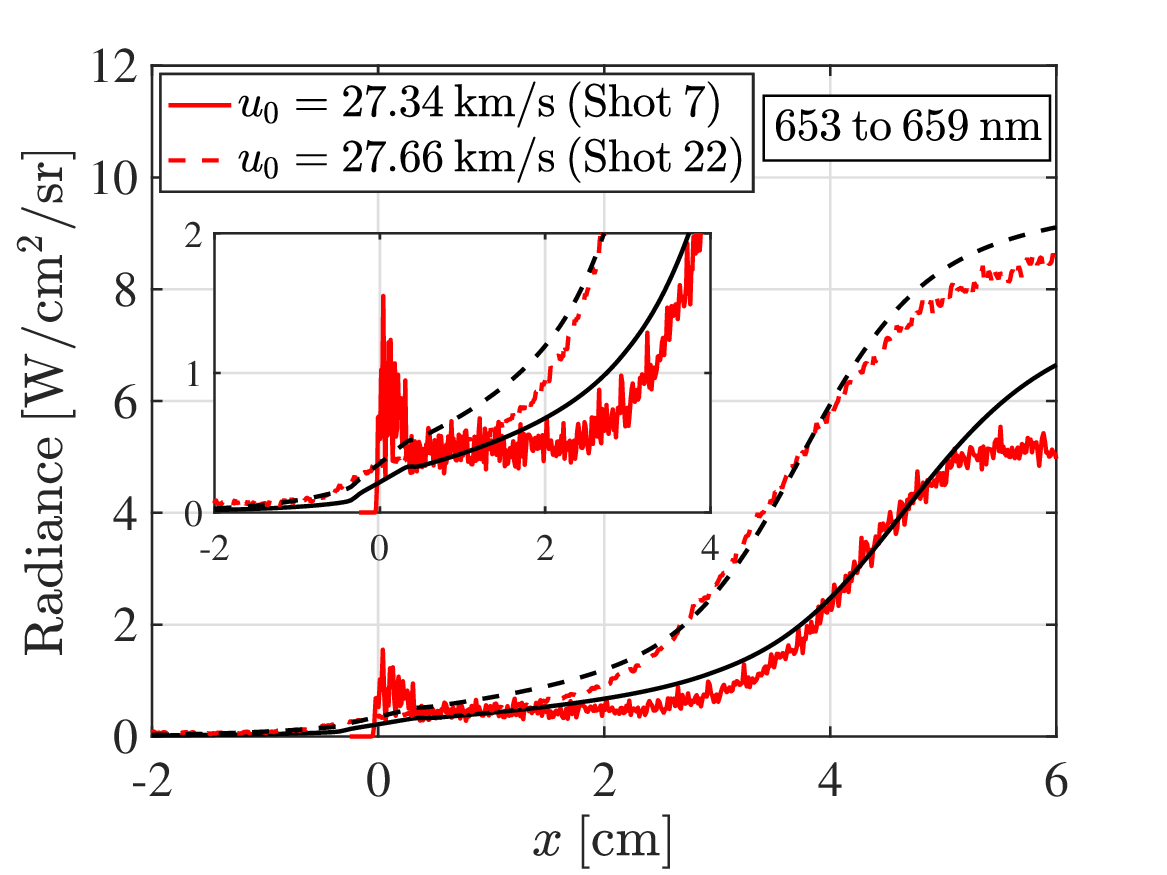}
     \end{subfigure}
     \begin{subfigure}[b]{0.33\textwidth}
         \centering
         \includegraphics[width=\textwidth,trim={0cm 0cm 1.2cm 0.5cm},clip]{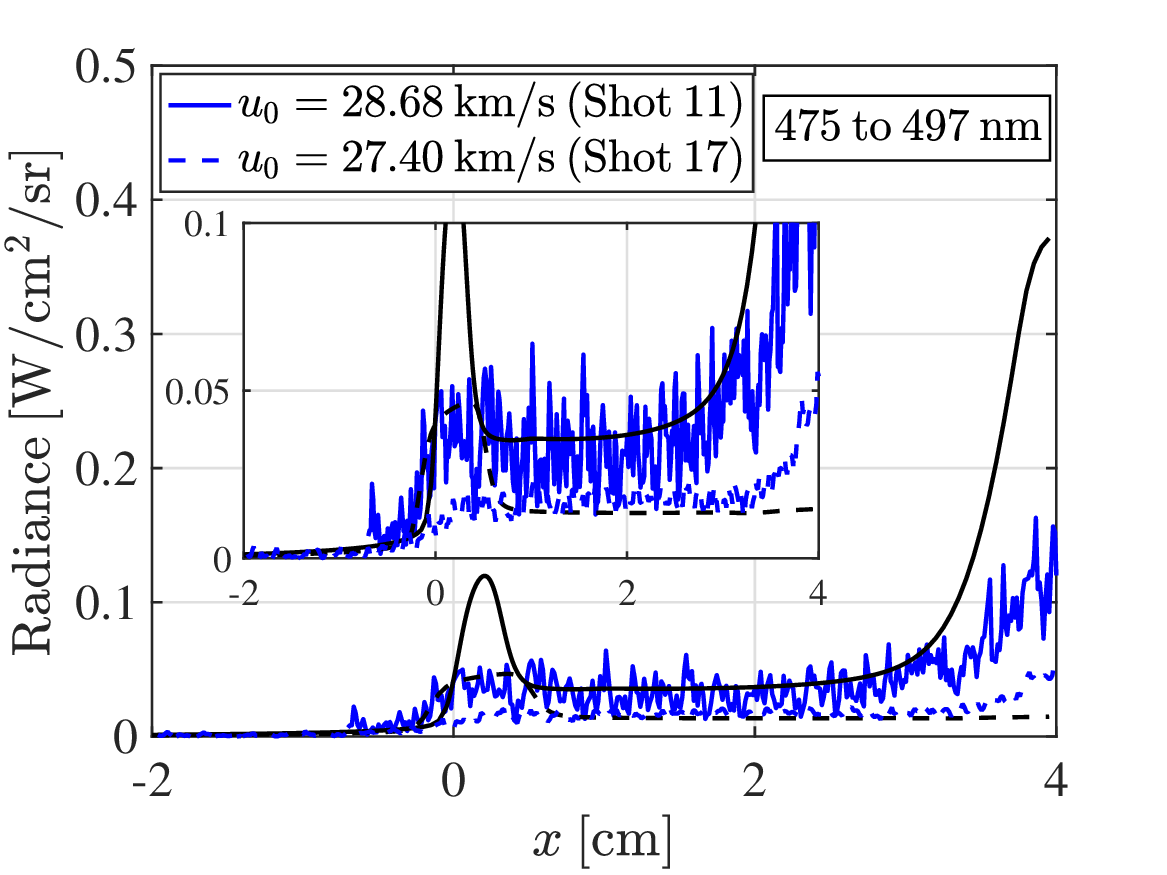}
         \caption{$\bm{P_0}$ = 0.1 torr}
     \end{subfigure}
     \hfill
     \begin{subfigure}[b]{0.33\textwidth}
         \centering
         \includegraphics[width=\textwidth,trim={0cm 0cm 1.2cm 0.5cm},clip]{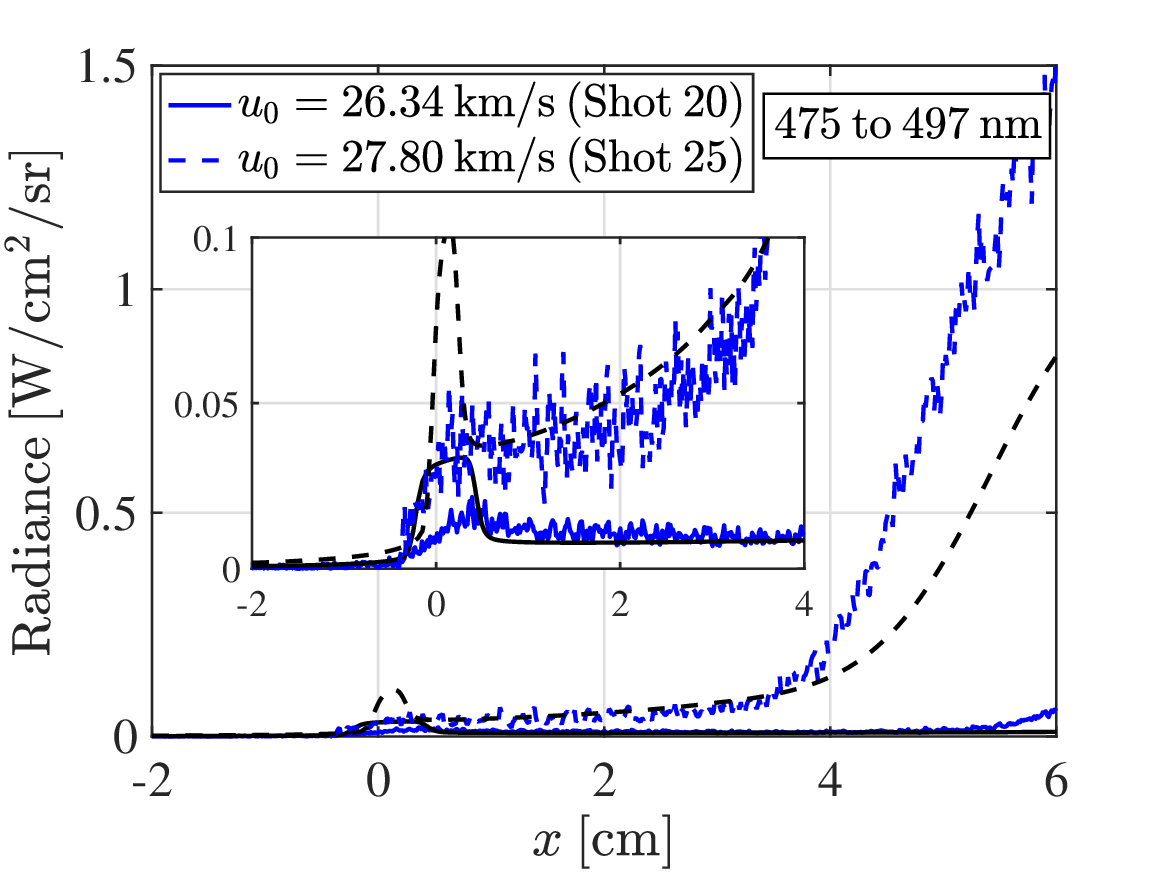}
         \caption{$\bm{P_0}$ = 0.2 torr}
     \end{subfigure}
     \hfill
     \begin{subfigure}[b]{0.33\textwidth}
         \centering
         \includegraphics[width=\textwidth,trim={0cm 0cm 1.2cm 0.5cm},clip]{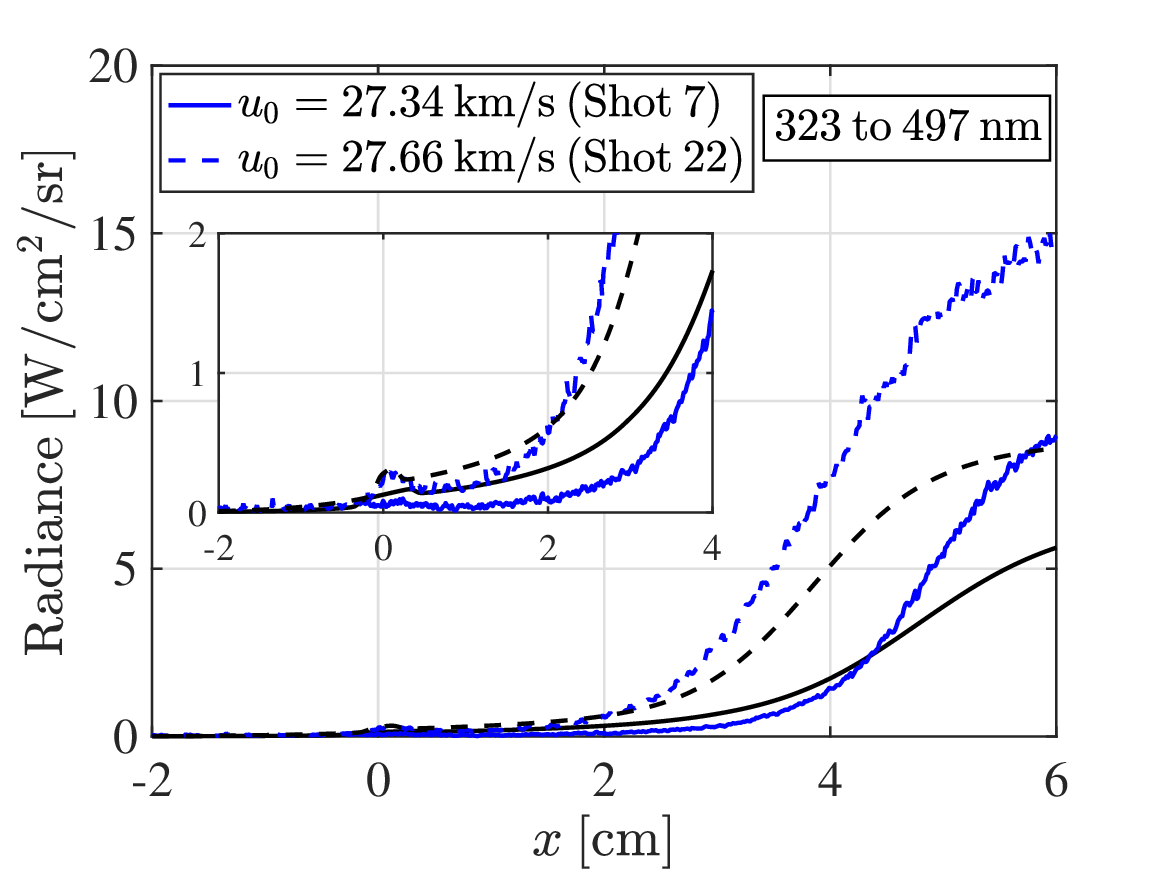}
         \caption{$\bm{P_0}$ = 0.5 torr}
     \end{subfigure}
    \caption{Integrated radiance for the Lyman-$\bm \alpha$ (top), Balmer-$\bm \alpha$ (center) and higher-order Balmer series (bottom) features at increasing freestream pressures (left to right). Colored and black lines correspond to measured and simulated radiance values, respectively. Integrated wavelength ranges are indicated on each figure.}
    \label{fig:H_allshots}
\end{figure}

The integrated radiance profiles for all other simulated shot conditions are given in Fig.~\ref{fig:H_allshots}. The Lyman-$\alpha$ feature was not measured for shots 7 and 11. For the higher-order Balmer series features at $P_0=$ 0.1 and 0.2 torr, the integrated wavelength range is limited to only include the Balmer-$\beta$ feature. This is done to minimize the impact of contaminants on the radiance profiles, which appear in the 330 to 475 nm range for these lower pressure conditions. While the radiance in the induction zones for the Lyman-$\alpha$ and Balmer-$\alpha$ features are overpredicted for some shot conditions, the locations and magnitudes for the post-induction zone radiance are well predicted for nearly all shot conditions. For the higher-order Balmer series features, the radiance in the induction zones are predicted accurately. However, the radiance in the far post-shock region is underpredicted for all shots besides shot 11.

All together, the results presented in Fig.~\ref{fig:S23_Lya} - \ref{fig:H_allshots} suggest that the proposed model of the present work gives relatively accurate predictions of both the trends and magnitudes of atomic H radiation for the simulated shock conditions. This implies that, using the kinetic model of the present work, the electron and heavy-particle-impact processes are sufficient for reproducing the majority of the trends in the EAST data for H radiation. This will be discussed further in the context of alternate kinetic models from the literature in section~\ref{sec:altmodels}.

\subsubsection{$H_2$ Lyman Band}
\label{sec:H2rad}

\begin{figure}[hbt!]
     \centering
     \hspace*{\fill}
     \begin{subfigure}[b]{0.4\textwidth}
         \centering
         \includegraphics[width=\textwidth,trim={0cm 0cm 1.2cm 0.5cm},clip]{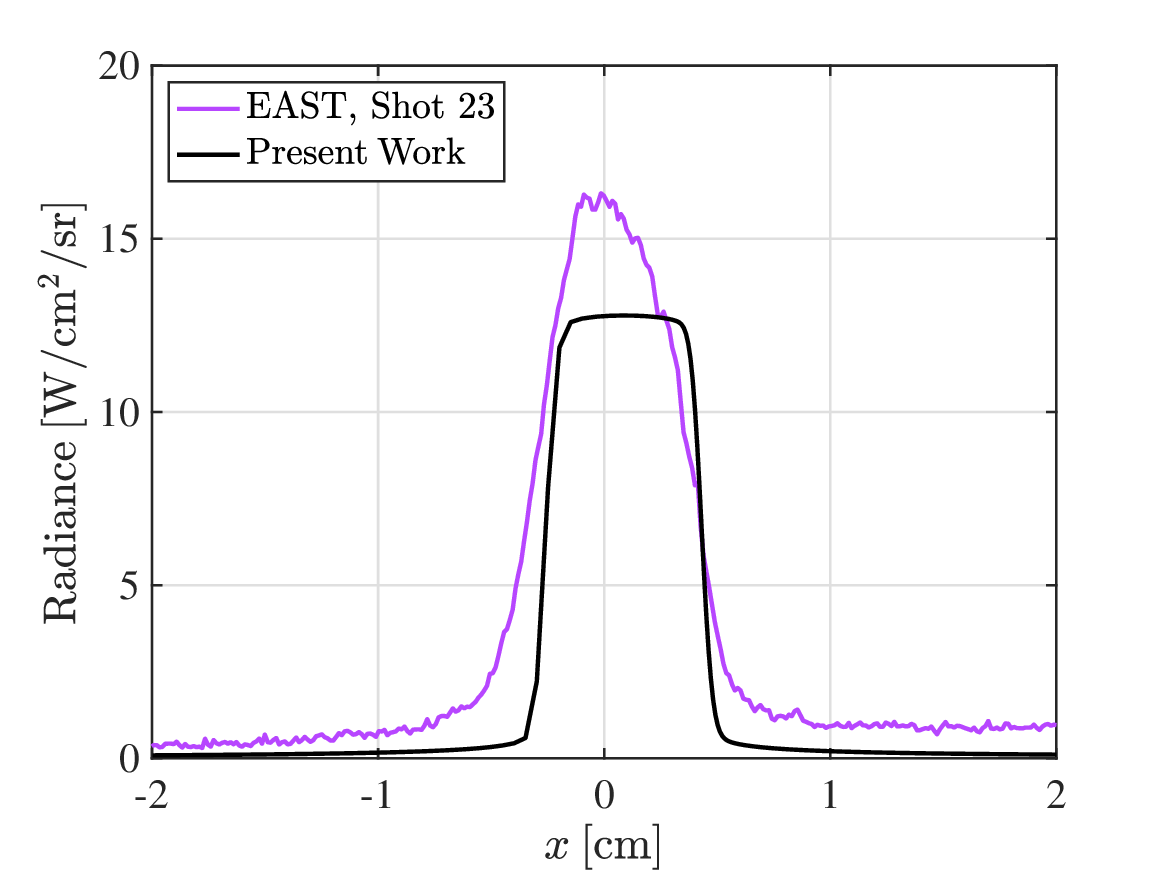}
     \end{subfigure}
     \hfill
     \begin{subfigure}[b]{0.4\textwidth}
         \centering
         \includegraphics[width=\textwidth,trim={0cm 0cm 1.2cm 0.5cm},clip]{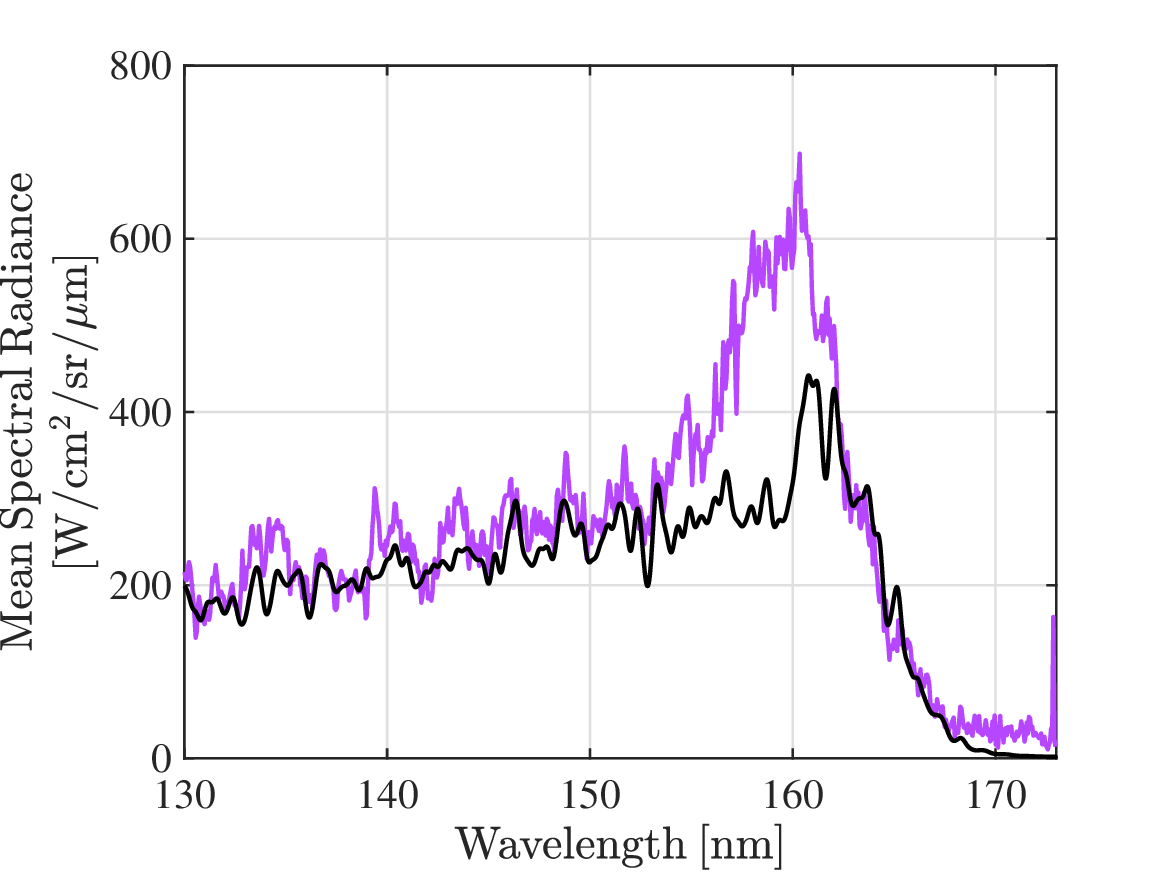}
     \end{subfigure}
     \hspace*{\fill}
    \caption{Integrated radiance over 130 to 173 nm (left) and mean spectral radiance over $\bm x$ = -0.5 to 0.5 cm (right) for the Lyman band feature.}
    \label{fig:S23_Lyband}
\end{figure}

The measured and computed radiance profiles for the $\rm H_2$ Lyman band for shot 23 are shown in Fig.~\ref{fig:S23_Lyband}. Recall that in the calculations, all $\rm H_2$ rovibrational and electronic states are assumed to follow Boltzmann distributions about $T_{\rm rv,H_2}$ and $T_{\rm t,e^-}$, respectively. Despite the simplicity of this assumption, the results of Fig.~\ref{fig:S23_Lyband} show that the majority of the features in the Lyman band are captured well by the computational model. In particular, there is excellent agreement in both the shape and magnitude of the radiance for most of the plotted spectral range. However, there is an underprediction of the highest magnitude peaks between approximately 154 and 162 nm, which results in a corresponding underprediction of the integrated radiance at the shock front when compared to the experiments. For the integrated radiance, the shape of the profile is impacted significantly by two effects: physical shock spreading due to viscosity, and the resolution of the measurements (represented by the SRF). As viscous effects are neglected in the calculations and the spatial motion is represented as a square, the shape of the simulated profile appears more rectangular than the experiments.

\begin{figure}[hbt!]
     \centering
     \begin{subfigure}[b]{0.33\textwidth}
         \centering
         \includegraphics[width=\textwidth,trim={0cm 0cm 1.2cm 0.5cm},clip]{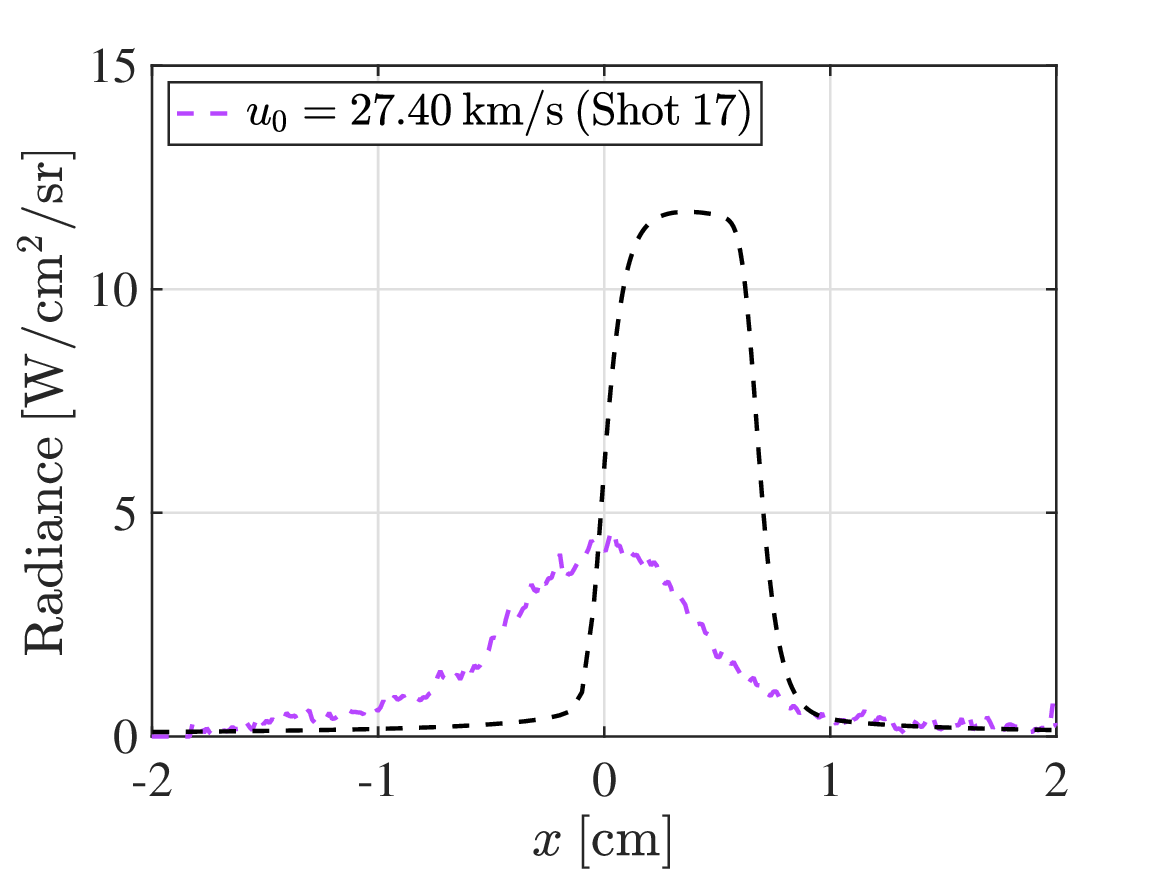}
         \caption{$\bm{P_0}$ = 0.1 torr}
     \end{subfigure}
     \hfill
     \begin{subfigure}[b]{0.33\textwidth}
         \centering
         \includegraphics[width=\textwidth,trim={0cm 0cm 1.2cm 0.5cm},clip]{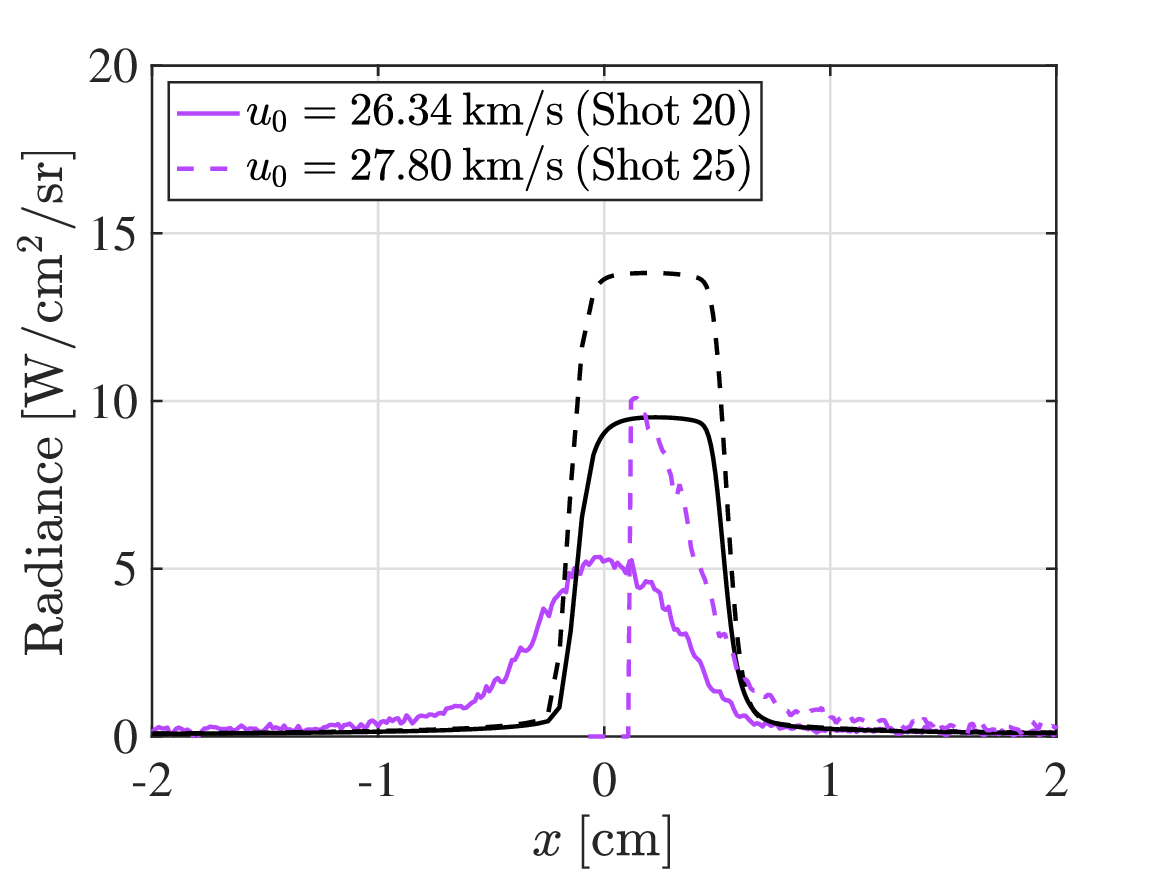}
         \caption{$\bm{P_0}$ = 0.2 torr}
     \end{subfigure}
     \hfill
     \begin{subfigure}[b]{0.33\textwidth}
         \centering
         \includegraphics[width=\textwidth,trim={0cm 0cm 1.2cm 0.5cm},clip]{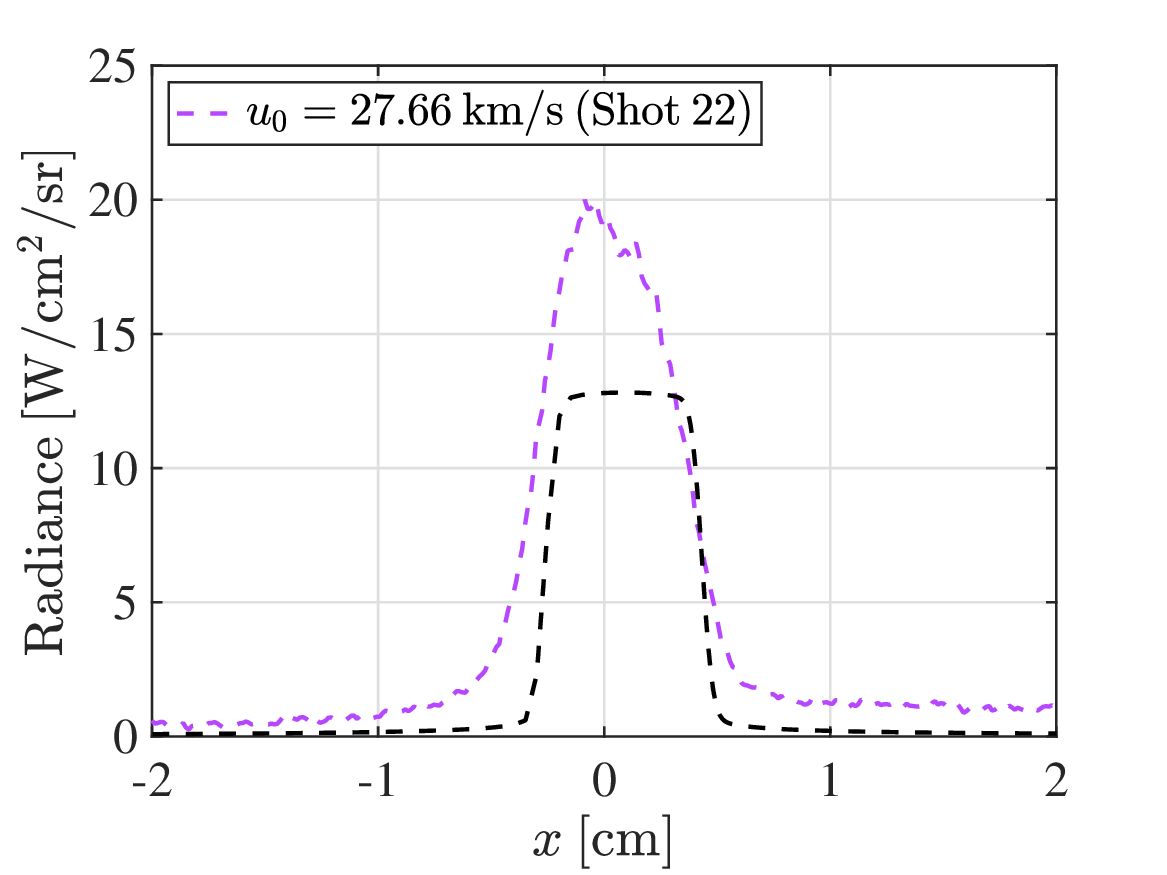}
         \caption{$\bm{P_0}$ = 0.5 torr}
     \end{subfigure}
    \caption{Integrated radiance for the Lyman band feature (130 to 173 nm) at increasing freestream pressures (left to right). Colored and black lines correspond to measured and simulated radiance values, respectively.}
    \label{fig:H2_allshots}
\end{figure}

The integrated radiance profiles for all other simulated shot conditions are shown in Fig.~\ref{fig:H2_allshots} (excluding shots 7 and 11 for which the Lyman band was not measured). In contrast to the results for shot 23, the integrated radiance at the shock front is overpredicted by the computational model at the lower pressure shot conditions ($P_0$ = 0.1 and 0.2 torr). Additionally, the predicted shock widths are noticeably smaller than the measured profiles at these conditions. This is likely due to the neglecting of viscosity/ diffusion in the calculations of the present work, as these effects will broaden the shock to have a non-zero thickness that is inversely proportional to the freestream pressure. For shot 22 at 0.5 torr, the same underprediction of the radiance at the shock front as was seen for shot 23 is observed.

While there are clearly some discrepancies between the measured and computed radiance values of the $\rm H_2$ Lyman band as shown in Fig.~\ref{fig:S23_Lyband} and \ref{fig:H2_allshots}, the integrated radiance is predicted within a factor of two for nearly all shot conditions using the very simple Boltzmann and inviscid core flow assumptions of the present work. Future work should incorporate viscous effects as well as a non-Boltzmann model for $\rm H_2$ electronic states to improve the agreement between the measured and simulated radiance profiles.

\subsection{Sensitivity Analysis}
\label{sec:sens}

\subsubsection{Kinetic Rates}
\label{sec:sens_kinetics}

\begin{figure}[hbt!]
     \centering
     \hspace*{\fill}
     \begin{subfigure}[b]{0.4\textwidth}
         \centering
         \includegraphics[width=\textwidth,trim={0cm 0cm 1.2cm 0.5cm},clip]{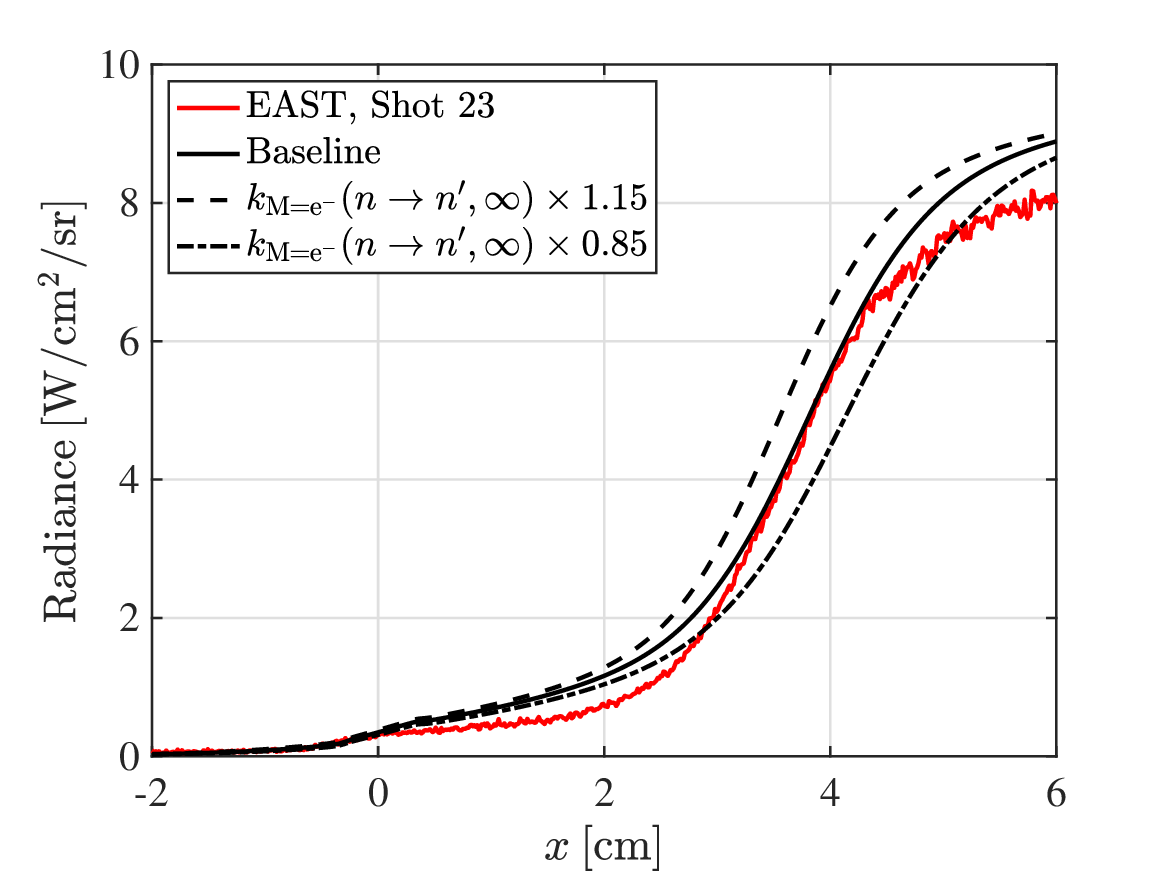}
     \end{subfigure}
     \hfill
     \begin{subfigure}[b]{0.4\textwidth}
         \centering
         \includegraphics[width=\textwidth,trim={0cm 0cm 1.2cm 0.5cm},clip]{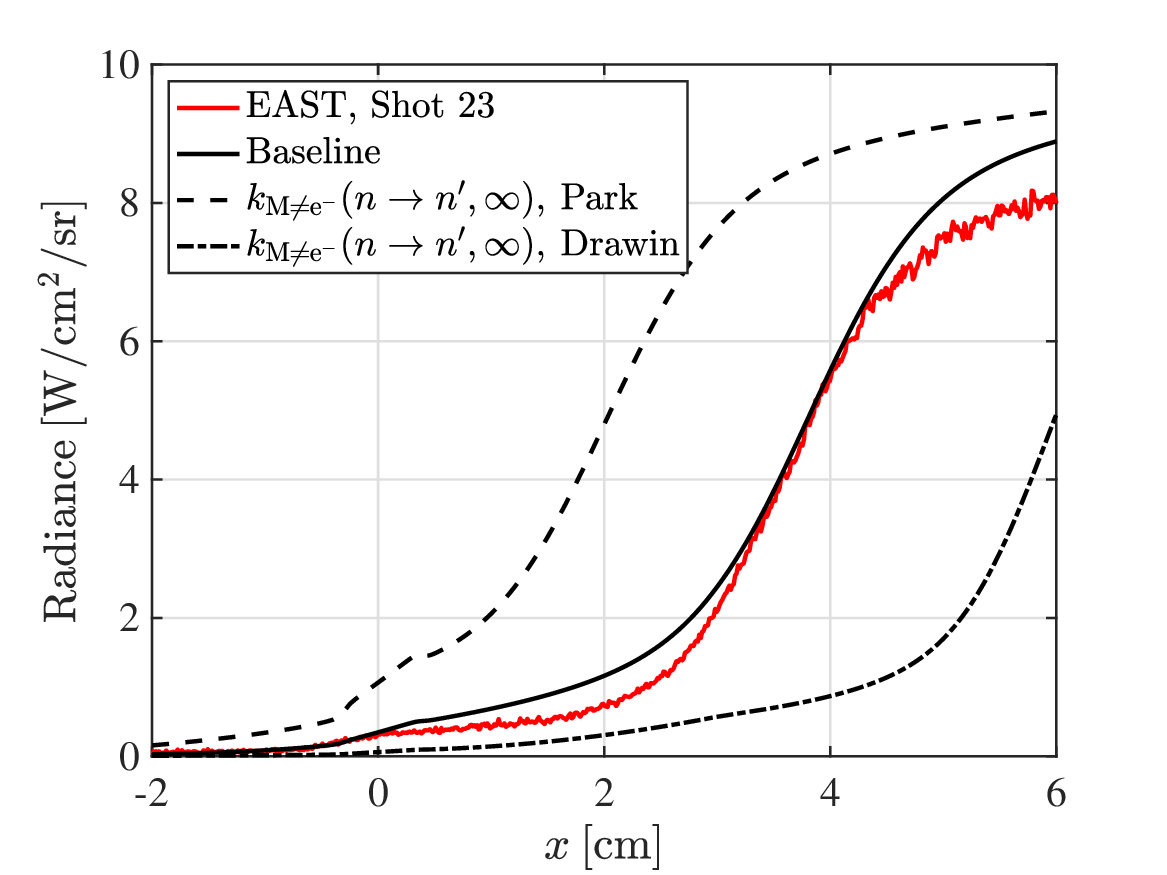}
     \end{subfigure}
     \hspace*{\fill}
    \caption{Sensitivity of the Balmer-$\bm \alpha$ integrated radiance over 653 to 659 nm to changes in the electron-impact (left) and heavy-particle-impact (right) excitation and ionization rates.}
    \label{fig:sens_elec}
\end{figure}

To illustrate the sensitivity of the computed radiance values to uncertainties in the kinetic rates, several calculations are performed for the shot 23 conditions with the rate constants modified from the baseline model presented in the previous sections. Figure~\ref{fig:sens_elec} shows the radiance for the Balmer-$\alpha$ feature, where either the electron-impact excitation and ionization rate constants have been increased or decreased by 15\% (based on the estimated uncertainty from section~\ref{sec:revEI}), or the heavy-particle-impact excitation and ionization rate constants have been replaced by those of Park~\cite{Park2012} or Drawin~\cite{Drawin1969}. In general, the significantly smaller uncertainty in the electron-impact rates leads to a much smaller variation in the radiance when compared to the heavy-particle-impact rates. For the heavy-particle-impact rates, the use of either the Park or Drawin rate constants leads to a considerably faster or slower equilibration process, respectively.
Additionally, the variation in the electron-impact rates leads to changes in the slope of the post-induction zone radiance but not the length of the induction zone itself, while the opposite is true for the heavy-particle-impact rates. This is consistent with the understanding that heavy-particle-impact ionization is responsible for producing the initial electrons in the kinetic model, so the heavy-particle-impact processes act as a rate-limiting step for the subsequent electron-impact processes. As the degree of ionization increases further downstream however, excitation and ionization become dominated by electron-impact collisions.

\begin{figure}[hbt!]
     \centering
     \begin{subfigure}[b]{0.33\textwidth}
         \centering
         \includegraphics[width=\textwidth,trim={0cm 0cm 1.2cm 0.5cm},clip]{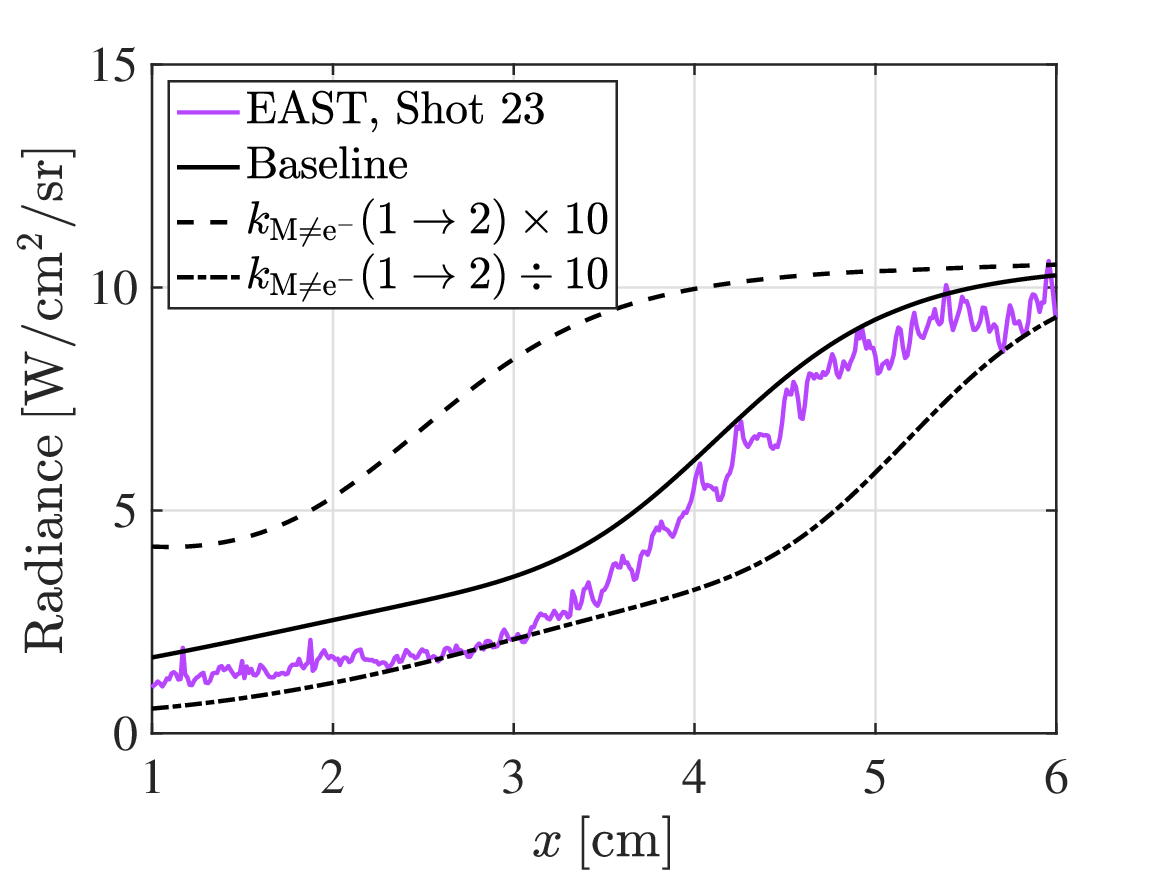}
     \end{subfigure}
     \hfill
     \begin{subfigure}[b]{0.33\textwidth}
         \centering
         \includegraphics[width=\textwidth,trim={0cm 0cm 1.2cm 0.5cm},clip]{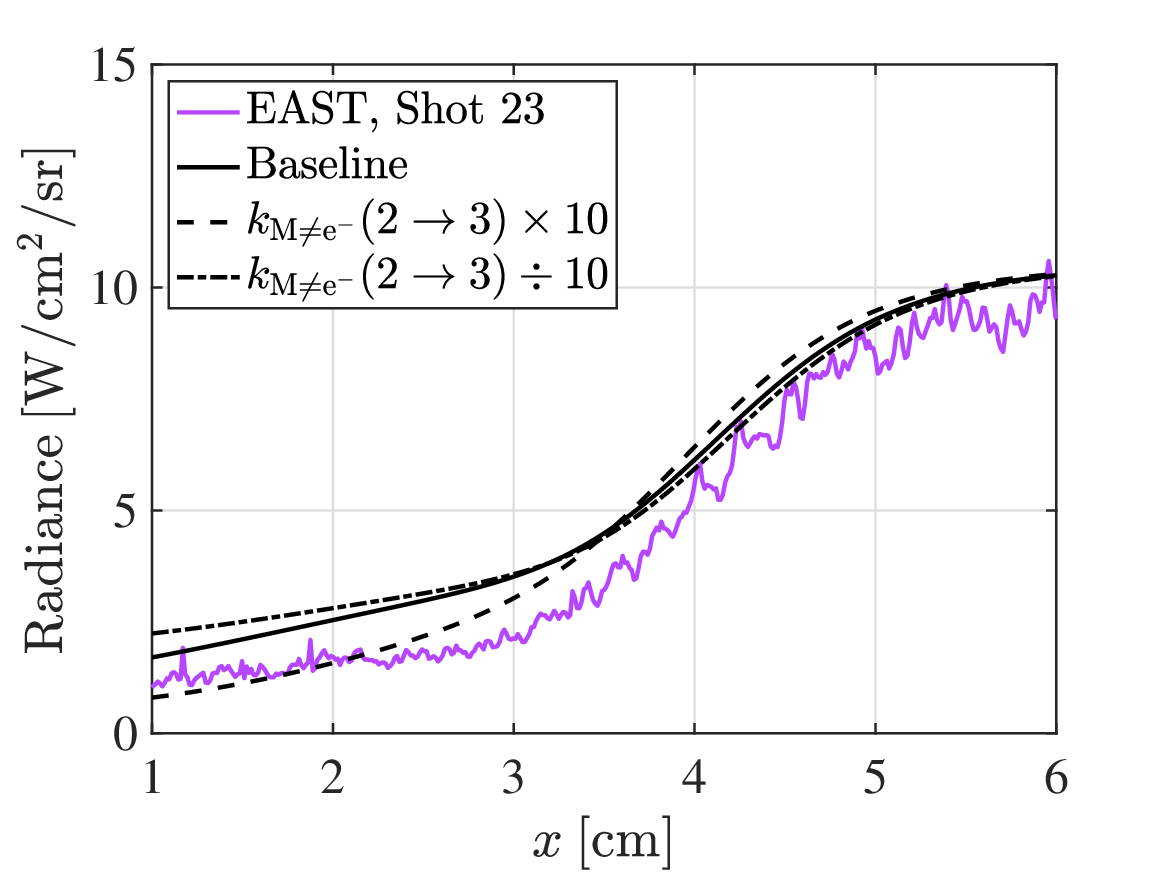}
     \end{subfigure}
     \hfill
     \begin{subfigure}[b]{0.33\textwidth}
         \centering
         \includegraphics[width=\textwidth,trim={0cm 0cm 1.2cm 0.5cm},clip]{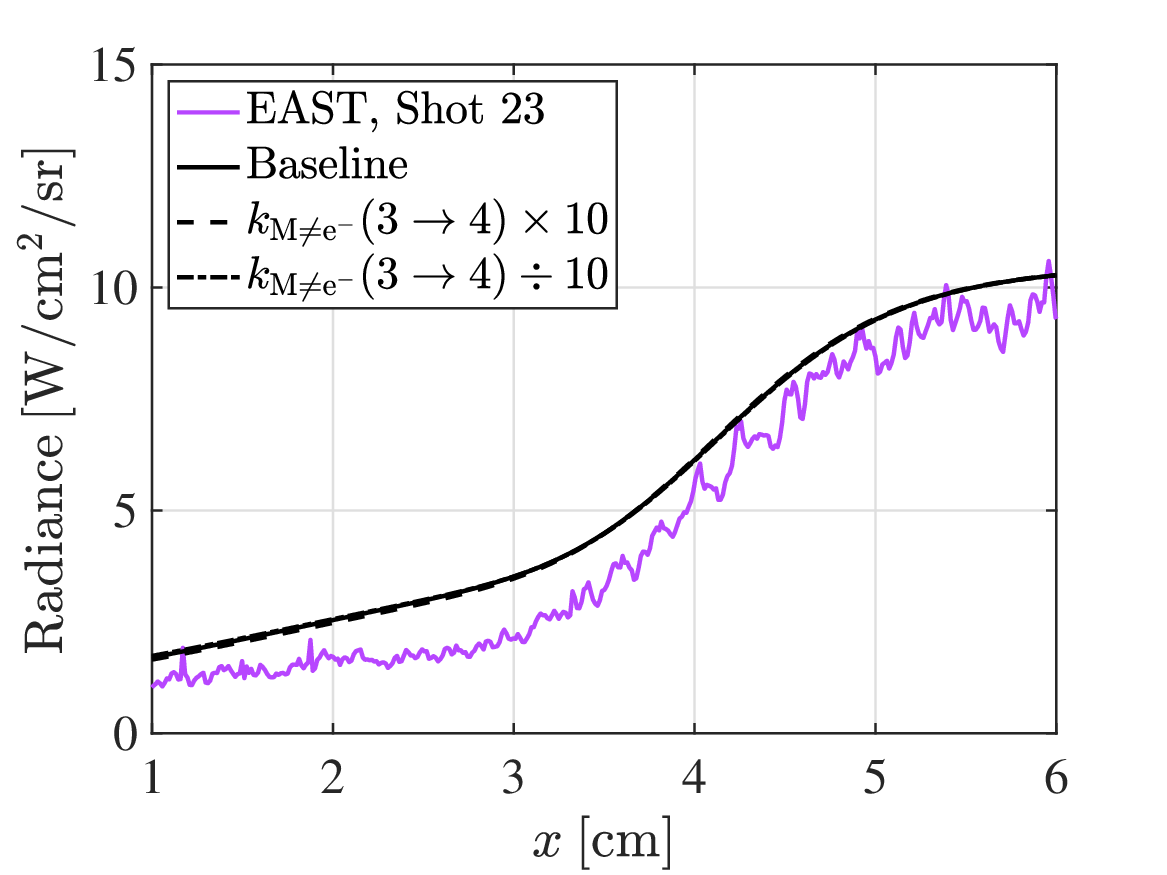}
     \end{subfigure}
     \begin{subfigure}[b]{0.33\textwidth}
         \centering
         \includegraphics[width=\textwidth,trim={0cm 0cm 1.2cm 0.5cm},clip]{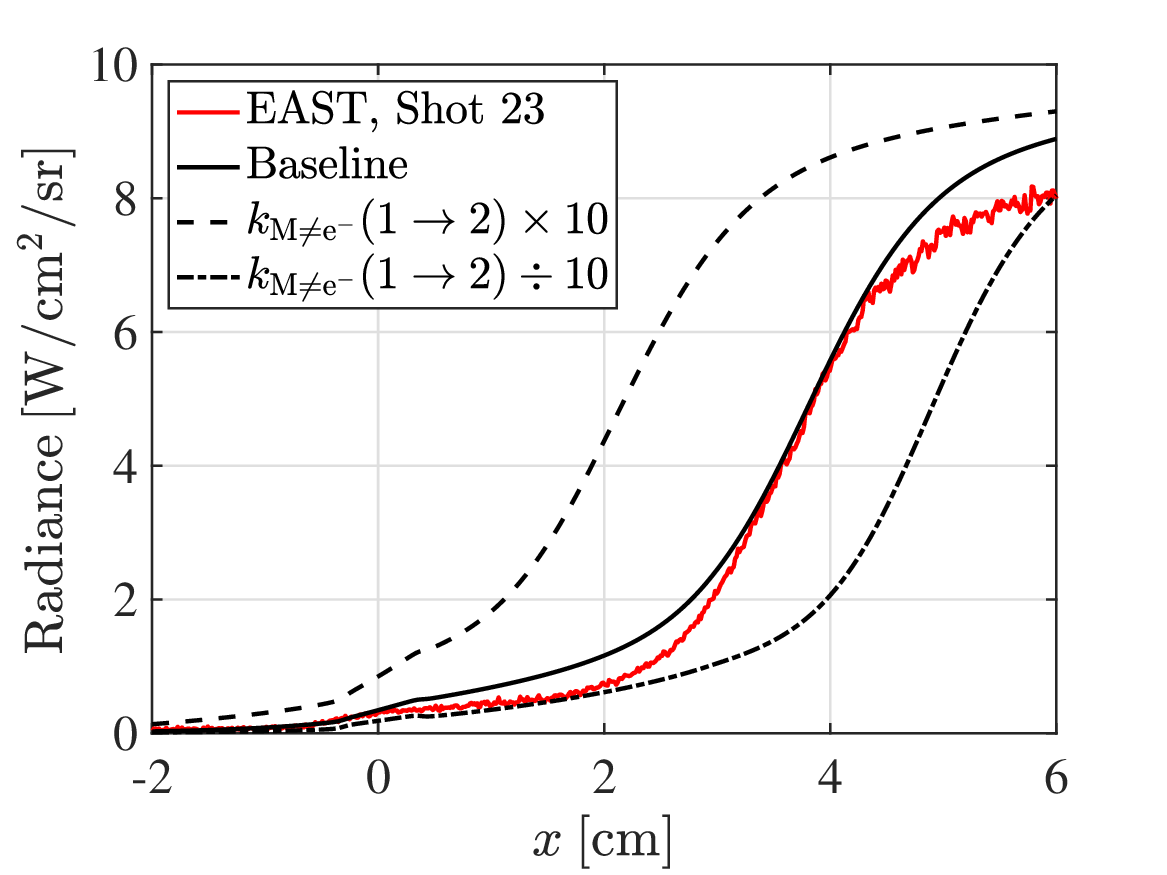}
         \caption{$\bm{k_{\rm M\neq e^-}(1\rightarrow2)}$}
     \end{subfigure}
     \hfill
     \begin{subfigure}[b]{0.33\textwidth}
         \centering
         \includegraphics[width=\textwidth,trim={0cm 0cm 1.2cm 0.5cm},clip]{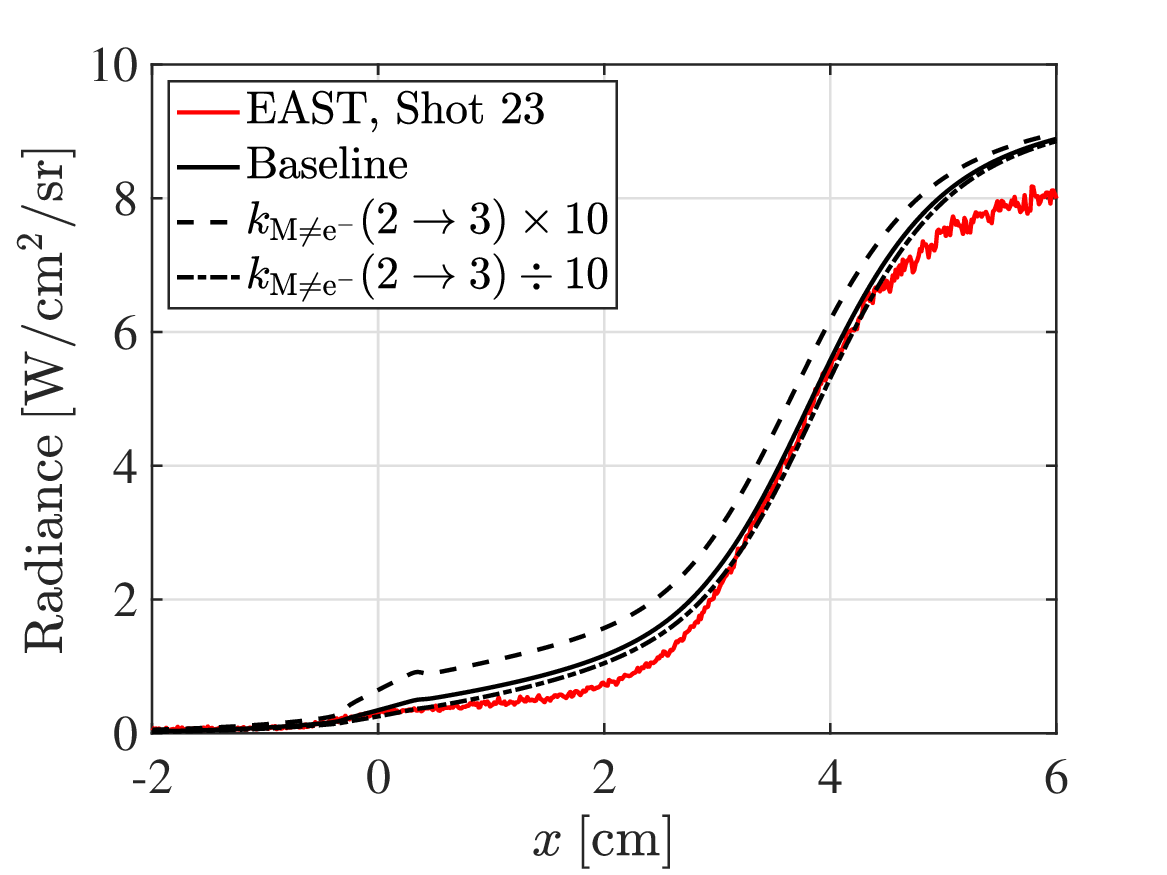}
         \caption{$\bm{k_{\rm M\neq e^-}(2\rightarrow3)}$}
     \end{subfigure}
     \hfill
     \begin{subfigure}[b]{0.33\textwidth}
         \centering
         \includegraphics[width=\textwidth,trim={0cm 0cm 1.2cm 0.5cm},clip]{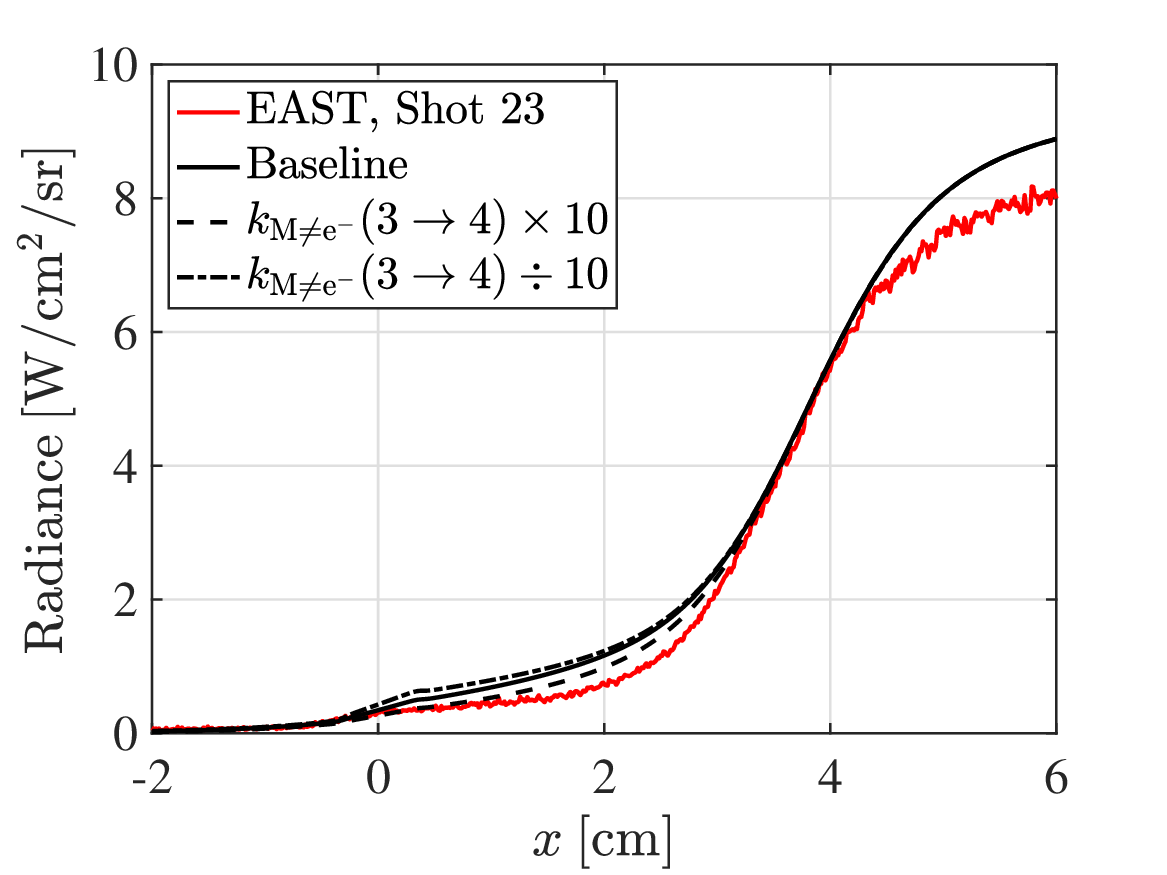}
         \caption{$\bm{k_{\rm M\neq e^-}(3\rightarrow4)}$}
     \end{subfigure}
    \caption{Sensitivity of the Lyman-$\bm \alpha$ integrated radiance over 117 to 130 nm (top) and Balmer-$\bm \alpha$ integrated radiance over 653 to 659 nm (bottom) to changes in the heavy-particle-impact excitation rate constants.}
    \label{fig:sens_hvy}
\end{figure}

To illustrate the sensitivity of the radiance profiles to the heavy-particle-impact rates in a more granular fashion, Fig.~\ref{fig:sens_hvy} shows the radiance for the Lyman-$\alpha$ and Balmer-$\alpha$ features, where only one of the heavy-particle-impact mono-quantum excitation rate constants has been modified. There are two key observations from these results. Firstly, there is a much larger sensitivity of the radiance to the $n=1\rightarrow2$ transition when compared to the higher-order transitions, such that the length of the induction zone is almost entirely controlled by the $n=1\rightarrow2$ rate constant. Secondly, the radiation in the induction zone for the Lyman-$\alpha$ and Balmer-$\alpha$ features can be reduced without significantly changing the post-induction zone radiance by increasing the $n=2\rightarrow3$ and $n=3\rightarrow4$ rate constants, respectively. This suggests that while the radiance predictions of the baseline model are reasonably accurate, the predictions in the induction zone could potentially be improved by modifying the baseline heavy-particle-impact rate constants.

\begin{figure}[hbt!]
     \centering
     \hspace*{\fill}
     \begin{subfigure}[b]{0.4\textwidth}
         \centering
         \includegraphics[width=\textwidth,trim={0cm 0cm 1.2cm 0.5cm},clip]{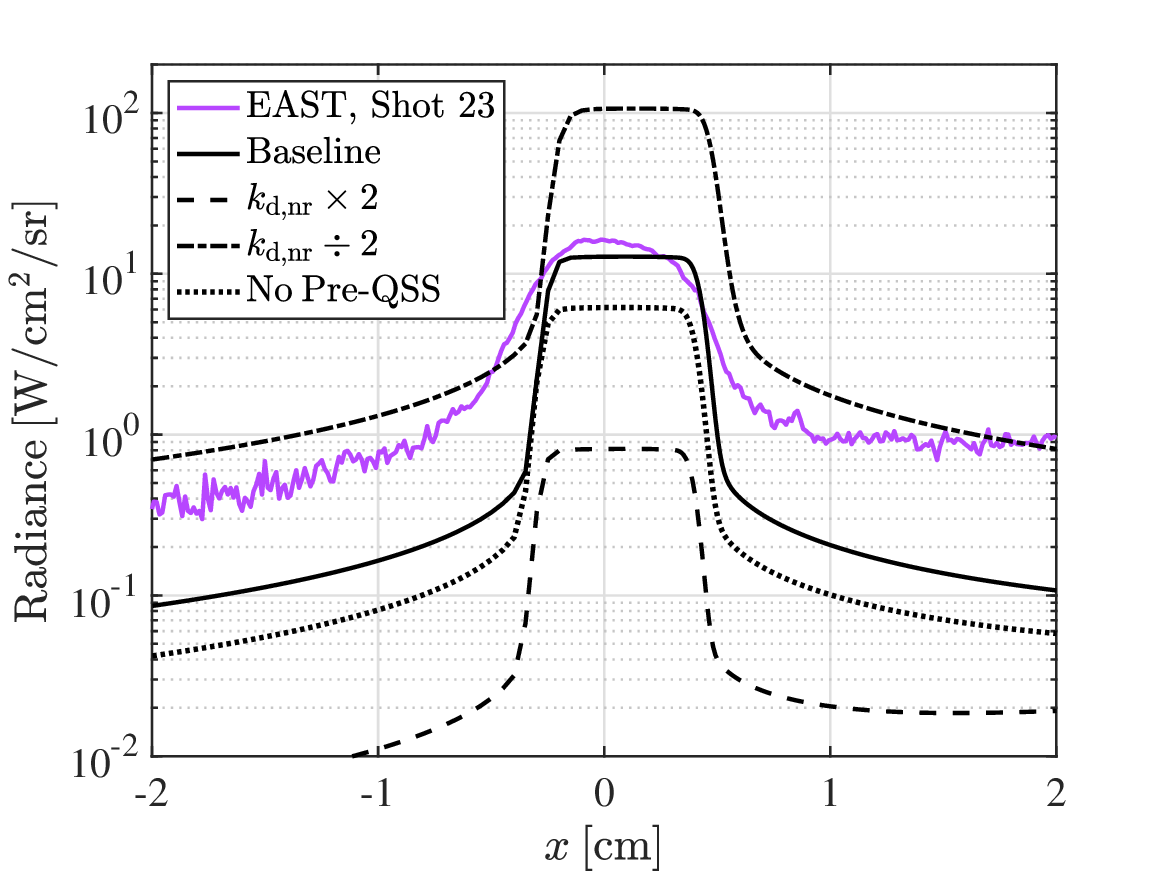}
     \end{subfigure}
     \hfill
     \begin{subfigure}[b]{0.4\textwidth}
         \centering
         \includegraphics[width=\textwidth,trim={0cm 0cm 1.2cm 0.5cm},clip]{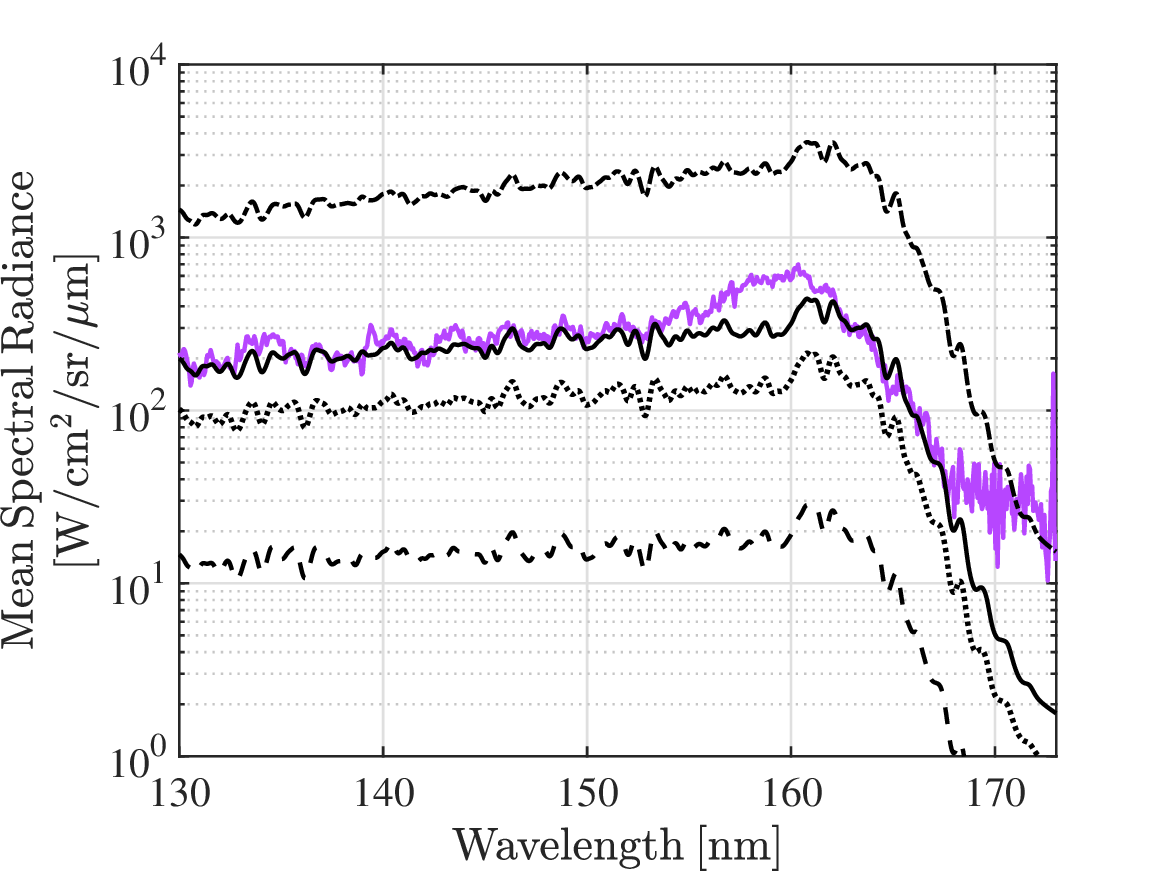}
     \end{subfigure}
     \hspace*{\fill}
    \caption{Sensitivity of the Lyman band integrated radiance over 130 to 173 nm (left) and mean spectral radiance over $\bm x$ = -0.5 to 0.5 cm (right) to changes in the dissociation rates.}
    \label{fig:sens_diss}
\end{figure}

Figure~\ref{fig:sens_diss} shows the radiance for the $\rm H_2$ Lyman band, where either the QSS non-recombining dissociation rate constants, $k_{\rm d,nr}$, for all third-bodies have been modified by a factor of two, or the pre-QSS correction in the dissociation model has been neglected. Carroll et al.~\cite{Carroll2026_diss2} found that the uncertainty in $k_{\rm d,nr}$ for all relevant third-bodies was approximately a factor of two. Only the results for the $\rm H_2$ Lyman band are shown, as minimal differences were seen in the atomic H features. When the rate of $\rm H_2$ dissociation is increased (either by increasing $k_{\rm d,nr}$ or by neglecting pre-QSS effects), $\rm H_2$ is consumed more quickly and hence the predicted radiance of the Lyman band is decreased, and vice versa. In particular, increasing or decreasing $k_{\rm d,nr}$ by a factor of two leads to over a magnitude decrease or increase, respectively, in the radiance, while neglecting pre-QSS effects reduces the radiance by approximately a factor of two. This large sensitivity implies that an accurate characterization of $\rm H_2$ dissociation rates, including rovibrational non-equilibrium and pre-QSS effects, is critical for predictions of the Lyman band radiance.

\subsubsection{Boundary Layer Treatment}
\label{sec:sens_BL}

\begin{figure}[hbt!]
     \centering
     \hspace*{\fill}
     \begin{subfigure}[b]{0.4\textwidth}
         \centering
         \includegraphics[width=\textwidth,trim={0cm 0cm 1.2cm 0.5cm},clip]{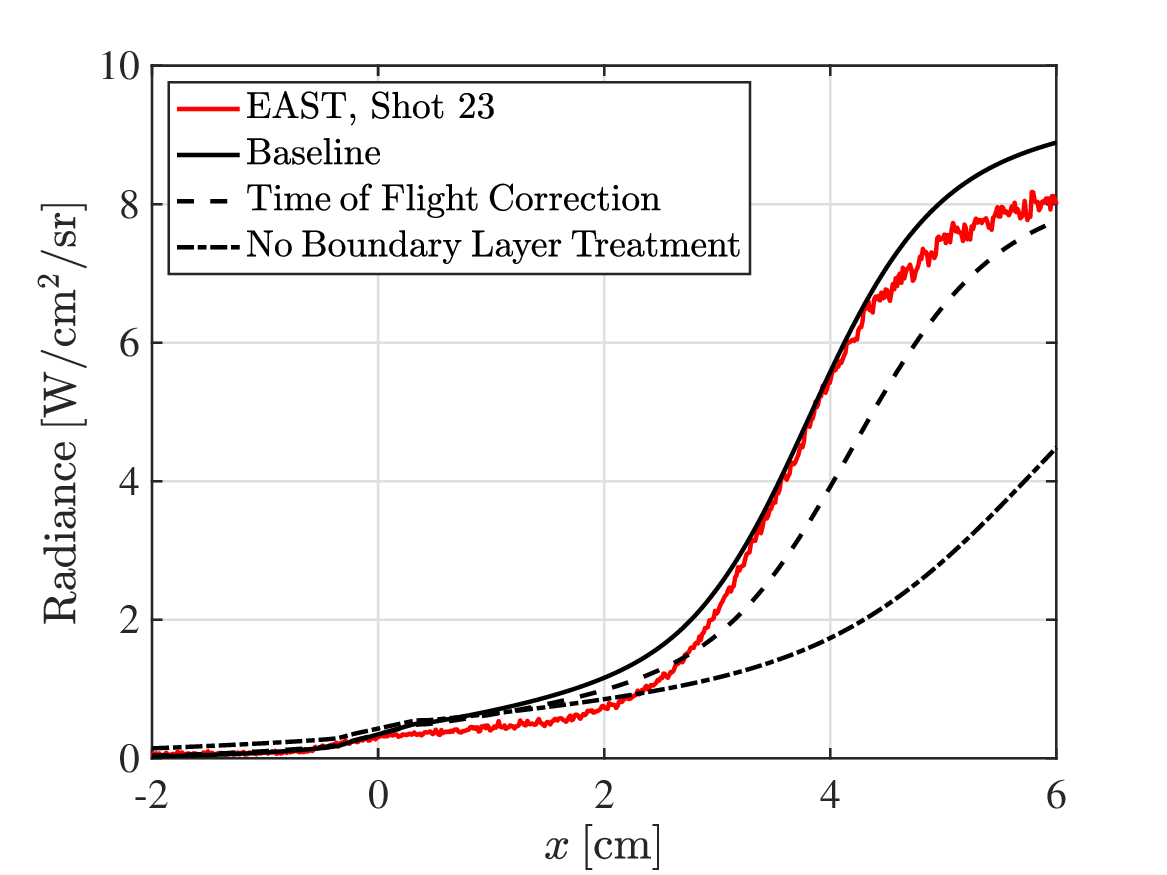}
     \end{subfigure}
     \hfill
     \begin{subfigure}[b]{0.4\textwidth}
         \centering
         \includegraphics[width=\textwidth,trim={0cm 0cm 1.2cm 0.5cm},clip]{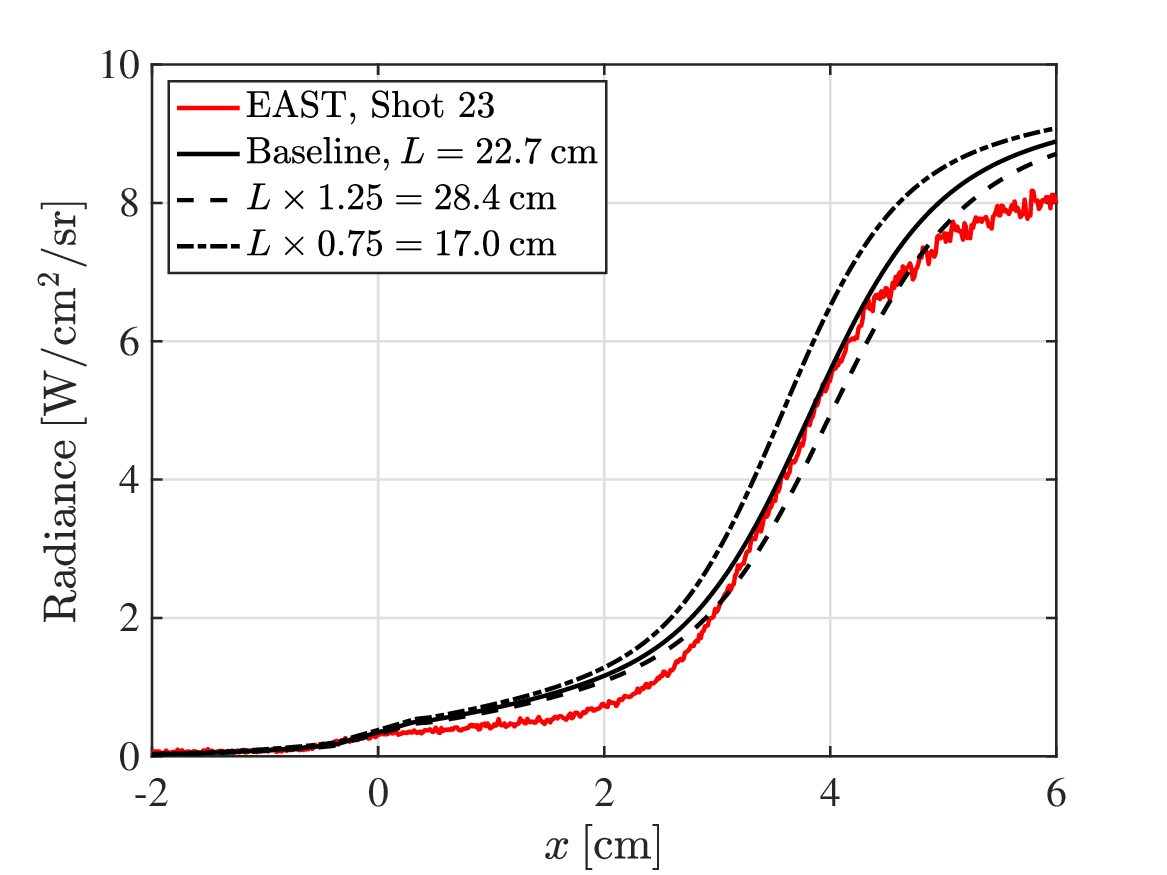}
     \end{subfigure}
     \hspace*{\fill}
    \caption{Sensitivity of the Lyman-$\bm \alpha$ integrated radiance over 117 to 130 nm to the boundary layer treatment method (left) or the test slug length (right).}
    \label{fig:sens_BL}
\end{figure}

The impact of varying either the treatment of the boundary layer in the governing equations or the test slug length on the Balmer-$\alpha$ radiance is shown in Fig.~\ref{fig:sens_BL}. In the ``No Boundary Layer Treatment'' case, the standard Rankine-Hugoniot governing equations are solved with no corrections applied for boundary layer effects; in the ``Time-of-Flight Correction'' case, the time-of-flight correction of Clarke et al.~\cite{Clarke2023} is applied a posteriori to the Rankine-Hugoniot solution to account for spatial transformation effects, but the coupled compression effects solved in Eq.~\eqref{eqn:species_march}-\eqref{eqn:electron_march} are neglected. While the time-of-flight correction significantly improves the agreement between the simulations and experiments, without the coupled compression effect, it is not sufficient to reproduce the slope in the post-induction zone radiance. This is especially true at the $P_0$ = 0.1 and 0.2 torr conditions (not shown), where the rise in the post-induction zone radiance is almost entirely due to boundary layer effects. Separately, varying the test slug length by $\pm$25\% shows that there is a relatively weak dependence of the radiance to uncertainties in $L$.

A key observation to make here is that of all of the tested uncertainties, only the electron-impact rates and the boundary layer treatment lead to significant changes in the slope of the post-induction zone radiance. Therefore, if a less physical treatment of the boundary layer was used, the electron-impact rate constants would have to be modified beyond their uncertainty bounds to compensate for this error. The agreement for the slope of the post-induction zone radiance, along with the relatively small uncertainties in the electron-impact rate constants, supports the validity of the boundary layer treatment in the present work.

\subsection{Comparison to Alternate Kinetic Models}
\label{sec:altmodels}

\begin{figure}[hbt!]
     \centering
     \hspace*{\fill}
     \begin{subfigure}[b]{0.4\textwidth}
         \centering
         \includegraphics[width=\textwidth,trim={0cm 0cm 1.2cm 0.5cm},clip]{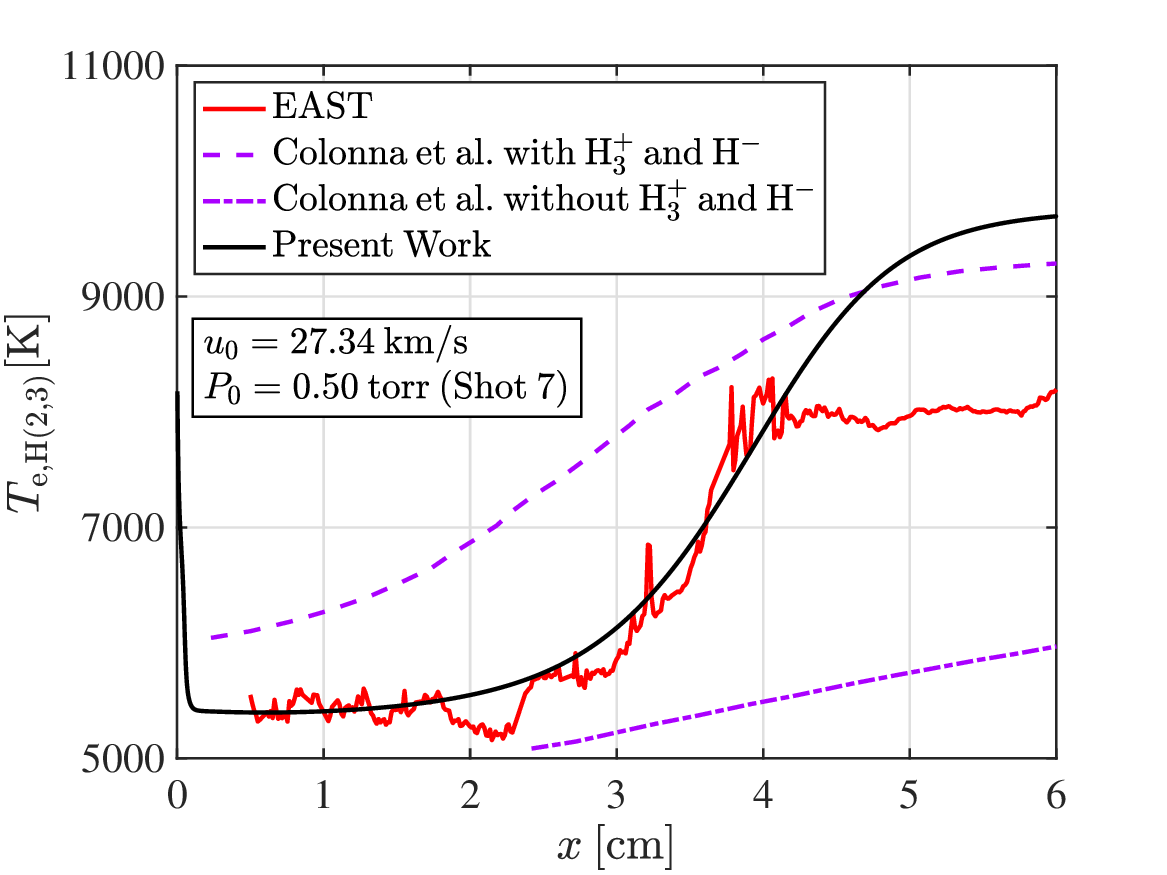}
     \end{subfigure}
     \hfill
     \begin{subfigure}[b]{0.4\textwidth}
         \centering
         \includegraphics[width=\textwidth,trim={0cm 0cm 1.2cm 0.5cm},clip]{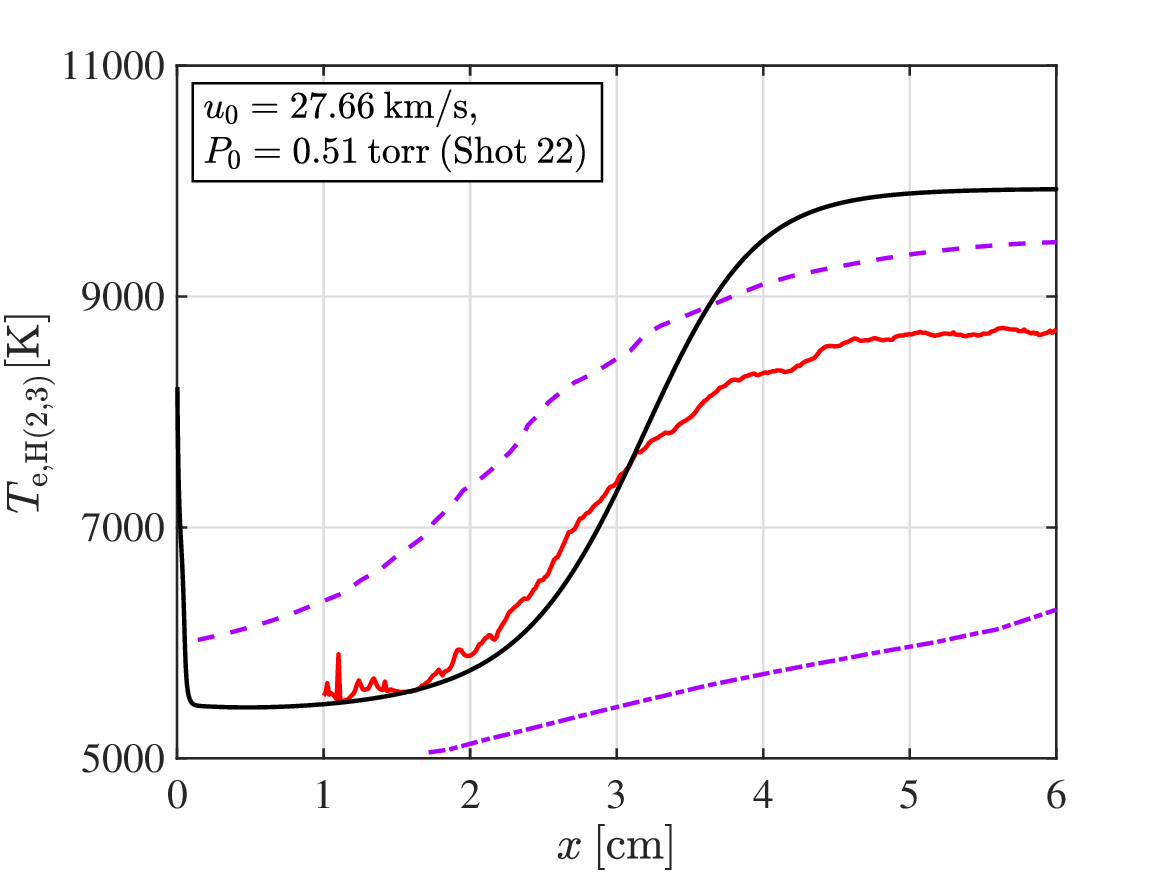}
     \end{subfigure}
     \hspace*{\fill}
    \caption{Second-level electronic temperatures of H for shot 7 (left) and shot 22 (right).}
    \label{fig:T23}
\end{figure}

Finally, the second-level electronic temperatures of H, computed as
\begin{equation}
    T_{\rm e,H(2,3)}=\frac{E_{{\rm H}(n=3)}-E_{{\rm H}(n=2)}}{k_{\rm B}\ln\left(\frac{N_{{\rm H(n=2)}}/g_{{\rm H(n=2)}}}{N_{{\rm H(n=3)}}/g_{{\rm H(n=3)}}}\right)},
\end{equation}
are shown in Fig.~\ref{fig:T23} alongside the predicted profiles by Colonna et al.~\cite{Colonna2020}, and the values extracted via Stark analysis of the Balmer-$\alpha$ line from the EAST experiments. There are two versions of the StS model by Colonna et al.; the first includes the species $\rm H_2$, $\rm H_2^+$, H, $\rm H^+$, He, $\rm He^+$, and $\rm e^-$ (referred to as their ``reduced'' model), while the only difference in the second is the inclusion of the additional species $\rm H_3^+$ and $\rm H^-$ (referred to as their ``full'' model). In both versions, the vibrational states of $\rm H_2$, the electronic states of $\rm H_2$, H, and He, and a non-Maxwellian electron energy distribution function are treated explicitly, such that over 100 pseudo-species are included. Both versions use the heavy-particle-impact excitation and ionization rate constants of Drawin~\cite{Drawin1969}.

Overall, the present model predicts $T_{\rm e,H(2,3)}$ more faithfully than either model of Colonna et al., despite only treating 11 species explicitly. The agreement in the induction zone temperatures is particularly satisfying, and indicates that the heavy-particle-impact rate constants of the low-lying states are reasonably accurate, as the H({\it n}) populations in the induction zone are effectively only a function of the heavy-particle-impact rates (as implied by the results presented in section~\ref{sec:sens_kinetics}). Without the inclusion of $\rm H_3^+$ and $\rm H^-$, the model of Colonna et al. predicts a significantly slower relaxation, which results in a $T_{\rm e,H(2,3)}$ profile that severely underpredicts the experiments. With the inclusion of $\rm H_3^+$ and $\rm H^-$, a faster relaxation process is predicted, with profiles that are more comparable, but still overpredict, the experiments. Based on these results, Colonna et al. argued that the inclusion of $\rm H_3^+$ and $\rm H^-$ is necessary, as these species provide an alternate pathway to ionization in the early post-shock region. However, the pathway that they postulate, namely $\rm H_2 + H_2 \rightarrow H_3^+ + H^- \rightarrow \dots \rightarrow H^+ + e^-+\dots$, begins in the immediate post-shock region.  As a result, it proceeds too quickly to reproduce the experimentally-observed induction zone behavior.

\begin{figure}[hbt!]
     \centering
     \begin{subfigure}[b]{0.33\textwidth}
         \centering
         \includegraphics[width=\textwidth,trim={0cm 0cm 1.2cm 0cm},clip]{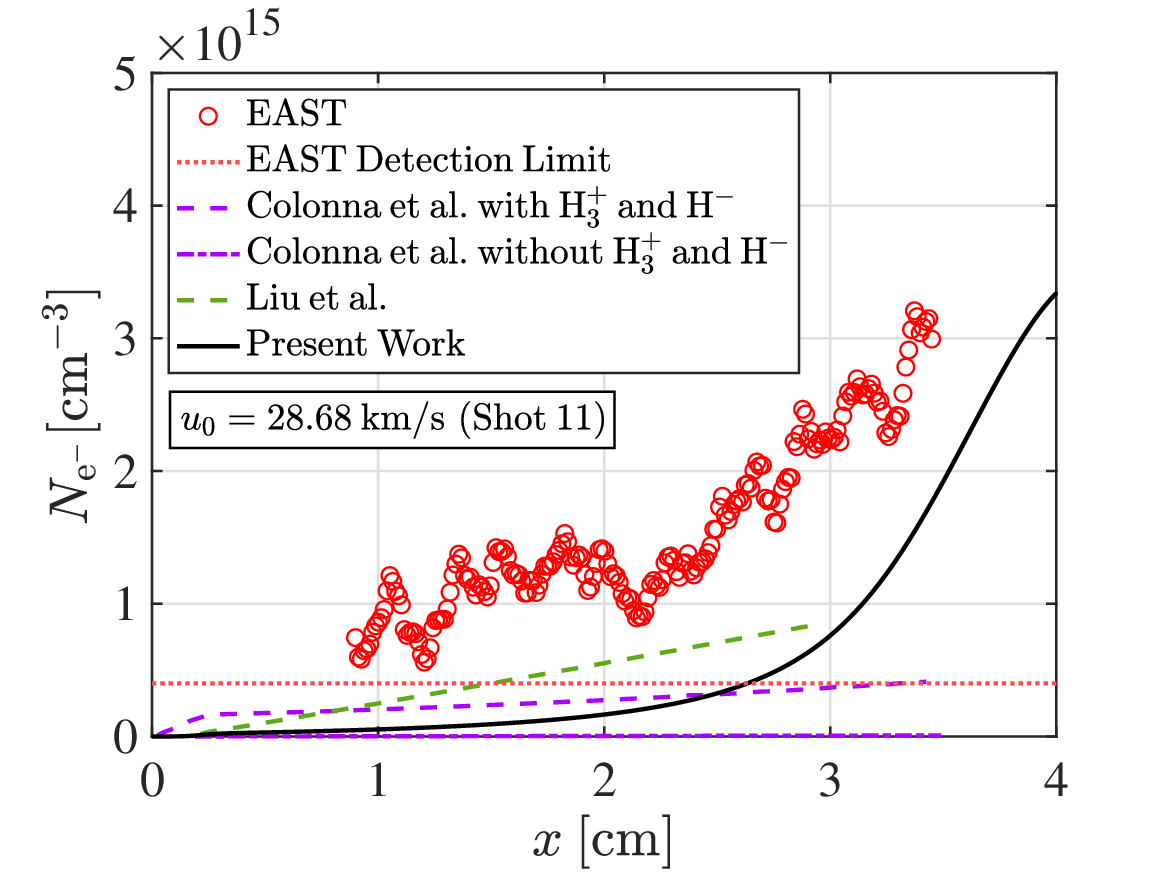}
     \end{subfigure}
     \hfill
     \begin{subfigure}[b]{0.33\textwidth}
         \centering
         \includegraphics[width=\textwidth,trim={0cm 0cm 1.2cm 0cm},clip]{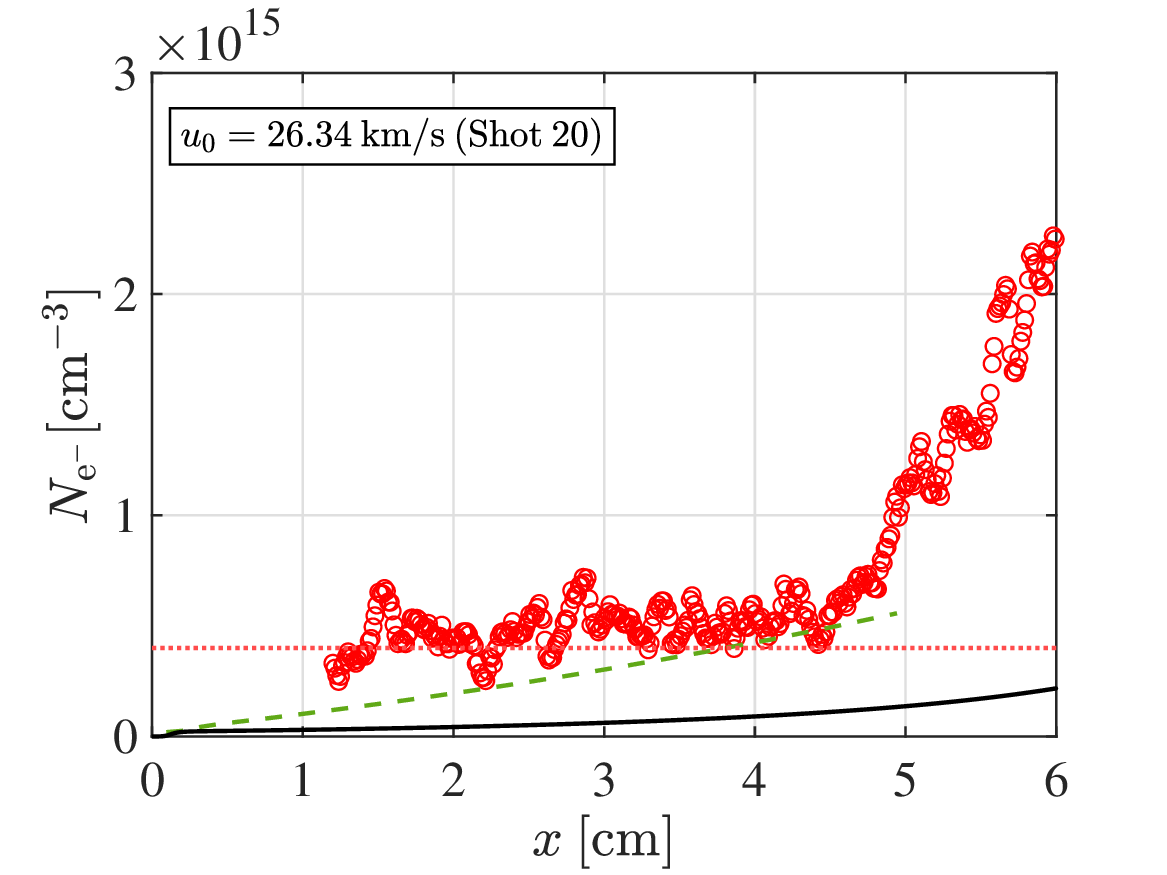}
     \end{subfigure}
     \hfill
     \begin{subfigure}[b]{0.33\textwidth}
         \centering
         \includegraphics[width=\textwidth,trim={0cm 0cm 1.2cm 0cm},clip]{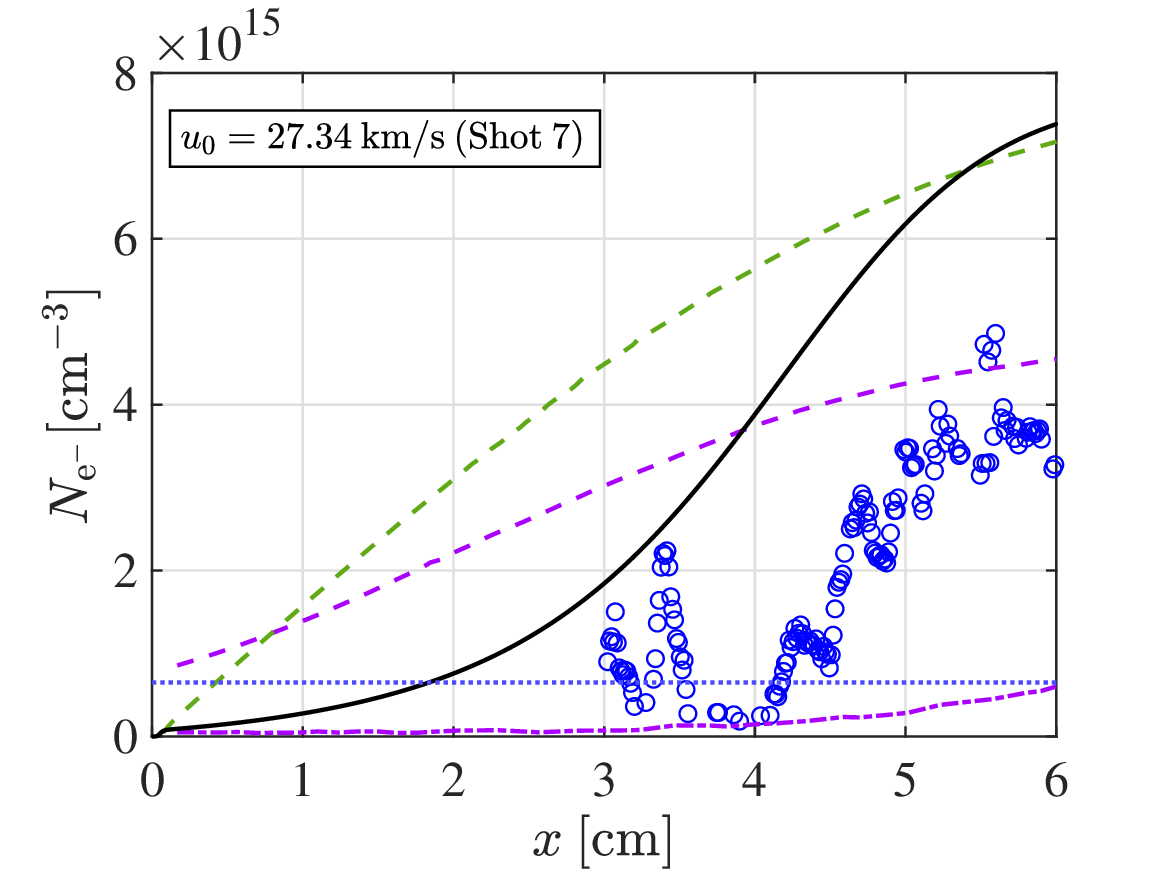}
     \end{subfigure}
     \begin{subfigure}[b]{0.33\textwidth}
         \centering
         \includegraphics[width=\textwidth,trim={0cm 0cm 1.2cm 0cm},clip]{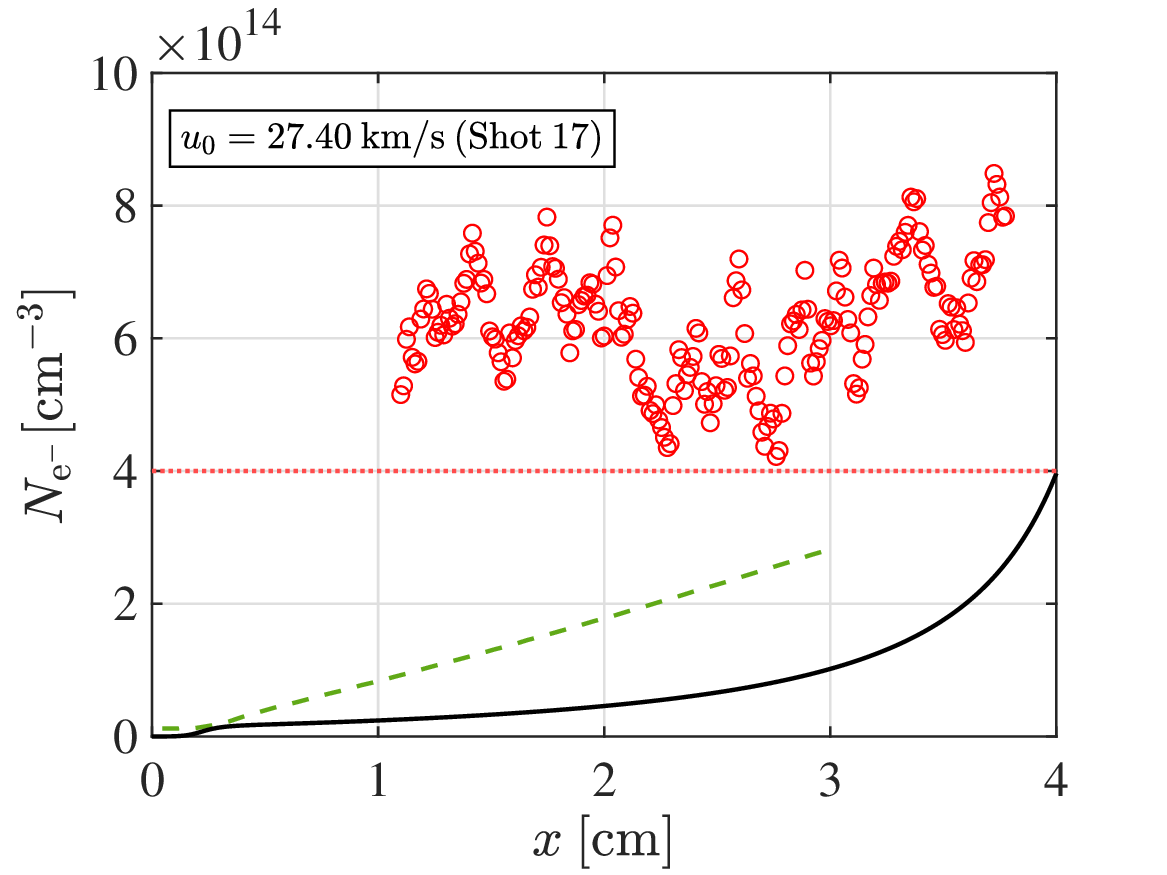}
         \caption{$\bm{P_0}$ = 0.1 torr}
     \end{subfigure}
     \hfill
     \begin{subfigure}[b]{0.33\textwidth}
         \centering
         \includegraphics[width=\textwidth,trim={0cm 0cm 1.2cm 0cm},clip]{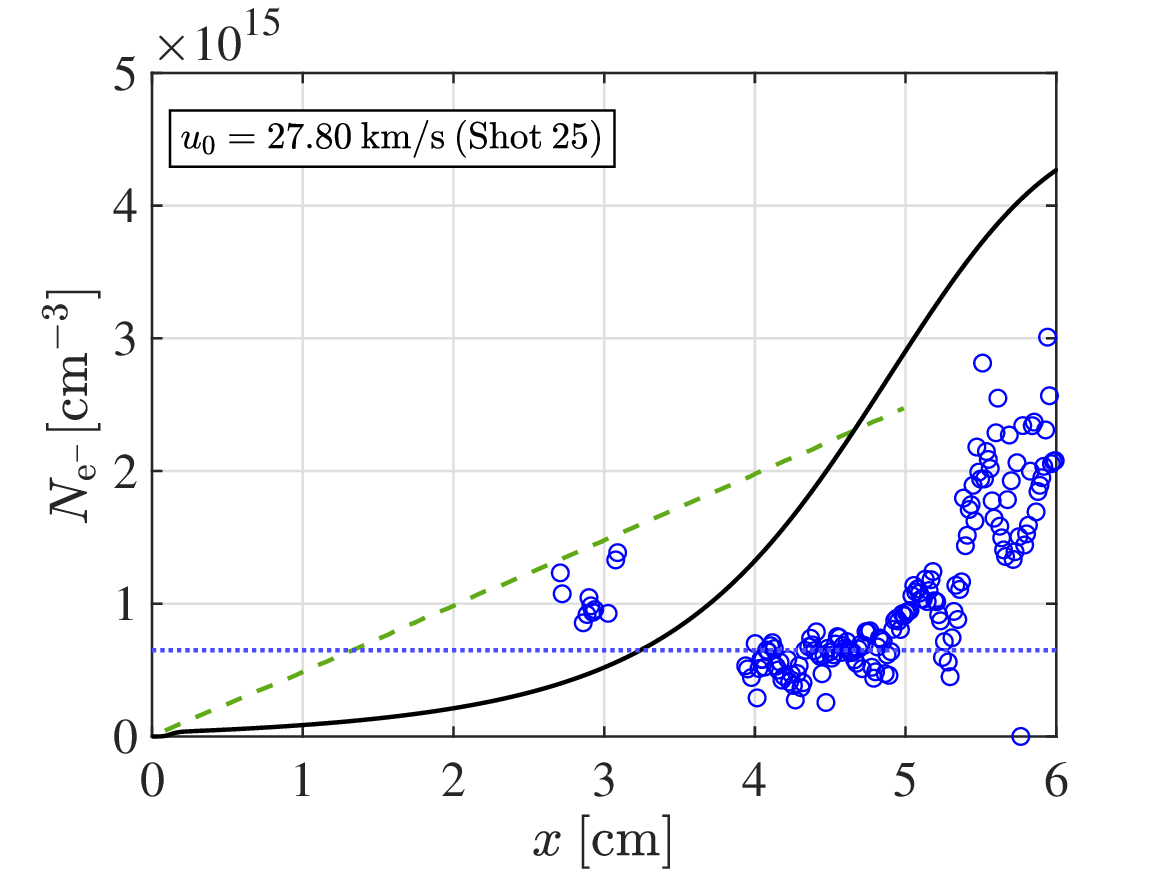}
         \caption{$\bm{P_0}$ = 0.2 torr}
     \end{subfigure}
     \hfill
     \begin{subfigure}[b]{0.33\textwidth}
         \centering
         \includegraphics[width=\textwidth,trim={0cm 0cm 1.2cm 0cm},clip]{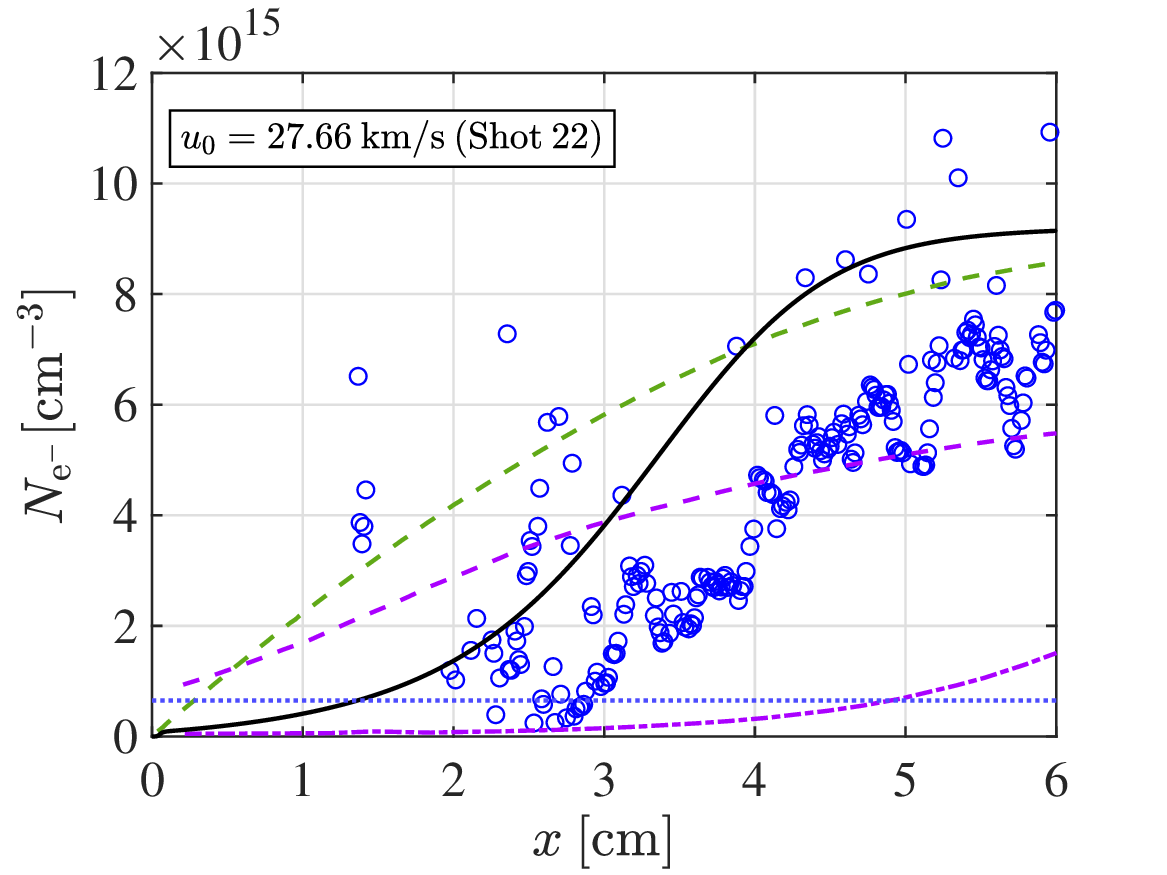}
         \caption{$\bm{P_0}$ = 0.5 torr}
     \end{subfigure}
    \caption{Electron number densities at increasing freestream pressures (left to right). Red and blue symbols indicate values extracted from the EAST experiments via the Balmer-$\bm \alpha$ and Balmer-$\bm \gamma$ lines, respectively.}
    \label{fig:ne}
\end{figure}

Figure~\ref{fig:ne} similarly shows the predicted electron number densities of the present work, alongside those of Colonna et al., Liu et al.~\cite{Liu2022}, and the experimentally-extracted values. The experimental values have been extracted via Stark analysis of the Balmer-$\alpha$ line for shots 11, 17, and 20 (shown as red symbols), and of the Balmer-$\gamma$ line for shots 7, 22, and 25 (shown as blue symbols)~\cite{Cruden2017}. The kinetic model of Liu et al. is based on a one-temperature framework which only treats the bulk species $\rm H_2$, H, $\rm H^+$, He, $\rm He^+$, and $\rm e^-$. Their rate constants are the same as those of Leibowitz~\cite{Leibowitz1973,Leibowitz1976}, except for the H electron and heavy-particle-impact ionization rate constants for which an ad hoc modification factor was applied.

For shots 11, 17, and 20, the simulated values of all models are smaller than the experimental values, and significant portions of the simulated profiles are below the detection limits of the experiments. For shot 11, the post-induction zone rise in $N_{\rm e^-}$ is best captured by the present work. As discussed in section~\ref{sec:sens_BL}, this is effectively entirely due to the treatment of boundary layer effects, which were neglected in the simulations of Colonna et al. and Liu et al. For shots 7, 22, and 25, the models of the present work, Liu et al., and Colonna et al. with $\rm H_3^+$ and $\rm H^-$ generally overpredict the experimental results (except Colonna et al. in the far post-shock region for shot 22). As in Fig.~\ref{fig:T23}, the model of Colonna et al. without $\rm H_3^+$ and $\rm H^-$ significantly underpredicts the experiments. Most notably however, the models of Colonna et al. and Liu et al. both predict a steady rise in $N_{\rm e^-}$ starting at the shock front and with no inflection point, and hence miss the induction zone behavior seen in the experiments and reproduced by the present work. As was shown in section~\ref{sec:sens_kinetics}, this induction zone behavior is controlled by the heavy-particle-impact processes, and thus suggests that the heavy-particle-impact rate constants are mischaracterized in these alternate models. Ultimately, the results of both Fig.~\ref{fig:T23} and \ref{fig:ne} imply that the present model gives improved predictions of the experimental profiles when compared to the models of Colonna et al. and Liu et al., despite not including the trace species $\rm H_3^+$, $\rm H^-$, or $\rm H_2^+$, and without the use of tuned rate constants.

\section{Conclusion}
\label{sec:conclusion}

In this work, the kinetics and radiation for high-speed $\rm H_2$/ He shocks were investigated using an 11-species thermochemical model, which explicitly treats the non-Boltzmann kinetics of atomic H in a state-specific framework. A detailed literature review of the electron and heavy-particle-impact kinetics of atomic H was conducted to construct this model. Then, a novel method for the treatment of coupled boundary layer effects in 1-D steady shocks was derived, and used to simulate $\rm H_2$/ He shocks corresponding to the experiments of Cruden and Bogdanoff~\cite{Cruden2017}. The simulated radiance profiles compared reasonably well to the experimental profiles, especially for the dominant mono-quantum Lyman-$\alpha$ and Balmer-$\alpha$ features of atomic H. To probe the impact of modeling uncertainties on the predicted radiance, an uncertainty analysis of the kinetics and boundary layer treatment was performed. From this, the sensitivity of the induction zone behavior to the heavy-particle-impact rates and boundary layer treatment method was revealed. Finally, the predictions of the present work were found to reproduce the experiments more accurately than the previous studies by Colonna et al.~\cite{Colonna2020} and Liu et al.~\cite{Liu2022}.

Overall, the present results suggest several key findings about the kinetic modeling and prediction of radiation in $\rm H_2$/ He shock environments. Firstly, there is currently a lack of reliable and consistent kinetic data available for heavy-particle-impact excitation and ionization of atomic H in the literature. Moreover, there is a large sensitivity for the predicted radiance in the induction zone to these heavy-particle-impact rates. Secondly, the correct treatment of boundary layer effects, including coupled compression effects, is critical for the accurate reproduction of experimental results, especially at the lower pressure conditions investigated in the present work ($P_0$ = 0.1 and 0.2 torr). Thirdly, while there are still some discrepancies remaining, the proposed kinetic model reproduces the experimental results more accurately than alternate models from the literature, despite not using any tuned rate constants, and by only explicitly treating the kinetics of $\rm H_2$ dissociation and H ionization.

To further improve radiance predictions for $\rm H_2$/ He shocks, future work should include the investigation of viscous effects and non-Boltzmann electronic state distributions for $\rm H_2$. Additionally, the studies by Park~\cite{park2011}, Cruden and Bogdanoff~\cite{Cruden2017}, and Coelho and Lino da Silva~\cite{Coelho2023} have all suggested that the trace methane ($\rm CH_4$) present in the atmospheres of the ice and gas giants may have a non-negligible impact on radiative heating. Therefore, $\rm CH_4$ and its associated dissociated/ ionized species should be incorporated into the kinetics, and the resulting model should be verified against the recent experiments of Steer et al.~\cite{Steer2023} and Cruden and Tibère-Inglesse~\cite{Cruden2024}.

\section*{Appendix A: Impact of Radiative Processes}

To simplify the solution procedure, radiative processes have been neglected in the present governing equations. This is equivalent to assuming that all radiation is completely self-absorbed/ optically thick. To assess the sensitivity of the results to this assumption, radiative transitions for H are incorporated into the kinetic model. Considering both bound-bound and free-bound (radiative recombination) transitions, the contribution of radiation to the chemical source term of H({\it n}) is given by
\begin{equation}
    \dot{\omega}_{\rm H({\it n}),rad} = \sum_{n'=n+1}^{n_{\rm max}} N_{\rm H({\it n'})}A_{n'\rightarrow n}\Lambda_{n'\rightarrow n}
    - \sum_{n'=1}^{n-1} N_{\rm H({\it n})}A_{n\rightarrow n'}\Lambda_{n\rightarrow n'}+N_{\rm H^+}N_{\rm e^-}A_{\infty\rightarrow n}\Lambda_{\infty\rightarrow n}.
\label{eqn:omega_rad}
\end{equation}
Here, the escape factor formulation is used to account for absorption, with values of $\Lambda=0$ and $\Lambda=1$ corresponding to the optically thick and thin limits, respectively. The corresponding total energy and electron energy radiative source terms are given by
\begin{equation}
    \dot{\Omega}_{\rm rad} = -\sum_{n=2}^{n_{\rm max}} \sum_{n'=1}^{n-1} N_{\rm H({\it n})}A_{n\rightarrow n'}\Lambda_{n\rightarrow n'}(E_{{\rm H}(n)}-E_{{\rm H}(n')}) -\sum_{n=1}^{n_{\rm max}}N_{\rm H^+}N_{\rm e^-}A_{\infty\rightarrow n}\Lambda_{\infty\rightarrow n}(E_{\rm H^+}-E_{{\rm H}(n)})
\label{eqn:Omega_rad}
\end{equation}
and
\begin{equation}
    \dot{\Omega}_{\rm e^-,rad} = -\sum_{n=1}^{n_{\rm max}}N_{\rm H^+}N_{\rm e^-}A_{\infty\rightarrow n}\Lambda_{\infty\rightarrow n}(E_{\rm H^+}-E_{{\rm H}(n)}),
\label{eqn:Omegae_rad}
\end{equation}
respectively. Bound-bound Einstein coefficients ($A_{n\rightarrow n'}$) are taken from the NIST database~\cite{NIST_ASD} while free-bound rate constants ($A_{\infty\rightarrow n}$) are taken from Janev et al.~\cite{Janev2003}. $\rm H_2$ and free-free (Bremsstrahlung) radiation are neglected.

\begin{figure}[hbt!]
     \centering
     \begin{subfigure}[b]{0.4\textwidth}
         \centering
         \includegraphics[width=\textwidth]{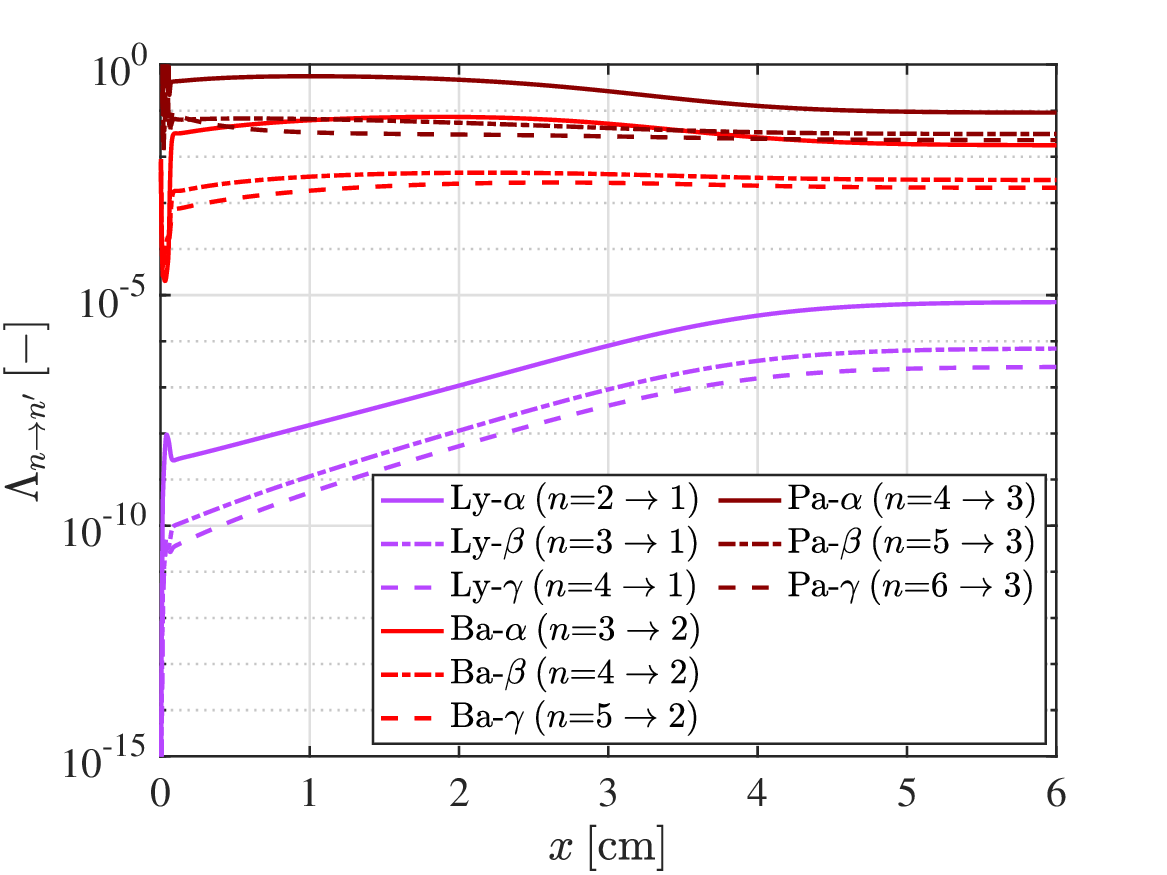}
     \end{subfigure}
        \caption{Escape factors as computed by NEQAIR for the shot 23 case.}
        \label{fig:rad_escf}
\end{figure}

To compute escape factors, the local temperature and number density values from the optically-thick calculations of section~\ref{sec:flowfield} are used as inputs into NEQAIR's escape factor routine (under the local approximation with a characteristic distance of 1.0 cm as recommended by~\cite{Brandis2019Neqair}). The resulting escape factors for the $\alpha$, $\beta$, and $\gamma$ lines of the Lyman, Balmer, and Paschen series for the shot 23 conditions are shown in Fig.~\ref{fig:rad_escf}. From these results, it is clear that the escape factor for each transition varies considerably with $x$. Therefore, to obtain a conservative estimate of radiative losses, the maximum value (in $x$) of each transition's escape factor is used in Eq.~\eqref{eqn:omega_rad}-\eqref{eqn:Omegae_rad} for the Lyman, Balmer, and Paschen series. All higher-order and free-bound transitions are assumed to be optically thin.

\begin{figure}[hbt!]
     \centering
     \hspace*{\fill}
     \begin{subfigure}[b]{0.4\textwidth}
         \centering
         \includegraphics[width=\textwidth,trim={0cm 0cm 1.2cm 0.5cm},clip]{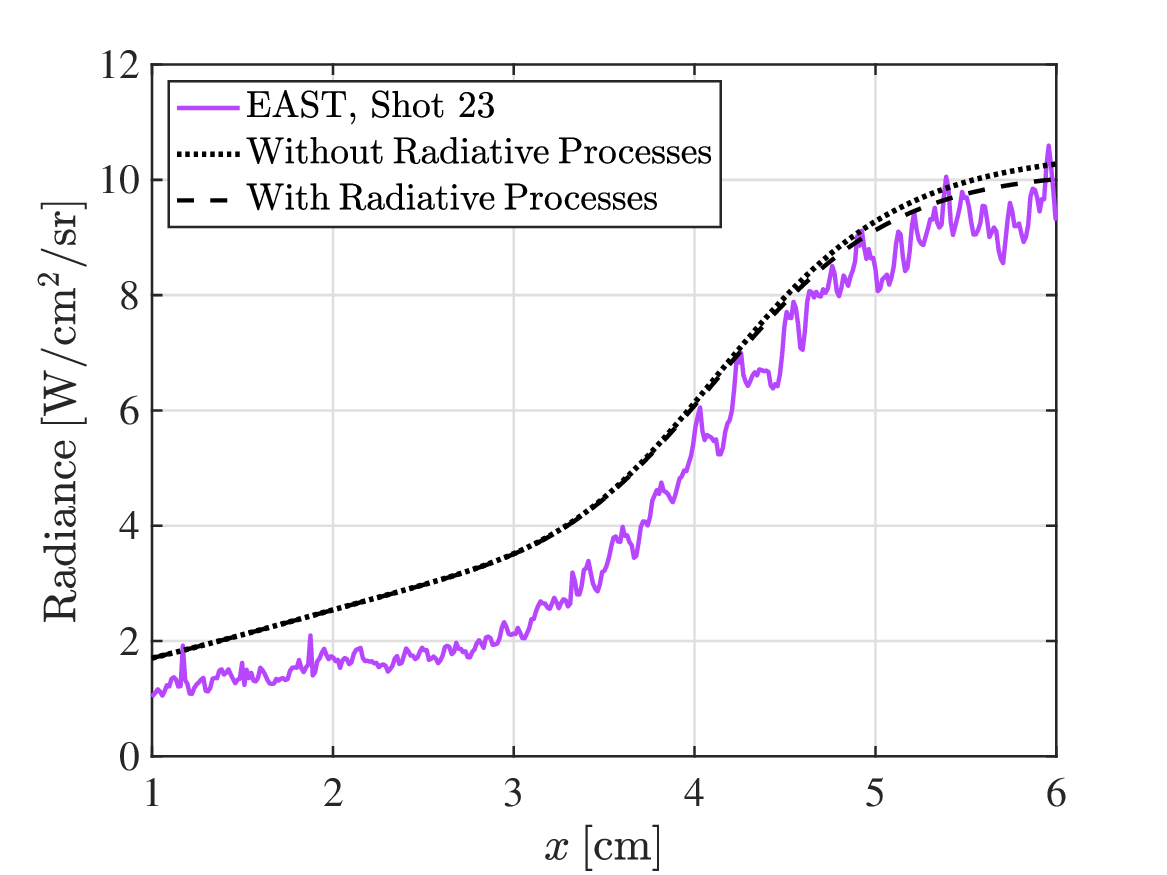}
     \end{subfigure}
     \hfill
     \begin{subfigure}[b]{0.4\textwidth}
         \centering
         \includegraphics[width=\textwidth,trim={0cm 0cm 1.2cm 0.5cm},clip]{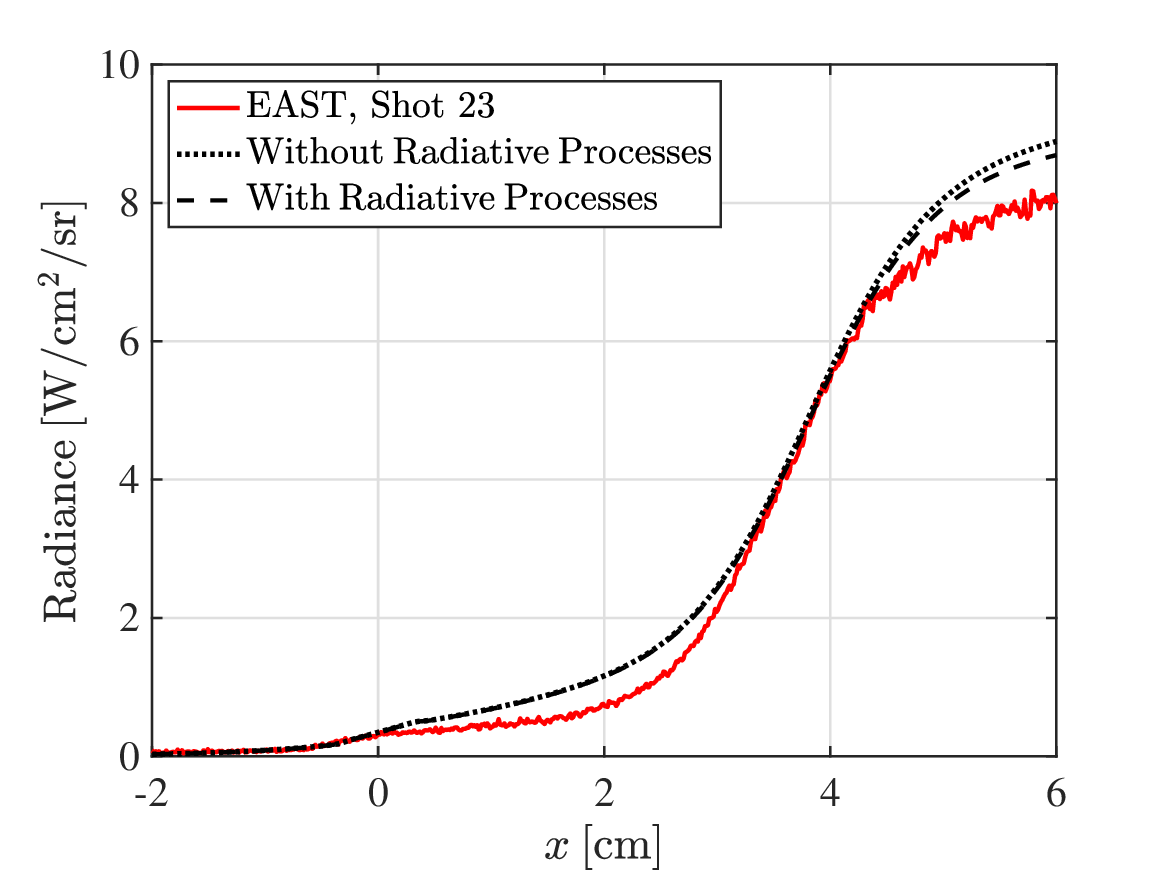}
     \end{subfigure}
     \hspace*{\fill}
    \caption{Integrated radiance profiles for the Lyman-$\bm\alpha$ (117 to 130 nm, left) and the Balmer-$\bm \alpha$ (653 to 659 nm, right) features with and without radiative processes.}
    \label{fig:rad}
\end{figure}

The resulting integrated radiance profiles for the Lyman-$\alpha$ and Balmer-$\alpha$ features for shot 23 are shown in Fig~\ref{fig:rad}. The impact of radiative processes are found to be negligible. Similar results were obtained for all shot conditions.

\section*{Appendix B: Impact of $\rm \mathbf{H_2^+}$}

Throughout the present work, the associative ionization reaction which results in the production of $\rm H_2^+$, i.e.,
\begin{equation}
    \rm H({\it n}\geq2) + H({\it n}=1) \leftrightarrow H_2^+ + e^-,
    \label{eqn:AI}
\end{equation}
has been neglected. However, for high-speed flows in air mixtures, associative ionization reactions are known to play a key role in the post-shock ionization process~\cite{Boyd2021,Aiken2025}. To assess the impact of associative ionization for $\rm H_2$/ He mixtures, reaction~\eqref{eqn:AI} is incorporated in the present kinetic model using the state-specific rate constants of Mihajlov et al.~\cite{Mihajlov2011} and Srećković et al.~\cite{Sreckovic2018}.

\begin{figure}[hbt!]
     \centering
     \begin{subfigure}[b]{0.33\textwidth}
         \centering
         \includegraphics[width=\textwidth,trim={0cm 0cm 1.2cm 0.5cm},clip]{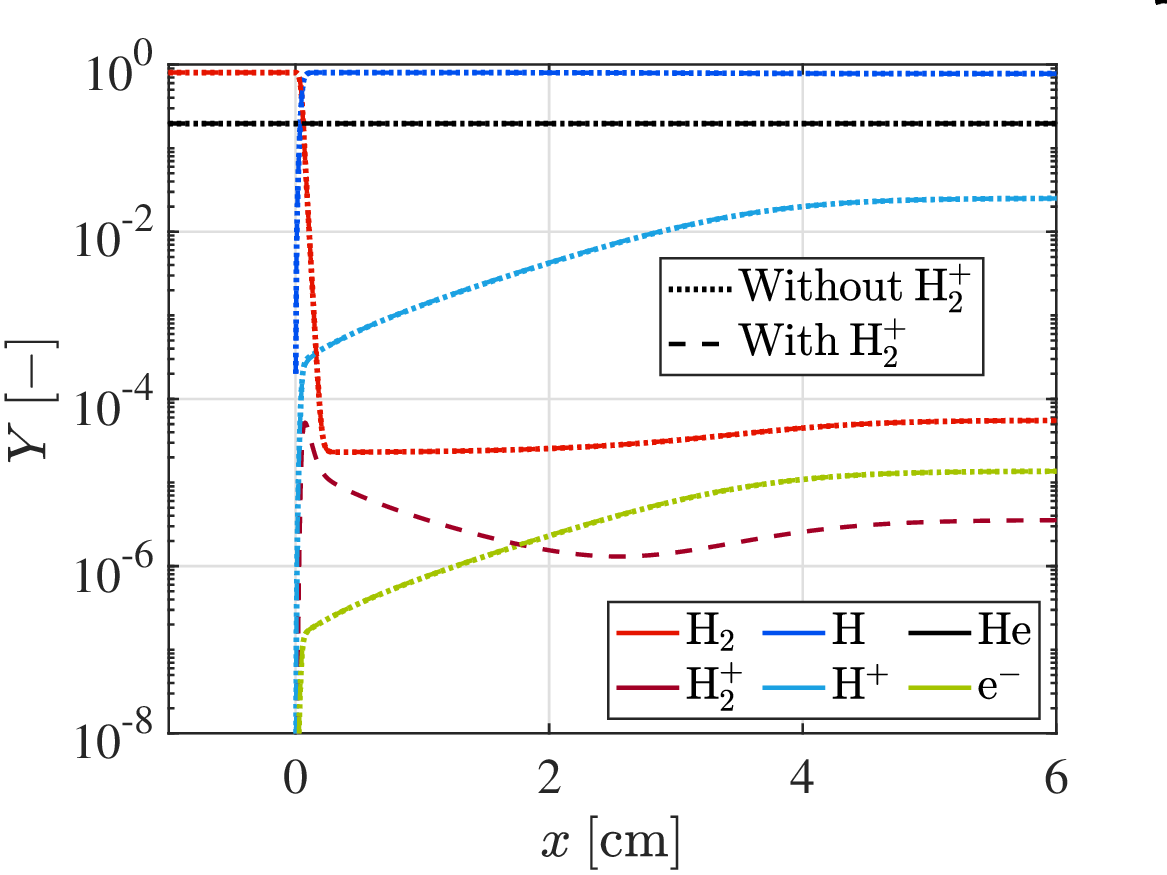}
     \end{subfigure}
     \hfill
     \begin{subfigure}[b]{0.33\textwidth}
         \centering
         \includegraphics[width=\textwidth,trim={0cm 0cm 1.2cm 0.5cm},clip]{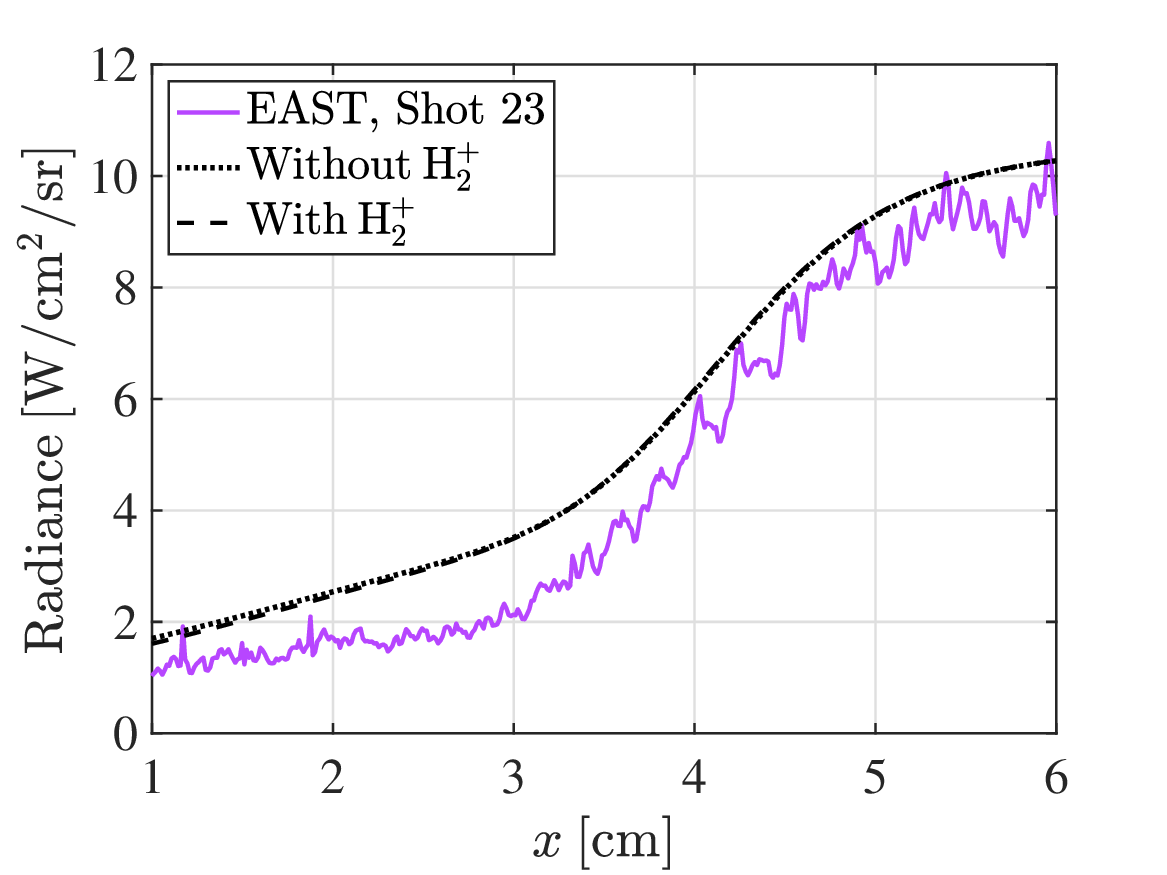}
     \end{subfigure}
     \hfill
     \begin{subfigure}[b]{0.33\textwidth}
         \centering
         \includegraphics[width=\textwidth,trim={0cm 0cm 1.2cm 0.5cm},clip]{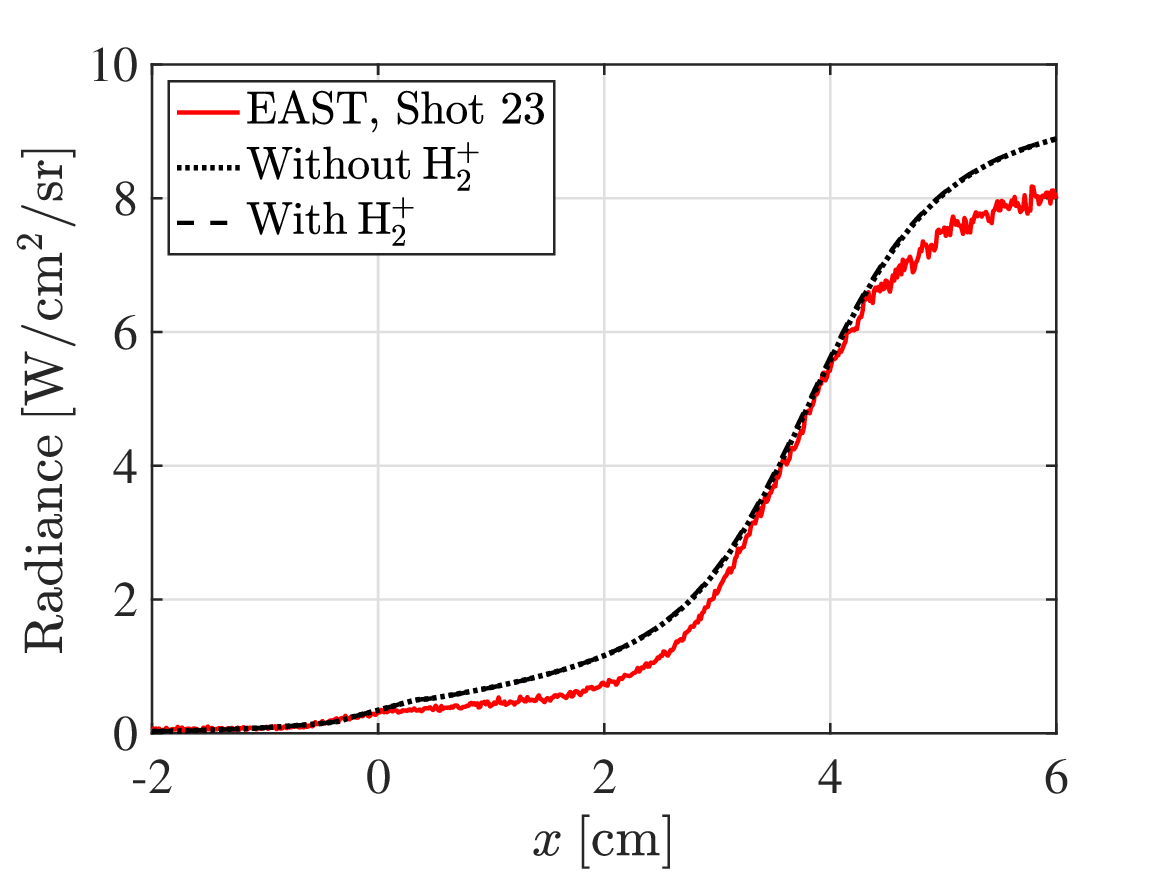}
     \end{subfigure}
    \caption{Simulated mass fractions (left) and integrated radiance profiles for the Lyman-$\bm\alpha$ (117 to 130 nm, middle) and the Balmer-$\bm \alpha$ (653 to 659 nm, right) features with and without $\bf H_2^+$ in the kinetic model.}
    \label{fig:H2+}
\end{figure}

Figure~\ref{fig:H2+} shows the computed mass fraction and Lyman-$\alpha$/ Balmer-$\alpha$ integrated radiance profiles for shot 23 with and without $\rm H_2^+$. Other than the non-zero mass fraction of $\rm H_2^+$, the inclusion of $\rm H_2^+$ has a negligible impact on all computed quantities of interest. The same result was found for all shot conditions. This suggests that $\rm H_2^+$ can be neglected in simulations of $\rm H_2$/ He shocks, at least at conditions similar to those considered in the present work.

\begin{figure}[hbt!]
     \centering
     \begin{subfigure}[b]{0.4\textwidth}
         \centering
         \includegraphics[width=\textwidth,trim={0cm 0cm 1cm 0cm},clip]{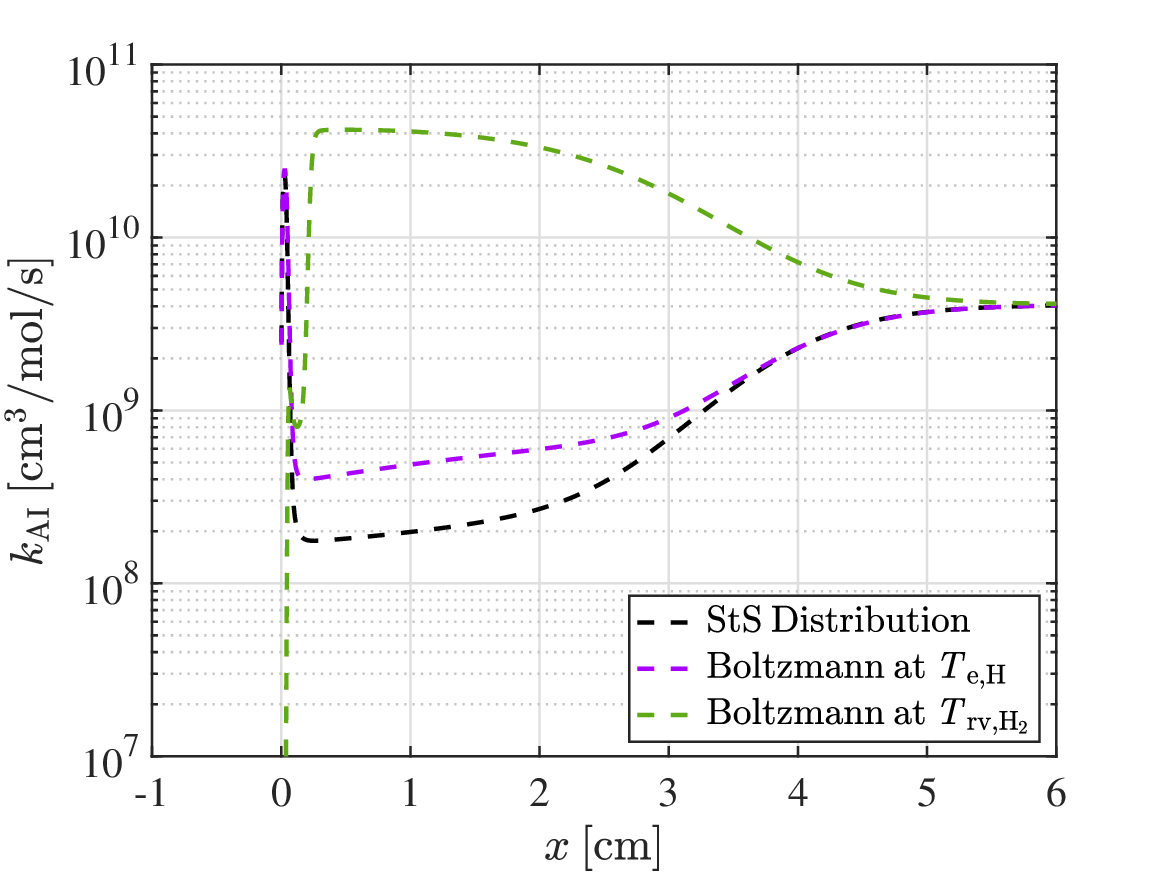}
     \end{subfigure}
    \caption{Macroscopic associative ionization rate constants computed a posteriori for the shot 23 case with $\bf H_2^+$.}
    \label{fig:kAI}
\end{figure}

This is the opposite conclusion to that reached in the recent ice giant aerothermal uncertainty analysis study of Rataczak et al.~\cite{Rataczak2025}. Specifically, Rataczak et al. found that the inclusion of $\rm H_2^+$ had a significant impact on ionization profiles and hence radiative heating predictions, despite using the same associative ionization rate constants as the present analysis. This seeming contradiction is likely due to the fact that Rataczak et al. used a traditional two-temperature kinetic model which assumed that the electronic states of H follow a Boltzmann distribution about a mixture-averaged vibrational-electronic temperature, $T_{\rm ve}$. To illustrate the impact of such an assumption, Fig.~\ref{fig:kAI} shows the effective macroscopic associative ionization rate constant for shot 23, where
\begin{equation}
    k_{\rm AI} = \frac{\sum_{n=1}^{n_{\rm max}}N_{{\rm H}(n)}k_{{\rm AI,H}(n)}(T_{\rm t,h})}{\sum_{n=1}^{n_{\rm max}}N_{{\rm H}(n)}}
\end{equation}
is computed a posteriori by either using the exact non-Boltzmann H({\it n}) distribution from the present StS simulation, or a ``reconstructed'' Boltzmann H({\it n}) distribution at $T_{\rm e,H}$ or $T_{\rm rv,H_2}$. While the $T_{\rm e,H}$ Boltzmann case overpredicts the StS rate constant by approximately a factor of two, the $T_{\rm rv,H_2}$ Boltzmann case overpredicts the StS rate constant by two orders of magnitude. These results further underscore the importance of treating non-Boltzmann H({\it n}) distributions in $\rm H_2$/ He shock simulations.

\section*{Acknowledgments}
This work is supported by the NASA entry systems modeling project under grant number 80NSSC21K1751.

\bibliography{main}

@article{Park2001,
  doi = {10.2514/2.6582},
  year = {2001},
  month = jan,
  publisher = {American Institute of Aeronautics and Astronautics ({AIAA})},
  volume = {15},
  number = {1},
  pages = {76--90},
  author = {Chul Park and Richard L. Jaffe and Harry Partridge},
  title = {Chemical-Kinetic Parameters of Hyperbolic Earth Entry},
  journal = {Journal of Thermophysics and Heat Transfer}
}

@article{Boyd1997,
  doi = {10.1063/1.869474},
  year = {1997},
  month = oct,
  publisher = {{AIP} Publishing},
  volume = {9},
  number = {10},
  pages = {3086--3095},
  author = {Iain D. Boyd},
  title = {Monte Carlo simulation of nonequilibrium flow in a low-power hydrogen arcjet},
  journal = {Physics of Fluids}
}

@article{Leibowitz1976,
  doi = {10.2514/3.61465},
  year = {1976},
  month = sep,
  publisher = {American Institute of Aeronautics and Astronautics ({AIAA})},
  volume = {14},
  number = {9},
  pages = {1324--1329},
  author = {Lewis P. Leibowitz and Ta-Jin Kuo},
  title = {Ionizational Nonequilibrium Heating During Outer Planetary Entries},
  journal = {{AIAA} Journal}
}

@article{Aggarwal1991,
  doi = {10.1088/0953-4075/24/6/024},
  year = {1991},
  month = mar,
  publisher = {{IOP} Publishing},
  volume = {24},
  number = {6},
  pages = {1385--1410},
  author = {K M Aggarwal and K A Berrington and P G Burke and A E Kingston and A Pathak},
  title = {Electron collision cross sections at low energies for all transitions between the n=1,  2,  3,  4 and 5 levels of atomic hydrogen},
  journal = {Journal of Physics B: Atomic,  Molecular and Optical Physics}
}

@article{Park1971,
  doi = {10.1016/0022-4073(71)90158-0},
  year = {1971},
  month = jan,
  publisher = {Elsevier {BV}},
  volume = {11},
  number = {1},
  pages = {7--36},
  author = {C. Park},
  title = {Electron impact excitation rate coefficients for hydrogen,  helium and alkali atoms},
  journal = {Journal of Quantitative Spectroscopy and Radiative Transfer}
}

@article{Park1989,
  doi = {10.2514/3.28771},
  year = {1989},
  month = jul,
  publisher = {American Institute of Aeronautics and Astronautics ({AIAA})},
  volume = {3},
  number = {3},
  pages = {233--244},
  author = {Chul Park},
  title = {Assessment of two-temperature kinetic model for ionizing air},
  journal = {Journal of Thermophysics and Heat Transfer}
}

@book{Park1989NonequilibriumHA,
  title={Nonequilibrium Hypersonic Aerothermodynamics},
  author={Chul Hi Park},
  year={1989}
}

@article{Park2011,
  doi = {10.2514/1.51810},
  year = {2011},
  month = nov,
  publisher = {American Institute of Aeronautics and Astronautics ({AIAA})},
  volume = {48},
  number = {6},
  pages = {897--903},
  author = {Chul Park},
  title = {Nonequilibrium Chemistry and Radiation for Neptune Entry},
  journal = {Journal of Spacecraft and Rockets}
}

@inproceedings{Steer2023,
  doi = {10.2514/6.2023-1729},
  year = {2023},
  month = jan,
  publisher = {American Institute of Aeronautics and Astronautics},
  author = {Joseph Steer and Peter L. Collen and Alex B. Glenn and Christopher Hambidge and Luke J. Doherty and Matthew McGilvray and Stefan Loehle and Louis Walpot},
  title = {Shock Radiation Tests for Ice Giant Entry Probes Including CH$_4$ in the T6 Free-Piston Driven Wind Tunnel},
  booktitle = {{AIAA} {SCITECH} 2023 Forum}
}

@article{Coelho2023,
  doi = {10.1016/j.asr.2022.12.024},
  year = {2023},
  month = apr,
  publisher = {Elsevier {BV}},
  volume = {71},
  number = {8},
  pages = {3408--3432},
  author = {Jo{\~{a}}o Coelho and M{\'{a}}rio Lino da Silva},
  title = {Aerothermodynamic analysis of Neptune ballistic entry and aerocapture flows},
  journal = {Advances in Space Research}
}

@inproceedings{Hansson2021,
  doi = {10.2514/6.2021-0706},
  year = {2021},
  month = jan,
  publisher = {American Institute of Aeronautics and Astronautics},
  author = {Kaelan Hansson and Alex T. Carroll and Savio J. Poovathingal and Iain D. Boyd},
  title = {Analysis of Chemical Kinetic Parameters for Hydrogen Atmospheres},
  booktitle = {{AIAA} Scitech 2021 Forum}
}

@book{Janev1987,
  doi = {10.1007/978-3-642-71935-6},
  year = {1987},
  publisher = {Springer Berlin Heidelberg},
  author = {Ratko K. Janev and William D. Langer and Douglass E. Post and Kenneth Evans},
  title = {Elementary Processes in Hydrogen-Helium Plasmas}
}

@article{Colonna2020,
  doi = {10.1016/j.ijheatmasstransfer.2020.119916},
  year = {2020},
  month = aug,
  publisher = {Elsevier {BV}},
  volume = {156},
  pages = {119916},
  author = {Gianpiero Colonna and Lucia Daniela Pietanza and Annarita Laricchiuta},
  title = {Ionization kinetic model for Hydrogen-Helium atmospheres in hypersonic shock tubes},
  journal = {International Journal of Heat and Mass Transfer}
}

@article{Colonna2018,
  doi = {10.1016/j.sab.2018.01.009},
  year = {2018},
  month = mar,
  publisher = {Elsevier {BV}},
  volume = {141},
  pages = {85--93},
  author = {Gianpiero Colonna and Annarita Laricchiuta and Lucia Daniela Pietanza},
  title = {Modeling plasma heating by ns laser pulse},
  journal = {Spectrochimica Acta Part B: Atomic Spectroscopy}
}

@article{Colonna2017,
  doi = {10.1140/epjd/e2017-80080-3},
  year = {2017},
  month = nov,
  publisher = {Springer Science and Business Media {LLC}},
  volume = {71},
  number = {11},
  author = {Gianpiero Colonna and Lucia D. Pietanza and Giuliano D'Ammando and Roberto Celiberto and Mario Capitelli and Annarita Laricchiuta},
  title = {Vibrational kinetics of electronically excited states in H$_2$ discharges},
  journal = {The European Physical Journal D}
}

@article{Cruden2017,
  doi = {10.2514/1.a33891},
  year = {2017},
  month = nov,
  publisher = {American Institute of Aeronautics and Astronautics ({AIAA})},
  volume = {54},
  number = {6},
  pages = {1246--1257},
  author = {Brett A. Cruden and David W. Bogdanoff},
  title = {Shock Radiation Tests for Saturn and Uranus Entry Probes},
  journal = {Journal of Spacecraft and Rockets}
}

@article{Higdon2018,
  doi = {10.2514/1.t5275},
  year = {2018},
  month = jul,
  publisher = {American Institute of Aeronautics and Astronautics ({AIAA})},
  volume = {32},
  number = {3},
  pages = {680--690},
  author = {Kyle J. Higdon and Brett A. Cruden and Aaron M. Brandis and Derek S. Liechty and David B. Goldstein and Philip L. Varghese},
  title = {Direct Simulation Monte Carlo Shock Simulation of Saturn Entry Probe Conditions},
  journal = {Journal of Thermophysics and Heat Transfer}
}

@article{Palmer2014,
  doi = {10.2514/1.a32768},
  year = {2014},
  month = may,
  publisher = {American Institute of Aeronautics and Astronautics ({AIAA})},
  volume = {51},
  number = {3},
  pages = {801--814},
  author = {Grant Palmer and Dinesh Prabhu and Brett A. Cruden},
  title = {Aeroheating Uncertainties in Uranus and Saturn Entries by the Monte Carlo Method},
  journal = {Journal of Spacecraft and Rockets}
}

@article{Mansbach1969,
  doi = {10.1103/physrev.181.275},
  year = {1969},
  month = may,
  publisher = {American Physical Society ({APS})},
  volume = {181},
  number = {1},
  pages = {275--289},
  author = {Peter Mansbach and James Keck},
  title = {Monte Carlo Trajectory Calculations of Atomic Excitation and Ionization by Thermal Electrons},
  journal = {Physical Review}
}

@article{Johnson1972,
  doi = {10.1086/151486},
  year = {1972},
  month = may,
  publisher = {American Astronomical Society},
  volume = {174},
  pages = {227},
  author = {L. C. Johnson},
  title = {Approximations for Collisional and Radiative Transition Rates in Atomic Hydrogen},
  journal = {The Astrophysical Journal}
}

@article{Vriens1980,
  doi = {10.1103/physreva.22.940},
  year = {1980},
  month = sep,
  publisher = {American Physical Society ({APS})},
  volume = {22},
  number = {3},
  pages = {940--951},
  author = {L. Vriens and A. H. M. Smeets},
  title = {Cross-section and rate formulas for electron-impact ionization,  excitation,  deexcitation,  and total depopulation of excited atoms},
  journal = {Physical Review A}
}

@article{Scholz1990,
  doi = {10.1093/mnras/242.4.692},
  year = {1990},
  month = aug,
  publisher = {Oxford University Press ({OUP})},
  volume = {242},
  number = {4},
  pages = {692--697},
  author = {T. T. Scholz and H. R. J. Walters and P. G. Burke and M. P. Scott},
  title = {Effective collision strengths for 1s{\textendash}2s and 1s{\textendash}2p electron{\textendash}hydrogen atom scattering},
  journal = {Monthly Notices of the Royal Astronomical Society}
}

@article{Callaway1994,
  doi = {10.1006/adnd.1994.1009},
  year = {1994},
  month = may,
  publisher = {Elsevier {BV}},
  volume = {57},
  number = {1-2},
  pages = {9--20},
  author = {J. Callaway},
  title = {Effective Collision Strengths for Hydrogen and Hydrogen-Like Ions},
  journal = {Atomic Data and Nuclear Data Tables}
}

@article{Anderson2000,
  doi = {10.1088/0953-4075/33/6/311},
  year = {2000},
  month = mar,
  publisher = {{IOP} Publishing},
  volume = {33},
  number = {6},
  pages = {1255--1262},
  author = {H Anderson and C P Ballance and N R Badnell and H P Summers},
  title = {An R-matrix with pseudo-states approach to the electron-impact excitation of H I for diagnostic applications in fusion plasmas},
  journal = {Journal of Physics B: Atomic,  Molecular and Optical Physics}
}

@article{Anderson2002,
  doi = {10.1088/0953-4075/35/6/701},
  year = {2002},
  month = mar,
  publisher = {{IOP} Publishing},
  volume = {35},
  number = {6},
  pages = {1613--1615},
  author = {H Anderson and C P Ballance and N R Badnell and H P Summers},
  title = {An R-matrix with pseudo-states approach to the electron-impact excitation of H I for diagnostic applications in fusion plasmas},
  journal = {Journal of Physics B: Atomic,  Molecular and Optical Physics}
}

@article{Przybilla2004,
  doi = {10.1086/421316},
  year = {2004},
  month = jul,
  publisher = {American Astronomical Society},
  volume = {609},
  number = {2},
  pages = {1181--1191},
  author = {Norbert Przybilla and Keith Butler},
  title = {Non-{LTE} Line Formation for Hydrogen Revisited},
  journal = {The Astrophysical Journal}
}

@book{Janev1993,
  author = {R. K. Janev and J. J. Smith},
  title = {Cross sections for collision processes of hydrogen atoms with electrons, protons and multiply charged ions},
  year = {1993},
  publisher = {International Atomic Energy Agency (IAEA)},
}

@techreport{Janev2003,
  author = {R. K. Janev and D. Reiter and U. Samm},
  title = {Collision processes in low-temperature hydrogen plasmas},
  institution = {Institut für Plasmaphysik EURATOM Association, Trilateral Euregio Cluster},
  year = {2003},
}

@article{Barklem2007,
  doi = {10.1051/0004-6361:20066686},
  year = {2007},
  month = feb,
  publisher = {{EDP} Sciences},
  volume = {466},
  number = {1},
  pages = {327--337},
  author = {P. S. Barklem},
  title = {Non-{LTE} Balmer line formation in late-type spectra: effects of atomic processes involving hydrogen atoms},
  journal = {Astronomy \& Astrophysics}
}

@article{Barklem2011,
  doi = {10.1051/0004-6361/201116745},
  year = {2011},
  month = may,
  publisher = {{EDP} Sciences},
  volume = {530},
  pages = {A94},
  author = {P. S. Barklem and A. K. Belyaev and M. Guitou and N. Feautrier and F. X. Gad{\'{e}}a and A. Spielfiedel},
  title = {On inelastic hydrogen atom collisions in stellar atmospheres},
  journal = {Astronomy \& Astrophysics}
}

@article{Mihajlov2011,
  doi = {10.1088/0067-0049/193/1/2},
  year = {2011},
  month = jan,
  publisher = {American Astronomical Society},
  volume = {193},
  number = {1},
  pages = {2},
  author = {Anatolij A. Mihajlov and Ljubinko M. Ignjatovi{\'{c}} and Vladimir A. Sre{\'{c}}kovi{\'{c}} and Milan S. Dimitrijevi{\'{c}}},
  title = {Chemi-Ionization in Solar Photosphere: Influence on the Hydrogen Atom Excited States Population},
  journal = {The Astrophysical Journal Supplement Series}
}

@article{Mihajlov2004,
  doi = {10.1088/0953-4075/37/22/008},
  year = {2004},
  month = nov,
  publisher = {{IOP} Publishing},
  volume = {37},
  number = {22},
  pages = {4493--4506},
  author = {A A Mihajlov and Lj M Ignjatović and Z Djurić and N N Ljepojević},
  title = {The rate coefficients for the processes of (n-n')-mixing in collisions of Rydberg atoms H*(n) with H(1s) atoms},
  journal = {Journal of Physics B: Atomic,  Molecular and Optical Physics}
}

@article{Dimitrijevic2021,
  doi = {10.1016/j.newast.2020.101529},
  year = {2021},
  month = apr,
  publisher = {Elsevier {BV}},
  volume = {84},
  pages = {101529},
  author = {Milan S. Dimitrijevi{\'{c}} and Vladimir A. Sre{\'{c}}kovi{\'{c}} and Ljubinko M. Ignjatovi{\'{c}} and Bratislav P. Marinkovi{\'{c}}},
  title = {The role of some collisional processes in {AGNs}: Rate coefficients needed for modeling},
  journal = {New Astronomy}
}

@article{Sreckovic2018,
  doi = {10.1093/mnras/sty2256},
  year = {2018},
  month = aug,
  publisher = {Oxford University Press ({OUP})},
  author = {V A Sre{\'{c}}kovi{\'{c}} and M S Dimitrijevi{\'{c}} and Lj M Ignjatovi{\'{c}}},
  title = {Atom-Rydberg atom chemi-ionization/recombination processes in the hydrogen clouds in Broad Line Region of {AGNs}},
  journal = {Monthly Notices of the Royal Astronomical Society}
}

@article{Drawin1969,
  doi = {10.1007/bf01392775},
  year = {1969},
  month = oct,
  publisher = {Springer Science and Business Media {LLC}},
  volume = {225},
  number = {5},
  pages = {483--493},
  author = {H. W. Drawin},
  title = {Influence of atom-atom collisions on the collisional-radiative ionization and recombination coefficients of hydrogen plasmas},
  journal = {Zeitschrift f\"{u}r Physik A Hadrons and nuclei}
}

@article{Drawin1973,
  doi = {10.1016/0375-9601(73)90331-9},
  year = {1973},
  month = mar,
  publisher = {Elsevier {BV}},
  volume = {43},
  number = {4},
  pages = {333--335},
  author = {H.W. Drawin and F. Emard},
  title = {Atom-atom excitation and ionization in shock waves of the noble gases},
  journal = {Physics Letters A}
}

@article{Thomson1912,
  doi = {10.1080/14786440408637241},
  year = {1912},
  month = apr,
  publisher = {Informa {UK} Limited},
  volume = {23},
  number = {136},
  pages = {449--457},
  author = {J.J. Thomson},
  title = {Ionization by moving electrified particles},
  journal = {The London,  Edinburgh,  and Dublin Philosophical Magazine and Journal of Science}
}

@article{Fleischmann1972,
  doi = {10.1007/bf01379737},
  year = {1972},
  month = oct,
  publisher = {Springer Science and Business Media {LLC}},
  volume = {252},
  number = {5},
  pages = {435--442},
  author = {H. H. Fleischmann and R. C. Dehmel},
  title = {On Drawin{\textquotesingle}s formula for ionization in atom-atom collisions},
  journal = {Zeitschrift f\"{u}r Physik}
}

@article{Lambert1993,
  doi = {10.1088/0031-8949/1993/t47/030},
  year = {1993},
  month = jan,
  publisher = {{IOP} Publishing},
  volume = {T47},
  pages = {186--198},
  author = {David L Lambert},
  title = {Quantitative stellar spectroscopy with large optical telescopes},
  journal = {Physica Scripta}
}

@ARTICLE{Steenbock1984,
       author = {{Steenbock}, W. and {Holweger}, H.},
        title = "{Statistical equilibrium of lithium in cool stars of different metallicity}",
      journal = {Astronomy \& Astrophysics},
         year = 1984,
        month = jan,
       volume = {130},
       number = {2},
        pages = {319-323},
       adsurl = {https://ui.adsabs.harvard.edu/abs/1984A&A...130..319S}
}

@article{Park2012,
  doi = {10.2514/1.t3689},
  year = {2012},
  month = apr,
  publisher = {American Institute of Aeronautics and Astronautics ({AIAA})},
  volume = {26},
  number = {2},
  pages = {231-243},
  author = {Chul Park},
  title = {Nonequilibrium Ionization and Radiation in Hydrogen-Helium Mixtures},
  journal = {Journal of Thermophysics and Heat Transfer}
}

@article{Livingston1976,
  doi = {10.2514/3.61466},
  year = {1976},
  month = sep,
  publisher = {American Institute of Aeronautics and Astronautics ({AIAA})},
  volume = {14},
  number = {9},
  pages = {1335--1337},
  author = {Floyd R. Livingston and P. T. Y. Poon},
  title = {Relaxation Distance and Equilibrium Electron Density Measurements in Hydrogen-Helium Plasmas},
  journal = {{AIAA} Journal}
}

@inproceedings{Carroll2023_conv,
  title = {Stagnation Point Convective Heating Correlations for Entry into H$_2$/He Atmospheres},
  DOI = {10.2514/6.2023-0208},
  booktitle = {AIAA SCITECH 2023 Forum},
  publisher = {American Institute of Aeronautics and Astronautics},
  author = {Carroll,  Alex T. and Brandis,  Aaron M.},
  year = {2023},
  month = jan 
}

@article{Liu2020,
  title = {Using Aerothermodynamic Similarity to Experimentally Study Nonequilibrium Giant Planet Entry},
  volume = {57},
  ISSN = {1533-6794},
  DOI = {10.2514/1.a34713},
  number = {5},
  journal = {Journal of Spacecraft and Rockets},
  publisher = {American Institute of Aeronautics and Astronautics (AIAA)},
  author = {Liu,  Yu and James,  Christopher M. and Morgan,  Richard G. and McIntyre,  Timothy J.},
  year = {2020},
  month = sep,
  pages = {1008–1020}
}

@inproceedings{Chaudhry2020,
  title = {Implementation of a Chemical Kinetics Model for Hypersonic Flows in Air for High-Performance {CFD}},
  DOI = {10.2514/6.2020-2191},
  booktitle = {AIAA Scitech 2020 Forum},
  publisher = {American Institute of Aeronautics and Astronautics},
  author = {Chaudhry,  Ross S. and Boyd,  Iain D. and Torres,  Erik and Schwartzentruber,  Thomas E. and Candler,  Graham V.},
  year = {2020},
  month = jan 
}

@article{Singh2020_1,
  title = {Consistent kinetic–continuum dissociation model {I.} Kinetic formulation},
  volume = {152},
  ISSN = {1089-7690},
  DOI = {10.1063/1.5142752},
  number = {22},
  journal = {The Journal of Chemical Physics},
  publisher = {AIP Publishing},
  author = {Singh,  Narendra and Schwartzentruber,  Thomas},
  year = {2020},
  month = jun 
}

@article{Magin2012,
  title = {Coarse-grain model for internal energy excitation and dissociation of molecular nitrogen},
  volume = {398},
  ISSN = {0301-0104},
  DOI = {10.1016/j.chemphys.2011.10.009},
  journal = {Chemical Physics},
  publisher = {Elsevier BV},
  author = {Magin,  Thierry E. and Panesi,  Marco and Bourdon,  Anne and Jaffe,  Richard L. and Schwenke,  David W.},
  year = {2012},
  month = apr,
  pages = {90–95}
}

@article{Liu2015,
  title = {General multi-group macroscopic modeling for thermo-chemical non-equilibrium gas mixtures},
  volume = {142},
  ISSN = {1089-7690},
  DOI = {10.1063/1.4915926},
  number = {13},
  journal = {The Journal of Chemical Physics},
  publisher = {AIP Publishing},
  author = {Liu,  Yen and Panesi,  Marco and Sahai,  Amal and Vinokur,  Marcel},
  year = {2015},
  month = apr 
}

@article{Venturi2020,
  title = {Data-Inspired and Physics-Driven Model Reduction for Dissociation: Application to the {$\rm O_2$} + {O} System},
  volume = {124},
  ISSN = {1520-5215},
  DOI = {10.1021/acs.jpca.0c04516},
  number = {41},
  journal = {The Journal of Physical Chemistry A},
  publisher = {American Chemical Society (ACS)},
  author = {Venturi,  S. and Sharma,  M. P. and Lopez,  B. and Panesi,  M.},
  year = {2020},
  month = sep,
  pages = {8359–8372}
}

@article{Shah1987,
  title = {Pulsed crossed-beam study of the ionisation of atomic hydrogen by electron impact},
  volume = {20},
  ISSN = {0022-3700},
  DOI = {10.1088/0022-3700/20/14/022},
  number = {14},
  journal = {Journal of Physics B: Atomic and Molecular Physics},
  publisher = {IOP Publishing},
  author = {Shah,  M B and Elliott,  D S and Gilbody,  H B},
  year = {1987},
  month = jul,
  pages = {3501–3514}
}

@article{Belyaev2003,
  title = {Cross sections for low-energy inelastic H + Li collisions},
  volume = {68},
  ISSN = {1094-1622},
  DOI = {10.1103/physreva.68.062703},
  number = {6},
  journal = {Physical Review A},
  publisher = {American Physical Society (APS)},
  author = {Belyaev, Andrey K. and Barklem, Paul S.},
  year = {2003},
  month = dec 
}

@article{Barklem2003,
  title = {Inelastic H+Li and H$^-$+Li$^+$ collisions and non-LTE Li I line formation in stellar atmospheres},
  volume = {409},
  ISSN = {1432-0746},
  DOI = {10.1051/0004-6361:20031199},
  number = {2},
  journal = {Astronomy \& Astrophysics},
  publisher = {EDP Sciences},
  author = {Barklem, P. S. and Belyaev, A. K. and Asplund, M.},
  year = {2003},
  month = oct,
  pages = {L1–L4}
}

@article{Belyaev2010,
  title = {Cross sections for low-energy inelastic H + Na collisions},
  volume = {81},
  ISSN = {1094-1622},
  DOI = {10.1103/physreva.81.032706},
  number = {3},
  journal = {Physical Review A},
  publisher = {American Physical Society (APS)},
  author = {Belyaev,  A. K. and Barklem,  P. S. and Dickinson,  A. S. and Gadéa,  F. X.},
  year = {2010},
  month = mar 
}

@article{Barklem2010,
  title = {Inelastic Na+H collision data for non-LTE applications in stellar atmospheres},
  volume = {519},
  ISSN = {1432-0746},
  DOI = {10.1051/0004-6361/201015152},
  journal = {Astronomy \& Astrophysics},
  publisher = {EDP Sciences},
  author = {Barklem,  P. S. and Belyaev,  A. K. and Dickinson,  A. S. and Gadéa,  F. X.},
  year = {2010},
  month = sep,
  pages = {A20}
}

@article{Guitou2011,
  title = {Inelastic Mg+H collision processes at low energies},
  volume = {44},
  ISSN = {1361-6455},
  DOI = {10.1088/0953-4075/44/3/035202},
  number = {3},
  journal = {Journal of Physics B: Atomic,  Molecular and Optical Physics},
  publisher = {IOP Publishing},
  author = {Guitou,  M and Belyaev,  A K and Barklem,  P S and Spielfiedel,  A and Feautrier,  N},
  year = {2011},
  month = jan,
  pages = {035202}
}

@article{Barklem2012,
  title = {Inelastic Mg+H collision data for non-LTE applications in stellar atmospheres},
  volume = {541},
  ISSN = {1432-0746},
  DOI = {10.1051/0004-6361/201219081},
  journal = {Astronomy \& Astrophysics},
  publisher = {EDP Sciences},
  author = {Barklem,  P. S. and Belyaev,  A. K. and Spielfiedel,  A. and Guitou,  M. and Feautrier,  N.},
  year = {2012},
  month = may,
  pages = {A80}
}

@article{Leibowitz1973,
  doi = {10.1063/1.1694174},
  year = {1973},
  month = jan,
  publisher = {{AIP} Publishing},
  volume = {16},
  number = {1},
  pages = {59--68},
  author = {Lewis P. Leibowitz},
  title = {Measurements of the structure of an ionizing shock wave in a hydrogen-helium mixture},
  journal = {The Physics of Fluids}
}

@article{Janev1979,
  title = {Excitation and deexcitation processes in slow collisions of Rydberg atoms with ground-state parent atoms},
  volume = {20},
  ISSN = {0556-2791},
  DOI = {10.1103/physreva.20.1890},
  number = {5},
  journal = {Physical Review A},
  publisher = {American Physical Society (APS)},
  author = {Janev,  R. K. and Mihajlov,  A. A.},
  year = {1979},
  month = nov,
  pages = {1890–1904}
}

@article{Birely1972,
  title = {Formation of $\mathrm{H}(2p)$ and $\mathrm{H}(2s)$ in Collisions of 1-25-keV Hydrogen Atoms with the Rare Gases},
  author = {Birely, J. H. and McNeal, R. J.},
  journal = {Phys. Rev. A},
  volume = {5},
  issue = {1},
  pages = {257--265},
  numpages = {0},
  year = {1972},
  month = {Jan},
  publisher = {American Physical Society},
  doi = {10.1103/PhysRevA.5.257}
}

@article{Sauers1974,
  title = {Small-angle scattering of metastable hydrogen formed by collisions of neutral hydrogen atoms with gas targets},
  author = {Sauers, I. and Thomas, E. W.},
  journal = {Phys. Rev. A},
  volume = {10},
  issue = {3},
  pages = {822--828},
  numpages = {0},
  year = {1974},
  month = {Sep},
  publisher = {American Physical Society},
  doi = {10.1103/PhysRevA.10.822}
}

@article{VanZyl1987,
  title = {Lyman-\ensuremath{\alpha} emission from low-energy H impact on rare-gas atoms},
  author = {Van Zyl, B. and Gealy, M. W.},
  journal = {Phys. Rev. A},
  volume = {35},
  issue = {9},
  pages = {3741--3748},
  numpages = {0},
  year = {1987},
  month = {May},
  publisher = {American Physical Society},
  doi = {10.1103/PhysRevA.35.3741}
}

@article{Grosser1984,
  title={Hydrogen 2s and 2p excitation in low energy H, D+, He collisions},
  author={Grosser, J and Kr{\"u}ger, W},
  journal={Zeitschrift f{\"u}r Physik. A, Atoms and nuclei},
  volume={318},
  number={1},
  pages={25--30},
  year={1984}
}

@techreport{Hunter1990,
  author       = {Hunter, H T and Kirkpatrick, M I and Alvarez, I and Cisneros, C and Phaneuf, R A and Barnett, C F},
  title        = {Atomic data for fusion},
  institution  = {Oak Ridge National Lab. (ORNL), Oak Ridge, TN (United States)},
  doi          = {10.2172/6570226},
  place        = {United States},
  year         = {1990},
  month        = {07}
}

@article{AlAtawneh2021,
  title = {Excitation cross sections in a collision between two ground-state hydrogen atoms},
  volume = {54},
  ISSN = {1361-6455},
  DOI = {10.1088/1361-6455/abece3},
  number = {6},
  journal = {Journal of Physics B: Atomic,  Molecular and Optical Physics},
  publisher = {IOP Publishing},
  author = {Al Atawneh,  Saed J and Tőkési,  K},
  year = {2021},
  month = mar,
  pages = {065202}
}

@article{Hill1980,
  title = {Excitation of fast metastable hydrogen atoms to the n=3 state in passage through gases},
  volume = {13},
  ISSN = {0022-3700},
  DOI = {10.1088/0022-3700/13/5/020},
  number = {5},
  journal = {Journal of Physics B: Atomic and Molecular Physics},
  publisher = {IOP Publishing},
  author = {Hill,  J and Geddes,  J and Gilbody,  H B},
  year = {1980},
  month = mar,
  pages = {951–958}
}

@article{Hill1979,
  title = {1s-2s excitation of fast hydrogen atoms in collisions with atomic and molecular hydrogen},
  volume = {12},
  ISSN = {0022-3700},
  DOI = {10.1088/0022-3700/12/17/016},
  number = {17},
  journal = {Journal of Physics B: Atomic and Molecular Physics},
  publisher = {IOP Publishing},
  author = {Hill,  J and Geddes,  J and Gilbody,  H B},
  year = {1979},
  month = sep,
  pages = {2875–2882}
}

@article{Riley1999,
  title = {Excitation and ionization in H(1s)-H(1s) collisions},
  volume = {32},
  ISSN = {1361-6455},
  DOI = {10.1088/0953-4075/32/22/306},
  number = {22},
  journal = {Journal of Physics B: Atomic,  Molecular and Optical Physics},
  publisher = {IOP Publishing},
  author = {Riley,  Merle E and Ritchie,  A Burke},
  year = {1999},
  month = nov,
  pages = {5279–5288}
}

@article{Stenrup2009,
  title = {Mutual neutralization in low-energy ${\mathrm{H}}^{+}+{\mathrm{H}}^{\ensuremath{-}}$ collisions: A quantum ab initio study},
  author = {Stenrup, Michael and Larson, \AA{}sa and Elander, Nils},
  journal = {Phys. Rev. A},
  volume = {79},
  issue = {1},
  pages = {012713},
  numpages = {12},
  year = {2009},
  month = {Jan},
  publisher = {American Physical Society},
  doi = {10.1103/PhysRevA.79.012713}
}

@article{Szucs1984,
  title = {Experimental study of the mutual neutralisation of H+and H-between 5 and 2000 eV},
  volume = {17},
  ISSN = {0022-3700},
  DOI = {10.1088/0022-3700/17/8/021},
  number = {8},
  journal = {Journal of Physics B: Atomic and Molecular Physics},
  publisher = {IOP Publishing},
  author = {Szucs,  S and Karemera,  M and Terao,  M and Brouillard,  F},
  year = {1984},
  month = apr,
  pages = {1613–1622}
}

@article{Bates1955,
  title = {Inelastic Heavy Particle Collisions Involving the Crossing of Potential Energy Curves III: Charge Transfer from Negative Ions of Atomic Hydrogen to Protons},
  volume = {68},
  ISSN = {0370-1298},
  DOI = {10.1088/0370-1298/68/3/306},
  number = {3},
  journal = {Proceedings of the Physical Society. Section A},
  publisher = {IOP Publishing},
  author = {Bates,  D R and Lewis,  J T},
  year = {1955},
  month = mar,
  pages = {173–180}
}

@inbook{Bray1995,
  title = {Calculation of Electron Scattering on Hydrogenic Targets},
  ISSN = {1049-250X},
  DOI = {10.1016/s1049-250x(08)60164-0},
  booktitle = {Advances In Atomic,  Molecular,  and Optical Physics},
  publisher = {Elsevier},
  author = {Bray,  Igor and Stelbovics,  Andris T.},
  year = {1995},
  pages = {209–254}
}

@article{Griffin2005,
  title = {The validity of classical trajectory and perturbative quantal methods for electron-impact ionization from excited states in H-like ions},
  volume = {38},
  ISSN = {1361-6455},
  DOI = {10.1088/0953-4075/38/12/l01},
  number = {12},
  journal = {Journal of Physics B: Atomic,  Molecular and Optical Physics},
  publisher = {IOP Publishing},
  author = {Griffin,  D C and Ballance,  C P and Pindzola,  M S and Robicheaux,  F and Loch,  S D and Ludlow,  J A and Witthoeft,  M C and Colgan,  J and Fontes,  C J and Schultz,  D R},
  year = {2005},
  month = jun,
  pages = {L199–L206}
}

@article{Witthoeft2004,
  title = {Time-dependent close-coupling theory for e + H elastic and inelastic scattering},
  volume = {70},
  ISSN = {1094-1622},
  DOI = {10.1103/physreva.70.022711},
  number = {2},
  journal = {Physical Review A},
  publisher = {American Physical Society (APS)},
  author = {Witthoeft,  M. C. and Loch,  S. D. and Pindzola,  M. S.},
  year = {2004},
  month = aug 
}

@article{Pindzola1996,
  title = {Total ionization cross section for electron-hydrogen scattering using a time-dependent close-coupling method},
  volume = {54},
  ISSN = {1094-1622},
  DOI = {10.1103/physreva.54.2142},
  number = {3},
  journal = {Physical Review A},
  publisher = {American Physical Society (APS)},
  author = {Pindzola,  M. S. and Robicheaux,  F.},
  year = {1996},
  month = sep,
  pages = {2142–2145}
}

@article{Defrance1981,
  title = {Electron impact ionisation of metastable atomic hydrogen},
  volume = {14},
  ISSN = {0022-3700},
  DOI = {10.1088/0022-3700/14/1/013},
  number = {1},
  journal = {Journal of Physics B: Atomic and Molecular Physics},
  publisher = {IOP Publishing},
  author = {Defrance,  P and Claeys,  W and Cornet,  A and Poulaert,  G},
  year = {1981},
  month = jan,
  pages = {111–117}
}

@article{Bartschat1996,
  title = {Electron-impact ionization of atomic hydrogen from the 1S and 2S states},
  volume = {29},
  ISSN = {1361-6455},
  DOI = {10.1088/0953-4075/29/15/005},
  number = {15},
  journal = {Journal of Physics B: Atomic,  Molecular and Optical Physics},
  publisher = {IOP Publishing},
  author = {Bartschat,  Klaus and Bray,  Igor},
  year = {1996},
  month = aug,
  pages = {L577–L583}
}

@article{Furudate2009,
  title = {Nonequilibrium Calculation of High-Temperature Radiating H$_2$-He Flowfield},
  volume = {23},
  ISSN = {1533-6808},
  DOI = {10.2514/1.43961},
  number = {4},
  journal = {Journal of Thermophysics and Heat Transfer},
  publisher = {American Institute of Aeronautics and Astronautics (AIAA)},
  author = {Furudate,  Michiko},
  year = {2009},
  month = oct,
  pages = {651–659}
}

@article{Clarke2023,
  title = {Spatial Transformations for Reacting Gas Shock Tube Experiments},
  volume = {61},
  ISSN = {1533-385X},
  DOI = {10.2514/1.j062604},
  number = {8},
  journal = {AIAA Journal},
  publisher = {American Institute of Aeronautics and Astronautics (AIAA)},
  author = {Clarke,  Justin and Di Mare,  Luca and McGilvray,  Matthew},
  year = {2023},
  month = aug,
  pages = {3365–3374}
}

@article{Clarke2024,
  title = {Quasi-one-dimensional non-equilibrium method for shock tube and stagnation line flows},
  volume = {36},
  ISSN = {1089-7666},
  DOI = {10.1063/5.0218676},
  number = {9},
  journal = {Physics of Fluids},
  publisher = {AIP Publishing},
  author = {Clarke,  J. and Brody,  S. and Steer,  J. and McGilvray,  M. and Di Mare,  L.},
  year = {2024},
  month = sep 
}

@article{Mirels1963,
  title = {Test Time in Low-Pressure Shock Tubes},
  volume = {6},
  ISSN = {0031-9171},
  DOI = {10.1063/1.1706887},
  number = {9},
  journal = {The Physics of Fluids},
  publisher = {AIP Publishing},
  author = {Mirels,  Harold},
  year = {1963},
  month = sep,
  pages = {1201–1214}
}

@article{Bartschat2013,
  title = {Computational methods for electron–atom collisions in plasma applications},
  volume = {46},
  ISSN = {1361-6463},
  DOI = {10.1088/0022-3727/46/33/334004},
  number = {33},
  journal = {Journal of Physics D: Applied Physics},
  publisher = {IOP Publishing},
  author = {Bartschat,  Klaus},
  year = {2013},
  month = aug,
  pages = {334004}
}

@article{Carroll2026_diss1,
  title = {Treatment of Thermal Nonequilibrium Dissociation: Theory and Application to H$_2$},
  ISSN = {1533-6808},
  DOI = {10.2514/1.t7242},
  journal = {Journal of Thermophysics and Heat Transfer},
  publisher = {American Institute of Aeronautics and Astronautics (AIAA)},
  author = {Carroll,  Alex T. and Blanquart,  Guillaume and Brandis,  Aaron M. and Cruden,  Brett A.},
  year = {2026},
  month = jan,
  pages = {1–16}
}

@article{Carroll2026_diss2,
  title = {Treatment of Thermal Nonequilibrium Dissociation: Rate Constant Review for H$_2$},
  author = {Carroll,  Alex T. and Wolmer,  Jacob and Blanquart,  Guillaume and Brandis,  Aaron M. and Cruden,  Brett A.},
  journal = {Journal of Thermophysics and Heat Transfer},
  publisher = {American Institute of Aeronautics and Astronautics (AIAA)},
  year = {2026 (in press)},
}

@article{Vegiri1998,
  title = {Theoretical investigation of metastable hydrogen de-excitation in collisions with He and Ne},
  volume = {31},
  ISSN = {1361-6455},
  DOI = {10.1088/0953-4075/31/3/015},
  number = {3},
  journal = {Journal of Physics B: Atomic,  Molecular and Optical Physics},
  publisher = {IOP Publishing},
  author = {Vegiri,  Aliki},
  year = {1998},
  month = feb,
  pages = {473–489}
}

@article{Appleton1964,
  title = {The conservation equations for a non-equilibrium plasma},
  volume = {20},
  ISSN = {1469-7645},
  DOI = {10.1017/s0022112064001458},
  number = {4},
  journal = {Journal of Fluid Mechanics},
  publisher = {Cambridge University Press (CUP)},
  author = {Appleton,  J. P. and Bray,  K. N. C.},
  year = {1964},
  month = dec,
  pages = {659–672}
}

@article{Mirels1966,
  title = {Flow Nonuniformity in Shock Tubes Operating at Maximum Test Times},
  volume = {9},
  ISSN = {0031-9171},
  DOI = {10.1063/1.1761542},
  number = {10},
  journal = {The Physics of Fluids},
  publisher = {AIP Publishing},
  author = {Mirels,  H.},
  year = {1966},
  month = oct,
  pages = {1907–1912}
}

@book{Oxenius1986,
  title = {Kinetic Theory of Particles and Photons},
  ISBN = {9783642707285},
  DOI = {10.1007/978-3-642-70728-5},
  publisher = {Springer Berlin Heidelberg},
  author = {Oxenius,  Joachim},
  year = {1986}
}

@article{Annaloro2017,
  title = {Non-uniqueness of the multi-temperature law of mass action. Application to 2T plasma composition calculation by means of a collisional-radiative model},
  volume = {71},
  ISSN = {1434-6079},
  DOI = {10.1140/epjd/e2017-80284-5},
  number = {12},
  journal = {The European Physical Journal D},
  publisher = {Springer Science and Business Media LLC},
  author = {Annaloro,  Julien and Teulet,  Philippe and Bultel,  Arnaud and Cressault,  Yann and Gleizes,  Alain},
  year = {2017},
  month = dec 
}

@article{Colonna2012,
  title = {Statistical thermodynamic description of {$\rm H_2$} molecules in normal ortho/para mixture},
  volume = {37},
  ISSN = {0360-3199},
  DOI = {10.1016/j.ijhydene.2012.03.103},
  number = {12},
  journal = {International Journal of Hydrogen Energy},
  publisher = {Elsevier BV},
  author = {Colonna,  Gianpiero and D’Angola,  Antonio and Capitelli,  Mario},
  year = {2012},
  month = jun,
  pages = {9656–9668}
}

@article{Popovas2016,
  doi = {10.1051/0004-6361/201527209},
  year = {2016},
  month = nov,
  publisher = {{EDP} Sciences},
  volume = {595},
  pages = {A130},
  author = {A. Popovas and U. G. J{\o}rgensen},
  title = {Partition functions - {I.} Improved partition functions and thermodynamic quantities for normal, equilibrium, and ortho and para molecular hydrogen},
  journal = {Astronomy \& Astrophysics}
}

@inproceedings{Cruden2024,
  title = {Impact of Trace CH$_4$ on Shock Layer Radiation in Outer Planet Entry},
  DOI = {10.2514/6.2024-2084},
  booktitle = {AIAA SCITECH 2024 Forum},
  publisher = {American Institute of Aeronautics and Astronautics},
  author = {Cruden,  Brett A. and Tibère-Inglesse,  Augustin C.},
  year = {2024},
  month = jan 
}

@article{Liu2022,
  title = {Electron Number Density Measurements in a Saturn Entry Condition},
  volume = {60},
  ISSN = {1533-385X},
  DOI = {10.2514/1.j060560},
  number = {3},
  journal = {AIAA Journal},
  publisher = {American Institute of Aeronautics and Astronautics (AIAA)},
  author = {Liu,  Yu and James,  Christopher M. and Morgan,  Richard G. and Jacobs,  Peter A. and Gollan,  Rowan and McIntyre,  Timothy J.},
  year = {2022},
  month = mar,
  pages = {1303–1315}
}

@inproceedings{Scoggins2024,
  title = {Aeroheating Environment of Aerocapture Systems for {Uranus} Orbiters},
  DOI = {10.2514/6.2024-0951},
  booktitle = {AIAA Scitech 2024 Forum},
  publisher = {American Institute of Aeronautics and Astronautics},
  author = {Scoggins,  James B. and Hinkle,  Andrew D. and Shellabarger,  Eli},
  year = {2024},
  month = jan 
}

@article{Steer2024,
  title = {Commissioning of Upgrades to {T6} to Study Giant Planet Entry},
  ISSN = {1533-6794},
  DOI = {10.2514/1.a35893},
  journal = {Journal of Spacecraft and Rockets},
  publisher = {American Institute of Aeronautics and Astronautics (AIAA)},
  author = {Steer,  Joseph and Collen,  Peter and Glenn,  Alex and Hambidge,  Christopher and Doherty,  Luke J and McGilvray,  Matthew and Sopek,  Tamara and Loehle,  Stefan and Walpot,  Louis},
  year = {2024},
  month = may,
  pages = {1–18}
}

@inproceedings{Steuer2024,
  title = {Investigation on the Usability of Arc-Jet Generators for the Experimental Simulation of Giant Planet Entry},
  DOI = {10.2514/6.2024-3552},
  booktitle = {AIAA Aviation FORUM AND ASCEND 2024},
  publisher = {American Institute of Aeronautics and Astronautics},
  author = {Steuer,  David C. and Donaldson,  Nathan and D\"{u}rnhofer,  Christian A. and Leiser,  David and Fasoulas,  Stefanos and Loehle,  Stefan},
  year = {2024},
  month = jul 
}

@BOOK{NAP13117,
  author    = "{National Research Council}",
  title     = "Vision and Voyages for Planetary Science in the Decade 2013-2022",
  isbn      = "978-0-309-22464-2",
  doi       = "10.17226/13117",
  year      = 2011,
  publisher = "The National Academies Press",
  address   = "Washington, DC"
}

@BOOK{NAP26522,
    author    = "{National Academies of Sciences, Engineering, and Medicine}",
    title     = "Origins, Worlds, and Life: A Decadal Strategy for Planetary Science and Astrobiology 2023-2032",
    isbn      = "978-0-309-47578-5",
    doi       = "10.17226/26522",
    year      = 2022,
    publisher = "The National Academies Press",
    address   = "Washington, DC"
}

@article{Rataczak2025,
  title = {Uncertainty Quantification and Sensitivity Analysis of Ice Giant Aerocapture Aerothermodynamics},
  ISSN = {1533-6808},
  DOI = {10.2514/1.t7134},
  journal = {Journal of Thermophysics and Heat Transfer},
  publisher = {American Institute of Aeronautics and Astronautics (AIAA)},
  author = {Rataczak,  Jens A. and Boyd,  Iain D. and McMahon,  Jay W.},
  year = {2025},
  month = jun,
  pages = {1–22}
}

@book{vincenti1965,
  title={Introduction to Physical Gas Dynamics},
  author={Vincenti, W.G. and Kruger, C.H.},
  isbn={9780471908357},
  lccn={65024297},
  year={1965},
  publisher={Wiley}
}

@article{Stalker1998,
  title = {Hypersonic Blunt-Body Flows in Hydrogen-Neon Mixtures},
  volume = {35},
  ISSN = {1533-6794},
  DOI = {10.2514/2.3399},
  number = {6},
  journal = {Journal of Spacecraft and Rockets},
  publisher = {American Institute of Aeronautics and Astronautics (AIAA)},
  author = {Stalker,  R. J. and Edwards,  B. P.},
  year = {1998},
  month = nov,
  pages = {729–735}
}

@inproceedings{Cruden2014AbsoluteRM,
  title={Absolute Radiation Measurements in Earth and Mars Entry Conditions},
  author={Brett A. Cruden},
  year={2014},
  booktitle={Radiation and Gas-surface Interaction Phenomena in High-Speed Reentry Lecture Series}
}

@techreport{Brandis2019Neqair,
    author = {Aaron M. Brandis and Brett A. Cruden},
    title = {NEQAIR v15.0 Release Notes: Nonequilibrium and Equilibrium Radiative Transport and Spectra Program},
    year = {2019},
    number = {ARC-E-DAA-TN72963},
    institution = {National Aeronautics and Space Administration},
}

@inproceedings{Cruden2014Neqair,
	Title = {Updates to the NEQAIR Radiation Solver},
	Author = {Brett A. Cruden and Aaron M. Brandis},
	Booktitle = {6th International Workshop on Radiation of High Temperature Gases in Atmospheric Entry},
	Year = {2014},
	Location ={St. Andrews, England, U.K.},
}

@article{Beardsell2020,
  title = {A cost-effective semi-implicit method for the time integration of fully compressible reacting flows with stiff chemistry},
  volume = {414},
  ISSN = {0021-9991},
  DOI = {10.1016/j.jcp.2020.109479},
  journal = {Journal of Computational Physics},
  publisher = {Elsevier BV},
  author = {Beardsell,  Guillaume and Blanquart,  Guillaume},
  year = {2020},
  month = aug,
  pages = {109479}
}

@article{Alves2014,
  title = {The IST-LISBON database on LXCat},
  volume = {565},
  ISSN = {1742-6596},
  DOI = {10.1088/1742-6596/565/1/012007},
  journal = {Journal of Physics: Conference Series},
  publisher = {IOP Publishing},
  author = {Alves,  L L},
  year = {2014},
  month = dec,
  pages = {012007}
}

@article{Zammit2016,
  title = {Complete Solution of Electronic Excitation and Ionization in Electron-Hydrogen Molecule Scattering},
  volume = {116},
  ISSN = {1079-7114},
  DOI = {10.1103/physrevlett.116.233201},
  number = {23},
  journal = {Physical Review Letters},
  publisher = {American Physical Society (APS)},
  author = {Zammit,  Mark C. and Savage,  Jeremy S. and Fursa,  Dmitry V. and Bray,  Igor},
  year = {2016},
  month = jun 
}

@article{Aiken2025,
  title = {Three-temperature collisional-radiative model of ionization and
                    recombination in hypersonic air plasmas},
  volume = {32},
  ISSN = {1089-7674},
  DOI = {10.1063/5.0294530},
  number = {10},
  journal = {Physics of Plasmas},
  publisher = {AIP Publishing},
  author = {Aiken,  Timothy T. and Boyd,  Iain D. and Adamovich,  Igor V.},
  year = {2025},
  month = oct 
}

@article{Boyd2021,
  title = {Analysis of Associative Ionization Rates for Hypersonic Flows},
  volume = {35},
  ISSN = {1533-6808},
  DOI = {10.2514/1.t6109},
  number = {3},
  journal = {Journal of Thermophysics and Heat Transfer},
  publisher = {American Institute of Aeronautics and Astronautics (AIAA)},
  author = {Boyd,  Iain D. and Josyula,  Eswar},
  year = {2021},
  month = jul,
  pages = {484–493}
}

@Misc{NIST_ASD,
author = {A.~Kramida and Yu.~Ralchenko and
J.~Reader and {NIST ASD Team}},
HOWPUBLISHED = {{NIST Atomic Spectra Database
(ver. 5.12).
National Institute of Standards and Technology,
Gaithersburg, MD.}},
year = {2024},
}

\end{document}